\documentclass[11pt,a4paper,oneside]{book}
\usepackage[hmargin={2.5cm,3cm},vmargin=3cm]{geometry}
\usepackage[british]{babel}  
\usepackage[latin1]{inputenc}
\usepackage{lmodern}
\usepackage[T1]{fontenc}
\usepackage{textcomp}
\usepackage{graphicx}
\usepackage{url}
\usepackage[font=small,%
            labelfont=bf,%
            labelsep=period]{caption}
\usepackage[explicit,clearempty]{titlesec} 
\usepackage{fancyhdr} 

\usepackage{color}
\usepackage{sidecap}
\usepackage{amssymb,amsmath,amsfonts,mathrsfs,verbatim,xkeyval,bm}
\numberwithin{equation}{chapter}
\usepackage{subfigure}
\usepackage[strict]{changepage}
\usepackage{tabu}

\fancypagestyle{plain}{
  \fancyhf{}
  
  \fancyfoot[C]{\thepage}
}

\newcommand{\df}{\textrm{d}}
\def\build#1_#2^#3{\mathrel{
    \mathop{\kern 0pt#1}\limits_{#2}^{#3}}}
\newcommand{\snot[1]}{{\scriptstyle \times 10}^{#1}}
\definecolor{myblue}{RGB}{0,138,230} 
\definecolor{mypink}{RGB}{224,11,122} 
\definecolor{mygray}{RGB}{194,194,194} 
\definecolor{mydarkgray}{RGB}{61,61,61} 
\definecolor{myyellowgreen}{RGB}{122,224,11}
\definecolor{mypurple}{RGB}{122,11,224}
\newtheorem{theorem}{Theorem}[section]
\newtheorem{lemma}[theorem]{Lemma}
\newtheorem{proposition}[theorem]{Proposition}

\newtheorem{definition}[theorem]{Definition}
\newtheorem{bookrule}[theorem]{Rule}

\begin{document}

\frontmatter
\thispagestyle{empty}
\begin{center}
    \begin{minipage}{\textwidth}
        \centering
        \includegraphics[totalheight=4cm]{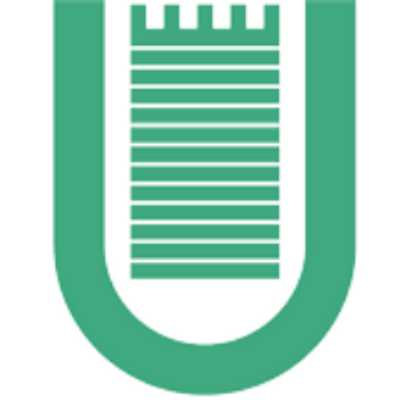}
    \end{minipage}
    \\[\intextsep]
    \vspace{0.8cm}
    \Large UNIVERSIT\`A DEGLI STUDI DI ROMA ``TOR VERGATA'' \\
    Dipartimento di Matematica

    \vspace{0.7cm}
    \Large Dottorato di Ricerca in Matematica\\
    \normalsize Ciclo XXVIII

    \vspace*{\stretch{1}}

    \LARGE  \textsc{ New  normal form approaches\\
    adapted to the Trojan problem} \\[0.8cm]
    \LARGE Roc\'io Isabel P\'AEZ \\

    \vspace*{\stretch{2}}

    \Large Relatore: Dr. Ugo LOCATELLI \\[0.2cm]
    \normalsize Coordinatore del corso di dottorato: \\
    \normalsize Prof. Carlo SINESTRARI \\
    
    \vspace*{\stretch{2}}    

    \normalsize \textsc{- Anno Accademico 2014/2015 -\\ Roma, Italia}
\end{center}
 
\chapter*{}\label{cha:dedicatoria}
\thispagestyle{empty}
\begin{flushright}
To my own personal Jesus
\end{flushright}

\chapter{Preface}\label{cha:prefacio}

The present thesis resumes the research carried out in the framework
of the author's Ph.D. program, supported by 
the Marie Curie Training Network 'Astronet-II The
Astrodynamics Network' (PITN-GA-2011-289240), under the supervision of
Dr. Ugo Locatelli. The results here presented are based on original research
developed by the author. Part of the results are
contained in the following publications:\\

{\small
\noindent
\underline{P\'aez, R.I.}, Locatelli, U., (2014)\\ Design
of maneuvers based on new normal form approximations: the case
study of the CPRTBP, \emph{ICNPAA 2014 Proceedings, AIP Conf. 
Proceedings} {\bf 1637}, p. 776.\\

\noindent
Efthymiopoulos, C., \underline{P\'aez, R.I.}, (2014)\\
Modeling resonant Trojan motions in planetary systems,
\emph{Complex Planetary Systems, Proceedings IAU Symposium 310, 
Z. Knezevic et A. Lemaitre, eds.}, p 70.\\

\noindent
\underline{P\'aez, R.I.}, Efthymiopoulos, C., (2014)\\
Modeling Trojan dynamics: diffusion mechanisms through resonances,
\emph{Complex Planetary Systems, Proceedings IAU Symposium 310, 
Z. Knezevic et A. Lemaitre, eds.}, p 96.\\

\noindent
\underline{P\'aez, R.I.}, Efthymiopoulos, C., (2015)\\ Trojan
resonant dynamics, stability and chaotic diffusion, for parameters
relevant to exoplanetary systems, \emph{Celest. Mech. Dyn. Astron.}
{\bf 121}, p. 776.\\

\noindent
\underline{P\'aez, R.I.}, Locatelli, U., (2015)\\ Trojan
dynamics well approximated by a new Hamiltonian normal form,
\emph{Mon. Not. R. Astron. Soc.} {\bf 453}, p. 2177.\\

\noindent
\underline{P\'aez, R.I.}, Locatelli, U., Efthymiopoulos, C.,
(2016)\\ The Trojan problem from a Hamiltonian perturbative
perspective, \emph{G\'omez G., Masdemont J. (eds) Astrodynamics
  Network AstroNet-II, Astrophysics and Space Science Proceedings}
{\bf 44}, p. 193.\\

\noindent
\underline{P\'aez, R.I.}, Locatelli, U.,  Efthymiopoulos, C., (2016)\\
New Hamiltonian expansions adapted to the Trojan problem
\emph{Celest. Mech. Dyn. Astron.}
{\bf 126}, p. 516.\\

\noindent
\underline{P\'aez, R.I.} (2017)\\ 
Analytical boundary for effective stability domains of Trojan motions
(\emph{preprint}).\\
}

\vspace*{1cm}

\begin{flushright}
Roc\'io Isabel Paez\\
e-mail: \url{paez@mat.uniroma2.it}\\
website: \url{http://www.mat.uniroma2.it/~paez}\\
Roma, February 2016. \\
\end{flushright}

\chapter{Abstract}\label{cha:resumen}

The main subject of this work is the study of the problem of the
Trojan orbits from a perturbative Hamiltonian perspective. We face
this problem by introducing first a novel Hamiltonian formulation,
exploiting the well-differentiated temporal scales of the Trojan
motion. The resulting Hamiltonian allows to separate the secular (very
slow) component of the motion from the librating and fast degrees of
freedom. This decompositon provides the foundation of a so-called
\emph{Basic Hamiltonian} model ($H_b$), i.e. the part of the
Hamiltonian for Trojan orbits independent of all secular
angles. Our study shows that, up to some extent, the model $H_b$
successfully represents the features of the motion under more complete
models, in a range of physical parameters relevant for dynamics in the
Solar System or in extrasolar planetary systems.

We propose, then, two novel normal form schemes in order to analytically
study the model $H_b$. The first scheme takes into account the existence
of a real singularity due to close encounters of the Trojan body with the
primary, by avoiding any polynomial or trigonometric expansion for the
librating angle. The second scheme exploits the fact that the Trojan orbits
are highly asymmetric with respect to the libration center. We then
analytically construct a so-called "asymmetric expansion", which extends
the domain of the normal form series' convergence with respect to the
usual polynomial expansions around the stable Lagrangian points L4
or L5.

Both schemes are tested in detail in the framework of the Circular and
Elliptic Restricted 3-Body Problems, focusing particularly on the analytical
derivation of the location of {\it secondary resonances} embedded within
the libration domain. Additionally, the second scheme provides an
analytical estimation of the width of such resonances.

Finally, the thesis analyses the key usefulness of the $H_b$ model,
pointing out the possibility for straightforward extensions allowing
to include additional bodies (Restricted Multi-Planet Problem), and/or
Trojan motions in 3D space.


\tableofcontents
\listoffigures

\mainmatter

\chapter{Introduction}\label{sec:1-Introduction}

This thesis is devoted to the study of the problem of the librational
motion in the neighborhood of the celebrated triangular points found
by Lagrange (1772) in the framework of the gravitational problem of
three bodies. This problem has a wide spectrum of applications. In our
Solar System, the moving bodies represent, for example, the so-called
{\it Trojan} asteroids located close to the Lagrangian points of the
Sun-Jupiter system. Trojan asteroids were also found near other
planets in our solar system.  It has been conjectured (see references
in Sect.~\ref{sec:1.4.works_on_Trojans}) that Trojan planets or
asteroids should exist in extrasolar planetary systems as well.

In his seminal work {\it Les M\'ethodes Nouvelles de la M\'ecanique
  Celeste}~\cite{Poincare-92}, H. Poincar\'{e} emphasized the use of
the {\it Hamiltonian} method in the problems of Celestial Mechanics
and in dynamical astronomy in general. The Hamiltonian formalism is
based on the use of sets of variables subject to a particular class of
transformations, called canonical or {\it symplectic}.
Such transformations lend themselves quite
conveniently to developing perturbative series solutions of the
equations of motion for, e.g., celestial bodies.  The associated
approach has become known in the mathematical literature as the method
of {\it Hamiltonian normal forms}.
In the present thesis, we exploit the
method of normal forms in the context of Trojan dynamics.

In this chapter, we briefly summarize some basic notions of the
canonical formalism as well as the relevant concepts of the problem of
Trojan motions, emphasizing those aspects which are needed in
subsequent chapters. Finally, we describe the goal and structure
of the present dissertation.


\section{Hamiltonian Mechanics}\label{sec:1.1-Ham_form}

\subsection{Hamilton's equations}\label{sec:1.1.X-ham-equations}

A system of ordinary differential equations of the 
type (for $\mathbf{x}=(p_1,\ldots,p_n,q_1,\ldots,q_n) \in \mathscr{D} 
\subset \mathbb{R}^{2n}$)
\begin{equation}\label{eq:dif_eqs}
\frac{\df \mathbf{x}}{\df t} = \mathbf{f}(\mathbf{x},t)
\end{equation}
is said to be in \emph{Hamiltonian form} if there exists a function
${\cal H}(p_1,\ldots,p_n,q_1,\ldots,q_n,t)$ such that the 
equations~\eqref{eq:dif_eqs} can be rewritten as
\begin{equation}\label{eq:Ham_eqs_gral}
\frac{\df p_i}{\df t} = -\frac{\partial {\cal H}}{\partial q_i}, \qquad
\frac{\df q_i}{\df t} = \frac{\partial {\cal H}}{\partial p_i}, \qquad
i=1,\ldots,n~.
\end{equation}
The function ${\cal H}$ is called the \emph{Hamiltonian} of the system
and the equations of motion in~\eqref{eq:Ham_eqs_gral} are
\emph{Hamilton's equations} (\cite{ArnKozNeish-06},
\cite{Goldstein-80}). The variables $p_1,\ldots,p_n,q_1,\ldots,q_n$ are
called canonical \emph{momenta} and \emph{coordinates} respectively.
The space spanned by the canonical variables is called \emph{phase space}. The 
system~\eqref{eq:Ham_eqs_gral} is said to be of $n$
\emph{degrees of freedom} (d.o.f.).  

We call \emph{cyclic} a
coordinate $q_j$ that does not appear explicitely in the Hamiltonian
function. In this case, the conjugate momentum $p_j$ is preserved. Taking
into account Eq.~\eqref{eq:Ham_eqs_gral}, the
total time derivative of the Hamiltonian is
\begin{equation}\label{eq:dHdt}
\frac{\df {\cal H}}{\df t} = \sum_{i=1}^{n} \left( \frac{\partial {\cal
    H}}{\partial q_i} \dot{q_i}+ \frac{\partial {\cal H}}{\partial
  p_i} \dot{p_i} \right) + \frac{\partial {\cal H}}{\partial t} =
\frac{\partial {\cal H}}{\partial t}~. 
\end{equation} 
Therefore, ${\cal H}$ is constant in time as
long as it does not depend explicitely on $t$ (\emph{autonomous
  Hamiltonian}). The \emph{non autonomous} systems are said to be of
$n+1/2$ d.o.f. In fact, it is possible to extend the phase-space, by including
an additional variable conjugate to the time $t$, that gives a new
Hamiltonian ${\cal H}'$ of $n+1$ d.o.f. preserved in time.
Thus, all the results applicable to autonomous systems may be generalized
to the non autonomous systems as well.

Let $f$ and $g$ be two generic functions of the generalized
coordinates $(\mathbf{p},\mathbf{q})$, $\mathbf{p}=p_1,\ldots,p_n$,
$\mathbf{q}=q_1,\ldots,q_n$. The Poisson bracket between $f$ and $g$
is defined by
\begin{equation}\label{eq:poisson_brack_def}
\{ f,g\}_{\mathbf{q},\mathbf{p}}= \sum_{i=1}^n \left( \frac{\partial f}{\partial q_i} \frac{\partial g}{\partial p_i} - \frac{\partial f }{\partial p_i} \frac{\partial g}{\partial q_i} \right)~.
\end{equation}
If we apply the Poisson bracket to $f$ and the Hamiltonian ${\cal
  H}$, we obtain
\begin{equation}\label{eq:Ham_induced_flow}
\{f, {\cal H}\}_{\mathbf{q},\mathbf{p}}= \sum_{i=1}^n \left( \frac{\partial f}{\partial q_i} \frac{\partial {\cal H}}{\partial p_i} - \frac{\partial f }{\partial p_i} \frac{\partial {\cal H}}{\partial q_i} \right) = \dot{f}~.
\end{equation}
In words, the computation of the Poisson bracket gives the time
evolution of any dynamical variable $f$, i.e. a differentiable
real function of the canonical coordinates
$(\mathbf{p},\mathbf{q})$, under the flow induced by the
Hamiltonian ${\cal H}$. Applying this to the generalized coordinates,
we can rewrite Hamilton's equations as follows:
\begin{equation}\label{eq:Ham_eq_poiss}
\dot{q_i} = \{q_i, {\cal H}\}_{\mathbf{q},\mathbf{p}}~, \qquad \dot{p_i}=\{p_i, {\cal H}\}_{\mathbf{q},\mathbf{p}}~.
\end{equation}

\subsection{Canonical transformations}\label{sec:1.1.1-can_trans}

Let us consider a time-independent transformation of variables defined by the
following (invertible) equations
\begin{equation}\label{eq:can_transf}
Q_{i} = Q_{i}(\mathbf{q},\mathbf{p}),\qquad P_{i} = P_{i}(\mathbf{q},\mathbf{p})~,
\quad i=1,\ldots,n~.
\end{equation}
If for any Hamiltonian ${\cal H} = {\cal H}(\mathbf{p},\mathbf{q})$
function of the canonical conjugated variables $(\mathbf{p},\mathbf{q})$,
the new set of variables $(\mathbf{P},\mathbf{Q})$,
 satisfies Hamilton's equations,
\begin{equation}\label{eq:Ham_eqs_QP}
\frac{\df P_i}{\df t} = -\frac{\partial {\cal K}}{\partial Q_i}, \qquad
\frac{\df Q_i}{\df t} = \frac{\partial {\cal K}}{\partial P_i}, \qquad
i=1,\ldots,n~
\end{equation}
where the new Hamiltonian function ${\cal K}$ is given by
\begin{equation}
{\cal K}={\cal H}(\mathbf{p}(\mathbf{P},\mathbf{Q}),\mathbf{q}(\mathbf{P},\mathbf{Q}))~,
\end{equation}
then the transformation~\eqref{eq:Ham_eqs_QP} is called
\emph{canonical}. In other words, a canonical transformation is a
change of coordinates that preserves the form of Hamilton's
equations.

A straightforward way to check whether a transformation of the
form~\eqref{eq:can_transf} is canonical is based on the following
property~\cite{Arnold-78}: the transformation~\eqref{eq:can_transf} is
canonical if and only if it preserves the fundamental Poisson
brackets,
\begin{equation}\label{eq:poiss_can_transf}
\{q_i,q_k\}_{\mathbf{Q},\mathbf{P}} = \{p_i,p_k\}_{\mathbf{Q},\mathbf{P}} = 0 ~, \qquad\{q_i,p_k\}_{\mathbf{Q},\mathbf{P}} = \delta_{ik}~,
\end{equation}
for $1\leq i\leq n$, $1\leq k\leq n$. 

Equations~\eqref{eq:poiss_can_transf} can be
used to check the canonical property when the transformation is given.
However, they cannot be used directly in order to construct canonical
transformations of the form~\eqref{eq:can_transf}. We now refer to two
methods that do allow to construct them explicitly.

The first method is based on the construction of a \emph{generating
  function} $S$. This is a function depending on a particular
combination of old/new momenta and coordinates. A common division distinguishes
four classes: \emph{1st class} $S_1$, if the generating functions
depends on the old and new coordinates,
i.e. $S_1(\mathbf{q},\mathbf{Q})$; \emph{2nd class} $S_2$, if it
depends on the old coordinates and new momenta,
i.e. $S_2(\mathbf{q},\mathbf{P})$; \emph{3rd class} if $S_3 \equiv
S_3(\mathbf{p},\mathbf{Q})$; \emph{4th class} if $S_4 \equiv
S_4(\mathbf{p},\mathbf{P})$.  The canonical transformation equations
are given by the corresponding derivatives of $S$. For example,
for an arbitrary choice of 
 generating function of 2nd class,
$S=S_2(\mathbf{q},\mathbf{P}) $, it can be shown that the transformation 
equations
\begin{equation}
p_i = \frac{\partial S_2}{\partial q_i}~, \qquad Q_i = \frac{\partial
  S_2}{\partial P_i}
\end{equation}
are canonical (see \S1.2a of~\cite{LichtLieb-92}), as long as they are
invertible, e.g.
\begin{equation} 
\mathrm{det}\left(\frac{\partial^2 S_2}{\partial q_j \partial P_i} \right) \neq 0~.
\end{equation}
In an analogous way, we derive the transformations for
the other classes of generating functions.

The second method consists of the use of
\emph{Lie generating functions} $\chi(\mathbf{p},\mathbf{q},s)$ of the original
canonical variables $\mathbf{q}$, $\mathbf{p}$ and a parameter
$s$. The function generates a canonical transformation from old
to new variables by means of Hamilton equations
\begin{equation}\label{eq:lie-1}
\frac{\df p_i}{\df s} = -\frac{\partial \chi }{\partial q_i }~,
\qquad \frac{\df q_i}{\df s} = \frac{\partial \chi }{\partial p_i }~,
\qquad i=1,\ldots,n~,
\end{equation}
and the new variables are related to the old variables by
\begin{equation}\label{eq:lie-2}
P_i=p_i(s)~,\quad Q_i=q_i(s)~,
\end{equation}
assuming that $p_i = p_i(0)$ and $q_i = q_i(0)$.  Since it is derived
from a Hamiltonian-like flow (induced by $\chi$), this transformation
must be canonical.  

The method of Lie generating functions is widely used in
perturbation theory for producing near to identity canonical
transformations (\cite{Hori-66},~\cite{Depri-69}). An advantage
of the Lie method is that it provides an explicit expression
of the transformations ${\mathbf P}({\mathbf p},{\mathbf q})$, 
${\mathbf Q}({\mathbf p},{\mathbf q})$. On the contrary, the method
based on the generating functions $S_1,\ldots,S_4$ provides
 implicit formul\ae~which need to be inverted in order to 
obtain the explicit form of the transformation equations.

\subsection{Action-angle variables}\label{sec:act-ang-var}

Let us consider a dynamical variable $\gimel$. If
\begin{equation}\label{eq:first-int}
\dot{\gimel} = \{ \gimel, {\cal H} \}_{\mathbf{q},\mathbf{p}} = 0
\end{equation}
then $\gimel$ is called \emph{first integral} of the Hamiltonan ${\cal
  H}$.  We can immediately notice that if two dynamical variables
$\gimel_1$ and $\gimel_2$ are first integrals, then $\{
\gimel_1,\gimel_2\}$ is a first integral as well. In particular, in
autonomous systems ${\cal H}$ is a first integral of the Hamiltonian
flow.

Let ${\cal H}$ be an autonomous Hamiltonian of $n$ d.o.f. possessing
a set of $n$ first integrals $(\gimel_1,\ldots,\gimel_n)$  which
satisfy the following properties: i. they are independent, i.e.
\begin{equation}\label{eq:indep-first-int}
\mathrm{rank} \left( \frac{\partial(\gimel_1,\ldots,\gimel_n)}{\partial 
(q_1,\ldots,q_n,p_1,\ldots,p_n)} \right) = n~,
\end{equation}
and ii. they form an involution system, i.e. they accomplish the property
\begin{equation}\label{}
\{ \gimel_i, \gimel_j \} = 0~\quad \mathrm{for} \quad i,j = 1,\ldots,n~.
\end{equation}
Assume also that the hypersurfaces of constant energy 
${\cal H}(\mathbf{p},\mathbf{q})$ 
are compact. 

\begin{SCfigure}
  \centering
  \includegraphics[width=.50\textwidth]{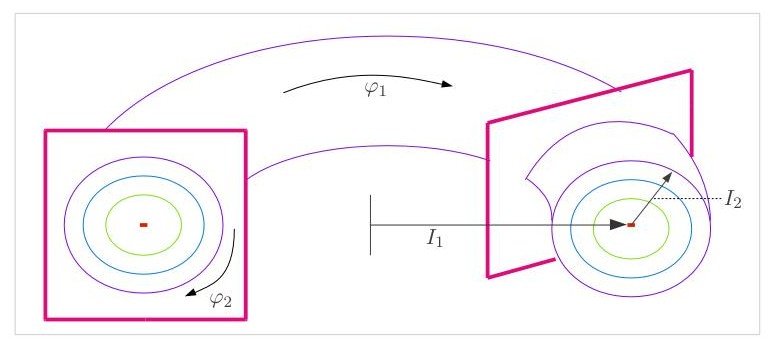}
  \caption[Invariant tori in the phase space of action-angle
    variable]{Schematic representation of a family invariant tori and a
    periodic orbit (red) in the phase space of action angle
    variables. The constant value of the action defines the radius of
    the torus. The three tori in this image have a similar value of
    the action $I_1$ (long dark arrow). The second action $I_2$
    increases its value from the greeen orbit to the purple orbit.}
  \label{fig:tori.jpg}
\end{SCfigure}

Under these conditions,
 the Liouville-Arnold-Jost theorem (\cite{Arnold-63a},\cite{Jost-68}) 
states that it is possible to \emph{locally} construct a canonical
transformation $(\mathbf{p},\mathbf{q}) \mapsto
(\mathbf{I},\boldsymbol{\varphi}) \in \mathbb{R}^{n}\times\mathbb{T}^{n}$ such
that in the new variables the Hamiltonian acquires the form
\begin{equation}\label{eq:intgr_Ham}
{\cal H} = {\cal H}(\mathbf{I})~.
\end{equation}
In Eq.~\eqref{eq:intgr_Ham} all the coordinates $\varphi_i$, called the
\emph{angles}, are cyclic. The solution of
Hamilton's equations is then trivial, given by
\begin{equation}\label{eq:acc_ang_var}
I_i = I_i|_{t=0} \equiv \, \textrm{const.}\,, \qquad \varphi_i =
\varphi_i|_{t=0} + \omega_i t~, \qquad i=1,\ldots,n~,
\end{equation}
where $\omega_i=\partial{\cal H}/\partial I_i$. 
The momenta $I_i$ are called the \emph{actions}\footnote{In general, the actions
are defined only locally, e.g the pendulum admits different actions inside
and outside the separatrix.}. Their value along any
particular orbit remains 
constant, while the temporal evolution of the \emph{angles} $\varphi_i$ is
linear with frequency $\omega_i$. A system of the above form is called
Arnold-Liouville integrable.
Figure~\ref{fig:tori.jpg} provides a schematic
representation of the orbits in the phase-space defined by the action-angle 
variables. Since the evolution of the motion is given strictly
by Eqs.~\eqref{eq:acc_ang_var}, the orbit lies on an invariant torus
$\mathbb{T}^{n}$, defined by the constant values of the actions
$I_i$. The uniform motion on the torus is given by the value of the
angles $\varphi_i$ according to Eq.~\eqref{eq:acc_ang_var}. 
The orbits
are called \emph{quasiperiodic} if the frequencies are such that
\begin{equation}\label{eq:non-res-torus}
k_1\,.\,\omega_1 + \ldots + k_n\,.\,\omega_n = 0 \iff k_i = 0~,\quad 
i=1,\ldots,n~.
\end{equation}
A torus on which \eqref{eq:non-res-torus} holds is called \emph{non resonant}. 
On the other hand, if there
exist one or more combinations of integer values $k_i$
such that
\begin{equation}\label{eq:res-torus}
k_1\,.\,\omega_1 + \ldots + k_n\,.\,\omega_n = 0 \quad \wedge \quad \sum_{i=1}^{n} |k_i| \neq 0~,
\end{equation}
the torus is called \emph{resonant}. The commensurability between
frequencies induces orbits lying on so-called \emph{lower-dimensional
  tori}, i.e. tori of dimension lower than $n$. If $m$ independent
relations of the form~\eqref{eq:res-torus} exist, the corresponding
orbits lie in a $(n-m)$-dimensional torus. Of particular importance is
the case $m=n-1$. Then, the corresponding orbits are $1$-dimensional
tori, called \emph{periodic orbits}. In
Fig.~\ref{fig:tori.jpg}, the red orbit represents a case of periodic orbit.
Every time when the orbit completes a period $T_1 = 2\pi/\omega_1$, its
position in terms of $I_2,\varphi_2$ is the same. Finally, a trivial consequence
of the Liouville-Arnold-Jost theorem is that an autonomous Hamiltonian of 
1 d.o.f. is integrable, since it has one first integral:
the Hamiltonian itself.

\subsection{Normal form theory for nearly-integrable systems}\label{sec:1.1.2-pert_theory}

Most dynamical systems in nature are non integrable. 
H. Poincar\'e~\cite{Poincare-92} emphasized the importance of systems for 
which the Hamiltonian function has the form
\begin{equation}\label{eq:quasi-int-ham}
{\cal H}(\mathbf{p},\mathbf{q})= H_0(\mathbf{p}) + \epsilon H_1(\mathbf{p},\mathbf{q})~,
\end{equation}
where $H_0$ is an integrable hamiltonian and $\epsilon H_1$ is a
fuction expandable as a convergent series in powers of $\epsilon$.
According to Poincar\'e, understanding the solutions of the systems of
the form~\eqref{eq:quasi-int-ham} constitutes the \emph{fundamental
  problem of dynamics}. In particular, if the size of the perturbation
$\epsilon$ to the integrable system $H_0$ is a small quantity, the
system is called \emph{nearly integrable}.

Although cases of strongly chaotic dynamics are easy to
identify, nearly integrable dynamics is the most frequent case encountered
in Solar System dynamics. In nearly integrable systems
the solutions cannot be found in such explicit form as in 
 integrable systems (like, for example, the Two-Body problem).
However, quite precise approximate solutions can be obtained by employing
the method of \emph{normal forms}.  

A normal form can be
defined as a Hamiltonian function yielding a simple-to-analyse
dynamics~\cite{LaPlata-11}. In general, this does not imply that we can
reduce a non-integrable Hamiltonian to an integrable one. However, we can
approximate it by a normal form with properties that render its study
simpler. This normal form is produced after implementing a set of
conveniently chosen canonical transformations to the variables of the original
Hamiltonian.

Let us consider a canonical transformation
of the form~\eqref{eq:can_transf}. 
In the normal form method, we look for a canonical transformation such 
that, after applying the change of variables, the Hamiltonian
is decomposed in the form
\begin{equation}\label{eq:Z-plus-R}
{\cal H}_{new}(\mathbf{P},\mathbf{Q}) = {\cal H}(\mathbf{p}(\mathbf{P},\mathbf{Q}),\mathbf{q}(\mathbf{P},\mathbf{Q})) = {\cal Z}(\mathbf{P},\mathbf{Q}) + 
{\cal R}(\mathbf{P},\mathbf{Q})~,
\end{equation}
with $|{\cal R}|<<|{\cal Z}|$. The term ${\cal Z}$
in~\eqref{eq:Z-plus-R} called the normal form, while ${\cal R}$ is
called the remainder. The importance of the latter lies on the fact
that it tells us how different is the dynamics of the original
Hamiltonian with respect to the normal form. In practice, while
applying a sequence of canonical transformations, we must control the
growth of the size of the remainder. Most so-called \emph{normalizing
  schemes}, i.e. sequences of canonical transformations, are designed
in order to keep appropriate control on this
growth~\cite{Pisa-03},~\cite{LaPlata-11}.

To obtain the normalized Hamiltonian, we need to produce a sequence of 
near-identity canonical transformations.
To this end, we discuss now
the technique of \emph{Lie series}. Let us consider
an arbitrary function $\chi(\mathbf{p},\mathbf{q})$.  As described
in~\eqref{eq:lie-1}, the flow produced by Hamilton's equations using
$\chi$ as the hamiltonian function, is given by
\begin{equation}\label{eq:lie-3}
\dot{p}_i = -\frac{\partial \chi}{\partial q_i}~,\quad \dot{q}_i = \frac{\partial \chi}{\partial p_i}~, \qquad i=1,\ldots,n~.
\end{equation}
Let $p_i(t)$ and $q_t(t)$ denote the solutions in time, with initial
conditions $p_i(0)$, $q_i(0)$. The transformation
$p_i(0),q_i(0)~\mapsto~p_i(t),q_i(t)$, viewed as a mapping,
is a canonical transformation for any arbitrary time
$t$~\cite{Arnold-78}. In this sense, $\chi(\mathbf{p},\mathbf{q})$ can
be considered as a function from which we can obtain infinitely many
canonical transformations, by solving Hamilton's equations for
differents values of the time $t$. Accordingly, $t$ can be thought 
of as
a parameter generating the whole family of canonical transformations
induced by $\chi$ (the $s$ parameter used in
Eq.~\ref{eq:lie-2}). 

In general, we are not able to explicititely integrate
Hamilton's equations~\eqref{eq:lie-3} for all times $t$.
However,
for $t$ small, it is possible to obtain a solution of the initial
value problem~\eqref{eq:lie-3} through \emph{Taylor}
expansions. The key remark is that, from the equations of the 
flow~\eqref{eq:lie-3} we can compute the time derivative of all
orders for the canonical variables $p_i$, $q_i$, $i=1,\ldots,n$,
as functions of the canonical variables themselves. For example,
\begin{equation}
\begin{aligned}\label{eq:lie-4}
\frac{\df^2 p_j}{\df t^2} = & \, \frac{\df}{\df t} \left(- \frac{\partial \chi}{\partial q_j}\right) = - \sum_{i=1}^{n} \left( \frac{\partial }{\partial q_i} \left(\frac{\partial \chi}{\partial q_j} \right) \dot{q_i} + \frac{\partial }{\partial p_i} \left(\frac{\partial \chi}{\partial q_j} \right) \dot{p_i} \right) \\
= & \, \sum_{i=1}^{n} \left(\frac{\partial^2 \chi}{ \partial p_i \partial q_j} \frac{\partial \chi}{\partial q_i} - \frac{\partial^2 \chi}{\partial q_i \partial q_j}\frac{\partial \chi}{\partial p_i}\right)~.
\end{aligned}
\end{equation}
Thus, the second order time derivative of $p_j$ is expressed in terms of
$\chi$ and partial derivatives of $\chi$. This process can be
generalized for all the variables and orders of the derivatives needed for
a truncated Taylor series.

Let us define the \emph{Lie operator} ${\cal L}_{\chi} \equiv
\{\cdot,\chi\}$. According to~\eqref{eq:Ham_induced_flow}, for a
generic dynamical variable $f$, we have
\begin{equation}\label{eq:lie-5}
\frac{\df f}{\df t} = \{f, \chi\} = {\cal L}_{\chi}f
\end{equation}
that corresponds to the time derivative of $f$ along a Hamiltonian
flow induced by $\chi$. Generalizing this notation for higher order
derivatives we have
\begin{equation}\label{eq:lie-6}
\frac{\df^m f}{\df t^m} = \{\ldots \{ \{ f,\chi\} \ldots \} \ldots \chi \} = 
{\cal L}^{m}_{\chi}f~.
\end{equation}
Now, we can construct the Taylor series for the solutions
$p_i(t),q_i(t)$,
\begin{equation}
\begin{aligned}\label{eq:lie-7}
p_i(t) &= p_i(0) + \frac{\df p_i(0)}{\df t} t + \frac{1}{2}
\frac{\df^2 q_i(0)}{\df t^2} t^2 + \ldots = \sum_{m=0}^{\infty}
\frac{1}{m!} \frac{\df^m p_i(0)}{\df t^m}
t^m~, \\ q_i(t) &= q_i(0) + \frac{\df q_i(0)}{\df t}
t + \frac{1}{2} \frac{\df^2 q_i(0)}{\df t^2} t^2 + \ldots =
\sum_{m=0}^{\infty} \frac{1}{m!} \frac{\df^m q_i(0)}{\df t^m}
t^m~, 
\end{aligned}
\end{equation}
where $\frac{\df^m q_i(0)}{\df t^m} = \frac{\df^m q_i}{\df
  t^m}|_{t=0}$ and $\frac{\df^m p_i(0)}{\df t^m} = \frac{\df^m
  p_i}{\df t^m}|_{t=0},\,m\in \mathbb{N}$ and $i=1,\ldots,n$. 
Replacing the \emph{Lie}
operator notation in Eqs.~\eqref{eq:lie-7}, we 
have
\begin{equation}
\begin{aligned}
p_i(t) = p_{i}^{(0)} + ({\cal L}_{\chi}p_{i}^{(0)})\,t + \frac{1}{2} ({\cal
  L}^2_{\chi}p_{i}^{(0)})\,t^2 + \ldots~, \label{eq:lie-series-1} \\
q_i(t) = q_{i}^{(0)} + ({\cal L}_{\chi}q_{i}^{(0)})\,t + \frac{1}{2} ({\cal
  L}^2_{\chi}q_{i}^{(0)})\,t^2 + \ldots~, 
\end{aligned}
\end{equation}
where $p_{i}^{(0)} = p_i(0)$ and $q_{i}^{(0)} =
q_i(0)$, and $i=1,\ldots,n$. The last equations correspond to the
formal definition of a \emph{Lie series}. As mentioned before, they
provide a family of canonical transformations for any value of the
time variable $t$ within their domain of convergence. The convergence
of the series is discussed, e.g., in~\cite{Pisa-03},~\cite{LaPlata-11}.
Let us
assume that $\chi$ and its derivatives are small enough
so that the series are convergent for $t=1$. Then, we have
\begin{equation}
\begin{aligned}
p^{(1)}_{i} = p^{(0)}_{i} + ({\cal L}_{\chi}p^{(0)}_{i}) + \frac{1}{2} ({\cal
  L}^2_{\chi}p^{(0)}_{i}) + \ldots~, \label{eq:lie-transform-1} \\
q^{(1)}_{i} = q^{(0)}_{i} + ({\cal L}_{\chi}q^{(0)}_{i}) + \frac{1}{2} ({\cal
  L}^2_{\chi}q^{(0)}_{i}) + \ldots~,
\end{aligned}
\end{equation}
which are the Lie canonical transformations from 
$(\mathbf{p}^{(0)},\mathbf{q}^{(0)}) \mapsto (\mathbf{p}^{(1)},\mathbf{q}^{(1)})$. 
Using the exponential
operator
\begin{equation}\label{eq:exp-op}
\mathrm{exp}\frac{\df \, \cdot}{\df t} = \frac{\df \,\cdot}{\df t} + \frac{1}{2} \frac{\df^2 \,\cdot}{\df t^2} +\ldots~,
\end{equation}
we re-express all the canonical transformations in a compact form
\begin{equation}\label{eq:exp-op-lie-trans}
  p^{(1)}_{i} = \mathrm{exp}({\cal L}_{\chi}) \,p^{(0)}_{i}, \quad q^{(1)}_{i} = \mathrm{exp}({\cal L}_{\chi}) \,q^{(0)}_{i}, \qquad i=1,\ldots,n~,
\end{equation}
with the \emph{Lie exponential operator} defined by
\begin{equation}\label{eq:Lie-exp-oper}
\exp\Big({\cal L}_{\chi}\Big)\, \cdot \, = \mathbb{I}\,\cdot\, 
 +  ({\cal L}_{\chi} \, \cdot \,) + \frac{1}{2}  ({\cal L}^2_{\chi} \, \cdot \,) + \ldots~.
\end{equation}
The following properties of the Lie series are relevant in practical 
computations:

\noindent
i. the function $\chi$ can be chosen in a completely
arbitrary way. This gives the freedom to choose $\chi$ in order to ensure
that the transformed Hamiltonian acquires the normal form properties we
look for;\\

\noindent
ii. in the computation of the
transformations~\eqref{eq:exp-op-lie-trans}, the only operations
involved are sums, products and derivatives, which are easy to
implement in an computer-algebraic program;\\

\noindent
iii. considering $\chi$ a small quantity, Eq.~\eqref{eq:Lie-exp-oper}
implies that the Lie series generates a \emph{near-identity}
transformation;\\

\noindent
iv. the time derivative of any function $f$ under the Hamiltonian flow
induced by $\chi$ is given by~\eqref{eq:lie-5}, hence
\begin{equation}\label{eq:lie-9}
f_1(\mathbf{p}^{(1)},\mathbf{q}^{(1)}) =
f(\mathbf{p}^{(0)}(\mathbf{p}^{(1)},\mathbf{q}^{(1)}),\mathbf{q}^{(0)}(\mathbf{p}^{(1)},\mathbf{q}^{(1)}))
= \mathrm{exp}({\cal L}_{\chi}) f(\mathbf{p}^{(1)},\mathbf{q}^{(1)})~.
\end{equation}
In other words, it is possible to find which form acquires the
function $f$ after applying the
transformation~\eqref{eq:exp-op-lie-trans} without performing function
compositions; instead we apply the corresponding Lie operator directly
on $f$~\cite{Grob-60},~\cite{Pisa-03} (property known as the 'Exchange
theorem'). This key argument allows to replace complicated
compositions of functions by just trivial operations.

We now briefly discuss the algorithm of computation of a normal form
via Lie series (for details see, e.g.,~\cite{LichtLieb-92}).
Following~\cite{LaPlata-11}, we introduce a convenient notation called
'book-keeping', which allows to expose the normal form algorithm in a
way easily transcribable to a computer-algebraic program. Let
$f(\mathbf{p},\mathbf{q})$ be a function of the canonical variables
depending on one or more small parameters, like $\epsilon$ in
Eq.~\eqref{eq:quasi-int-ham}. In some problems of Celestial Mechanics, such
parameters can be the masses of the planets (divided by the mass of
the Sun), eccentricities or inclinations, amplitudes of libration
around particular equilibrium solutions, etc. Depending on the size of
these parameters, in perturbation theory we encounter series
expansions of $f(\mathbf{p},\mathbf{q})$ of the form
$f(\mathbf{p},\mathbf{q}) = f_0(\mathbf{p},\mathbf{q}) +
f_1(\mathbf{p},\mathbf{q}) + f_2(\mathbf{p},\mathbf{q}) \ldots$, where
$f_r(\mathbf{p},\mathbf{q})$ represents terms estimated as of ''$r$-th
order of smallness''. The value of $r$ can be chosen to be connected
to the exponents of various small parameters by a specific rule, that
we hereafter call the \emph{book-keeping rule}. To formally account
for this, we introduce in the notation a symbol $\lambda$, with
numerical value equal to $\lambda=1$.  Thus, in the sequel, a term
with a factor $\lambda^r$ in front is meant as a term estimated to be
of $r$-th order of smallness.

Returning to the normal form algorithm, let ${\cal H}^{(0)}$ denote the
initial Hamiltonian. At the first normalization step, we separate the
terms of ${\cal H}^{(0)}$ as follows
\begin{equation}\label{eq:1stordHam}
{\cal H}^{(0)} = {\cal Z}_0 + \lambda {\cal H}_1^{(0)} + \lambda^2 {\cal H}_2^{(0)}
+ \ldots
\end{equation}
${\cal Z}_0$ denotes the terms of the Hamiltonian which we choose to 
be in normal form at the zero-th order. All remaining terms should be
assigned book-keeping order $1$ or higher. The choice of ${\cal Z}_0$
can have a certain degree of arbitrariness and different choices
lead to different normalization schemes.

We now seek to determine a canonical transformation 
$(\mathbf{p}^{(0)},\mathbf{q}^{(0)}) \mapsto (\mathbf{p}^{(1)},\mathbf{q}^{(1)})$ 
such that the Hamiltonian function, expressed in the new variables
$(\mathbf{p}^{(1)},\mathbf{q}^{(1)})$, is in normal form up to order 
${\cal O}(\lambda)$. Let $\chi_1$ be the generating function accomplishing
this transformation. Let us consider for a moment
that $\chi_1$ was given. The Hamiltonian after applying the
transformation reads
\begin{equation}\label{eq:transformed_H_lie}
{\cal H}^{(1)}(\mathbf{p}^{(1)},\mathbf{q}^{(1)}) = \mathrm{exp}({\cal L}_{\chi_1}) 
{\cal H}^{(0)}(\mathbf{p}^{(1)},\mathbf{q}^{(1)})~,
\end{equation}
whose lowest order terms are
\begin{equation}\label{eq:lanecesitoabajo}
{\cal H}^{(1)} = {\cal H}^{(0)} + {\cal L}_{\chi_1} {\cal H}^{(0)} + \frac{1}{2} {\cal L}^2_{\chi_1} {\cal H}^{(0)} + \ldots
\end{equation}
Replacing Eq.~\eqref{eq:lanecesitoabajo} in Eq.~\eqref{eq:1stordHam}, we obtain
\begin{equation}
\begin{aligned}
{\cal H}^{(1)} = & {\cal Z}_0 + \lambda {\cal H}_1^{(0)} + \{ {\cal Z}_0,\chi_1\}
 +\lambda \{ {\cal H}_1^{(0)}, \chi_1\} \\
+ & \frac{1}{2} \{ \{ {\cal Z}_0,
\chi_1\},\chi_1\} + \lambda \frac{1}{2} \{ \{ {\cal H}_1^{(0)},
\chi_1\},\chi_1\} +  \ldots~,
\end{aligned}
\end{equation} 
If we define $\chi_1$ as a quantity of first order in $\lambda$ 
($\cal O(\lambda)$), we have
\begin{align}\label{eq:orders_lambda}
\lambda {\cal H}_1^{(0)}+ \{ {\cal Z}_0,\chi_1\} &\rightarrow {\cal O}(\lambda)~, & \nonumber\\
\lambda \{ {\cal H}_1^{(0)}, \chi_1\} + \frac{1}{2} \{ \{ {\cal Z}_0,
\chi_1\},\chi_1\} &\rightarrow   {\cal O}(\lambda^2)~, &\\
\lambda \frac{1}{2} {\cal L}^2_{\chi_1} {\cal H}^{(0)} &\rightarrow   {\cal O}(\lambda^3)~.& \nonumber
\end{align}
Now we aim to produce a new Hamiltonian ${\cal H}^{(1)}$ such that,
at order ${\cal O}(\lambda)$,
it contains only terms
 in normal form. To achieve so, the generating function $\chi_1$ must be
defined so as to satisfy the equation
\begin{equation}\label{eq:gen_homol_eq_1}
\lambda \phantom{i}^{\ast}{\cal H}_1^{(0)} + {\{\cal Z}_0, \chi_1 \} = 0~,
\end{equation}
where $\phantom{i}^{\ast}{\cal H}_1^{(0)}$ are the terms of ${\cal
  H}^{(0)}_1$ which are \emph{not} in normal form, i.e. those we do not
want to keep in ${\cal H}^{(1)}$. An equation
like~\eqref{eq:gen_homol_eq_1} is known as the \emph{homological
  equation}. Its solution specifies the generating function
$\chi_1$. Then, implementing Eq.~\eqref{eq:transformed_H_lie}, we also
find the now transformed Hamiltonian ${\cal H}^{(1)}$.

The solution of the homological equation can be found in a straighforward
way, if the term ${\cal Z}_0$ is chosen in the initial step as
\begin{equation}\label{eq:normal_form_Z}
{\cal Z}_0 = \sum_{i=1}^n \, \omega_i \, p^{(0)}_{i} ~,
\end{equation}
with $\omega_i =\,\mathrm{const.}$, $i=1,\ldots,n$. A choice of the
form~\eqref{eq:normal_form_Z} is possible when the variables
$\mathbf{p},\mathbf{q}$ are action-angle variables, and the problem
under study has the structure of coupled nonlinear oscillators.
By replacing each trigonometric function of the angles by
\begin{equation}
\cos q_{i,0} = \frac{(e^{{\rm i}q_{i,0}} + e^{-{\rm i}q_{i,0}})}{2}~, 
\end{equation}
for $i =1,\ldots,n$ (and similarly for $\sin q_{i,0}$),
the function $\phantom{i}^{\ast}{\cal H}_1^{(0)}$ can be written in the form
\begin{equation}\label{eq:h(1)}
\phantom{i}^{\ast}{\cal H}_1^{(0)} = \sum_{\mathbf{k}_i}
b(\mathbf{p}^{(0)}) \, e^{{\rm i}(\mathbf{k}_i \cdot \mathbf{q}^{(0)})}~,
\end{equation}
with $\mathbf{k}_i \in \mathbb{Z}^n$. Substituting
Eqs.~\eqref{eq:h(1)} and~\eqref{eq:normal_form_Z} in the homological
equation~\eqref{eq:gen_homol_eq_1}, we obtain the solution for the
generating function
\begin{equation}\label{eq:gral_gen_funct}
\chi_1 = \lambda \sum_{\mathbf{k}_i} 
\frac{b(\mathbf{p}^{(0)})}{{\rm i}(\mathbf{k}_i \cdot \boldsymbol{\omega})}
 e^{{\rm i}(\mathbf{k}_i \cdot \mathbf{q}^{(0)})}~,
\end{equation}
where $\boldsymbol{\omega} = (\omega_1,\ldots,\omega_n)$.  For the
solution to exist, all denominators $\mathbf{k}_i \cdot
\boldsymbol{\omega}$ must be different from zero.  This restriction
limits our possible choice of functions $\phantom{i}^{\ast}{\cal
  H}_1^{(0)}$, i.e. $\phantom{i}^{\ast}{\cal H}_1^{(0)}$ should
contain no Fourier terms with wavevectors $\mathbf{k}_i$ satisfying
$\mathbf{k}_i\cdot \boldsymbol{\omega} = 0$. Albeit not explicitely
needed, in practice we exclude also the terms
satisfying 
\begin{displaymath}
\mathbf{k}_i\cdot \boldsymbol{\omega} \simeq 0
\end{displaymath}
 from
$\phantom{i}^{\ast}{\cal H}_1^{(0)}$. In fact, the question of how to properly
identify terms that generate small divisors is one of the most
importants in normal form theory.  

With the expression of the
generating function $\chi_1$, we finally obtain the new transformed
Hamiltonian by applying the Lie series operator,
\begin{equation}
{\cal H}^{(1)} = \exp({\cal L}_{\chi_1}) {\cal H}^{(0)}~,
\end{equation}
which, by construction, is in normal form up to order ${\cal O}(\lambda)$.

The above normalization step of the Hamiltonian ${\cal H}^{(0)}$ can
be generalized to any higher order.  Let us consider the Hamiltonian
${\cal H}^{(r)}$, which we assume is in normal form up to order ${\cal
  O}(\lambda^r)$:
\begin{equation}\label{eq:Hr}
{\cal H}^{(r)} = {\cal Z}_{0} + \lambda {\cal Z}_1 + \ldots + 
\lambda^r {\cal Z}_{r} + \lambda^{r+1} {\cal H}_{r+1}^{(r)} + 
\lambda^{r+2} {\cal H}_{r+2}^{(r)} + \ldots~.
\end{equation}
Some of the terms of ${\cal H}_{r+1}^{(r)}$, denoted as
$\phantom{i}^{\ast}{\cal H}_{r+1}^{(r)}$, are not in normal form, thus
we want to eliminate them. 
As in Eq.~\eqref{eq:orders_lambda}, from the Lie series $\exp ({\cal
  L}_{\chi_{r+1}}) H^{(r)}$, we see that the only terms of order
$\lambda^{r+1}$ are those coming from $\lambda^{r+1} {\cal
  H}_{r+1}^{(r)}$ and ${\cal L}_{\chi_{r+1}} {\cal Z}_0$.  Therefore,
the corresponding homological equation for this step reads
\begin{equation}\label{eq:gen_homol_eq_r}
\lambda^{r+1} \phantom{i}^{\ast}{\cal H}_{r+1}^{(r)} + {\{\cal Z}_0,
\chi_{r+1} \} = 0~,
\end{equation}
which can be solved in the same way as Eq.~\eqref{eq:gen_homol_eq_1}.
Applying the Lie series operator to $H^{(r)}$
\begin{equation}\label{eq:H_rp1_oper}
{\cal H}^{(r+1)} = \exp({\cal L}_{\chi_{r+1}}) {\cal H}^{(r)}~,
\end{equation}
we obtain $H^{(r+1)}$ which, by construction, is in normal form up to 
${\cal O}(\lambda^{r+1})$, i.e.
\begin{equation}\label{eq:H_rp1_exp}
{\cal H}^{(r+1)} = {\cal Z}_0 + \lambda {\cal Z}_1 + \ldots + 
\lambda^r {\cal Z}_r + \lambda^{r+1} {\cal Z}_{r+1} + 
\lambda^{r+2} {\cal H}_{r+2}^{(r+1)} + \ldots~,
\end{equation}
where ${\cal Z}_{r+1} = H_{r+1}^{(r)}-\!\phantom{i}^{\ast}{\cal H}_{r+1}^{(r)}$.

\subsection{Linear and non-linear stability}\label{sec:1.4.X.non_stab}

Let us consider a system represented by a Hamiltonian function
${\cal H}$. Let $(\mathbf{p}_0,\mathbf{q}_0)$ be an equilibrium point
 of ${\cal H}$. Hamilton's equations
\eqref{eq:Ham_eqs_gral} for the equilibrium point yield
\begin{equation}\label{eq:equil_pos_ham}
\left. \frac{\partial {\cal H}}{\partial q_i} \right|_{\mathbf{p}_0,\mathbf{q}_0} = \left. \frac{\partial {\cal H}}{\partial p_i} \right|_{\mathbf{p}_0,\mathbf{q}_0} = 0~, \quad i=1,\ldots,n~. 
\end{equation}
We define an orbit slightly displaced with respect to
the equilibrium as
\begin{equation}\label{eq:small_displ_1}
(\mathbf{q},\mathbf{p})~ =
  (\mathbf{q}_0+\boldsymbol{\delta}\mathbf{q},
  \mathbf{p}_0+\boldsymbol{\delta} \mathbf{p})~, 
\end{equation}
where $\mathbf{X}\equiv (\boldsymbol{\delta} \mathbf{q},
\boldsymbol{\delta} \mathbf{p}) \in \mathbb{R}^{2n}$ is a vector whose
components are all small quantities. If we replace
Eq.~\eqref{eq:small_displ_1} in Eqs.~\eqref{eq:Ham_eqs_gral}, and we
keep only terms of first order around the equilibrium position,
we produce the \emph{linearized} system,
\begin{equation}\label{eq:linear_matx_form}
\dot{\mathbf{X}} = \mathbf{A} \mathbf{X},
\end{equation}
where the constant matrix $\mathbf{A}$ contains the second
derivatives of ${\cal H}$ evaluated at the equilibrium point. 
The solutions to
Eq.~\eqref{eq:linear_matx_form} are given by
\begin{equation}\label{eq:sol_X}
\mathbf{X} = \mathbf{B}^{-1} \mathbf{Y} = \mathbf{B}^{-1} 
\begin{pmatrix}
\, c_1 \mathrm{e}^{\lambda_1 t} \,\\
\, c_2 \mathrm{e}^{\lambda_2 t} \,\\
\, \vdots \,\\
\, c_n \mathrm{e}^{\lambda_n t} \,\\
\end{pmatrix}~, 
\end{equation}
where $\lambda_i$ are the $n$ eigenvalues corresponding to
$\mathbf{A}$, $\mathbf{B}$ is the $n \times n$ matrix whose columns
are the eigenvectors of $\mathbf{A}$, and $c_i$ are constants of
integration, in practice derived by the initial conditions
\cite{Lang-87}. 

According to the nature of the eigenvalues $\lambda_i$, the solutions
are stable oscillations around the equilibrium point (all $\lambda_i$
are imaginary), or present one or more unstable components (at least
one of the $\lambda_i$ has $\mathrm{Re}(\lambda_i) \neq 0$). Thus, the
linear stability of the motion in the vicinity of
$(\mathbf{p}_0,\mathbf{q}_0)$ can be concluded only by the eigenvalues
of the matrix $\mathbf{A}$. The latter are computed by solving the
characteristic equation
\begin{equation}\label{eq:charact_eq}
\mathrm{det}(\mathbf{A} - \lambda \mathbf{I} ) = 0~,
\end{equation}
with $\mathbf{I}$ the identity matrix.
 
Linear stability around an equilibrium point does not guarantee also
stability when terms of order higher than linear are retained in the
variational equations. Higher order perturbations to the linear
system may raise unstable trajectories, causing escapes. Additionally,
some orbits that are quasiperiodic in the linear approach may not
persist as such in the non-linear model.

We now briefly refer to some well-known approaches to the problem of
nonlinear stability of the orbits in systems of the
form~\eqref{eq:quasi-int-ham}.  The so-called Kolmogorov-Arnold-Moser
(KAM) theorem (\cite{Kolmo-54},~\cite{Moser-62},~\cite{Arnold-63})
examines the existence of quasiperiodic orbits when an integrable
Hamiltonian system is disturbed by a sufficiently small Hamiltonian
perturbation.  Let us consider a system represented by
Eq.~\eqref{eq:quasi-int-ham}, where the integrable part $H_0$
generates solutions of the form~\eqref{eq:acc_ang_var}, with
$\omega_i=\partial H_0/\partial p_i$.  Let us assume also that $H_0$
satisfies an appropriate non-degeneracy condition (see,
e.g.~\cite{Arnold-78}). The simplest such condition is that the
gradients $\nabla_{\mathbf{p}}\omega_i(\mathbf{p})$ for $i=1,\ldots,n$,
are linearly independent.  The theorem states that, for a $\epsilon$
sufficiently small, a large (of order $1-{\cal O}(\sqrt{\epsilon})$) measure
of the non resonant invariant tori of the unperturbed problem $H_0$
survive as deformed invariant tori, with the original frequencies, in
the perturbed problem ${\cal H} = H_0 + \epsilon H_1$ (\S 6.3
of~\cite{ArnKozNeish-06}, \S 2.3.5. of~\cite{Contopoulos-02}). The
preserved invariant tori are called \emph{KAM tori}, and they are
characterized by frequencies which are 'far from being resonant', typically
described by the \emph{diophantine} condition
\begin{equation}\label{dioff}
\exists \, \gamma, \tau > 0 \quad \mathrm{such\:that} \quad  | {\textstyle \sum_{i=1}^n} k_i\,\omega_i|
  \geq \frac{\gamma}{\|\mathbf{k}\|^{\tau}}~,\: \forall \, \mathbf{k} \in 
\mathbb{Z}^n\setminus \{0\}~,
\end{equation}
where $\mathbf{k}$ is the integer vector $(k_1,\ldots,k_n)$.

In the case of a system of 2 d.o.f, the phase space is
4-dimensional and any trajectory evolves on an 3D isoenergetic
surface. Thus, a KAM torus divides the
phase space in two non-communicating parts, restricting the motion to one
part or the other. This ensures the non linear stability for all the 
orbits confined in the interior of an invariant KAM torus. However,
in systems of more than 2 d.o.f., the invariant tori do not isolate
the orbits in their interior. Thus, the non linear stability of such
orbits has to be examined by other methods.

On the other hand, the Nekhoroshev theorem
(\cite{Nekh-77},~\cite{Nekh-79}), whose analytic part was suggested in
earlier works (\cite{Little-59a},~\cite{Little-59b}), establishes the
stability of the orbits on a finite, albeit exponentially long time.
Let us assume that we have a nearly-integrable Hamiltonian system of
the form~\eqref{eq:quasi-int-ham}, where $H_0$ satisfies some
so-called conditions of non-degeneracy, steepness (or convexity) and
analyticity (see~\cite{Pisa-03}). Then, the theorem states that the
actions $\mathbf{p}$ are bounded according to
\begin{equation}\label{eq:Nekh_est}
\|\mathbf{p}(t) -\mathbf{p}(0) \| < \epsilon^{\alpha} \qquad \mathrm{for\:all}\: \: t \leq T \,, \: \mathrm{with} \: T = {\cal O}(\mathrm{e}^{(\frac{\epsilon_o}{\epsilon})^b}) ~,
\end{equation}
where $\alpha$ and $b$ are parameters depending on the number of
degrees of freedom and the particular form of $H_0$ and $H_1$, and 
 $T$, called the Nekhoroshev time, gives a minimum of the time of practical
stability of the orbits. 

As mentioned already, the existence of invariant Kolmogorov tori in
systems of more than 2 d.o.f.  does not suffice to isolate open
regions in the phase space.  However, in regions where the KAM tori
have a large measure, they create partial blocking structures, which
practically ensure the stability of all the orbits in their
neighborhood for extremely long times. In particular, the joint use of
the KAM theorem with Nekhoroshev's theorem ensures that the diffusion
at very small distances from the KAM tori is super-exponentially slow,
yielding stability for times exceeding by far even those found by the
Nekhoroshev theorem~\cite{MorbGio-95}. Although the normal form
construction involved in the theorem of super-exponential stability is
local (attached to one KAM torus, see~\cite{MorbGio-95}), the theorem
provides a lower bound of the true time of effective stability for a
large measure of orbits in cases of systems of three or higher number
of d.o.f. possesing a large measure of invariant KAM tori.


\section{The Two-Body Problem (2BP)}\label{sec:1.2-2bodyprob}

\subsection{Main features}\label{sec:1.2.1-feat2bp}

\begin{SCfigure}
  \centering
  \includegraphics[width=.50\textwidth]{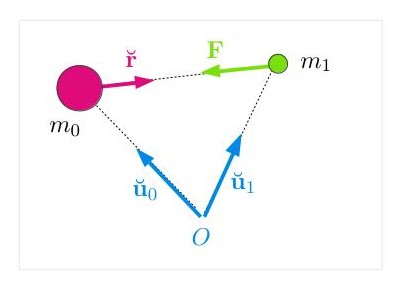}
  \caption[Schematic representation of the Problem of 2
    bodies]{Schematic representation of the problems of 2 bodies, of
    masses $m_0$ and $m_1$, located at positions $\mathbf{u}_0$ and
    $\mathbf{u}_1$ in an inertial frame $O$. The attraction force that
    $m_0$ exerts on $m_1$ is represented by vector $\mathbf{F}$, with
    magnituded given in~\eqref{eq:Newt_grav_law}. The distance that
    separates the two masses is $r$, with its vectorial representation
    $\mathbf{r}$ in heliocentric coordinates.
  \vspace{1.2cm}}
  \label{fig:2body_prob.jpg}
\end{SCfigure}

The most basic problem in Celestial Mechanics is Kepler's problem,
i.e. the motion of a pair of bodies under their mutual gravitational
interaction. It represents the motion of many diverse systems, as
binary stars, satellites moving around planets or planets around the
stars. We now present some main elements of the Keplerian motion,
mostly introducing formul\ae~needed in the rest of the thesis.

Kepler's three laws summarize the
results of his observations of the planetary motion: the planets move
in ellipses with the Sun at one focus (1st. law), in orbits whose
periods are proportional to the cube of their major semiaxis (3rd. law),
and the radial vector which connects the Sun with the planet spans
equal areas in equal times (2nd. law). Later on, Newton provided the
mathematical relation for these empirical laws (see \S2.3
of~\cite{MurrDerm-99}): the attraction force between any two
masses, $m_0$ and $m_1$ separated by a distance $r$ is given by
\begin{equation}\label{eq:Newt_grav_law}
F = {\cal G} \frac{m_0 m_1}{r^2}~,
\end{equation}
where ${\cal G}$ is the universal constant of gravitation, ${\cal G}=
6.67260\times 10^{-11} Nm^2 kg^{-2}$. 
\begin{SCfigure}
  \centering
  \caption[Orbital plane of the 2BP]{Representation of the orbital
    plane (the colored ellipse) defined by the motion of $m_1$ around
    $m_0$. Since the angular momentum $\mathbf{L} = \mathbf{r}\times
    \dot{\mathbf{r}}$ is constant, the plane of the motion remains
    always the same and normal to the vector $L$
  \vspace{1.5cm}}
  \includegraphics[width=.70\textwidth]{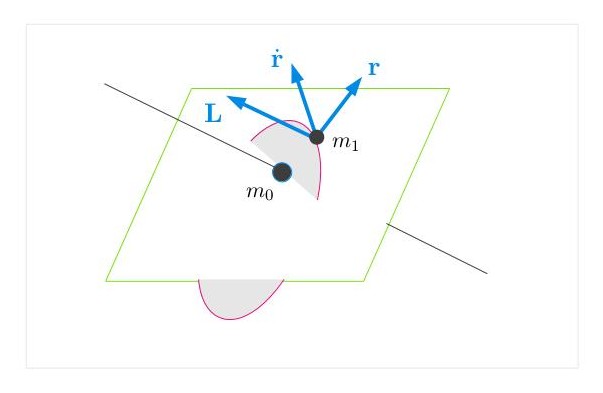}
  \label{fig:ang_mom_plot.jpg}
\end{SCfigure}

Let us consider the motion of the two masses, in an inertial frame,
such that the attraction force between them is given
by~\eqref{eq:Newt_grav_law}. 
Under such assumptions, the center of mass results stationary or in
constant rectilinear motion. Therefore, it is possible to reduce the
problem to the relative motion of $m_1$ with respect to $m_0$. If
$m_0$ corresponds to the Sun, the new coordinates are called
\emph{heliocentric}. Considering the relative vector
$\mathbf{r}=\mathbf{u}_1-\mathbf{u}_0$, where $\mathbf{u}_i$ are the
position vectors in the inertial frame for each mass, we obtain
the equation of relative
  motion of $m_1$ around $m_0$ (see Fig.~\ref{fig:2body_prob.jpg}):
\begin{equation}\label{eq:Newt_eq_hel_2BP}
\frac{\df^2 \mathbf{r}}{\df t^2}=-\frac{{\cal
    G}(m_0+m_1)}{\|\mathbf{r}\|^3}\mathbf{r}~.
\end{equation}
Thus, the gravitational interaction between masses is represented by a
central field, characterized by the inverse square dependence on the
distance. Any central field (of arbitrary dependence on $r$) provides
certain useful symmetries (see \S3-2 of~\cite{Goldstein-80}), related
to the conservation of two important quantities: the total energy of
the system and the angular momentum vector.  If we take the vector
product of $\mathbf{r}$ with $\ddot{\mathbf{r}}$, by
Eq.~\eqref{eq:Newt_eq_hel_2BP}, we get
\begin{equation}\label{eq:ang_momentum}
\mathbf{r} \times \ddot{\mathbf{r}}=0, \quad \textrm{and\,integrating}\quad
\mathbf{r} \times \dot{\mathbf{r}} = \mathbf{L}~, 
\end{equation}
where $\mathbf{L}$ is a constant vector perpendicular to both
$\mathbf{r}$ and $\dot{\mathbf{r}}$. Thus,
the position vector $\mathbf{r}$ and the velocity vector
$\dot{\mathbf{r}}$ always lie in a plane perpendicular to the direction
defined by $\mathbf{L}$, the \emph{angular momentum vector} (per unit mass).  
In other
words, the motion of $m_1$ around $m_0$ always takes place in
the same plane (see Fig.~\ref{fig:ang_mom_plot.jpg}), called the
\emph{orbital plane}. In polar coordinates $(r,\theta)$ in the orbital
plane, the preserved modulus of $\mathbf{L}$
is given by $L=r^2\dot{\theta}$. The total energy per unit
mass provides an additional first integral of motion
\begin{equation}\label{eq:Energy_2BP}
{\cal E}=\frac{1}{2}\|\mathbf{v}\|^2- \frac{{\cal G}(m_0+m_1)}{r}
 = T(\mathbf{v})+V(\mathbf{r}) \equiv const~.
\end{equation}
Thus, the 2BP has four independent integrals of motion: the energy
${\cal E}$, and the three components of the angular momentum vector
$\mathbf{L}$. Three of them (${\cal E}$, $L = |\mathbf{L}|$ and $L_z$) are
in involution. Therefore the 2BP is an integrable
system.

\subsection{Orbital elements}\label{sec:1.2.2-orb-elem}

\begin{SCfigure}
  \centering
  \includegraphics[width=.60\textwidth]{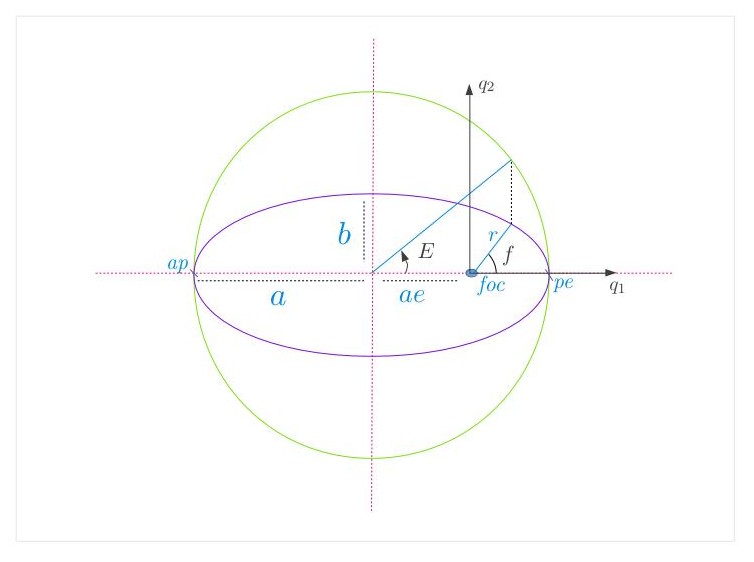}
  \caption[Orbital elements 1]{Graphical representation of an ellipse and the
    quantities that define it: major semiaxis $a$, minor semiaxis $b$
    and eccentricity $e$.  Considering the attracting mass at the
    focus labeled as \emph{foc}, $r$ is the distance
    between the two masses. Accordingly, the labels \emph{pe} and
    \emph{ap} define the positions of the pericenter and apocenter,
    respectively. The angles $E$ and $f$ denote the eccentric and true
    anomalies.
  \vspace{0.5cm}}
  \label{fig:ellipse_elem.jpg}
\end{SCfigure}

Taking into account that the motion in the 2BP always takes place on a
planar ellipse, it is convenient to characterize the motion in
terms of quantities that: i) describe the geometrical properties of
the ellipse, ii) orient the ellipse in space, and iii) define the
actual position of the body on the ellipse. These quantities are the
\emph{orbital elements}.  Every ellipse is described by two different
quantities: its major semiaxis $a$ and its minor semiaxis $b$, that
give the shape of the ellipse, as in Fig.~\ref{fig:ellipse_elem.jpg}.
In an equivalent way, the ellipse can be characterized by the major
semiaxis and the eccentricity $e$, i.e. the ratio between the distance
from the center of the ellipse to the focus ($ae$ in
Fig.~\ref{fig:ellipse_elem.jpg}) and $a$. This quantity indicates how
much the orbit differs from a perfect circle ($e=0$ for circular
orbits and $e=1$ for a segment of length $2a$). For a fixed value of
the major semiaxis, the range $0<e<1$ gives any possible ellipse. Let
us assume that the star is located at the focus labeled as
\emph{foc} in Fig.~\ref{fig:ellipse_elem.jpg} and the instantaneous
distance between $m_0$ and $m_1$ is given by $r$. Thus, $r$ is bounded
between $a(1-e)$ (minimum) and $a(1+e)$ (maximum). The shortest
possible distance defines the \emph{pericenter} (also
\emph{perihelion} or \emph{perigee}, according to the central body),
labeled in the figure as \emph{pe}. Equivalently, the largest distance
defines the \emph{apocenter} (\emph{aphelion}, \emph{apogee})
\emph{ap}. 

To locate the instantaneous position of the body on the
ellipse, we define an orthogonal reference frame $q_1,q_2$ with origin
at the position of $m_0$. One of the axis ($q_1$) is defined in the
direction of the line
that connects the focus with the pericenter. In this frame, we introduce
\emph{polar coordinates} ($r$,$\theta$). It is customary to refer to
the angle $\theta$ as the \emph{true anomaly} $f$.
Additionally, we define a second angle $E$ called \emph{eccentric anomaly},
which corresponds to the angle subtended at the center of the ellipse
by the projection of the position of the body on a circle with radius
equal to $a$ and tangent to the ellipse at the pericenter and
apocenter (see Fig.~\ref{fig:ellipse_elem.jpg}). At a
certain time $t$, the polar coordinates for $m_1$ are
given by
\begin{equation}\label{eq:r_polar}
r = a(1-e \cos E)~,
\end{equation}
and
\begin{equation}\label{eq:true_anomaly}
\cos f = \frac{\cos E - e}{1 - e \cos E}~.
\end{equation}
On the other hand, the instantaneous position in the $(q_1,q_2)$ frame is given
by
\begin{equation}\label{eq:q1q2}
q_1 = a (\cos E -e),\qquad q_2 = a \sqrt{1-e^2} \sin E~~.
\end{equation}
Since it is enough to know $a$, $e$ and $E$ for obtaining either
$r$ and $f$, or $q_1$ and $q_2$, the instantaneous position of the
body at time $t$ on the ellipse is completely determined by
these three quantities.

\begin{SCfigure}
  \centering
  \includegraphics[width=.65\textwidth]{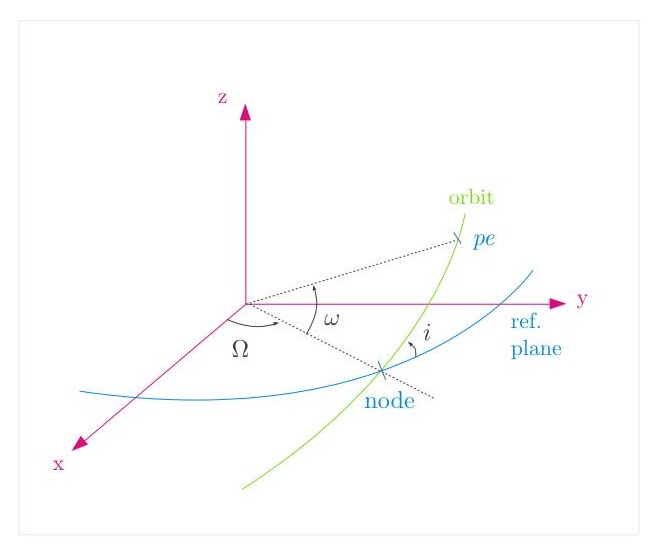}
  \caption[Orbital elements 2]{Orbital plane and reference plane in
    the orthogonal system of reference $(x,y,z)$. The angles $\Omega$,
    $\omega$ and $i$ define the position and orientation of the
    ellipse in the reference frame. The position of the ascending node
    is denoted by the label \emph{node} and the pericenter of the
    ellipse by $pe$.}
  \label{fig:ang_elem.jpg}
\end{SCfigure}

According to Kepler's second law, the motion of $m_1$ on
the ellipse is such that the position vector $(q_1,q_2)$ spans equal areas
of the orbital plane in equal times. In other words, the transverse
velocity of the test particle varies along the ellipse, 
maximum at the pericenter and minimum at the apocenter. Since $E$
corresponds to a projection of this motion, it does not evolve
linearly with time. We then introduce a new angle 
\begin{equation}\label{eq:mean_anomaly}
M=n(t-t_0)~,
\end{equation}
where 
\begin{equation}\label{eq:mean_motion}
n=\frac{\sqrt{{\cal G}(m_0+m_1)}}{a^{3/2}}
\end{equation}
is the body's orbital frequency, also called \emph{mean motion}, and
$t_0$ is the time of passage at pericenter. Unlike the eccentric
anomaly, the \emph{mean anomaly} $M$ changes linearly with time. From
the corresponding equation of motion~\eqref{eq:Newt_eq_hel_2BP} and
the relations in Eqs.~\eqref{eq:q1q2}, \eqref{eq:true_anomaly} and
\eqref{eq:r_polar}, it is possible to obtain an equation that relates
these anomalies.  The so-called \emph{Kepler equation}, given by
\begin{equation}\label{eq:Kep_equation}
E - e \sin E = M = n(t-t_0)~,
\end{equation}
gives the relation between $E$ and $M$ and shows the non-linear
dependence of $E$ on time. Since $E$ is one-to-one to $M$, the set
$a$, $e$ and $M$ also defines completely the instantaneous position of
$m_1$ on the ellipse.

Finally, we define the position and orientation of the ellipse in
space. Let us consider an orthogonal inertial system of reference
$(x,y,z)$, centered at the central body $m_0$. In this frame, the
location of the ellipse is based on three different angles. First, we
define the inclination $i$ of the orbital plane with respect to the
reference plane ($x,y$). Except for the particular case when the
orbital plane coincides with the reference plane ($i=0$), the orbit
intersects the plane ($x,y$) at two different points, called
\emph{nodes}.  We differentiate them by considering the
\emph{ascending} node, i.e.  when the test particle crosses from
negative values of $z$ to positive values (see
Fig.~\ref{fig:ang_elem.jpg}), and the \emph{descending} node (from
positive to negative values).  Two additional angles define the
orientation of the ellipse in this system of reference: the
\emph{longitude of the node} $\Omega$, i.e. the angular position of
the ascending node measured from the $x$ axis, completes the
orientation of the orbital plane; the \emph{argument of the
  pericenter} $\omega$, i.e. the angular position of the pericenter
with respect to the line that joins the central body and the ascending
node (measured on the orbital plane), characterizes the orientation of
the ellipse in its own plane.

Summarizing, the quantities $a$, $e$, $i$, $\omega$ and $\Omega$ give
the shape, position and orientation of the ellipse in
space. Additionally, the mean anomaly $M$ gives the position of the
body in the ellipse. The whole set, known as \emph{orbital elements},
completely defines the position and velocity of the body in the
inertial frame centered at the central mass. The correspondance
between the inertial positions and velocities ($x$, $y$, $z$, $\dot{x}$,
$\dot{y}$,$\dot{z}$) and the orbital elements can be found, e.g., in \S2.8
of~\cite{MurrDerm-99}.

In the case $i=0$, the position of the nodes
is not determined and therefore $\omega$ and $M$ are not defined.
In addition to this, for $i\neq 0$ but $e=0$, the mean anomaly is not
defined either, since in this case the position of the pericenter
is not determined. In order to remove these inconsistencies,
it is preferable to consider new well-defined angles. 
Thus, we introduce the \emph{longitude
of the pericenter}, the angle between the x-axis and the pericenter,
\begin{equation}
\varpi = \omega+\Omega ~,
\end{equation}
and the \emph{mean longitude} 
\begin{equation}\label{mean_long}
\lambda = M + \varpi~.
\end{equation}
The first angle is well defined when $i=0$, while the second one is
well defined when $i$ or $e$ (or both) are equal to zero. Thus, this
alternative set completely defines the positions and velocities in
the inertial frame.

It is possible to obtain an expression of
the preserved quantities of the 2BP (${\cal E}$ and $\mathbf{L}$) in
terms of the orbital elements. 
Considering, from Section~\ref{sec:1.2.1-feat2bp}, that
$L=r^2\dot{f}$, thus
\begin{equation}\label{eq:area_diff}
\frac{\df A}{\df t} = \frac{1}{2} L~,
\end{equation}
where $A$ is the area scanned by the radial vector in a time $t$.
For $t=T$, the period of a revolution on the ellipse, Eq.~\eqref{eq:area_diff}
reads
\begin{equation}\label{eq:area_diff_2}
A_{ellipse} = \pi a \, b \, = \frac{1}{2} \, L\, T~.
\end{equation}
Replacing $T = \frac{2\pi}{n}$ and $b = a\sqrt{1-e^2}$ in 
Eq.~\eqref{eq:area_diff_2}, we obtain
\begin{equation}\label{eq:mod_ang_mom_elem}
 L=n a^2\sqrt{1-e^2}~,
\end{equation}
which gives the modulus of the angular momentum in terms of orbital elements.

From equations~\eqref{eq:r_polar}
and~\eqref{eq:true_anomaly}, we obtain
\begin{equation}\label{eq:r_elem}
r=\frac{a(1-e^2)}{1+e\cos f}~.
\end{equation}
Differentiating~\eqref{eq:r_elem}, we obtain 
\begin{equation}\label{eq:r_elem_dot}
\dot{r}= \frac{r \dot{f} e \sin f}{1+ e \cos f}~.
\end{equation}
Thus, from Eqs.~\eqref{eq:mod_ang_mom_elem} and~\eqref{eq:r_elem_dot}, we have
\begin{equation}
\dot{r}=\frac{n a}{\sqrt{1-e^2}}e \sin f\, \qquad r \dot{f} =
\frac{na}{{\sqrt{1-e^2}}}(1+e \cos f)
\end{equation}
and then, considering the expression of the velocity in terms of orbital
elements,
\begin{equation}\label{eq:vel_elem}
\|\mathbf{v}\|^2 = v^2 = \dot{r}^2 + r^2 \dot{f}^2 = \frac{n^2
  a^2}{1-e^2} (1+2 e \cos f + e^2) = \frac{n^2 a^2}{1- e^2} \left(
\frac{2 a (1-e^2)}{r} - (1-e^2) \right)~.
\end{equation}
Hence
\begin{equation}\label{eq:otra_que_va}
v^2 = n^2 a^2  \left( \frac{2}{r} - \frac{1}{a} \right) = 
{\cal G}(m_0+m_1) \left( \frac{2}{r} - \frac{1}{a} \right)~.
\end{equation}
Replacing Eq.~\eqref{eq:otra_que_va} in the expression for the
energy~\eqref{eq:Energy_2BP} per unit mass, we finally obtain
\begin{equation}\label{eq:Energy_2BP_elem}
{\cal E}=\frac{1}{2}\|\mathbf{v}\|^2- \frac{{\cal G}(m_0+m_1)}{r}
=\frac{1}{2} {\cal G}(m_0+m_1) \left( \frac{2}{r} - \frac{1}{a} \right)
- \frac{{\cal G}(m_0+m_1)}{r} = -\frac{{\cal G}(m_0+m_1)}{2a}~.
\end{equation}
Therefore, it is possible to write the Hamiltonian for 
the 2BP in terms of orbital elements, as
\begin{equation}\label{eq:Ham_2BP_elem}
H = {\cal E} = -\frac{{\cal G}(m_0+m_1)}{2a}~,
\end{equation}
a function depending only on the major semiaxis $a$.
But we should emphasize here that none of the set of orbital elements
conforms a set of \emph{canonical coordinates}. Such variables are
introduced in the next subsection.

\subsection{Delaunay coordinates}\label{sec:1.2.3-del_coord}

The Hamiltonian representation of the 2BP, as well as its
perturbations (like the Three-Body problem, see below), require to
define a suitable set of canonical coordinates.  We now introduce a
set of such coordinates, in the form of action-angle variables of the
2BP (see \S1.9.1. of~\cite{Morbi-11} for details on the construction).
The set of canonical \emph{Delaunay} action-angles variables
reads\footnote{We use the same symbols for the Delaunay variables $L$,
  $H$, as before for the angular momentum $\mathbf{L}$ and the
  Hamiltonian H because they are traditionally defined this way in the
  bibliographic references. In cases where this might be confusing, we
  introduce additional labels in order to properly distinguish them.}
\begin{equation}
\begin{aligned}\label{eq:delau_vari}
L=&\, \sqrt{{\cal G}(m_0+m_1)a}~, &l &= M~,  \\ 
G=&\, L\sqrt{1-e^2}~, &g &=\omega~,  \\ 
H=&\, G \cos i~, &h &= \Omega ~.
\end{aligned}
\end{equation}
The value of the action $G$ coincides with the modulus of the angular
momentum vector. Additionally, as discussed in
Section~\ref{sec:1.2.2-orb-elem}, we can introduce new angles for
those cases when $M$ and $\omega$ are not well defined. Thus, 
the \emph{modified Delaunay variables} are given by
\begin{equation}
\begin{aligned}\label{eq:mod_delau_vari}
\Lambda =& L=\, \sqrt{{\cal G}(m_0+m_1)a}~, & \lambda &= M+\varpi~,
 \\ 
\Gamma =&\, L(1-\sqrt{1-e^2})~, &\gamma &= -\varpi~, \\ 
Z=& \,\Gamma (1-\cos i)~, &\zeta &= -\Omega
~. 
\end{aligned}
\end{equation}
The action variables ($\Lambda$,$\Gamma$,$Z$) are a measure of the
major semi-axis, the eccentricity and the inclination, respectively.
This last set incorporates a common problem of the description in polar
variables: the angles $\gamma$ and $\zeta$ are multi-valued whenever their
corresponding actions are null. In order to remove this
fictitious singularity, we define an additional set of variables
known as \emph{Poincar\'e variables}
\begin{equation}
\begin{aligned}\label{eq:mod_delau_vari}
\Lambda &~, &\lambda &~, \\ 
\xi &= \sqrt{2\, \Gamma}\cos \gamma~, &\eta &= \sqrt{2\,\Gamma}\sin \gamma~\\
\varsigma &= \sqrt{2\,Z}\cos \zeta~, &\vartheta &= \sqrt{2\,Z}\sin \zeta~
\end{aligned}
\end{equation}
From~\eqref{eq:Ham_2BP_elem}, we can express the Hamiltonian of the 2BP in
terms of any of the above canonical sets. For example,
\begin{equation}\label{eq:Ham_2BP_del}
H_{\mathrm{2BP}} = -\frac{{\cal G}^2(m_0+m_1)^2}{2\Lambda^2}~.
\end{equation}


\section{The Restricted Three-Body Problem (R3BP)}\label{sec:1.3-RTBP}

\subsection{The Hamiltonian of the Restricted Three Body problem}\label{sec:1.3.3-ham_r3bp}

\begin{figure}[t]
  \centering
  \includegraphics[width=0.99\textwidth]{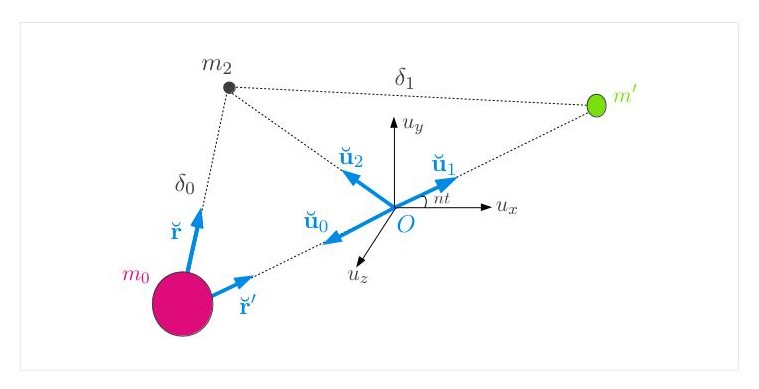}
  \caption[Schematic representation of the Problem of 3
    bodies]{Schematic representation of the problem of 3 bodies, of
    masses $m_0$, $m'$ and $m_2$ (negligible), located at positions
    $\mathbf{u}_0$, $\mathbf{u}_1$ and $\mathbf{u}_2$ in an inertial
    frame $O$. In heliocentric coordinates, the positions of $m'$ and
    $m_2$ relative to $m_0$ are $\mathbf{r}'$ and $\mathbf{r}$,
    respectively. The distance that separates the test particle from
    $m_0$ is given by $\delta_0$, and the one up to $m'$ is
    $\delta_1$.  The line that connects $m_0$ and $m'$ rotates with
    respect to the line defining the $u_x$ axis with angular velocity
    $nt$, $n$ corresponds to the mean motion of the planet.}
  \label{fig:r3bodyprob.jpg}
\end{figure}

After setting the features of the 2BP, we now pass to the more complex
Restricted Three-Body Problem (R3BP). We consider a system of three bodies: a
dominant mass $m_0$, and two additional smaller bodies, with masses
$m_1 \equiv m',\,m_2$, as in Fig.~\ref{fig:r3bodyprob.jpg}. In a barycentric
inertial reference frame, Newton's equations of motion
are
\begin{equation}\label{eq:Newt_eq_baryc_TBP}
\frac{\df^2\mathbf{u}_i}{\df t^2 } = -{\cal G} \sum_{j\neq i} m_j \frac{\mathbf{u}_i-\mathbf{u}_j}{\|\mathbf{u}_i-\mathbf{u}_j\|^3}~.
\end{equation}
Introducing the heliocentric
positions of the smaller bodies
$\mathbf{r}=\mathbf{u}_2-\mathbf{u}_0$ and $\mathbf{r}' \equiv \mathbf{r}_1 =\mathbf{u}_1-\mathbf{u}_0$, the equations above can be
rewritten as
\begin{align}
\frac{\df^2\mathbf{r}'}{\df t^2} = - \frac{{\cal G}\left(m_0+m' \right)}{\|\mathbf{r}'\|^3} \mathbf{r}' + {\cal G}m_2 \left( \frac{\mathbf{r}-\mathbf{r}'}{\|\mathbf{r}-\mathbf{r}'\|^3} - \frac{\mathbf{r}}{\|\mathbf{r}\|^3} \right) \quad \text{for} \: m'~,\label{eq:Newt_eq_hel_TBP-A}\\
\frac{\df^2\mathbf{r}}{\df t^2} = - \frac{{\cal G}\left(m_0+m_2 \right)}{\|\mathbf{r}\|^3} \mathbf{r} + {\cal G}m' \left( \frac{\mathbf{r}'-\mathbf{r}}{\|\mathbf{r}'-\mathbf{r}\|^3} - \frac{\mathbf{r}'}{\|\mathbf{r}'\|^3}\right) \quad \text{for} \: m_2~. \label{eq:Newt_eq_hel_TBP-B}
\end{align}
The motion of $m_0$ is given by $\mathbf{u}_0=
  -(m'\mathbf{r'}+m_2\mathbf{r})/(m_0+m'+m_2)$ and needs not be
explicitely considered.  From Eqs.~\eqref{eq:Newt_eq_hel_TBP-A}
and~\eqref{eq:Newt_eq_hel_TBP-B}, we can construct the heliocentric
equations of motion of a body of \emph{negligible} mass $m_2 = 0$,
under the influence of the two massive bodies $m_0$ and $m'$ ($ m'<
m_0$). From this point on we refer to $m_0$ as \emph{the Sun} or
\emph{star}, to $m'$ as \emph{planet} or \emph{primary} and to $m_2$
as the \emph{massless body} or \emph{test particle}.  These terms are
used just for brevity reasons, since the same formulation can be applied
to any restricted problem (e.g.  Earth-Moon-spacecraft system), as
long as the same physical assumptions are made.
Setting $m_2=0$ in Eqs.~~\eqref{eq:Newt_eq_hel_TBP-A}
and~\eqref{eq:Newt_eq_hel_TBP-B}, we recover the Keplerian equations of
motion for $\mathbf{r}'$, i.e. the motion of the planet around the Sun is
described by a fixed Keplerian ellipse, that we assume
\emph{given}. Then, the equation of motion for the massless body (the
so-called Restricted Three-Body problem) reads
\begin{equation}\label{eq:Newt_eq_hel_RTBP}
\frac{\df^2\mathbf{r}}{\df t^2} = -\frac{{\cal
    G}\,m_0}{\|\mathbf{r}\|^3} \mathbf{r}+ {\cal G}m'
\left( \frac{\mathbf{r}'-\mathbf{r}}{\|\mathbf{r}'-\mathbf{r}\|^3} -
\frac{\mathbf{r}'}{\|\mathbf{r}'\|^3}\right)~,
\end{equation}
where $\mathbf{r}$ and $\mathbf{r}'$ are the heliocentric position
vector of the massless body and of the planet, respectively.  The term
$\frac{{\cal G}\,m_0}{\|\mathbf{r}\|^3} \mathbf{r}$ has the form
Eq.~\eqref{eq:Newt_eq_hel_2BP} of the 2BP, but with $m_1$ replaced by
$m_2=0$ corresponding to the test particle. If $m'\ll m_0$ and
$\|\mathbf{r}'-\mathbf{r}\| \gg 0$, the remaining terms in the
r.h.s. of Eq.~\eqref{eq:Newt_eq_hel_RTBP}, depending on the mass of
the planet, play the role of a small perturbation with respect to the
influence of the star. Therefore, we infer that the resulting motion
for the massless body is close to (although not exactly) a Keplerian
orbit.

From~\eqref{eq:Newt_eq_hel_RTBP}, we can define a
scalar potential $U(\mathbf{r})$ of the form
\begin{equation}\label{eq:potential_U}
U(\mathbf{r}) = -\frac{{\cal G}m_0}{\|\mathbf{r}\|} - {\cal G} m'
\left( \frac{1}{\Delta} - \frac{\mathbf{r} \cdot
  \mathbf{r}'}{\|\mathbf{r}'\|^3} \right)~,
\end{equation}
where $\Delta=\|\mathbf{r}-\mathbf{r}'\|$. Then,
Eq.~\eqref{eq:Newt_eq_hel_RTBP} acquires the form
\begin{equation}\label{eq:grad_U}
\frac{\df^2\mathbf{r}}{\df t^2} = -\nabla_{\mathbf{r}} U(\mathbf{r})~.
\end{equation}
Associating $\mathbf{q}$ with $\mathbf{r}$ and $\mathbf{p}$ with
$\dot{\mathbf{r}}$, it is simple to demonstrate that
Eq.~\eqref{eq:grad_U} accomplishes the conditions
in~\eqref{eq:Ham_eqs_gral} for being in Hamiltonian form, with a
Hamiltonian function
\begin{equation}\label{eq:Ham_hel_RTBP_1}
H = \frac{\|\mathbf{p}\|^2}{2} + U(\mathbf{r}) ~.
\end{equation}
Including the expression
of the potential, we have the Hamiltonian function of the R3BP
\begin{equation}\label{eq:Ham_hel_RTBP_2}
H = \frac{\|\mathbf{p}\|^2}{2}-\frac{{\cal G}m_0}{\|\mathbf{r}\|} -
{\cal G} m' \left( \frac{1}{\Delta} - \frac{\mathbf{r} \cdot
  \mathbf{r}'}{\|\mathbf{r}'\|^3} \right)~.
\end{equation}
We note here that the previous formula is composed by two different
contributions. The first two terms
\begin{equation}\label{eq:kepl_part}
K = \frac{\|\mathbf{p}\|^2}{2}-\frac{{\cal G}m_0}{\|\mathbf{r}\|}~~,
\end{equation}
the so-called \emph{Keplerian part}, correspond to the Hamiltonian
function associated to the system of
equations~\eqref{eq:Newt_eq_hel_2BP} of the Two-Body problem (with
$m_1=0$). Thus, this part of the Hamiltonian induces a Keplerian-like
solution. The second part is called the \emph{disturbing function} of
the R3BP. It represents a small perturbation, whose size relative to
$K$ is proportional to $m'/m_0$. Since the Keplerian part is 
integrable, it is the disturbing function which introduces all
the interesting dynamical features of the problem.

\subsection{Rotating frame and Lagrangian equilibrium points}\label{sec:1.3.2-rot_fram_Lag}

The R3BP exhibits some well known features of which we make use along
this thesis. In particular, for introducing the Trojan problem, we
need to refer to the properties of the so-called \emph{Lagrangian equilibrium
points}.

Let $(u_x,u_y,u_z)$ be the inertial frame as in
Fig.~\ref{fig:r3bodyprob.jpg}. We recall that the position vectors in
this system are $\mathbf{u}_i=(u_{x,i},u_{y,i},u_{z,i})$, with $i=0$
(for the star), $1$ (planet), $2$ (massless body), and the equations
of motion are given by Eq.~\eqref{eq:Newt_eq_baryc_TBP}.  Taking
Eq.~\eqref{eq:Newt_eq_baryc_TBP} for the test particle in components,
we have
\begin{equation}
\begin{aligned}\label{eq:iner_eq_comp_r3bp}
\ddot{u}_{x,2} &= {\cal G} m_0 \, \frac{u_{x,0}-u_{x,2}}{\delta_0^3} +
     {\cal G} m' \, \frac{u_{x,1}-u_{x,2}}{\delta_1^3}~,
     \nonumber\\ \ddot{u}_{y,2} &= {\cal G} m_0 \,
     \frac{u_{y,0}-u_{y,2}}{\delta_0^3} + {\cal G} m' \,
     \frac{u_{y,1}-u_{y,2}}{\delta_1^3}~, \\ \ddot{u}_{z,2}
     &= {\cal G} m_0 \, \frac{u_{z,0}-u_{z,2}}{\delta_0^3} + {\cal G}
     m' \, \frac{u_{z,1}-u_{z,2}}{\delta_1^3}~,\nonumber
\end{aligned}
\end{equation}
where, as it can be seen in Fig.~\ref{fig:r3bodyprob.jpg}, 
\begin{equation}
\begin{aligned}\label{eq:iner_eq_comp_r3bp-1}
\delta_0 &=
\sqrt{(u_{x,0}-u_{x,2})^2+(u_{y,0}-u_{y,2})^2+(u_{z,0}-u_{z,2})^2}~, \\ 
\delta_1 &= \sqrt{(u_{x,1}-u_{x,2})^2+(u_{y,1}-u_{y,2})^2+(u_{z,1}-u_{z,2})^2}~.
\end{aligned}
\end{equation}

For the rest of Sect.~\ref{sec:1.3-RTBP}, we consider the case in
which $m_0$ and $m'$ have \emph{circular} orbits around their common
center of mass. This aproximation of the R3BP is known as the
\emph{Circular Restricted Three-Body problem} (CR3BP). In the CR3BP,
the bodies $m_0$ and $m'$ have a constant separation at every moment,
and their motion on the orbital circle has constant angular frequency
$n$, given by the mean motion.  We introduce the following measure
units: we assume that the longitude, time and mass units are such that
the distance that separates $m_0$ from $m'$, the value of ${\cal
  G}(m_0+m')$ and the mean motion $n$ are all equal to $1$.  We define
now the mass parameter
\begin{equation}\label{eq:mass_param}
\mu = \frac{m'}{m'+m_0}~.
\end{equation}
In these units, the revolution period of the planet around the star (or
of any of them around the mass center) is
$T=2\pi$. Furthermore, ${\cal G}m_0 = 1-\mu$ and ${\cal G}m' = \mu$.

\begin{SCfigure}
  \centering
  \includegraphics[width=.50\textwidth]{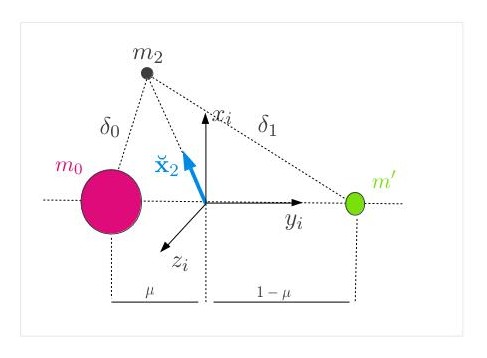}
  \caption[Position of the masses in the synodic rotating
    frame]{Position of the masses in the synodic rotating frame
    $(x_i,y_i,z_i)$. The position of $m_0$ is $(x_0,y_0,z_0) =
    (-\mu,0,0)$, the one of $m'$ is $(x_1,y_1,z_1) = (1-\mu,0,0)$ and
    the one of $m_2$ is $(x_2,y_2,z_2)$. The distance that separates
    the test particle from $m_0$ is given by $\delta_0$, and the one
    up to $m'$ is $\delta_1$.
  \vspace{1.5cm}}
  \label{fig:r3bodyprob_2.jpg}
\end{SCfigure}

Let us consider a non-inertial \emph{synodic} system of reference
$(x,y,z)$ that rotates with angular velocity $n = 1$, its origin is
located at the center of mass of the system and its $x$ axis is
defined by the line connecting $m_0$ and $m'$, with the positive
values on the direction of $m'$ (Fig.~\ref{fig:r3bodyprob_2.jpg}). The
positions of $m_0$ and $m'$ in this system are fixed, given by
\begin{equation}\label{eq:prim_position}
(x_0,y_0,z_0) = (-\mu,0,0)~,\qquad (x_1,y_1,z_1) = (1-\mu,0,0)~.
\end{equation}
The distances $\delta_0$ and $\delta_1$ turn out to be
\begin{align}
\delta_0 &= \sqrt{(x_2+\mu)^2+y_2^2+z_2^2}~, \label{eq:iner_eq_comp_r3bp-a} \\
\delta_1 &= \sqrt{(x_2-(1-\mu))^2+y_2^2+z_2^2}~.\label{eq:iner_eq_comp_r3bp-b}
\end{align}
Regarding $m_2$, the transformation passing
its coordinates from the synodic system $(x,y,z)$ to the inertial
barycentric system $(u_x,u_y,u_z)$ reads
\begin{equation}\label{eq:syn_inert_trans}
\begin{pmatrix}
\,u_{x,2}\, \\
\,u_{y,2}\, \\
\,u_{z,2}\, \\
\end{pmatrix}
=
\begin{pmatrix}
\,\cos nt & -\sin nt & 0 \,\\
\,\sin nt & \cos nt & 0 \,\\
\, 0 & 0 & 1 \,\\
\end{pmatrix}
\begin{pmatrix}
x_2 \\
y_2 \\
z_2 \\
\end{pmatrix}~.
\end{equation}
By means of Eq.~\eqref{eq:syn_inert_trans}, it is possible
to compute and replace the derivatives of $u_{x,2}$, $u_{y,2}$ and $u_{z,2}$, in Eq.~\eqref{eq:iner_eq_comp_r3bp}. Thus, the equations of motion of the test
particle in synodic coordinates are given by
\begin{align}\label{eq:eq_mot_syn_frame}
\ddot{x}_2 - 2n \dot{y}_2 - n^2 x_{2} &= - \left[ (1-\mu) \, \frac{x_2 + \mu}{\delta_0^3} + \mu \, \frac{x_2 - (1-\mu)}{\delta_1^3} \right]~,\\
\ddot{y}_2 + 2n \dot{x}_2 - n^2 y_{2} &= - \left[ \frac{(1-\mu)}{\delta_0^3} + \frac{\mu}{\delta_1^3} \right] y_2 ~,\\
\ddot{z}_2 &= - \left[ \frac{(1-\mu)}{\delta_0^3}+ \frac{\mu}{\delta_1^3} \right] z_2 ~.
\end{align}
These accelerations can be written also in terms of the gradient of a certain scalar function $U$:
\begin{align}
\ddot{x}_2 - 2n \dot{y}_2 &= \frac{\partial U}{\partial x_2}\, \label{eq:eq_mot_syn_a}\\
\ddot{y}_2 + 2n \dot{x}_2 &= \frac{\partial U}{\partial y_2}\, \label{eq:eq_mot_syn_b}\\
\ddot{z}_2 &= \frac{\partial U}{\partial z_2}\,\label{eq:eq_mot_syn_c}
\end{align}
where $U=U(x_2,y_2,z_2)$ is given by
\begin{equation}\label{eq:pseudo-pot}
U= \frac{n^2}{2}(x_2^2+y_2^2)+\frac{1-\mu}{\delta_0}+\frac{\mu}{\delta_1}~.
\end{equation}
Due to the passage to the rotating system of reference,
Eqs.\eqref{eq:eq_mot_syn_a}-\eqref{eq:eq_mot_syn_c} contain not only
the terms of the gravitational potential
$\frac{1-\mu}{\delta_0}+\frac{\mu}{\delta_1}$, but also the terms
producing the centrifugal acceleration, $n^2(x_2^2+y_2^2)/2$, and the
Coriolis terms $-2n\dot{y}$ and $2n\dot{x}$, that depend on the
velocity of the particle and are also proportional to the angular
velocity of the frame.

In terms of the synodic variables, we introduce the \emph{Jacobi
  integral} (or Jacobi constant) $C_{\mathrm{j}}$.  It corresponds to
the only first integral of the CR3BP, since the total energy and the
angular momentum vector are not preserved.  Therefore, unlike the 2BP,
the CR3BP is not integrable in the Liouville sense. Nevertheless,
important information about the behavior of the orbits can be obtained
from the level curves od the Jacobi constant
$C_{\mathrm{j}}$. Multiplying Eq.~\eqref{eq:eq_mot_syn_a} by
$\dot{x}_2$, Eq.~\eqref{eq:eq_mot_syn_b} by $\dot{y}_2$ and
Eq.~\eqref{eq:eq_mot_syn_c} by $\dot{z}_2$, and adding the three
terms, we have
\begin{equation}\label{eq:Jaco_const_1}
\dot{x}_2\ddot{x}_2 + \dot{y}_2\ddot{y}_2 +\dot{z}_2\ddot{z}_2= \frac{\partial U}{\partial x_2}\dot{x}_2 + \frac{\partial U}{\partial y_2}\dot{y}_2 + \frac{\partial U}{\partial z_2}\dot{z}_2~.
\end{equation}
This equation can be integrated, yielding
\begin{equation}\label{eq:Jaco_const_2}
\dot{x}_2^2+\dot{y}_2^2+\dot{z}_2^2 = 2U - C_{\mathrm{j}}
\end{equation}
where the Jacobi integral $C_{\mathrm{j}}$ enters as a constant of integration.
Since $\dot{x}_2^2+\dot{y}_2^2+\dot{z}_2^2 = v_2^2$, the square of the
velocity of the massless particle in the rotating frame, we have
\begin{equation}\label{eq:Jaco_const_3}
v_2^2 = 2U - C_{\mathrm{j}}~.
\end{equation}
Replacing with the expression for the potential $U$ in
Eq.~\eqref{eq:pseudo-pot}, and isolating the term of $C_{\mathrm{j}}$,
we find
\begin{equation}\label{eq:Jaco_const_4}
C_{\mathrm{j}} = n^2 (x_2^2 + y_2^2) + 2 \left( \frac{1-\mu}{\delta_0} +\frac{\mu}{\delta_1}\right) - \dot{x}_2^2 - \dot{y}_2^2 - \dot{z}_2^2  ~.
\end{equation}
We can re-express the Jacobi constant in terms of the set of inertial
variables through the transformation~\eqref{eq:syn_inert_trans}. Thus,
$C_{\mathrm{j}}$ as function of $(u_{x,2},u_{y,2},u_{z,2})$ reads
\begin{equation}\label{eq:Jaco_const_5}
C_{\mathrm{j}} = 2 \left( \frac{1-\mu}{\delta_0} +\frac{\mu}{\delta_1}
\right) + 2n\,(u_{x,2}\,\dot{u}_{y,2} - u_{y,2}\,\dot{u}_{x,2}) -
\dot{u}_{x,2}^2 - \dot{u}_{y,2}^2 - \dot{u}_{z,2}^2~.
\end{equation}
From the vectorial expression of the angular momentum 
\begin{displaymath}
\mathbf{L} =
\mathbf{r}\times \dot{\mathbf{r}} = (u_{x,2},u_{y,2},u_{z,2}) \times
(\dot{u}_{x,2},\dot{u}_{y,2},\dot{u}_{z,2})~,
\end{displaymath}
we find 
$\mathbf{L}_z=u_{x,2} \dot{u}_{y,2} - \dot{u}_{x,2} u_{y,2}$. Thus, from
Eq.~\eqref{eq:Jaco_const_5}, we obtain
\begin{equation}\label{eq:Jaco_const_6}
\frac{1}{2} \left( \dot{u}_{x,2}^2 + \dot{u}_{y,2}^2 + \dot{u}_{z,2}^2 \right)
- \left( \frac{1-\mu}{\delta_0} +\frac{\mu}{\delta_1}
\right) = \mathbf{L}\cdot \mathbf{n} - \frac{1}{2} C_{\mathrm{j}}~,
\end{equation}
with $\mathbf{n}= (0,0,n)$. The left-hand side of Eq.~\eqref{eq:Jaco_const_6}
corresponds to the total energy per unit mass of the massless
particle. Thus, since $\mathbf{L}\cdot
\mathbf{n}$ is not constant, the total energy is not conserved either.

\begin{figure}
  \centering
  \includegraphics[height=0.90\textheight]{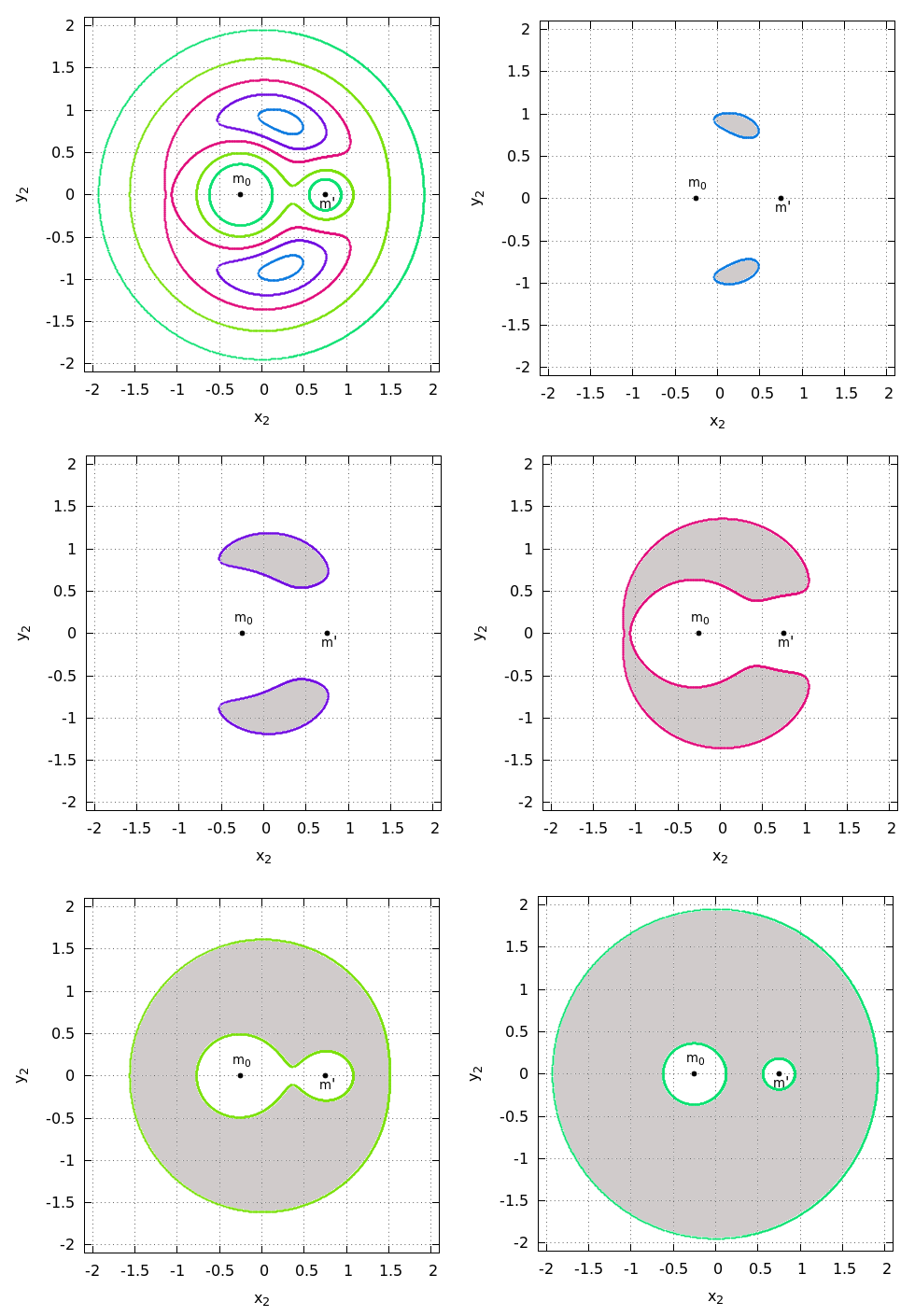}
  \caption[Examples of zero velocity curves of the CR3BP]{The
    \emph{zero velocity curves} of the CR3BP for a fixed value of the
    mass parameter $\mu=0.25$ and $5$ different Jacobi integral
    values: $C_{\mathrm{j}} = 2.85$ (blue, upper right),
    $C_{\mathrm{j}} = 3.00$ (purple, central left), $C_{\mathrm{j}} =
    3.25$ (pink, central right), $C_{\mathrm{j}} = 3.80$
    (yellow-green, lower left) and $C_{\mathrm{j}} = 4.80$ (green,
    lower right). Forbidden regions appear shadowed. The upper left plot
    shows all five curves together. The black dots denote the position of the
    primary masses $m_0$ at $(-\mu,0.0)$ and $m'$ at $(1-\mu,0.0)$.}
  \label{fig:plotCJs.png}
\end{figure}

Despite the lack of preservation of the angular
momentum and total energy that renders impossible to compute
an exact solution, the Jacobi integral still allows to
discriminate regions where the motion is \emph{allowed} from those
where it is not. From $C_{\mathrm{j}}$ we obtain
the so-called \emph{zero velocity surfaces}, i.e. a set of surfaces
that bound the motion in the system. If we consider only the
intersection of these surfaces with the plane $x,y$ ($z=0$), we
reduce the surfaces to zero velocity curves.  For the
definition of the latter, we just assume that $v = 0$ in the expression of
$C_{\mathrm{j}}$~\eqref{eq:Jaco_const_4}, thus
\begin{equation}\label{eq:zero_vel_curv}
2U = C_{\mathrm{j}} \quad \Longrightarrow \quad n^2 (x_{2}^2 + y_{2}^2) + 
2  \left( \frac{1-\mu}{\delta_1} +\frac{\mu}{\delta_2}
\right) = C_{\mathrm{j}}~,
\end{equation}
where $x_{2}$, $y_{2}$ are the coordinates of the test particle.
Since the velocity of the test particle must be real, the motion is
limited to those regions where $2U \geq C_{\mathrm{j}}$ (see
Eq.~\ref{eq:Jaco_const_3}), and the zero velocity curves trace the
analytic border of these regions. In Fig.~\ref{fig:plotCJs.png}, we
provide a few examples of the computation of these delimiting
borders. We consider a value of the mass parameter $\mu=0.25$ and $5$
different Jacobi integral values. From top to bottom, left to right,
the values for the Jacobi constant are $C_{\mathrm{j}} = 2.85$ (blue),
$3.00$ (purple), $3.25$ (pink), $3.80$ (yellow-green), $4.80$
(green). The shadowed areas correspond to the forbidden regions. The
plot in the upper left panel shows together all five curves. The
positions of the masses $m_0$ and $m'$ are represented by black dots,
according to Eq.~\eqref{eq:prim_position}.

Even if the Jacobi integral looks
not so restrictive, there are several conclusions that can be raised
from its analysis. For instance, in the last panel of
Fig~\ref{fig:plotCJs.png}, we present the zero velocity curve for the
particular case of $C_{\mathrm{j}}= 4.80$. If the test particle is
located in the region enclosing $m_0$, it is impossible for the
particle to escape from the system or to orbit around $m'$, since this
would require crossing the forbidden region. For a different (smaller)
value of $C_{\mathrm{j}}$, in the fifth panel, the permitted area
surrounds both primary masses, so the test particle may be transfered
from an orbit enclosing $m_0$ to one enclosing $m'$, but it is not
allowed to escape from their common domain. Also,
a test particle with initial conditions outside
the external zero velocity curve cannot orbit one of the two masses
individually. On the other hand, as the value of the Jacobi
integral decreases, the forbidden regions become smaller (first panel,
Fig.~\ref{fig:plotCJs.png}), and for values smaller than $C_j \sim 3 -
\mu$, they disappear.

\begin{SCfigure}
  \centering
  \caption[Zero velocity curves around the Lagrangian points $L_4$ and
    $L_5$]{Zero velocity curves around the Lagrangian points $L_4$ and
    $L_5$. The mass parameter considered is $\mu =0.35$ and the Jacobi
    constant for each curve is $C_{\mathrm{j}} = 2.8$ (blue), $2.90$
    (purple), $3.0$ (pink), $3.15$ (yellow-green), $3.34$ (green).
    Positions of the primary masses and lagrangian points $L_3$, $L_4$
    and $L_5$ are denoted with black dots.
  \vspace{1.5cm}}
  \includegraphics[width=.55\textwidth]{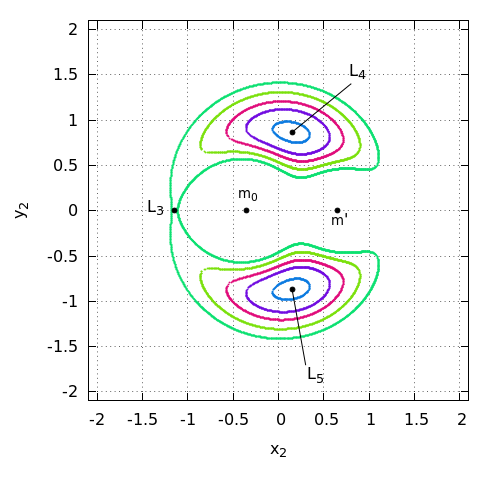}
  \label{fig:L4L5cjs.png}
\end{SCfigure}

The Lagrangian equilibrium points are those points in the synodic frame
where the centrifugal and
gravitational forces are balanced (see \S3.5 of \cite{MurrDerm-99}).
Their computation can be done
 directly from the equations of
motion~\eqref{eq:eq_mot_syn_a}, \eqref{eq:eq_mot_syn_b}
and~\eqref{eq:eq_mot_syn_c}. As in~\cite{Brouw-Clem-61}, we
 adapt the expression of the potential $U$, in order to
facilitate the computation of the derivatives involved.  Considering
the definitions of the distances $\delta_0$ and $\delta_1$
in~\eqref{eq:iner_eq_comp_r3bp-a} and~\eqref{eq:iner_eq_comp_r3bp-b},
we have
\begin{equation}\label{eq:una_eq_mas_1}
(1-\mu) \, \delta_0^{2} + \mu \, \delta_1^2 = x_{2}^2 + y_2^2 + (1-\mu)\mu~,
 \end{equation}
therefore the potential $U$ can be expressed as
\begin{equation}\label{eq:una_eq_mas_2}
U = (1-\mu) \left( \frac{1}{\delta_0} + \frac{\delta_0^2}{2} \right)
+ \mu \left( \frac{1}{\delta_1} + \frac{\delta_1^2}{2} \right) - \frac{1}{2}
(1-\mu)\mu~.
\end{equation}
This version of $U$ lacks an expression of the direct dependence
 on $x_{2}$ and $y_{2}$, substituted by $\delta_0$ and
$\delta_1$. Thus, the partial derivatives of $U$ are computed as
follows
\begin{equation}
\begin{aligned}\label{eq:partialU-1a}
\frac{\partial U}{\partial x_2} &= \frac{\partial U}{\partial \delta_0} \frac{\partial \delta_0}{\partial x_2} + \frac{\partial U}{\partial \delta_1} \frac{\partial \delta_1}{\partial x_2} = 0~,  \\
\frac{\partial U}{\partial y_2} &= \frac{\partial U}{\partial \delta_0} \frac{\partial \delta_0}{\partial y_2} + \frac{\partial U}{\partial \delta_1} \frac{\partial \delta_1}{\partial y_2} = 0~. 
\end{aligned}
\end{equation}
For the stationary condition, we consider that both the
acceleration and velocity are null ($\dot{x}_2 = \dot{y}_2 =
\ddot{x}_2 = \ddot{y}_2 = 0$). So, by Eqs.~\eqref{eq:eq_mot_syn_a},
\eqref{eq:eq_mot_syn_b}, \eqref{eq:eq_mot_syn_c}, we have that the
points for which Eqs.~\eqref{eq:partialU-1a}
are equal to zero are those corresponding to the equilibrium positions.
Replacing the partial derivatives with the corresponding 
formul\ae~derived from~\eqref{eq:una_eq_mas_2}, we find the equations for
the equilibrium positions
\begin{align}
(1-\mu) \left( -\frac{1}{\delta_0^2} + \delta_0 \right) \frac{x_2 +
    \mu}{\delta_0} + \mu \left( -\frac{1}{\delta_1^2} + \delta_1
  \right) \frac{x_2 - (1-\mu)}{\delta_1} & =
  0~, \label{eq:partialU-3}\\ (1-\mu) \left( -\frac{1}{\delta_0^2} +
  \delta_0 \right) \frac{y_2}{\delta_0} + \mu \left(
  -\frac{1}{\delta_1^2} + \delta_1 \right) \frac{y_2}{\delta_1} = 0~.
\label{eq:partialU-4}
\end{align}
If we set $y_2 = 0$, Eq.~\eqref{eq:partialU-4} is trivially satisfied.
Also, we get $\delta_0 = |x_2 + \mu|$, $\delta_1 = |x_2 -1 + \mu|$.
Substituting to Eq.~\eqref{eq:partialU-3} allows to fine the
x-coordinate of the \emph{collinear} Lagrangian
points $L_1$, $L_2$ and $L_3$ (see~\cite{Brouw-Clem-61}). These points
are unstable and lie all on the line connecting the star with the
planet. Of particular importance below is the point $L_3$, shown in
Fig.~\ref{fig:L4L5cjs.png}. On the other hand, inspecting
Eqs.~\eqref{eq:partialU-3} and~\eqref{eq:partialU-4}, we see that
$\delta_0 = \delta_1 = 1$ corresponds also to a solution.  Considering
Eqs.~\eqref{eq:iner_eq_comp_r3bp-a}
and~\eqref{eq:iner_eq_comp_r3bp-b}, we obtain
\begin{equation}
(x_2 + \mu)^2 + y_2^2 = 1~, \qquad (x_2 - (1-\mu))^2 + y_2^2 = 1~,
\end{equation}
that provides two different solutions
\begin{equation}\label{eq:equil_post_point}
x_2 = \frac{1}{2} - \mu, \qquad y_2= \pm \frac{\sqrt{3}}{2}~.
\end{equation}
Since $\delta_0 = \delta_1 = 1$, the positions of $L_4$ or $L_5$ 
and those
 of the two main masses $m_0$ and $m'$
lie at the vertices of
 two \emph{equilateral triangles}. The 
solution~\eqref{eq:equil_post_point}
defines the \emph{equilateral equilibrium points} $L_4$ and
$L_5$.  We note that these points represent
equilibria in the synodic rotating frame. Thus, a
particle at rest in $L_4$ or $L_5$ has a circular orbit with
angular frequency $n$ in the inertial system of reference.

An important point regarding the Lagrangian points is the
characterization of the motion in the vicinity of the equilibrium
position. We compute the associated value of the Jacobi
integral for particles at rest in such positions.
We find 
\begin{equation}\label{eq:Cl4-l5}
C_{\mathrm{j},L_4} = C_{\mathrm{j},L_5} = 3 -\mu + \mu^2~.
\end{equation}
Figure~\ref{fig:L4L5cjs.png} shows the zero velocities curves
corresponding to $C_{\mathrm{j}} = 2.80,\, 2.90,\, 3.00,\, 3.15,\,
3.34,$ for $\mu=0.35$. The first value is very close to that
corresponding to a particle at rest in $L_4/L_5$, while the last value
is close to that of $L_3$, $C_{\mathrm{j},L_3}\simeq 3.3362$. If
$C_{\mathrm{j},L_4} < C_j < C_{\mathrm{j},L_3}$, the motion in the
vicinity of $L_4/L_5$ is energetically allowed to take place
surrounding the Lagrangian points $L_4$ or $L_5$ only (although it is
not energetically restricted to do so). On the other hand, if $C_j >
C_{\mathrm{j},L_3}$, the orbits necessarily surround all three points
$L_3,\,L_4,\,L_5$. This distinction raises two different kinds of
motion in the neighbourhood of the equilateral points, known as
\emph{tadpole} and \emph{horseshoe} orbits. Nevertheless, this study
of the motion is inconclusive as far as a stability study has not been done.


\subsection{Linear stability around the $L_4$ and $L_5$}\label{sec:1.4.X.stab_equil_pt}

Let us consider the position of the massless
body as a small displacement with respect to the position of $L_4$ (or
$L_5$)
\begin{equation}\label{eq:little_disp}
x_2 = x_{2,0} + X_2 \qquad y_2 = y_{2,0} + Y_2~,
\end{equation}
where $X_2$ and $Y_2$ are the two small displacements and
$x_{2,0}$ and $y_{2,0}$ are given in Eq.~\eqref{eq:equil_post_point}.
We now replace Eq.~\eqref{eq:little_disp} in the equations of motion
for the test particle~\eqref{eq:eq_mot_syn_a}
and~\eqref{eq:eq_mot_syn_b}\footnote{Eq.~\eqref{eq:eq_mot_syn_c} is
  not considered because we focus on planar motions (on the primaries'
  orbital plane). The solution for the vertical component of the
  spatial case is simply an oscillator~\cite{MurrDerm-99}.}. The
corresponding linearized equations of motion (see
Sect.~\ref{sec:1.4.X.non_stab}) are
\begin{equation}\label{eq:lin_dif_eq}
\ddot{X}_2 - 2 \dot{Y}_2 = X_2 U_{xx} + Y_2 U_{xy}, \qquad
\ddot{Y}_2 + 2 \dot{X}_2 = X_2 U_{xy} + Y_2 U_{yy},
\end{equation}
where
\begin{equation}\label{eq:deriv_Us}
U_{xx} = \left(\frac{\partial^2 U}{\partial x_2^2} \right)_{0},\:\:
U_{xy} = \left(\frac{\partial^2 U}{\partial x_2 \partial y_2} \right)_0, \:\:
U_{yy} = \left(\frac{\partial^2 U}{\partial y_2^2} \right)_0~, 
\end{equation}
represent the partial derivatives of the potential evaluated at
($x_{2,0}$,$y_{2,0}$).  The linearized system
can be represented in a matrix form as in Eq.~\eqref{eq:linear_matx_form},
by considering
\begin{equation}\label{eq:vector_x_matrix_a}
\mathbf{X} =
\begin{pmatrix}
\, X_2 \,\\
\, Y_2 \,\\
\,\dot{X}_2 \,\\
\,\dot{Y}_2 \,\\
\end{pmatrix}\: \:
\mathrm{and}\: \:
\mathbf{A} = 
\begin{pmatrix}
\, 0 & 0 & 1 & 0 \,\\
\, 0 & 0 & 0 & 1 \,\\
\, U_{xx} & U_{xy} & 0 & 2 \,\\
\, U_{xy} & U_{yy} & -2 & 0 \,\\
\end{pmatrix}~,
\end{equation}
where $\mathbf{A}$ is a square matrix of constant coefficients.

For the particular case of the stability around the equilateral
Lagrangian points $L_4/L_5$, we have
\begin{equation}\label{eq:deriv_Us_L4L5}
U_{xx} = \frac{3}{4},\:\:
U_{xy} = \pm 3 \frac{\sqrt{3}(1 -2 \mu)}{4}~, \:\:
U_{yy} = \frac{9}{4} ~, 
\end{equation}
Hence, the characteristic equation~\eqref{eq:charact_eq} in this case reads
\begin{equation}\label{eq:caract_eq_L4L5}
\lambda^4 + \lambda^2 + \frac{27}{4} \mu (1-\mu) = 0~,
\end{equation}
whose $4$ solutions are given by
\begin{equation}\label{eq:eig_val_l1l2}
\lambda_{1,2} = \pm \frac{\sqrt{-1-\sqrt{1 - 27(1 -\mu) \mu}}}{\sqrt{2}}~,
\end{equation}
\begin{equation}\label{eq:eig_val_l3l4}
\lambda_{3,4} = \pm \frac{\sqrt{-1+\sqrt{1 - 27(1 -\mu) \mu}}}{\sqrt{2}}~.
\end{equation}
We can see that the eigenvalues are strictly imaginary if the condition
\begin{equation}\label{eq:routh_1}
1-27(1 -\mu) \mu \geq 0
\end{equation}
holds. This implies that the linear stability is guaranteed only for
\begin{equation}\label{eq:routh_2}
\mu \leq \mu_{R} = \frac{27 - \sqrt{621}}{54} \approx 0.0385~.
\end{equation}
Equation~\eqref{eq:routh_2} is known as the Routh criterion \cite{Gascheau-1843}.  If we
limit ourselves to the cases within the Solar System, all the systems
well represented by the R3BP as, for example, the Sun-Jupiter-asteroid or
the Earth-Moon-spacecraft systems, accomplish
condition~\eqref{eq:routh_2}\footnote{The only relevant exception is
  the system Pluto-Charon-asteroid. This is a case of a binary rather than
a hierarchical system.}.  Thus, the stable motions around the
equilateral points provide an interesting and common feature of the
dynamics in the Solar System and most probably in other planetary
systems as well. Nevertheless, these results are applicable in very
small domains, where the linearization of the system is a valid 
approximation. More general results can be provided only by the use of 
non-linear stability theorems (Sect.~\ref{sec:1.4.X.non_stab})
or by numerical analysis of the orbits.

\subsection{Numerical examples}\label{sec:1.4.X.motionL4L5}

\begin{SCfigure}
  \centering
  \caption[Epicyclic motion and guiding center]{Decomposition of the
    librational motion of the test particle around the equilateral
    Lagrangian point $L_4$. The motion is decomposed into a guiding
    center motion (large blue ellipse surrounding $L_4$) and an epicyclic
    motion (small purple ellipse). We refer to the position of $L_4$ with
    the pink star, the position of the epicenter with the dark dot and the
    position of the particle with the green circle.}
  \includegraphics[width=.50\textwidth]{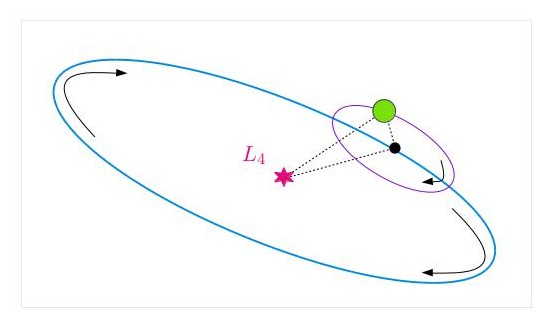}
  \label{fig:epicenter.jpg}
\end{SCfigure}

From Eq.~\eqref{eq:eig_val_l1l2} and 
Eq.~\eqref{eq:eig_val_l3l4}, and assuming that $\mu$ is small, we can
obtain
\begin{equation}\label{eq:lambdas_approx}
\lambda_{1,2} = \pm \,\mathrm{i}\, (1-\frac{27}{8} \mu )  + {\cal O}(\mu^2) =  
\pm\, \mathrm{i}\, \omega_{f}
\qquad
\lambda_{3,4} = \pm \,\mathrm{i}\, \sqrt{\frac{27}{4} \mu} + {\cal O}(\mu)
= \pm\, \mathrm{i}\, \omega_{s} ~,
\end{equation}
where the subscripts $f$ and $s$ stand for 'fast' and 'synodic'
respectively.  Equation~\eqref{eq:lambdas_approx} corresponds to the
lowest order (in $\mu$) approximation of the eigenvalues.  If we
consider that the solution is a linear combination of the modes of
$\mathrm{e}^{\pm i\omega_{f}}$ and $\mathrm{e}^{\pm i\omega_{s}}$,
then the motion results
\begin{equation}\label{eq:solut_harm_X}
X_{2}(t) = \mathsf{a}_1 \cos \omega_{f} t + \mathsf{a}_2 \cos
\omega_{s} t + \mathsf{a}_3 \sin \omega_{f} t+ \mathsf{a}_4
\sin \omega_{s} t~,
\end{equation}
and
\begin{equation}\label{eq:solut_harm_Y}
Y_{2}(t) = \mathsf{b}_1 \cos \omega_{f} t + \mathsf{b}_2 \cos
\omega_{s} t + \mathsf{b}_3 \sin \omega_{f} t+ \mathsf{b}_4
\sin \omega_{s} t~,
\end{equation}
where the coefficients $\mathsf{a}_i$ and $\mathsf{b}_i$ come from the
expression of the corresponding matrix $\mathbf{B}$ in
Eq.~\eqref{eq:sol_X}. Equations~\eqref{eq:solut_harm_X}
and~\eqref{eq:solut_harm_Y} clearly show the oscillatory nature of the
motion, when a small displacement $X_2$, $Y_2$ is considered with
respect to the position of the equilibrium point. The frequencies of
the oscillations are given by $\omega_{f}$ and $\omega_{s}$. The two
oscillations have very different
timescales~\eqref{eq:lambdas_approx}. On one hand, $\omega_{f}$ is
approximately equal to $1$, i.e.  it gives an oscillation of period
similar to that of the motion of the primary around the star.  On the
other hand, $\omega_{s}$ is proportional to the square root of the
small parameter $\mu$, which is a small parameter itself.  Thus, the
motion of the test particle can be decomposed in two different
contributions: the slow motion, associated to the motion of an
\emph{guiding center} around the position of equilibrium, with long
period $2\pi/\omega_{s}$ and known as the \emph{synodic libration},
and the fast one, attributed to the short period motion of the
particle around the guiding center.  The two motions correspond to two
different ellipses, whose dimensions
are associated to the amplitudes $\mathsf{a}_i$ and
$\mathsf{b}_i$. Figure~\ref{fig:epicenter.jpg} represents
schematically the decomposition given by Eqs.~\eqref{eq:solut_harm_X}
and~\eqref{eq:solut_harm_Y}. The small ellipse in the figure
corresponds to the fast motion of the particle, while the big ellipse
gives the libration of the epicenter around the Lagrangian point.  The
elongated ellipse describing the motion of the guiding center is
roughly determined by the size of a corresponding zero velocity
curve. The rate of the semiaxes if approximately $b/a \sim \sqrt{3
  \mu}$ (\S 3.10 of~\cite{MurrDerm-99}). For the ellipse that
describes the motion of the particle around the guiding center, it can
be proved that $b/a \sim 0.5$ (\S 3.8 of~\cite{MurrDerm-99}).

\begin{figure}[h]
  \centering
  \includegraphics[width=.95\textwidth]{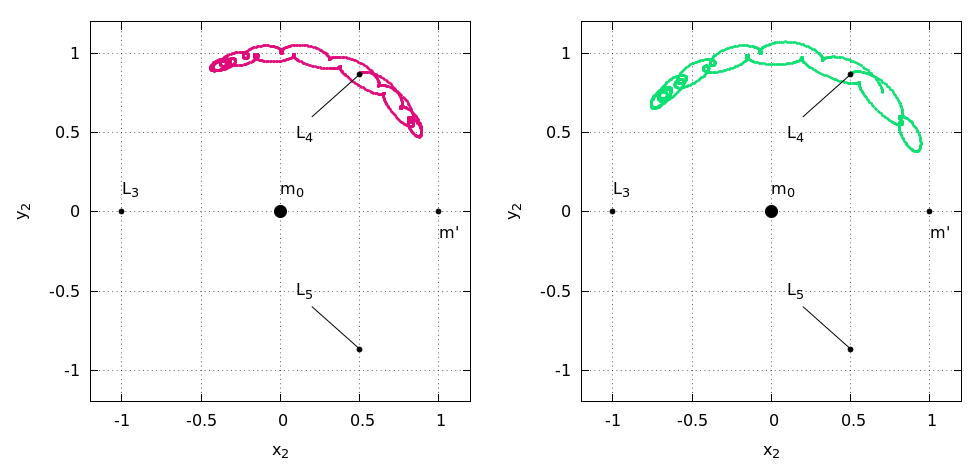}
  \caption[Examples of tadpole orbits]{Two different examples of
    tadpole orbits for $\mu = 0.001$.  The positions of the two
    primary masses and of the Lagrangian points are denoted with black
    points. The orbit on the left (pink points) corresponds to the
    initial condition $(x_2,y_2,\dot{x}_2,\dot{y}_2) =
    (0.5055,0.87252,0,0)$. The orbit on the right (green points)
    corresponds to the initial condition
    $(x_2,y_2,\dot{x}_2,\dot{y}_2) = (0.507,0.87402,0,0)$. }
  \label{fig:tadpoles.png}
\end{figure}

Figure~\ref{fig:tadpoles.png} shows two examples of numerical
integrations for these orbits.  We integrate the equations of motion
for the test particle in the synodic system of
reference~\eqref{eq:eq_mot_syn_frame}, for a time equal to several
revolutions of the primaries, for two different orbits. The initial
conditions considered for the first orbit (pink) are
$(x_2,y_2,\dot{x}_2,\dot{y}_2) = (0.5055,0.87252,0,0)$.  The initial
conditions for the second orbit (green),
$(x_2,y_2,\dot{x}_2,\dot{y}_2) = (0.507,0.87402,0,0)$, are chosen so
that the second orbit presents a larger displacement with respect to
the equilibrium point with respect to the first one. In both cases,
the chosen mass parameter corresponds to $\mu=0.001$.  In the figure,
the decomposition of the motion in the synodic and epicyclic
oscillation is clear: these yield the elongated tadpole-shaped
deformed ellipse enclosing $L_4$ and the smaller ellipse around the
guiding center. For larger displacements (as for the green orbit), the
synodic libration of the guiding center turns to be larger as
well. In this case, the angular excursion of the test particle with
respect to the position of $m'$ is more than
$90^{\circ}$. Additionally, the curves show that the angular extension
of the orbits is not symmetric, extending (in the case of the green
orbit) up to twice the distance in the direction of $L_3$ than in the
direction towards the primary. This characteristic pattern of the
curves around the equilateral points gives the name of \emph{tadpole}
to this kind of orbits.

It is possible to check, by means of numerical integrations,
what happens for orbits more and more displaced with respect to the
Lagrangian points, and to compare with the tadpole cases considered
before.
\begin{figure}[h]
  \centering
  \includegraphics[width=.85\textwidth]{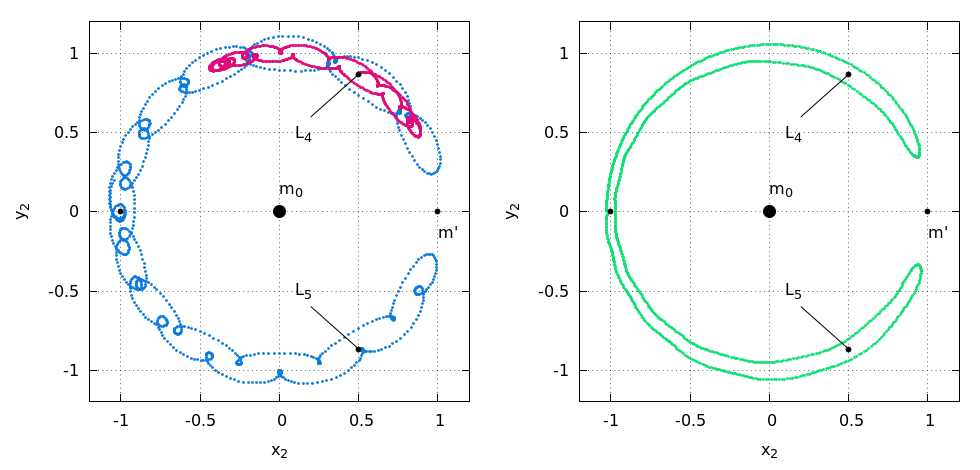}
  \caption[Examples of horseshoe orbits]{Two different examples of
    horseshoe orbits for $\mu=0.001$.  The positions of the two
    primary masses and the Lagrangian points are denoted with black
    points. The blue orbit (left panel) corresponds to the initial
    conditions $(x_2,y_2,\dot{x}_2,\dot{y}_2) =
    (-0.97668,0,0,-0.06118)$. For comparison with a tadpole orbit, an
    orbit of Figure~\ref{fig:tadpoles.png} is shown in pink. The green
    orbit (right panel) corresponds to the initial conditions
    $(x_2,y_2,\dot{x}_2,\dot{y}_2) = (-1.02745,0,0,0.04032)$.}
  \label{fig:horseshoes.png}
\end{figure}
Figure~\ref{fig:horseshoes.png} represents the orbits of two
additional numerical integrations performed for the same system as in
Fig.~\ref{fig:tadpoles.png}, with $\mu=0.001$. The initial conditions
used for these plots are highly displaced with respect to $L_4$: we
consider $(x_2,y_2,\dot{x}_2,\dot{y}_2)=(-0.97668,0,0,-0.06118)$ for
the blue orbit and $(x_2,y_2,\dot{x}_2,\dot{y}_2) =
(-1.02745,0,0,0.04032)$ for the green orbit. As expected, due to the
shape of the zero velocity curves (Fig.~\ref{fig:plotCJs.png}), the
orbits now enclose both lagrangian points $L_4$ and $L_5$, describing
a \emph{horseshoe} shape. The second orbit (green, right panel
Fig.~\ref{fig:horseshoes.png}) is chosen in order to reduce the fast
oscillations around the epicenter, rendering possible to visualize the
libration: the orbit resembles an extremely elongated deformed
ellipse. For comparison purposes, we expose in the figure the curve
corresponding to the first (pink) orbit of
Fig.~\ref{fig:tadpoles.png}. We recall here that for a fixed value of
the Jacobi integral, an orbit cannot get closer to the Lagrangian
points than the corresponding zero-velocity curve.  Thus, although not
entirely limiting the motion, the zero-velocity curve can rule the
inner boundary and thus, effectively, the shape of the orbit.

\section{The Trojan problem}\label{sec:1.4.works_on_Trojans}

The description given so far is only accurate if the heliocentric
orbit of the primary (e.g. planet) is circular (since the solutions
come from such an approximation), and the motion of the test particle
is planar and takes place in the vicinity of the equilateral
Lagrangian point. The study of the motion under a non-circular approximation
needs a more elaborate analysis and it is deferred to
later in this work.  Let us note here that in real systems of
astronomical interest, the motion of the planet can itself be quite
complicated, since in general it can be influenced by other planets.
 We can thus consider a hierarchy of problems, passing from the
circular to the elliptic R3BP, the secularly evolving R3BP with one or
more additional disturbing bodies, etc. As we will see, in all these
problems, for small perturbations, we can still have tadpole-like
motions of the massless body, whose analytical description becomes,
however, more and more cumbersome. In the sequel we will refer to the
general problem of the study of such tadpole motions as the {\it
  Trojan problem}.

The study of the Trojan problem has a long history in the
literature. Various authors have emphasized different aspects of the
problem, and have introduced a variety of sets of variables and/or
techniques in order to facilitate its analytical or numerical
study. We now briefly refer to some (non-exhaustive) literature on the
subject.

Focusing on the case of the CR3BP, early works (\cite{WoltjerJr-24},
\cite{DepDelie-65}, \cite{DepHenrRom-65}, \cite{Garfinkel-77},
\cite{Garfinkel-78}, etc.)  emphasized two main aspects of the
problem, namely i) the form and nature of the {\it periodic orbits},
and ii) the development of approximative \emph{series solutions} representing
the orbits in the tadpole domain.

In the CR3BP, it can be shown that a set of periodic Lyapunov orbits (in the
synodic frame) bifurcates from the equilibrium points L4 or L5
(\cite{Rabe-61}, \cite{Rabe-62}, \cite{Rabe-67}, \cite{DepHenr-69},
\cite{Garfinkel-77}, \cite{Garfinkel-83}).  For $C_J<C_{J,L4}$ these
are \emph{short period} orbits, i.e. orbits forming a small circle around
the equilibrium points. This circle represents the epicyclic motion
along a Keplerian ellipse as viewed in the synodic frame. The motion
takes place with a period $\approx 2\pi/\omega_f$. On the other hand,
for $C_{J,L4}<C_J<C_{L3}$ we have the family of {\it long period
  orbits}, i.e. elongated orbits surrounding again L4 or L5 but with
period $\approx 2\pi/\omega_s$, i.e. much longer than the short
one. In the limit of the linearized solutions of 
Eqs.~\eqref{eq:solut_harm_X} and~\eqref{eq:solut_harm_Y}, these periodic
orbits correspond to the cases where the terms associated to one of
the two frequencies vanish. Physically, this means that the short
period family consists of orbits with a null guiding center
oscillation, while the long period family consists of orbits with a
null epicyclic oscillation. When nonlinear terms are also taken into
account in the orbital equations of motion, each of the two families
can be constructed by a formal elimination, in analytical series
solutions, of the latters' dependence on either the short or the long
(synodic) frequencies~\cite{DepHenrRom-65}.  Then, one finds that the
periods turn to have a dependence on the Jacobi constant $Cj$ and on the
mass parameter $\mu$~\cite{Garfinkel-80}.  

Furthermore, there exist
more complicated periodic orbits which form 'bridges' connecting the
long with the short period family (\cite{Rabe-67b}, \cite{Rabe-68},
\cite{DepRabe-69}). These more complicated periodic orbits correspond
to a commensurability between the fast and the synodic
frequency. Namely, if $\omega_{f}\approx n \omega_{s}$, with $n$ integer,
there exist periodic orbits forming $n$ epicyclic oscillations
(e.g. loops, like in Fig.~\ref{fig:tadpoles.png}) while they
accomplish one full synodic libration. Varying $C_J$ and $\mu$, such
orbits can be traced down to their bifurcation either from a short
period or a long period orbit. Also, the phase space structure is
influenced by such orbits, which give rise to domains of so-called
'secondary resonances' (\cite{Garfinkel-77},~\cite{Rabe-67b}). The
study of these resonances is a principal part of the present thesis
(see Chapters 3-5).

Along a different line of approach, several authors presented series
expansions for the Trojan orbits in general, i.e., not restricted to
the periodic cases. Among the first of such attempts,
in~\cite{DepDelie-65} quasiperiodic orbital solutions in terms of the
two main frequencies $\omega_f$ and $\omega_s$ were calculated via
trigonometric expansions called `d'Alembert series'. The presence of
secondary resonances affects also these series, as it gives rise to
so-called critical terms, i.e., terms depending on a resonant
combination of the angles. In normal form constructions, such terms
cannot be \emph{averaged} (i.e. eliminated by the usual canonical
transformations as in subsection~\ref{sec:1.1.2-pert_theory}). Thus,
their presence prevents the splitting of the normal form Hamiltonian
in a `short period' and a `secular' (i.e. long period) part
and requires a special treatment
(e.g.~\cite{Garfinkel-77}, see also Chapter 5 below). On the other
hand, in the Trojan problem it can be shown that their effect on the
slow (secular) motions is rather limited~\cite{Namouni-99}.

Let us mention, finally, that the use of averaging techniques allows
to simplify the study of the synodic librations by finding a
simplified form of the equation of motion for the so-called critical
argument $\tau=\lambda-\lambda'$, with $\lambda$, $\lambda'$ the mean
longitude of the test particle and the planet, respectively.  The
libration of $\tau$ around the $L_4$ ($L_5$) value $\pi/3$ ($5\pi/3$)
can be represented by a Newton-equation using a so-called
`ponderomotive potential' $V(\tau)$~\cite{NamMurr-00}.  Studying the
properties of this function, it is possible, for example, to express
the center of the libration as a function of the eccentricity,
inclination and longitude of the perihelion of the Trojan body. It is
found that the position of the center in some cases may be shifted
considerably from the position of the Lagrangian points $L_4$ and
$L_5$~\cite{NamMurr-00}. Some new analytical expressions for this
shift are given in Chapter 3.

Whereas the CR3BP may be a good first model for developing the theory
of Trojan orbits, it clearly does not suffice to represent more
realistic problems. As a natural extension, there exist several
approximations to the analytical solution of the Trojan problem in the
framework of the Elliptic Restricted 3-Body problem (ER3BP). This
generalization brings new interesting features to the formulation.
While most works on the CR3BP consider two time scales (associated to
$\omega_f$ and $\omega_s$), in the ER3BP \emph{three} times scales are
necessary~\cite{Erdi-77}, associated to the fast, synodic and secular 
frequency.
From the physical point of view, these three scales are associated to
the epicyclic oscillation (fast, ${\cal O}(1)$), the
libration around the libration center (synodic, ${\cal O}(\sqrt{\mu})$) 
and the slow precession of the perihelion of the orbit of the
Trojan body (secular, ${\cal O}(\mu)$)~\cite{Erdi-78}.  

Starting from
an analytic solution up to second order in the mass parameter $\mu$
in~\cite{Erdi-77}, for the planar case, and in~\cite{Erdi-78}, for the
spatial case, it is possible to derive explicit formul\ae~for the
variation of the libration angle $\tau$ and of the major semiaxis
$a$ of the Trojan body as series in a paramater $\ell$,
called the \emph{proper libration}~\cite{Erdi-81}. This corresponds
to the amplitude of the synodic librations which represents an 
approximate constant of motion. We find: 
\begin{align}
\tau &= \left( \frac{\pi}{3} + \frac{\sqrt{3}}{4} \ell^2 \right) +
\left( \ell + \frac{5}{64} \ell^3 \right) \cos( \omega_s
t)+\ldots~, \label{eq:variation-u} \\ 
a &= a ' + a' \sqrt{\mu}\,\sqrt{3}\, \ell
\left(1 - \frac{3}{64} \ell^2 \right) \, \sin( \omega_s
t) + \ldots~, \label{eq:variation-a}
\end{align}
Regarding the secular motions, in~\cite{Erdi-79} approximate formul\ae~were
given for the time evolution of 
the components of the so-called \emph{eccentricity vector} 
$(e \cos \varpi, e \sin \varpi)$,
\begin{equation}
\begin{aligned}\label{eq:var_ecc_vector_sin}
e \sin \varpi = \alpha \,-\, c \,\sin \phi~,\\
e \cos \varpi = \beta \,-\, c \,\cos \phi~,
\end{aligned}
\end{equation}
with
\begin{align*}
\omega_s &= \sqrt{\frac{27}{4} \left(1 - \frac{3}{8} \ell^2 - \frac{97}{512} \ell^4\right)}~, &\phi &= \left(\frac{27}{8}+ \frac{129}{64} \ell^2 \right)\,\mu\, t~, \\ 
\alpha &= e' \left(\frac{\sqrt{3}}{2} - \frac{73 \sqrt{3}}{144} \ell^2\right)~, &\beta &= e' \left(\frac{1}{2} + \frac{17}{48} \ell^2 \right)~.
\end{align*}
Again $\ell$ and $c$ are positive constants of integration, $a'$ is
the major semi-axis of the primary $m'$, $e'$ its eccentricity and
most of the expressions are up to order ${\cal
  O}(\ell^2)$. Equations~\eqref{eq:variation-u}
and~\eqref{eq:variation-a} represent the shift in the position of the
libration center with respect to $\ell$~\cite{Erdi-81}. This result
yields a further correction, depending on $\ell$, for the shift of the
center with the orbital elements $e$, $i$ and $\varpi$ found
in~\cite{NamMurr-00}.

On the other hand, according to Eqs.~\eqref{eq:var_ecc_vector_sin},
 in the elliptic problem ($e'\neq 0$) the
center of the eccentricity vector appears displaced from the origin,
due to the (constant) terms depending on $e'$ and $\ell$ in the
expressions of $\alpha$ and $\beta$. A generalization of
Eqs.~\eqref{eq:var_ecc_vector_sin}
for the spatial ER3BP is presented in~\cite{Erdi-88}. The two slow
time scales (${\cal O}(\sqrt{\mu})$ and ${\cal O}(\mu)$) affecting
different orbital elements appear clearly depicted in the equations
above.

\begin{figure}
  \includegraphics[width=.98\textwidth]{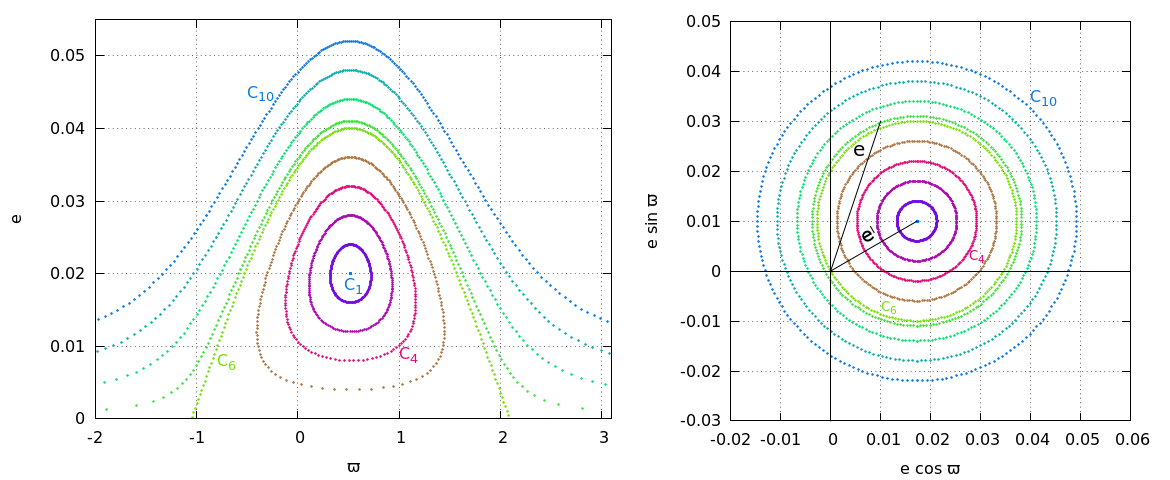}
  \caption[Paradoxal libration and the eccentricity vector
    evolution]{Two different regimes of the evolution of the
    eccentricity and longitude of perihelion for a set of ten orbits
    with different values of $c$ ($\ell=0)$. In the left panel, the
    two quantities are represented independently, while in the right
    panel their evolution is in terms of the eccentricity vector
    $(e\cos\varpi,e\sin\varpi)$. The displacement of the center
    represents the 'forced equilibrium' induced by the eccentric orbit
    of $m_1$. The two apparently different regions in the left panel
    are due to an apparent division of the orbits for which $c > e'$
    (circulation) from those for which $c < e'$ (paradoxal
    libration), while clearly both sets correspond to a unique regime
    in the right panel.}
  \label{fig:paradoxeccvec.png}
\end{figure}
In Fig.~\ref{fig:paradoxeccvec.png} we present some examples of
numerical computations of Eqs.~\eqref{eq:var_ecc_vector_sin}. 
We fix the value of the libration
amplitude $\ell = 0$ and the primary's eccentricity $e'=0.02$. Ten
different values of $c$ are used for the generation of orbits. The
minimum value $c_1 = 0$ corresponds to the innermost and the maximum
$c_{10} = 0.035$ to the outermost orbit. In the left panel, the
initial conditions are apparently divided in two regions according to
whether the angle $\varpi$ oscillates between two extremes or takes all
possible values in the interval $\left[0,2\pi\right]$. The limit curve
labeled $c_6$ corresponds to the orbit with $c = e' = 0.02$. That
curve is not associated with any real
structure of separatrix. The motions for $c < c_6$ are called
\emph{paradoxal} in~\cite{Beugetal-07}. They imply a regime of libration of
$\varpi$, which is a consequence of the non-zero eccentricity of the
primary's orbit, since, for $e'=0$, $\varpi$ could only circulate.
However, this libration regime does not stem from a separatrix 
structure, as made clear from the right panel of 
Fig.~\ref{fig:paradoxeccvec.png}, which shows that the paradoxal
and non paradoxal motions are topologically equivalent in the 
eccentricity vector plane.


While the previous solutions apply to the cases where the motion of
the primary is a given fixed ellipse, it is known that the influence
of additional bodies perturbs the motion of $m_1$. The so-called
\emph{Laplace-Lagrange} theory~(\cite{Bretagn-74}, for an extension up
to second order in the masses, see~\cite{LibSans-13}) allows the
remove the short-term behaviors in the motion of the planets under the
assumption of no mean motion resonances.  In this case, the
eccentricity vector of the primary ($e'\cos\varpi',e'\sin\varpi'$) can
be expressed as a secularly varying quantity, depending periodically
on the secular frequencies associated with the other bodies. By
replacing this behavior in the solutions of
Eqs.~\eqref{eq:var_ecc_vector_sin},
we can recover the effect of a \emph{secularly precessing primary} on
the orbital elements of the Trojan~\cite{Erdi-96}.  Since, in general,
the time-scale of the secular precession of the primary is long
compared with the synodic timescale of the libration of the Trojan,
this effect is not evident for just one libration
period. Nevertheless, for the long-term behavior, the precessing
primary induces widenings of the structures depicted in
Fig.~\ref{fig:paradoxeccvec.png}, by shifting the position of the
center of libration on a very long periodic timescale. An orbit which
for a big value of $e'$ is paradoxal (librating), may become
non-paradoxal (circulating) for a different (smaller) $e'$. Thus, the
variations of $e'$ may induce a change of regime in the long-term
evolution of the orbit, passing from perihelion circulation to
libration or vice versa~\cite{Erdi-96},~\cite{SanErd-03} (but see also
Chapter 4). In addition, the introduction of additional frequencies in
the problem may induce an increment of chaos and affect the borders of
the domain of stability~\cite{SanErd-03}. Also, if the Trojan particle
is not limited to the plane of $m_0$ and $m'$, additional effects are
produced by the variations in the Trojan body's
inclination~\cite{Erdi-96}.

Finally, under the assumption that the additional bodies are not in
mean motion resonance with the Trojan (nor the primary), it is
possible to average the equations of motion with respect to the fast
angles and obtain a general secular theory approximation for the
Trojan motion, where only the long-term behavior is
highlighted~\cite{Morais-99},~\cite{Morais-01}.

The above approaches emphasize the study of individual orbits under
various configurations of the disturbing bodies (planets, etc.). However,
a different line of research deals with the characterization of 
the \emph{long-term stability} of the orbits.

The stability of the motion around the center of libration in the
CR3BP was studies in several contexts. Besides linear
stability~\cite{Gascheau-1843}, which holds approximately only in an
extremely small domain in which the linear approximation is valid, the
theorems of non-linear stability (see Sect.~\ref{sec:1.4.X.non_stab})
have been also used. Regarding, in particular, the periodic orbits
associated to $L_4$ and $L_5$, invariant tori delimiting the
motion around them were constructed in the planar~\cite{Deprix2-67}
and in the spatial~\cite{Markeev-72} approximation. In addition, for a
small set of the Trojan asteroids of the Sun-Jupiter system, it has
become possible to approximate their orbits by KAM tori, by use of the
Kolmogorov normalization~\cite{GabJorLoc-05}.

In a different approach, it can be shown theoretically that, in the
framework of the CR3BP, the necessary conditions for applying
Nekhoroshev's theorem in the vicinity of the Lagrangian points hold
true for every value of $\mu<\mu_{R}$, except three isolated
values~\cite{BenFasGuz-98}.  As introduced in
Sec.~\ref{sec:1.4.X.non_stab}, the Nekhoroshev theorem does not
provide a division of the phase-space in systems of 2 d.o.f, as the
KAM theorem does. However, by fixing the Nekhoroshev time $T$ in
Eq.~\eqref{eq:Nekh_est} to an appropriately large time, it is possible
to determine the size of a domain around the elliptic fixed point
within which the orbits remain practically stable up to the time
$T$. This concept is known as \emph{effective
  stability}~\cite{Gioretal-89}, and it allows to prove that there
exist (small) regions that are Nekhoroshev-stable around the
equilateral points during the expected lifetime of the Solar System in
the case of Jupiter's Trojan asteroids, for the CR3BP
(\cite{Gioretal-89},
\cite{CellGior-91},~\cite{GiorSko-97},~\cite{SkoDok-01},
\cite{Efthy-05},~\cite{EfthySan-05}) and in the elliptic approximation
(\cite{LhotEfthyDvo-08}). Further generalizations in a Hamiltonian
formalism for an elliptic problem different to that of Jupiter's
asteroids are presented in~\cite{Efthy-13}.

So far, we summarized some important results provided by analytical
developments of the theory of Trojan motion. Although these provide
some explicit approaches to the
solutions, they are also quite limited by the simplifications of
the models considered. More complex and realistic approaches can be
studied by means of numerical integrations. Several of these studies
are devoted in particular to the understading of the dynamics of the
Trojan asteroids in the Solar System, where basic models such 
as the CR3BP or the ER3BP do not
reproduce accurately enough the true dynamics~\cite{GabJobRob-04}.

The first Trojan body ever observed was the asteroid Achilles,
coorbital to Jupiter. Nowadays, in the libration regions of Jupiter
there are more than 6000 bodies observed and classified. When modern
computers allowed to produce large simulations, it was possible to
test the long-term stability numerically, including also the direct
effect of additional bodies. It was found that the orbits in the
Trojan swarm are not indefinitely stable. As a consequence, the swarm
suffers a slow process of 'erosion' in the borders of the Trojan
domain, due to the gravitational effect of the giant
planets~\cite{Levetal-97}. About 5\%-20\% of the observed Trojans are
not stable in a scale of 4 Gyrs~\cite{Tsig-05}.  The main responsable
factor for the chaos induced in the Trojan swarm are
resonances. In~\cite{Robetal-05} and~\cite{RobGab-06}, the families of
the most prominent resonances that take place within the Trojan
libration domain are identified. The secular interaction with Saturn
(which is slightly out of the 5:2 MMR with Jupiter) turns to produce
remarkable effects. In particular, the family of the resonances
involving the secular frequency of the Trojan and the difference $5
n_J - 2 n_S$ (where $n_J$ and $n_S$ are the mean motions of Jupiter
and Saturn respectively) seems to be the main cause of the 'erosion'
of the Trojan swarm detected in~\cite{Levetal-97}.

One of the benefits from the study of this set of asteroids is that
the large number of identified objects makes statistical studies
feasible. The theoretical analysis of the Trojan motion evidences the
existence of approximate integrals of motion in the secular solutions,
called \emph{proper elements}. They are: the amplitude of libration
$D_p$ (directly related to the constant $\ell$ discussed above), the
proper eccentricy $e_{p,0} = c$ and the proper inclination $i_{p,0}$.
In more general models (or longer numerical integrations) these
quantities are not exactly constant, but secularly changing, thus they
remain \emph{quasi-invariant} for long periods of
time~\cite{BienSchu-87},\cite{SchuBien-87}.

\begin{figure}
  \includegraphics[width=.98\textwidth]{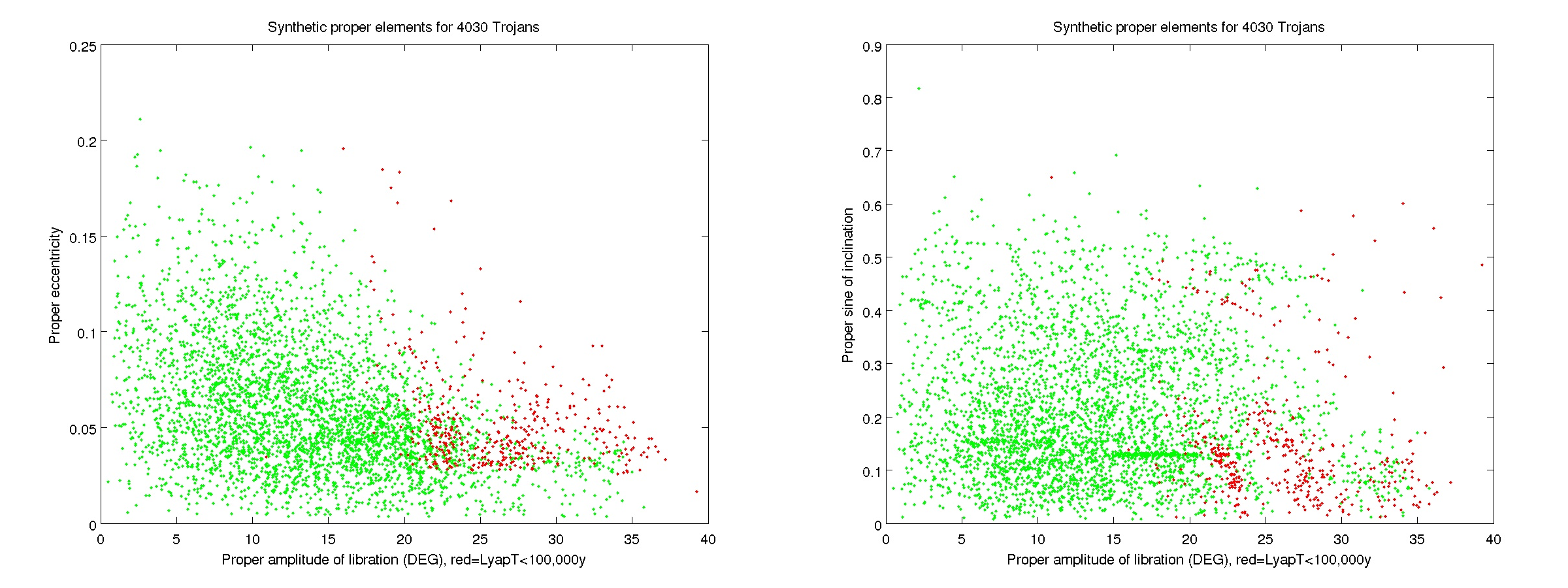}
  \caption[Classification of proper elements for Trojan
    asteroids]{Classification of proper elements for Trojan
    asteroids. The red orbits indicate strongly chaotic orbits, the
    green points indicate proper elements degraded by secular
    resonances. The plots and the computation of synthetic proper
    elements are provided by the site
    \textsc{http://hamilton.dm.unipi.it/astdys/}.}
  \label{fig:propelements_plots.png}
\end{figure}

The fast variation of the proper elements is an indication of highly
chaotic orbits. However, there exist several cases of \emph{stable
  chaos} in orbits for which the Lyapunov times are small (i.e., the
Lyapunov exponents are big), but the variation of proper elements is
negligible~\cite{Milani-93}.  Figure~\ref{fig:propelements_plots.png}
presents an updated computation of syntethic proper
elements~\cite{Milani-93}, done by using the semianalytical model
of~\cite{BeauRoig-01}, where the different timescales presented in the
case of Jupiter's Trojan asteroids are exploited by means of adiabatic
invariance theory.
It is conjectured that groups of asteroids with very similar proper
elements could have been generated by the collisional disruption of a
bigger body.  The numerical computation of proper elements for a big
number of asteroids allows their classification into
families~\cite{Milani-93}.

For several years, it was thought that the population of Jupiter's
Trojan domain was sustained by the accretion, captures or collisions
\emph{in situ}~\cite{Yoder}, although the mechanisms were proven not
to justify properly the features of the real
Trojans~\cite{MarzSch-98}. In fact, numerical simulations of planetary
migration have shown that a slow change of the position in the orbits
of Jupiter and Saturn, resulting in their crossing the mutual 2:1 MMR,
(the so-called \emph{Nice model}) could lead to a complete depletion
of Jupiter's Trojan domain~\cite{Michtetal-01},~\cite{MarzSch-07},
even if without such crossing the migration itself could be
harmless~\cite{Gomes-98},~\cite{RobBod-09}. On the other hand,
although the crossing of one or more planetary MMR induces a
considerable amount of chaos, it also makes possible the chaotic
capture of new Trojans~\cite{Morbietal-05}. A reformulated version of
this mechanism is presented in~\cite{NesVokr-09}, where the hypothesis
of a scattering between Jupiter and a lost icy planet is introduced
(the so-called \emph{Jumping Jupiter model}). The latter models are so
far the only able to reproduce the inclinations of the observed Trojan
asteroids and, perhaps, the asymmetry presented in the number of the
objects observed in the two domains.

Regarding the remaining planets in the Solar System, nowadays it is known 
that there exist coorbital companions at least for
Mars, Neptune, Uranus and the Earth. Neptune presents the second most
populated Trojan domain, after Jupiter, with 12 components, with
characteristically high inclinations. The stability in the equilateral
libration regions is proven to be high, thus Neptune may have kept about the
50\% of its Trojan population after the migration
process~\cite{NesDon-02},\cite{Lyketal-11}. The high inclination
values give hints that the main contributions to the population may be
due to captures of bodies rather than collisions~\cite{Lyketal-09}. As
happens with Jupiter and Saturn, the crossing of the MMR with Uranus
depletes the region but may trigger also the capture
mechanism~\cite{Lyketal-09}. Studies of the stability
 were performed by numerical N-body simulations for Mars 
(\cite{MikInn-94},~\cite{TabEv-99}) and 
Uranus (\cite{NesDon-02},~\cite{Dvoetal-10}). In the case of the
Earth, its only observed Trojan companion $2010\:\mathrm{TK}_{7}$ 
has been shown to be only temporarly
trapped in the libration region~\cite{Conetal-11}. 

Regarding Saturn, numerical simulations show that the secular
interaction with Jupiter suffices for destabilizing the coorbital
domain~\cite{InnMik-89},~\cite{NesDon-02}, thus Trojan companions are
not expected for Saturn. Nevertheless, in Saturn's system of moons,
there are two cases of coorbital resonances: Telesto and Calipso are
librating Tethys's $L_4$ and $L_5$ domains respectively, and Helene
and Polydeuces are librating Dione's $L_4$ and $L_5$.

With the advent of extrasolar planets observations, the spectrum of
architechtures for planetary systems has been widened considerably.
Extrasolar systems reveal planetary configurations that differ from
that of the Solar System, and therefore, the question of the existence
of Trojan bodies in these new enviroments takes importance. Such an
existence would require examining at least two different topics:
i. the possibility of formation or capture of a Trojan body, and
ii. the long-term stability of such bodies in the 1:1 libration region
of the coorbital planet. Regarding the first point, some authors have
studied the possibility of formation of a Trojan planet within the
libration region by accretion of material.  In general, accretion can
produce only small size planets \cite{Beugetal-07}. If two initial
coorbital bodies compete for the accretion of material, the process is
dominated by the body more massive initially~\cite{CressNel-09}. On
the other hand, if the process is individualized, i.e. the dominant
body already finished its accretion process, bigger Trojan bodies are
feasible to obtain~\cite{Lyraetal-09}.

Regarding the stability of the extrasolar
Trojans, some parametric studies were performed for different
planetary masses and configurations~\cite{Nauen-02}, in some cases
applied to observed systems~\cite{Schwarzetal-09}. Of different
coorbital configurations, the libration around the equilateral
points seems to be the most stable~\cite{SmithLissa-10}. The
architecture of the extrasolar system is determinant for the
stability of a Trojan coorbital: the stability regions are more
dependent on the physical and orbital parameters of the primary,
as well as their correlation with additional planets, than on the mass
of the Trojan body itself~\cite{Erdietal-07}.

Despite the predictions based on the simulations about formation and
stability, there are no observed extrasolar Trojan bodies so far.
Although different authors examinated the possibility that the
libration of Trojan bodies may generate characteristic signatures in
the observations of planetary transits~\cite{Haghietal-13} or stellar
radial velocities~\cite{Nauen-02}, these signatures may be highly
misinterpreted by the usual techniques of
identification~\cite{FordGau-06},~\cite{Dobro-13},~\cite{Giupetal-12}.
Nevertheless, tests on new methods specifically designed for the
detection of Trojan bodies have given promising
results~\cite{LaughCham-02},~\cite{Leleuetal-15}\footnote{For more
  extense reviews on the Trojan problem, see~\cite{Erdi-97}
  or~\cite{RobSouc-10}.}.


\section{Summary}\label{sec:1.5-summary}

From what was exposed in the previous introductory paragraphs, it
becomes clear that the dynamical richness of the Trojan problem
represents a challenge for its theoretical study based on some
form of perturbation theory. In this thesis, we attempt to address this
problem by: i) introducing a novel Hamiltonian formulation and ii)
proposing novel normal form schemes able to deal with the particular
characteristics of the Trojan motion. Such characteristics are the
existence of differentiated temporal scales, the asymmetry of the
motion with respect to the libration center, and the singular behavior
of the equations of motion at large libration amplitudes.

In \emph{Chapter 2}, we define a new normal form algorithm implemented
to the CR3BP.  We use modified Delaunay-Poincar\'e variables which
make possible a separation of the fast from the synodic d.o.f. The
only real singularity of the Trojan problem is due to close encounters
of the Trojan body with the primary $m'$. Such a singularity implies
that expansions in the critical argument $\tau = \lambda - \lambda'$
around the $L_4$ value $\pi/3$ converge only in the domain $|\tau -
\pi/3| < \pi/3$ (and similarly for $L_5$).  In Chapter 2 we propose a
novel normal form construction which avoids completely expansions in
the variable $\tau$. Applying the Lie series technique for averaging
the fast degree of freedom (independent of $\tau$) renders possible to
construct a normal form unaffected by the functional form of all series
quantities on $\tau$.  We provide a complete description of the
expansions and of the proposed normalizing scheme. We finally test the
normal form obtained by this method by means of numerical experiments.

In \emph{Chapter 3} we revisit the main features of the
ER3BP. Starting again from a representation in terms of modified
Delaunay variables, we propose a sequence of canonical transformations
leading to a Hamiltonian decomposition in the fast, synodic and
secular d.o.f. From the latter, we introduce a model called the 'basic
Hamiltonian' $H_b$. This corresponds to the part of the Hamiltonian
independent of the secular angle. The three d.o.f. interact through
different resonances, that we classify according to the frequency
relation which they involve.  These resonances may destabilize the
orbits, inducing escapes.  By means of numerical experiments, we
depict the resonance web in the ER3BP in terms of stability maps and
phase portraits. We finally show that there exists a correlation
between escapes, sticky regions of the phase space and resonant
dynamics.

\emph{Chapter 4} is devoted to the study of the dynamics under $H_b$.
The basic Hamiltonian $H_b$ is a model representing the short period
and synodic components of the Trojan motion. Averaging over the fast
angle, the $\langle H_b \rangle$ turns to be an integrable
Hamiltonian, yet depending on the value of $e'$. Thus, it allows to
formally define action-angle variables for the synodic d.o.f., even
when $e'\neq 0$. In addition, by means of a trivial reinterpretation
of the canonical transformation, it can be proven that the functions
$H_b$ derived from the ER3BP and the one derived from a more complex
model involving more disturbing planets, called the Restricted
Multi-Planet Problem (RMPP), are formally the same.  
Based on this formal equivalence, we study numerically some properties
of the model $H_b$, using for convenience the ER3BP as a complete
model allowing numerical comparisons, although the results are 
expected to hold in the RMPP as well.
In addition, we introduce a method
for locating the position of the secondary resonances 
based on the use of the normalized $\langle H_b\rangle$. We show that
the combination of the normalizing scheme of Chapter 2 (adapted here
to the elliptic approximation) and the representation by the $H_b$ is
efficient enough so as to allow to locate the so-called
\emph{transverse} resonances involving also very slow secular
frequencies.

\emph{Chapter 5} deals in detail with the problem of Trojan
\emph{secondary resonances}. We formulate yet one more novel expansion
allowing to predict the size and the location of secondary resonances
in the phase space.  This is possible by implementing a
\emph{resonant} normal form on the basic Hamiltonian model $H_b$. To
this end, we face the issue of expansions in the critical angle $\tau$
(Chapter 2). This is addressed by introducing a new expansion called
\emph{asymmetric}. The latter stems from exploting the natural
asymmetry of the Trojan orbits in their angular excursion away from
$L_4$ or $L_5$.  We make a comparison of the asymmetric and symmetric
expansions performed around the Lagrangian points. The symmetric
expansion proves to be inadequate for a correct representation of the
problem, inducing spurious dynamics in particular cases of resonances.
The use of the asymmetric expansion partially remedies this issue.

\emph{Chapter 6} summarizes the conclusions of the present thesis, and
gives some perspectives on possible extensions for future work. 

The Appendixes~\ref{sec_app:expansions},
\ref{sec_app:expan_del_ellip}, \ref{appex:theHb}, \ref{app:asymm_exp},
\ref{Appe:res-norm-form} provide detailed explanations of the
technical aspects of the expansions and some examples up to low
orders.

\chapter{Novel normalizing scheme of extended convergence}\label{sec:2-CRTBP}

\section{Motivation}\label{sec:motivation_nf}

As already emphasized in the previous chapter, the Trojan motion in
the ER3BP has three components, with well separated temporal scales: a
fast (epicyclic) motion with frequency of order ${\cal O}(1)$, a
synodic motion with frequency of order ${\cal O}(\sqrt{\mu})$ and a
secular motion with frequency of order ${\cal O}(\mu)$. Due to the
decoupling of the secular d.o.f., in the CR3BP the motion can be
expressed in terms of the fast and the synodic frequencies only, as in
Sect.~\ref{sec:1.4.X.motionL4L5}.

As customary in Celestial Mechanics, the most basic form of
Hamiltonian normalization stems from averaging the Hamiltonian over
the fast angles. Independently of the formalism used, what remains 
after such averaging gives the synodic motion around the libration center. The
use of a suitable set of variables is necessary
for performing the averaging. A good choice for this purpose are
Delaunay variables (see Sect.~\ref{2.1-expansion} below).

\begin{SCfigure}
  \centering 
  \includegraphics[width=.5\textwidth]{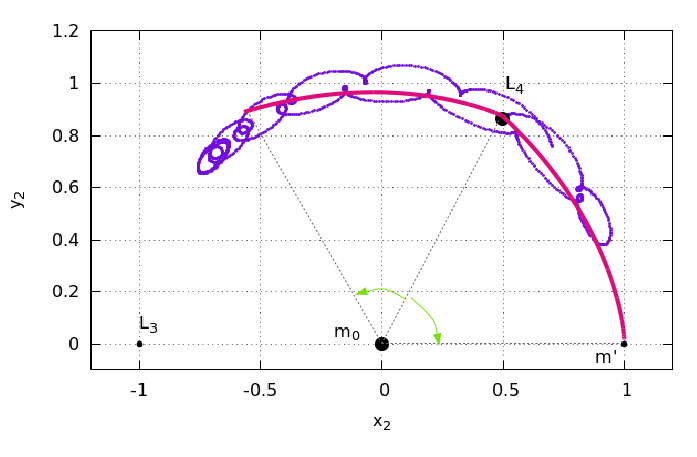}
  \caption[Representation of the convergence domain of series
    expansion around $L_4$]{{\footnotesize Representation of the
      domain of $\tau$ where polynomial series are convergent if the
      expansion takes place around $L_4$. The radius of convergence
      (thick pink line) of the series is given by the distance between
      $L_4$ and the primary, namely $60^{\circ}$ ($\tau=\pi/3$). While
      this does not induce any problem in the direction towards the
      primary, it does limit the convergence in the opposite
      direction. As shown by the purple curve (same as the green curve
      of Fig.~\ref{fig:tadpoles.png}), a typical Trojan orbit may
      greatly exceed the leftward angular limit of $60^{\circ}$ from
      $L_4$.}
\vspace{0.5cm}}
\label{fig:convdomain1.png}
\end{SCfigure}

Regarding the above, we point out a crucial remark.  The Hamiltonian
of the CR3BP has a real singularity corresponding to close encounters
of the massless body with the primary $m'$. This singularity
takes place at $ a = a'$, $\tau = \lambda-\lambda'=0$. The key
remark is that any \emph{polynomial} series expansion of the equations
of motion (or the Hamiltonian) with respect to $\tau$ around a fixed
value is convergent in a disk of radius equal to the distance between
the fixed point and the singularity. In the literature, it has been
common to consider such polynomial expansions around the position of
equilibrium ($\tau_{L_{4},L_{5}}=\pm\frac{\pi}{3}$).  Due to the
asymmetry of the librations (Sect.~\ref{sec:1.4.works_on_Trojans}), it
is easy to see that the above limited convergence affects severely the
representation of the orbits mainly in the opposite direction to the
primary (see Fig.~\ref{fig:convdomain1.png}).

We will now present a Hamiltonian scheme allowing to average the
Hamiltonian over short period terms \emph{without} making expansions
affected by the singular behavior of the Hamiltonian at
$|\tau-\pi/3|=\pi/3$.  In the framework of the CR3BP, the so found
averaged Hamiltonian is an integrable system of one d.o.f. able to
describe the synodic librations in the whole domain of allowed Trojan
motion. We then test the degree of approximation of our analytical
method against numerical experiments.

\section{Delaunay-Poincar\'e expansion for the pCR3BP}\label{2.1-expansion}

We start from the Hamiltonian of the planar R3BP
in heliocentric cartesian coordinates, given in Eq.~\eqref{eq:Ham_hel_RTBP_2}.
By the system of units defined, we have ${\cal G}(m_0+m') = 1$. Thus,
\begin{equation}
\begin{aligned}\label{eq:Ham_hel_RTBP_3}
H =\,& \frac{\|\mathbf{p}\|^2}{2} -\frac{{\cal G}(m_0+m')}{\|\mathbf{r}\|}
+\frac{{\cal G} m'}{\|\mathbf{r}\|}
 - {\cal G} m' \left( \frac{1}{\Delta} - \frac{\mathbf{r} \cdot
  \mathbf{r}'}{\|\mathbf{r}'\|^3} \right)~ \\
 =\,& \frac{\|\mathbf{p}\|^2}{2} -\frac{1}{\|\mathbf{r}\|}
 - {\cal G} m' \left( \frac{1}{\Delta} - \frac{1}{\|\mathbf{r}\|} -\frac{\mathbf{r} \cdot
  \mathbf{r}'}{\|\mathbf{r}'\|^3} \right)~,
\end{aligned}
\end{equation}
where $\Delta=\|\mathbf{r}-\mathbf{r}'\|$, $\mathbf{r}'$
is the heliocentric position vector for the planet, $\mathbf{r}$ for
the Trojan and $\mathbf{p} = \dot{\mathbf{r}}$. This decomposition of
$H$ (including a term $1/r$ in the disturbing function) allows
to define Delaunay variables independent of the mass parameter
$\mu$~\cite{BrownShook-33}. The heliocentric vectors in polar coordinates are
\begin{equation}\label{eq:simply}
\|\mathbf{r}\| = r~,\quad \mathbf{r}= (r \cos \theta, r \sin \theta)~,
\end{equation}
and (considering the circular approximation for the pR3BP)
\begin{equation}\label{eq:simply_2}
\|\mathbf{r'}\| = 1~,\quad \mathbf{r'}= (\cos f', \sin f')~.
\end{equation}

\begin{SCfigure}
\includegraphics[width=.5\textwidth]{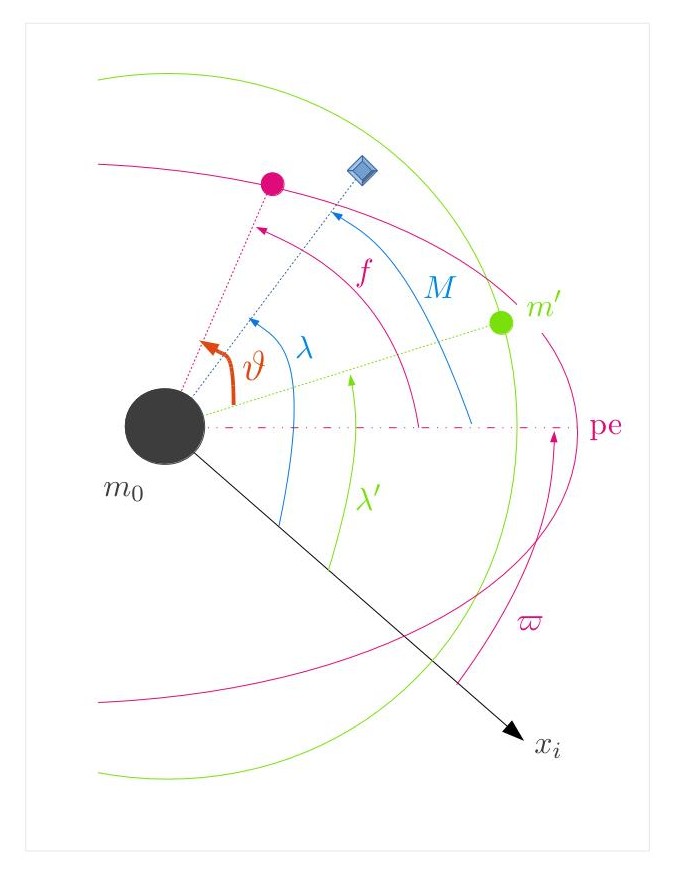} 
\caption[Angular variables in the pCR3BP]{{\small The angular
    variables used for the representation of the Hamiltonian, seen on
    the plane $(x,y)$ of the inertial reference frame. The black
    circle denotes the position of $m_0$ at the center of the circular
    orbit of $m'$ (yellow-green) and at one of the foci of the
    elliptical orbit of the test particle (pink, $pe$ for the position
    of the pericenter). $\vartheta$ is the instantaneous angular
    distance between $m'$ and the test particle (thick orange line).
    The mean longitudes $\lambda$ (blue), $\lambda'$ (green) and the
    longitude of the pericenter $\varpi$ (pink) are measured from the
    $x$-axis.  The true anomaly $f$ (pink) and mean anomaly $M$ (blue)
    are measured from the line that joins $m_0$ with the pericenter of
    the elliptic orbit.  Without loss of generality, we define
    $\varpi'=0$. The blue diamond indicates the position of a
    fictitious body that moves on a circular orbit of $a$ and $n$
    equal to those of the massless body's orbit. Thus, $\vartheta =
    \lambda - M + f - \lambda'$.}}
\label{fig:vartheta.jpg}
\end{SCfigure}

We can re-express the terms of the disturbing function as
\begin{align}\label{eq:rdotrp}
\mathbf{r}\cdot \mathbf{r'} & = r \cos \theta \, \cos \theta' + r \sin
\theta \, \sin \theta' \nonumber\\ & = r \left( \cos \theta \, \cos
\theta' + \sin \theta \, \sin \theta' \right) \\ & = r \cos (\theta - \theta') =
r \cos \vartheta~, \nonumber
\end{align}
\begin{equation}
\begin{aligned}\label{eq:delta_sol}
\Delta & = \|\mathbf{r}-\mathbf{r}'\| = \| r \cos \theta - \cos \theta', r \sin f - \sin \theta' \| \\
& = \sqrt{r^2 +1 - 2r \cos (\theta -\theta')} =  \sqrt{r^2 +1 - 2r \cos \vartheta}~, 
\end{aligned}
\end{equation}
where $\vartheta = \theta - \theta'$ (see
Fig.~\ref{fig:vartheta.jpg}). Replacing Eqs.~\eqref{eq:rdotrp}
and~\eqref{eq:delta_sol} in Eq.~\eqref{eq:Ham_hel_RTBP_3}, we find
\begin{equation}\label{eq:Ham_hel_RTBP_4}
H = \frac{p^2}{2} - \frac{1}{r} - \mu \left( \frac{1}{\sqrt{r^2+1-2 r \cos\vartheta}} - \frac{1}{r} - r \cos \vartheta \right)~,
\end{equation}
with $\mu$ defined in Eq.~\eqref{eq:mass_param} and $\|\mathbf{p}\| = p$.
The gravitational
potential defining the disturbing function depends only on the 
\emph{relative} configuration between the bodies, thus the Hamiltonian 
is independent of the orientation of the inertial system where
the coordinates $(r,\theta)$ are defined.

At this point, we introduce
the first canonical transformation to Delaunay-like variables, as
\begin{equation}
\begin{aligned}\label{eq:Delaunay-coord_cR3BP}
\qquad \Gamma & = \sqrt{a}(1-\sqrt{1-e^2})~, &   M & \, ~,\qquad\\
\qquad G & = \sqrt{a(1-e^2)}~, &  \lambda & = \varpi+M~,\qquad 
\end{aligned}
\end{equation}
where $a$, $e$, $M$, $\varpi$, $\lambda$ are the major semiaxis,
eccentricity, mean anomaly, longitude of the perihelion and mean
longitude of the Trojan body. The Keplerian part of the
Hamiltonian~\eqref{eq:Delaunay-coord_cR3BP} (see
Eq.~\ref{eq:Energy_2BP}) in terms of Delaunay variables reads
\begin{equation}\label{eq:Kep_part_Del}
\frac{p^2}{2} - \frac{1}{r} = - \frac{1}{2 a} = -\frac{1}{2 (G + \Gamma)^2}~.
\end{equation}
Due to the splitting performed in Eq.~\eqref{eq:Ham_hel_RTBP_3}, it is
possible to define the actions of the Delaunay set independently of $\mu$
(unlike in Eqs.~\ref{eq:mod_delau_vari}).

The disturbing function in~\eqref{eq:Ham_hel_RTBP_3} contains the quantities
\begin{equation*}
\frac{1}{r}~, \: \: r \cos \vartheta~, \: \: \mathrm{and} \: \: r^2~.
\end{equation*}
For the expansion of $1/r$, we have
\begin{equation}\label{eq:r_solo}
r = \frac{p}{1+ e \cos f} = \frac{a (1-e^2)}{1 + e \cos f}
= \frac{G^2}{1+e \cos f}~.
\end{equation}
Thus, 
\begin{equation}\label{eq:g_I}
\frac{1}{r} = \frac{1}{G^2} (1 + e \cos f(e,M)) =  \sum_{j=0}^{\infty} e^j {\cal P}_{j}^{(I)} (M)~,
\end{equation}
where ${\cal P}_{j}^{(I)}$ are trigonometric (finite) polynomials of degree $j$
in the mean anomaly. The construction of this expansion is based
on the expansions for the true anomaly $f$ in terms of the mean anomaly
$M$ (see, e.g., \S 2.5 of~\cite{MurrDerm-99})
\begin{equation}
\begin{aligned}\label{eq:cosv-series-expan}
\cos f &= -e +
\frac{2\big(1-e^2\big)}{e}\sum_{n=1}^{\infty}\big[J_n(ne)\cos(nM)\big]~,\\
\sin f &= 2\sqrt{1-e^2}\sum_{n=1}^{\infty}\big[J_{n}^{\prime}(ne)\sin(nM)\big]~, 
\end{aligned}
\end{equation}
where
\begin{equation} \label{eq:def-Bessel-Jn-and-derivative-Chap2}
J_{n}(x) =
\sum_{j=0}^{\infty}\frac{(-1)^j}{j!\,(j+n)!}\left(\frac{x}{2}\right)^{n+2j}
\ \quad{\rm and}\ \quad
J_{n}^{\prime}(x) =
\sum_{j=0}^{\infty}\frac{(-1)^j(2j+n)}{2\big(j!\,(j+n)!\big)}
\left(\frac{x}{2}\right)^{n+2j-1}
\end{equation}
are the Bessel functions of first kind and their derivatives.
Substituting
with~\eqref{eq:cosv-series-expan}--\eqref{eq:def-Bessel-Jn-and-derivative-Chap2}
in~\eqref{eq:g_I} we obtain the trigonometric polynomials ${\cal
  P}_{j}^{(I)}$. Up to order ${\cal O}(e^2)$:
\begin{equation*}\label{eq:expan_cosf}
\cos f = -e + \cos M + e\, \cos 2M - \frac{9}{8}\, (e^2 \, cos M - e^2
\cos 3M) + \ldots~,
\end{equation*}
\begin{equation*}\label{eq:expan_sinf}
\sin f= \sin M - \frac{7}{8}\, e^2 \sin M + e\, \sin 2M + 
\frac{9}{8}\, e^2 \sin 3M +\ldots,
\end{equation*}
\begin{equation*}\label{eq:expan_1/r}
\frac{1}{r} = \frac{1}{G^2} \left( 1 - e^2 + e \cos M + e^2 \cos 2M
\right) +\ldots~.
\end{equation*}\\

\noindent
For the expansion of $r^2$, from  Eq.~\eqref{eq:r_solo}, we have
\begin{equation}\label{eq:r_ov_G}
r = G^2 \left(\frac{1}{1+e \cos f} \right) =
G^2 \left( \sum_{j=0}^{\infty} (-1)^j e^j \cos^j f  \right)~.
\end{equation}
Replacing Eq.~\eqref{eq:cosv-series-expan} into Eq.~\eqref{eq:r_ov_G},
it is possible to compute $r^2$.  Up to order ${\cal O}(e^2)$:
\begin{equation*}\label{eq:expan_r}
r = G^2 \left( 1 + \frac{3}{2} e^2 - e \cos M -\frac{1}{2} e^2 \cos 2M \right)
+\ldots~,
\end{equation*}
\begin{equation*}\label{eq:expan_r2}
r^2 = G^4 \left( 1 + \frac{7}{2} e^2 - 2 e \cos M -\frac{1}{2} e^2 \cos 2M \right)+\ldots~.
\end{equation*}

\noindent
For the expansion of $r \cos \vartheta$, we first expand $\vartheta$,
considering that $\vartheta = \lambda - \lambda' + f - M$ (Fig.~\ref{fig:vartheta.jpg}). Hence,
\begin{equation}
\begin{aligned}\label{eq:cosvarthet1}
\cos \vartheta = &\, \cos (\lambda - \lambda' + f - M)~,\\ 
= &\, \cos M \cos f \cos \lambda \cos \lambda' 
+ \cos \lambda \cos \lambda' \sin M \sin f
- \cos f \cos \lambda' \sin M \sin \lambda \\ 
- & \cos M \cos \lambda' \sin f \sin \lambda 
- \cos f \cos \lambda \sin M \sin \lambda' 
+ \cos M \cos \lambda \sin f \sin \lambda'\\ 
+ & \cos M \cos f \sin \lambda \sin \lambda' 
+ \sin M \sin f \sin \lambda \sin \lambda'~.
\end{aligned}
\end{equation}

Replacing ~\eqref{eq:cosv-series-expan}
 into Eq.~\eqref{eq:cosvarthet1}, we obtain
\begin{equation}\label{eq:cosvarthet3}
\cos \vartheta = \, {\cal G}^{(II)}(e,M,\lambda,\lambda')~.
\end{equation}
Up to order ${\cal O}(e^2)$
\begin{align*}
\cos \vartheta = & \cos (\lambda - \lambda') - e^2 \cos (\lambda - \lambda') +
e \cos (M + \lambda - \lambda')\\ 
+ & \frac{9}{8} e^2 \cos (2M + \lambda - \lambda' )-
e \cos (M - \lambda + \lambda') - 
\frac{1}{8} e^2 \cos (2M - \lambda + \lambda') + \ldots~.
\end{align*}

Replacing all the above expressions in~\eqref{eq:Ham_hel_RTBP_4}, and
setting $e =\sqrt{1 - (G/L)^2}$, the Hamiltonian is given by
\begin{equation}\label{eq:Ham_hel_RTBP_5}
H(G,\Gamma,\lambda,M,\lambda') = -\frac{1}{2(G+\Gamma)^2} -\mu
R(G,\Gamma,\lambda,\lambda',M)~.
\end{equation}
The Hamiltonian in Eq.~\eqref{eq:Ham_hel_RTBP_5} is a system of 
 $2+\frac{1}{2}$ d.o.f., since the action conjugated to  
 $\lambda'$ does not appear explicitely. In order to produce an
autonomous Hamiltonian, we add a dummy action $G'$, with no
other physical meaning than to complete the set of canonical
variables. We define $G'$ through Hamilton's equations, 
$\dot{G}'= -\frac{\partial H}{\partial\lambda'}$,
$\dot{\lambda}'= \frac{\partial H}{\partial G'}$,
so the extended Hamiltonian is
\begin{equation}\label{eq:Ham_hel_RTBP_6}
H(G,\Gamma,G',\lambda,M,\lambda') = -\frac{1}{2(G+\Gamma)^2} + G' -\mu
R(G,\Gamma,\lambda,\lambda',M)~.
\end{equation}
Equation~\eqref{eq:Ham_hel_RTBP_6} corresponds to a sytem of
3 d.o.f.  Nevertheless, it is possible to trivially 
reduce one of them, by the following remark: all the 
trigonometric terms in the disturbing function have
arguments of the form
\begin{equation}\label{eq:arg_d'alemb_rul}
\varkappa = k_1 \, \lambda' + k_2 \, \lambda + k_3 \, \varpi' + k_4 \, \varpi + k_5 \, \Omega' + k_6 \, \Omega~,
\end{equation}
with $k_i \in \mathbb{N}$, $i=1,\ldots,6$.  As a
consequence of the invariance of the Hamiltonian to rotation
transformations, the coefficients $k_i$ must satisfy the following
condition
\begin{equation}\label{eq:d'alem_rule}
k_1 + k_2 + k_3 + k_4 + k_5 + k_6 = 0 ~,
\end{equation}
known as the \emph{first d'Alembert rule}. In the pCR3BP, the variables
$\Omega$ and $\Omega'$ do not appear, hence $k_5 = k_6 = 0$. Also, since 
$e'=0$, the longitude of the
perihelion of the circular orbit of $m'$ does not appear in the
disturbing function (second d'Alembert rule), thus $k_3 = 0$. Hence,
\begin{equation}\label{eq:d'alem_rule}
k_1 + k_2 +  k_4 = 0 ~,\: \:\mathrm{or,\:equivalently\:\:} k_2+k_4 = - k_1~.
\end{equation}
If we replace $\varpi = \lambda - M$~(Eq.~\eqref{eq:Delaunay-coord_cR3BP}) 
in Eq.~\eqref{eq:arg_d'alemb_rul} for the pCR3BP, we obtain
\begin{align}\label{eq:arg_circ_2}
\varkappa = & \, k_1 \, \lambda' + k_2 \, \lambda + k_4 \, (\lambda -
M)~,\nonumber\\ = & \, k_1 \, \lambda' + (k_2+k_4) \, \lambda -k_4 \,
M ~, \\ = & \, k_1 \, \lambda' - k_1 \, \lambda -k_4 \, M ~,
\nonumber\\ = & \, -k_1 (\lambda-\lambda') - k_4 \, M ~. \nonumber
\end{align}
Thus, as a consequence of the d'Alembert rules, $\lambda$ and
$\lambda'$ appear in the Hamiltonian exclusively through the
difference $\lambda-\lambda'$. This fact can be checked explicitely,
for instance, in the expansions up to order ${\cal O}(e^2)$ of
Eq.~\eqref{eq:cosvarthet3}.

We thus introduce a canonical transformation
$(G,\Gamma,G',\lambda,M,\lambda') \mapsto
(\Upsilon_1,\Upsilon_2,\Upsilon_3,\tau_1,\tau_2,\tau_3)\,$ through the
generating function  
$S_1(\lambda,\lambda',M,\Upsilon_1,\Upsilon_2,\Upsilon_3,)$ of second class (old
angles and new actions), defined by
\begin{equation}\label{eq:1st-can-trans}
S_1 = (\lambda-\lambda') \Upsilon_1 + \lambda' \Upsilon_2 + M \Upsilon_3
\end{equation}
that generates the transformation equations
\begin{equation}
\begin{aligned}\label{eq:1s-can-trans-eqs}
&\tau_1 = \frac{\partial S_1}{\partial \Upsilon_1} = \lambda-\lambda'~,~~
&\tau_2& = \frac{\partial S_1}{\partial \Upsilon_2} = \lambda'~,~~
&\tau_3 = \frac{\partial S_1}{\partial \Upsilon_3} = M~, \\
& G = \frac{\partial S_1}{\partial \lambda} = \Upsilon_1~, ~~
& G'& = \frac{\partial S_1}{\partial \lambda'} = \Upsilon_2 - \Upsilon_1~,~~ 
&\Gamma = \frac{\partial S_1}{\partial M} = \Upsilon_3~.
\end{aligned}
\end{equation}
Applying this transformation to~\eqref{eq:Ham_hel_RTBP_5}, we obtain
\begin{equation}\label{eq:Ham_hel_RTBP_7}
H(G,\Gamma,\tau,M) = -\frac{1}{2(G+\Gamma)^2} + \Upsilon_2 - G -\mu R(G,\Gamma,\tau,M)~,
\end{equation}
where $\tau \equiv \tau_1 = \lambda -\lambda'$ .  This form of
$H$ is independent of $\tau_2$, thus $\Upsilon_2$ remains constant in
time. Since a constant term in the Hamiltonian does not induce any
effect on the dynamics of the problem, we directly neglect it.  We
keep the original symbol for those variables that were transformed by the
identity transformation.

In order to avoid the fictitious singularity that takes place if $e=0$
(Sect.~\ref{sec:1.2.3-del_coord}), we introduce Poincar\'e-like
variables, through the transformation $(\rho,\xi,\lambda,\eta) \mapsto
(G,\Gamma,\tau,M)$ \footnote{We symbolize with
  $\mathrm{arctan}\,(a,b)$ the function
  $\mathrm{tan}^{-1}(a/b):\mathbb{R}^2\to \mathbb{T}^1$, of two variables, that
  maps the value of the arctangent to the corresponding quadrant
in the coordinate system with $b$ as the abscissa and $a$ as the ordinate.}:
\begin{equation}
\begin{aligned}\label{eq:Garfinkel-coord}
\qquad G &=  \rho+1~, & \tau &=  \tau~, \\
\qquad \Gamma &= \frac{\xi^2+\eta^2}{2}, &  M &= \mathrm{arctan}\,(\eta,\xi)~.
\end{aligned}
\end{equation}
A similar set of coordinates but dependent on $\mu$
is used in~\cite{Garfinkel-77}. By the computation of the
Poisson bracket, it is straightforward to check that the 
transformation~\eqref{eq:Garfinkel-coord} is canonical.
In these new variables, the transformed Hamiltonian reads
\begin{equation}\label{eq:Garfinkel-Ham}
H(\rho,\xi,\lambda,\eta)=
-\frac{1}{2\left[1+\rho+\frac{1}{2}(\xi^2+\eta^2)\right]^2}-1-\rho
-\mu R(\rho,\xi,\lambda,\eta)~.
\end{equation}

In Delaunay-Poincar\'e variables, the fast and synodic dynamics are
represented by the two independent pairs of canonical coordinates.
The first terms of the Keplerian part of
Hamiltonian~\eqref{eq:Ham_hel_RTBP_7} read
\begin{equation}
\begin{aligned}\label{eq:quadr-part-Kepl-appr}
- \frac{1}{2[1+\rho+\frac{1}{2}(\xi^2+\eta^2)]^2} -1 - \rho
 = &\, \frac{3}{2}
 +\frac{\xi^2+\eta^2}{2} - 
\frac{3}{2}\left[\rho+\frac{\xi^2+\eta^2}{2}\right]^2 + \ldots\, \\
= & \, \frac{3}{2}+\Gamma-\frac{3}{2}(\rho+\Gamma)^2
 + \ldots\ ,
\end{aligned}
\end{equation}
where $\xi$ and $\eta$ are back-transformed according to 
$\xi=\sqrt{2\Gamma}\cos M$ and $\eta=\sqrt{2\Gamma}\sin M$. 
Thus, the frequencies associated to each angle are
\begin{equation}\label{eq:slow-fast-ang-velocities}
\dot{\tau}=\frac{\partial\,H}{\partial\rho}\simeq 0 + \ldots~,
\qquad
\dot{M}=\frac{\partial\,H}{\partial\Gamma}\simeq 1 + \ldots~,
\end{equation}
where we assume that $\mu$, $\rho$ and $\Gamma$ are small quantities
in the neighborhood of the Lagrangian points.  Hence, the normalizing
scheme only has to remove the dependence of the Hamiltonian on $M$ or,
in Poincar\'e variables, it has to remove any combination of $\xi$ and
$\eta$ different from $\frac{\xi^2+\eta^2}{2}$.

We will now show that, by means of canonical transformations based on
Lie series, it is possible to introduce a normalizing scheme that
removes $M$ and on which $\tau$ is not involved. This allows to keep a
complicated dependence of the Hamiltonian on $\tau$, avoinding to
introduce expansions in the synodic variables ($\rho,\tau$) with
convergence problems due to singularities, as mentioned before.  We
notice that the term that induces this singularity is $1/\Delta$, 
which contains the factor
\begin{equation}\label{eq:def_beta}
\beta(\tau) = \frac{1}{\sqrt{2-2\cos\tau}}~.
\end{equation}

Expanding the Hamiltonian in Eq.~\eqref{eq:Garfinkel-Ham} with respect
to $\rho$, $\xi$ and $\eta$, we obtain the polynomial expression of H,
given by
\begin{equation}\label{eq:initial-Ham-rewritten}
\vcenter{\openup1\jot\halign{ \hbox {\hfil $\displaystyle {#}$} &\hbox
    {\hfil $\displaystyle {#}$\hfil} &\hbox {$\displaystyle
      {#}$\hfil}\cr H(\rho,\xi,\tau,\eta) &= &
      -\frac{1}{2} \sum_{j=0}^{\infty} (-1)^{j}
      (j+1) \left(\rho+\frac{\xi^2+\eta^2}{2} \right)^{j} \, -1
      -\rho \,+\,\mu\big(1+\cos(\tau)-\beta(\tau)\big) \cr &+ \mu
      &\sum_{l=1}^{\infty}\,\sum_{{\scriptstyle{m_1+m_2}}\atop{\scriptstyle
      {+m_3=l}}} \ \sum_{{\scriptstyle{k_1+k_2\le
      l}}\atop{\scriptstyle {j\le
      2l+1}}} \mathsf{e}_{m_1,m_2,m_3,k_1,k_2,j}
\,\rho^{m_1}\xi^{m_2}\eta^{m_3}
    \,\cos^{k_1}(\tau)\sin^{k_2}(\tau)\,\beta^j(\tau)\ , \cr
}}
\end{equation}
where $\beta(\tau)$ is defined in Eq.~\eqref{eq:def_beta} and the
coefficients $\mathsf{e}_{m_1,m_2,m_3,k_1,k_2,j}$ are rational
numbers. The Hamiltonian in Eq.~\eqref{eq:initial-Ham-rewritten} is
the function that we use for initializing the normalization.

In practice, these expansions are easier to perform if the
intermediate step associated to the
variables~\eqref{eq:Delaunay-coord_cR3BP} is avoided, and we pass
directly from the orbital elements $e$, $M$, to Poincar\'e-like
variables.  In the Appendix~\ref{sec_app:expansions}, we provide
explicit formul\ae~for these expansions, including the construction of
the terms with dependence on $\beta$.

\section{Normalization scheme}\label{2.2-normalization}

The elimination of the fast angle $M$, in our variables, implies to
obtain a Hamiltonian that depends on $\xi$ and $\eta$ only through
powers of $\frac{\xi^2+\eta^2}{2} \, = \, \Gamma$
(Eq.~\ref{eq:Garfinkel-coord}). Such terms are said to be in \emph{normal
  form}.  Each step of the algorithm, based on Lie series, eliminates
terms depending on $M$ of certain order in two small
parameters: $\mu$ and a given combination of $\rho$,
$\xi$ and $\eta$.

Let us first introduce the following definition:
\begin{definition}\label{definition1}
A generic function $g=g(\rho,\xi,\tau,\eta)$ belongs to the class
${\cal P}_{l,s}\,$, if its expansion is of the type:
\begin{equation*}
\sum_{2m_1+m_2+m_3=l}
\ \sum_{{\scriptstyle{k_1+k_2\le l+4s-3}}\atop{\scriptstyle {j\le 2l+7s-6}}}
c_{m_1,m_2,m_3,k_1,k_2,j}\, \mu^s\,\rho^{m_1}\xi^{m_2}\eta^{m_3}
\,(\cos\tau)^{k_1}(\sin\tau)^{k_2}\,\big(\beta(\tau)\big)^j\ ,
\end{equation*}
where $c_{m_1,m_2,m_3,k_1,k_2,j}$ are real coefficients.
\end{definition}

Definition~\ref{definition1} classifies the terms appearing in the
algorithm according to their dependence on the powers $s$
and $l$ ruling the iteration step. The relation between the exponents
of the variables is a consequence of the application of Poisson
brackets of during the algorithm (see
Proposition~\ref{lem:alg-prop-Pois-brackets} below) and make the
formal algorithm consistent. Let $r_1$ and $r_2\,$ be two integer
counters, in the intervals $[1,R_1]$ and $[1,R_2]$ respectively, with
$R_1\,,\,R_2\in\mathbb{N}$ the fixed maximum orders of
normalization. Let us assume that the Hamiltonian expansion at the
$(r_1,r_2-1)$-th step is such that
{\small
\begin{equation}\label{eq:H(r1,r2-1)}
\vcenter{\openup1\jot\halign{
 \hbox {\hfil $\displaystyle {#}$}
&\hbox {\hfil $\displaystyle {#}$\hfil}
&\hbox {$\displaystyle {#}$\hfil}\cr
H^{(r_1,r_2-1)}(\rho,\xi,\tau,\eta) 
&= &\frac{\xi^2+\eta^2}{2}+
\sum_{l\ge 4}Z_l^{(0)}\Big(\rho,\frac{\xi^2+\eta^2}{2}\Big)
\cr
&+ &\sum_{s=1}^{r_1-1}\left(\sum_{l=0}^{R_2}
  \mu^s Z_l^{(s)}\Big(\rho,\frac{\xi^2+\eta^2}{2},\tau\Big)
+ \sum_{l>R_2}\mu^{s}f_l^{(r_1,r_2-1;s)}(\rho,\xi,\tau,\eta)\right)
\cr
&+ &\sum_{l=0}^{r_2-1}\mu^{r_1}
 Z_l^{(r_1)}\Big(\rho,\frac{\xi^2+\eta^2}{2},\tau\Big)
\,+\,\sum_{l\ge r_2}\mu^{r_1}f_l^{(r_1,r_2-1;r_1)}(\rho,\xi,\tau,\eta)
\cr
&+ &\sum_{s>r_1}\sum_{l\ge 0}\mu^sf_l^{(r_1,r_2-1;s)}(\rho,\xi,\tau,\eta)\ ,
\cr
}}
\end{equation}}
where $Z_l^{(0)}\in {\cal P}_{l,0}$ $\forall\ l\ge 4$, $Z_l^{(s)}\in
{\cal P}_{l,s}$ $\forall\ 0\le l\le R_2\,,\ 1\le s<r_1\,$,
$Z_l^{(r_1)}\in {\cal P}_{l,r_1}$ $\forall\ 0\le l<r_2\,$,
$f_l^{(r_1,r_2-1;r_1)}\in {\cal P}_{l,r_1}$ $\forall\ l\ge r_2\,$,
$f_l^{(r_1,r_2-1;s)}\in {\cal P}_{l,s}$ $\forall\ l>R_2\,,\ 1\le
s<r_1\,$ and $\forall\ l\ge 0,\ s>r_1\,$. In words, $H^{(r_1,r_2-1)}$
contains terms in normal form up to order $r_1$ in $\mu$, and order
$r_2-1$ in $l$, $l=2 m_1 +m_2 +m_2$, with $m_1$, $m_2$, $m_3$ the
exponents of $\rho$, $\xi$ and $\eta$ respectively. 
For greater orders, the dependence on $\xi$ and $\eta$ is polynomial
but arbitrary. 

The $r_1,r_2$-th normalization step requires to compute 
\begin{equation}\label{eq:def-funzionale-H(r1,r2)}
H^{(r_1,r_2)}=\exp\Big({\cal L}_{\mu^{r_1}\chi_{r_2}^{(r_1)}}\Big)H^{(r_1,r_2-1)}~,
\end{equation}
with the operator $\exp\Big({\cal
  L}_{\mu^{r_1}\chi_{r_2}^{(r_1)}}\Big) \cdot \,$, in
Eq.~\eqref{eq:Lie-exp-oper} and the generating function
$\chi_{r_2}^{(r_1)}$ to be determined. With this aim, we introduce the
following propositions:
\begin{proposition}\label{lem:sol_homol_eq}
If $Z_2^{(0)}=(\xi^2+\eta^2)/2$ and
$f_{r_2}^{(r_1,r_2-1;r_1)}\in {\cal P}_{r_2,r_1}\,$, then there exists a
generating function $\chi_{r_2}^{(r_1)}\in {\cal P}_{r_2,r_1}$ and a
normal form term $Z_{r_2}^{(r_1)}\in {\cal P}_{r_2,r_1}$  satisfying
  the homological equation
\begin{equation}\label{eq:chi(r1,r2-1)}
{\cal L}_{\chi_{r_2}^{(r_1)}}Z_2^{(0)}+f_{r_2}^{(r_1,r_2-1;r_1)}=Z_{r_2}^{(r_1)}\ .
\end{equation}
\end{proposition}
\begin{proposition}\label{lem:alg-prop-Pois-brackets}
Let $f$ and $g$ be two generic functions such that $f\in {\cal P}_{r,s}$
and $g\in {\cal P}_{r^{\prime},s^{\prime}}\,$, then
\begin{equation*}
{\rm if}\ r+r^{\prime}\ge 2\ \ \Rightarrow
\ \ \{f,g\} \in {\cal P}_{r+r^{\prime}-2,s+s^{\prime}}\ ,
\qquad
{\rm else}\ \ \Rightarrow\ \ \{f,g\} =0 ~.
\end{equation*}
\end{proposition}
We just sketch the procedure to follow so as to  determine a
solution of~\eqref{eq:chi(r1,r2-1)}. First, we replace the fast coordinates
$(\xi,\eta)$ with the complex conjugate canonical variables $(z,{\rm
i}{\overline z})$, such that $\xi=(z-{\overline z})/\sqrt{2}$ and
$\eta=(z+{\overline z})/\sqrt{2}$.  Since the kernel of the
homological equation $Z_2^{(0)}$ does not depend on slow coordinates
$(\rho,\tau)$, the Poisson bracket ${\cal L}_{\chi_{r_2}^{(r_1)}}Z_2^{(0)}$ 
does not affect them. Thus, we
expand the homological equation~\eqref{eq:chi(r1,r2-1)} in Taylor
series with respect to $(z,{\rm i}{\overline z})$, using $(\rho,\tau)$
as fixed parameters.  We solve term-by-term the
equation~\eqref{eq:chi(r1,r2-1)} in the unknown coefficients
$x_{m_1,m_2,m_3,k_1,k_2,j}$ and $\zeta_{m_1,m_2,m_2,k_1,k_2,j}$ such
that
{\small
\begin{equation*}
\vcenter{\openup1\jot\halign{
 \hbox {\hfil $\displaystyle {#}$}
&\hbox {\hfil $\displaystyle {#}$\hfil}
&\hbox {$\displaystyle {#}$\hfil}\cr
\chi_{r_2}^{(r_1)}(\rho,z,\lambda,{\rm i}{\overline z}) &=
&
\sum_{\scriptstyle{2m_1+m_2}\atop{\scriptstyle {+m_3=l}}}
\ \sum_{{\scriptstyle{k_1+k_2\le l+4r_1-3}}\atop{\scriptstyle {j\le 2l+7r_1-6}}}
x_{m_1,m_2,m_3,k_1,k_2,j}\,\rho^{m_1}z^{m_2}({\rm i}{\overline z})^{m_3}
\,(\cos\tau)^{k_1}(\sin\lambda)^{k_2}\,\big(\beta(\lambda)\big)^j
\cr
}}
\end{equation*}}
and
{\small
\begin{equation*}
\vcenter{\openup1\jot\halign{
 \hbox {\hfil $\displaystyle {#}$}
&\hbox {\hfil $\displaystyle {#}$\hfil}
&\hbox {$\displaystyle {#}$\hfil}\cr
Z_{r_2}^{(r_1)}(\rho,z,\tau,{\rm i}{\overline z}) &=
&
\sum_{\scriptstyle{2m_1+{\phantom{l}}}\atop{\scriptstyle {2m_2=l}}}
\ \sum_{{\scriptstyle{k_1+k_2\le l+4r_1-3}}\atop{\scriptstyle {j\le 2l+7r_1-6}}}
\zeta_{m_1,m_2,m_2,k_1,k_2,j}\,\rho^{m_1}(z\cdot {\rm i}{\overline z})^{m_2}
\,(\cos\tau)^{k_1}(\sin\tau)^{k_2}\,\big(\beta(\tau)\big)^j\ .
\cr
}}
\end{equation*}}
Finally, we express the expansions above by replacing
$(z,{\rm i}{\overline z})$ with $(\xi,\eta)$, and we obtain the final
solutions in the form
$\chi_{r_2}^{(r_1)}=\chi_{r_2}^{(r_1)}(\rho,\xi,\tau,\eta)$ and
$Z_{r_2}^{(r_1)}=Z_{r_2}^{(r_1)}\big(\rho,(\xi^2+\eta^2)/2,\tau\big)$.
Regarding Prop.~\ref{lem:alg-prop-Pois-brackets}, its proof just
requires long but basically trivial computations, thus it is
omitted.

By construction, the generating function $\chi_{r_2}^{(r_1)}$ satisfying
Prop.~\ref{lem:sol_homol_eq}, generates a Hamiltonian $H^{(r_1,r_2)}$,
through the transformation~\eqref{eq:def-funzionale-H(r1,r2)}, that is
in normal form up to order $(r_1,r_2)$, i.e.,
\begin{equation}\label{eq:H(r1,r2)}
\vcenter{\openup1\jot\halign{
 \hbox {\hfil $\displaystyle {#}$}
&\hbox {\hfil $\displaystyle {#}$\hfil}
&\hbox {$\displaystyle {#}$\hfil}\cr
H^{(r_1,r_2)}(\rho,\xi,\tau,\eta) &=
&\frac{\xi^2+\eta^2}{2}+
\sum_{l\ge 4}Z_l^{(0)}\Big(\rho,\frac{\xi^2+\eta^2}{2}\Big)
\cr
&+ & \sum_{s=1}^{r_1-1}\left(\sum_{l=0}^{R_2}
  \mu^s Z_l^{(s)}\Big(\rho,\frac{\xi^2+\eta^2}{2},\tau\Big)
  +\sum_{l>R_2}\mu^{s}f_l^{(r_1,r_2;s)}(\rho,\xi,\tau,\eta)\right)
\cr
&+ &\sum_{l=0}^{r_2}\mu^{r_1}
 Z_l^{(r_1)}\Big(\rho,\frac{\xi^2+\eta^2}{2},\tau\Big)
\,+\,\sum_{l>r_2}\mu^{r_1}f_l^{(r_1,r_2;r_1)}(\rho,\xi,\tau,\eta)
\cr
&+ &\sum_{s>r_1}\sum_{l\ge 0}\mu^sf_l^{(r_1,r_2;s)}(\rho,\xi,\tau,\eta)\ ,
\cr
}}
\end{equation}
where $Z_l^{(0)}\in {\cal P}_{l,0}$ $\forall\ l\ge 4$, $Z_l^{(s)}\in
{\cal P}_{l,s}$ $\forall\ 0\le l\le R_2\,,\ 1\le s<r_1\,$,
$Z_l^{(r_1)}\in {\cal P}_{l,r_1}$ $\forall\ 0\le l\le r_2\,$,
$f_l^{(r_1,r_2;r_1)}\in {\cal P}_{l,r_1}$ $\forall\ l> r_2\,$,
$f_l^{(r_1,r_2;s)}\in {\cal P}_{l,s}$ $\forall\ l>R_2\,,\ 1\le
s<r_1\,$ and $\forall\ l\ge 0,\ s>r_1\,$.  The additional terms
induced by the transformation of Eq.~\ref{lem:alg-prop-Pois-brackets}
correspond to higher orders in $l$ and $s$, and they are normalized in
subsequent steps of the algorithm.\footnote{It is straightforward to
prove that the above normalizing scheme corresponds to an algorithm of the
kind described in Sect.~\ref{sec:1.1.2-pert_theory} with the following
book-keeping rule:
\begin{bookrule}\label{bkp:rule-1}
To every monomial of the type 
\begin{equation*}
\mathsf{e}_{m_1,m_2,m_3,k_1,k_2,j} \mu^s
\,\rho^{m_1}\xi^{m_2}\eta^{m_3}
    \,\cos^{k_1}(\tau)\sin^{k_2}(\tau)\,\beta^j(\tau)\
\end{equation*}
there corresponds a book-keeping parameter of type $\lambda^{r(s,m_1,m_2,m_3)}$,
given by
\begin{equation*}
 r(s,m_1,m_2,m_3) = R_2(s-1)+ 2m_1 + m_2 + m_3~,     
\end{equation*}
\end{bookrule}
where $R_2$ is the maximum order of normalization for $l$.}

The Hamiltonian in~\eqref{eq:initial-Ham-rewritten} is in a
suitable form for starting the normalization step $r_1=r_2~=~1$
($H^{(1,0)}=H$). Thus, the whole algorithm consists of the
subsequent determination of the Hamiltonians
\begin{equation}\label{eq:norm_steps}
H^{(1,0)}=H,\ H^{(1,1)},\ \ldots\,,\ H^{(1,R_2)},\ \ldots\,,
\ H^{(R_1,0)},\ H^{(R_1,1)},\ \ldots\,,\ H^{(R_2,R_1)}~~,
\end{equation}
up to the maximum orders $R_1,\,R_2$.
Nevertheless, since this process is finite, the last computed Hamiltonian
\begin{equation}\label{eq:H(R1,R2)}
H^{(R_1,R_2)}(\rho,\xi,\tau,\eta) =
{\cal Z}^{(R_1,R_2)}\big(\rho,(\xi^2+\eta^2)/2,\tau \big)
+{\cal R}^{(R_1,R_2)}(\rho,\xi,\tau,\eta)~,
\end{equation}
which is in normal form up to orders $R_1$ and $R_2$, differs from the
original Hamiltonian by a remainder term ${\cal R}^{(R_1,R_2)}$.  The
best choice for the truncation orders $R_1$ and $R_2$ are those values
that make this remainder minimum. In practice, we encounter
computational limitations enforcing to fix \emph{a priori} the values
of $R_1,R_2$. In the sequel, we check that the truncation order is not
very inaccurate by numerical tests showing how well the normal form
${\cal Z}^{(R_1,R_2)}$ reproduces the dynamics of the original
Hamiltonian $H$.

\section{Numerical studies on the normal form}\label{2.3-numeric}

\subsection{Semi-analytical integration scheme}\label{2.3.X-intscheme}

Let us denote by
$\big(\rho^{(r_1,r_2)},\xi^{(r_1,r_2)},\tau^{(r_1,r_2)},\eta^{(r_1,r_2)}\big)$
the set of canonical coordinates introduced at the $(r_1,r_2)$--th
normalization step. Let $\varphi$ be
{\small
\begin{equation}\label{eq:coord-change-(r1,r2)} 
\varphi^{(r_1,r_2)} \big(\rho^{(r_1,r_2)},\xi^{(r_1,r_2)},
\tau^{(r_1,r_2)},\eta^{(r_1,r_2)}\big) =
   \exp\Big({\cal L}_{\mu^{r_1}\chi_{r_2}^{(r_1)}}\Big)
   \big(\rho^{(r_1,r_2)},\xi^{(r_1,r_2)},\tau^{(r_1,r_2)},\eta^{(r_1,r_2)}\big)~.
\end{equation}}
By the 'Exchange theorem' (Eq.~\ref{eq:lie-9}), we have that
{\footnotesize
\begin{equation}\label{eq:exchange-theorem}
 H^{(r_1,r_2)} \big(\rho^{{\scriptscriptstyle
     (r_1,r_2)}},\xi^{{\scriptscriptstyle
     (r_1,r_2)}}, \tau^{{\scriptscriptstyle
     (r_1,r_2)}},\eta^{{\scriptscriptstyle (r_1,r_2)}}\big)
     =H^{(r_1,r_2-1)}\Big(\varphi^{(r_1,r_2)} \big(\rho^{{\scriptscriptstyle
     (r_1,r_2)}},\xi^{{\scriptscriptstyle
     (r_1,r_2)}}, \tau^{{\scriptscriptstyle
     (r_1,r_2)}},\eta^{{\scriptscriptstyle
     (r_1,r_2)}}\big)\Big)~.
\end{equation}}
Thus, the whole
normalization procedure is described by the total canonical transformation
\begin{equation}\label{eq:def-total-canonical-transf}
{\cal C}^{(R_1,R_2)}=\varphi^{(1,1)}\circ\ldots\circ\varphi^{(1,R_2)}\circ
\varphi^{(2,1)}\circ\ldots
\circ\varphi^{(R_1,1)}\ldots\circ\varphi^{(R_1,R_2)} ~,
\end{equation}
connecting the normalized variables
$\big(\rho^{(R_1,R_2)},\xi^{(R_1,R_2)},\tau^{(R_1,R_2)},\eta^{(R_1,R_2)}\big)$, 
with the non normalized variables
$\big(\rho^{(0,0)},\xi^{(0,0)},\tau^{(0,0)},\eta^{(0,0)}\big)$.

With the total canonical transformation,
we provide the following semi-analytical integration scheme
{\small
\begin{equation}\label{semi-analytical_scheme}
\vcenter{\openup1\jot\halign{
  \hbox to 34 ex{\hfil $\displaystyle {#}$\hfil}
&\hbox to 11 ex{\hfil $\displaystyle {#}$\hfil}
&\hbox to 34 ex{\hfil $\displaystyle {#}$\hfil}\cr
\left(\rho^{(0,0)}(0),\xi^{(0,0)}(0),
\tau^{(0,0)}(0),\eta^{(0,0)}(0)\right)
&\build{\longrightarrow}_{}^{{{\scriptstyle
\big({\cal C}^{(R_1,R_2)}\big)^{-1}}
\atop \phantom{0}}}
&\left(\rho^{(R_1,R_2)}(0),\xi^{(R_1,R_2)}(0),
\tau^{(R_1,R_2)}(0),\eta^{(R_1,R_2)}(0)\right)
\cr
& &\big\downarrow \build{\Phi_{{\cal Z}^{(R_1,R_2)}}^{t}}_{}^{}
\cr
\left(\rho^{(0,0)}(t),\xi^{(0,0)}(t),
\tau^{(0,0)}(t),\eta^{(0,0)}(t)\right)
&\build{\longleftarrow}_{}^{{{\scriptstyle {\cal C}^{(R_1,R_2)}} \atop \phantom{0}}}
&\left(\rho^{(R_1,R_2)}(t),\xi^{(R_1,R_2)}(t),
\tau^{(R_1,R_2)}(t),\eta^{(R_1,R_2)}(t)\right)
\cr
}}
\qquad\qquad\ 
\end{equation}}
where $\Phi_{{\cal K}}^{t}$ denotes the flow induced by a generic
Hamiltonian ${\cal K}$ on the canonical coordinates, for an interval
of time equal to $t$. In words, we estimate the time evolution of the
non-normalized system by the study of the normalized system, provided
we perform the transformations accordingly between initial and final
conditions. This integration scheme provides just an {\it approximate}
solution, which is more accurate as smaller the remainder part ${\cal
  R}^{(R_1,R_2)}$ is with respect to ${\cal Z}^{(R_1,R_2)}$.

We note here that ${\cal Z}^{(R_1,R_2)}$ is {\it integrable} and its
flow is easy to compute. To this end, we introduce the temporary
action--angle variables $\big(\Gamma^{(R_1,R_2)},M^{(R_1,R_2)}\big)$,
such that $\xi^{(R_1,R_2)}=\sqrt{2\Gamma^{(R_1,R_2)}}\cos
M^{(R_1,R_2)}$ and $\eta^{(R_1,R_2)}=\sqrt{2\Gamma^{(R_1,R_2)}}\sin
M^{(R_1,R_2)}$. $\Gamma^{(R_1,R_2)}$ is a constant of motion for the
normal form ${\cal Z}^{(R_1,R_2)}={\cal
  Z}^{(R_1,R_2)}\big(\rho^{(R_1,R_2)},\Gamma^{(R_1,R_2)},\tau^{(R_1,R_2)}\big)$.
Thus, we compute $\rho^{(R_1,R_2)}(t)$ and $\tau^{(R_1,R_2)}(t)$ at
any time $t\,$ by the quadrature method, with $\Gamma^{(R_1,R_2)}$ as
a fixed parameter.  Regarding $M^{(R_1,R_2)}(t)$, we compute the
integral corresponding to the differential equation ${\dot
  M}^{(R_1,R_2)}=\frac{\partial\,{\cal
    Z}^{(R_1,R_2)}}{\partial\Gamma^{(R_1,R_2)}}\,$. Finally, the
values of $\xi^{(R_1,R_2)}(t)$ and $\eta^{(R_1,R_2)}(t)$ are obtained
from $\Gamma^{(R_1,R_2)}(t)$ and $M^{(R_1,R_2)}(t)$.  For practical
purposes, the application of the classical quadrature method can be
replaced by any numerical integrator. Due to the available computational
resources, in the numerical experiments that follow, the
maximum truncation powers are $R_1=3$, $R_2=5$, for $s$ and $l$
respectively.

\subsection{Numerical surfaces of section and semi-analytical level curves}\label{2.3.X-suf.vs.levelc}

\begin{figure}
 \includegraphics[width=.63\textwidth]{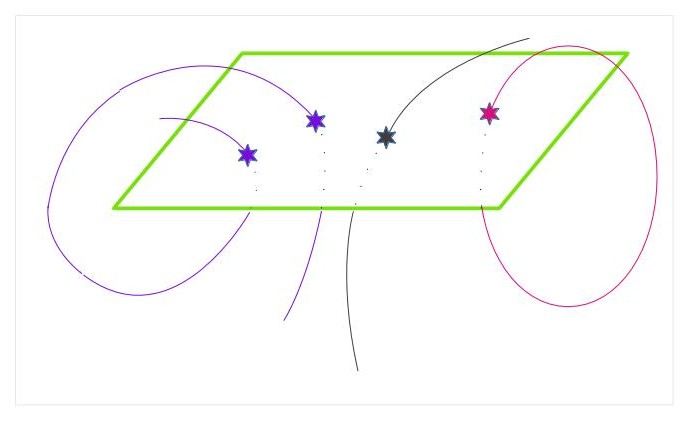} 
 \includegraphics[width=.37\textwidth]{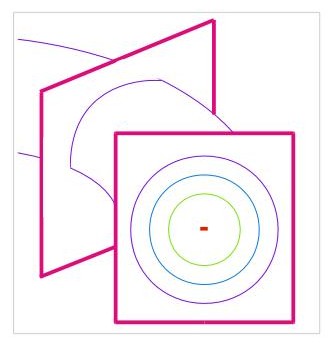} 
 \caption[Schematic Poincar\'e surface of section]{On the left panel,
   three different orbits (schematic) cross an arbitrary surface of
   section at the points denoted with stars. If the system is
   described in the action-angle phase space (right panel), the orbits
   lie on invariant tori. In a surface of section perpendicular to the
   flow, the quasi--periodic orbits look like closed curves (green,
   blue, purple circles), while the periodic orbits are isolated sets
   of fixed points (red).}
\label{fig:poincareSoS.jpg}
\end{figure}
The comparison between the
normalized Hamiltonian and the complete pCR3BP is done by means 
of numerical computations. Since the former is a system of 1 d.o.f.
and the latter is a system of 2 d.o.f., the illustration of the 
pCR3BP dynamics is made through phase portraits, i.e. so-called 
\emph{Poincar\'e surfaces of section}.

This method consists of retaining the points of the orbits that
intersects a particular surface, defined by a \emph{section
  condition}. A suitable (transverse to the flow) surface allows to
distinguish the dynamics of the problem, roughly discriminating
between periodic, quasi-periodic and chaotic orbits (see
Fig.~\ref{fig:poincareSoS.jpg}). In action-angle variables, the
quasi-periodic orbits lie on invariant tori
(Sec.~\ref{sec:act-ang-var}). Back transforming to the original
variables, the portrait under the full Hamiltonian resembles a
deformed version of the portrait under the normalized Hamiltonian,
with the addition of possible chaotic orbits.

\begin{SCfigure}
  \includegraphics[height=.40\textwidth]{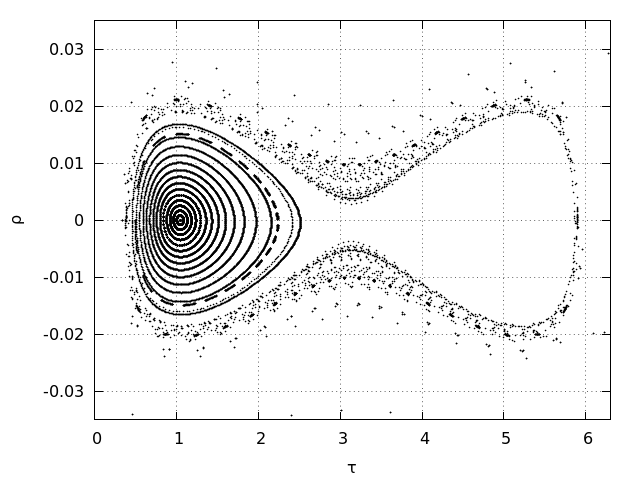}
 \caption[Numerical example of a surface of section]{An example of
   surface of section for the pCR3BP, for $\mu = 0.0005$.
\vspace{1cm}}
\label{fig:cirsurfsec.png}
\end{SCfigure}

In Fig.~\ref{fig:cirsurfsec.png} we show an
example of a \emph{pericenter} surface of section for the pCR3BP, defined by 
the condition
\begin{equation}\label{eq:peric-surf-sect}
M (\xi,\eta) = 0 \quad \mathrm{or,\:equivalently,} \quad \eta = 0 ~~.
\end{equation}

We obtain the phase portraits as follows. We fix the value of the mass
parameter and compute the value of the energy $H$
in~\eqref{eq:Garfinkel-Ham} for the point $L_4$ ($\tau =
\pi/3$, $\rho=\xi=\eta =0$). We define initial conditions on the
surface of section ($\eta = 0$), by varying $\tau$ and choosing $\xi$
such that $H$ is equal for all the initial conditions of the set. We
translate the initial conditions to cartesian variables
($u_{x,i},u_{y,i},\dot{u}_{x,i},\dot{u}_{y,i}$) and we integrate
Eqs.~\eqref{eq:iner_eq_comp_r3bp} with a Runge-Kutta
$7\>$-$\>8^{\>{\rm th}}$ order integrator, along 500 periods of the
primaries, with time-step equal to $2\pi/100$. During this integration
we collect the points contained in the surface of section and finally
the output data is again translated to Delaunay variables and plotted.
Since the surface of section condition corresponds to exactly one period
of the fast angle $M$, the portrait represents the dynamics of
the synodic d.o.f., as in the normal form ${\cal Z}^{(R_1,R_2)}$.
Thus, the orbits computed by the normal form and the phase portraits
are possible to compare.

The example presented in Fig.~\ref{fig:cirsurfsec.png}, for $\mu =
0.0005$, gives an estimation of how the libration regime of the pCR3BP
looks like. One of its most characteristic features is the large,
asymmetric variations in $\tau$. For this value of $\mu$, it is
possible to recover also some horseshoe orbits, librating around the
two equilateral points. A successful normal form must be able
to recover these properties of the Trojan motion. Hence, in the
experiments that follow, we test whether the normal form: i. efficiently
reproduces the large variations of the angle $\tau$, in particular for
orbits close to the border of stability, and ii. distinguishes between
tadpole and horseshoe orbits, in the cases where stable orbits exist
in both domains.

The first test is a graphical comparison between the orbits provided
for each Hamiltonian. The initial conditions used are fictitious, but
each set is derived from the catalogued position of a real Trojan
body, called the generating body. A whole portrait corresponds to a
value of the Jacobi constant equal to the one of the generating
body. We show below examples for two generating bodies, namely 2010
TK$_7$, Trojan asteroid of the Sun-Earth system and the asteroid 1872
Helenos of the Sun-Jupiter system.

From a catalogue, we obtain the coordinates (rotated to the plane of
primaries) of each generating body for a certain epoch, that we
transform to $(\rho_{gb},\tau_{gb},\xi_{gb},\eta_{gb})$. This initial
condition provides the Jacobi constant $C_{J_{gb}}$ for the body.  The
set of 10 initial conditions are generated by $\rho=\rho_{gb}$,
$\eta=\eta_{gb}$, a variable value for $\tau$ and $\xi$ such that
$C_J(\rho,\tau,\xi,\eta)= C_{J_{gb}}$ (isoenergetic orbits). These
orbits are numerically integrated for a short time, up to
accomplishing $M(\xi,\eta) = 0$. The final values used for the
integrations, gathered in $\mathscr{S}_{gb}$, lie on the surface of
section $M=0$.

We first compute the numerical surfaces of section of the pCR3BP for
the initial conditions in $\mathscr{S}_{gb}$ in the same way as in
Fig.~\ref{fig:cirsurfsec.png}. Passing now to the normal form, the
invariant curves of the numerical surface of section correspond to
level curves of ${\cal Z}^{(3,5)}$. To compute the latter, we first
translate the initial conditions of ${\cal S}_{gb}$ to normalized
variables, applying the inverse total canonical transformation $({\cal
  C}^{(3,5)})^{-1}$.  From ${\cal Z}^{(3,5)}$, we derive Hamilton's
equations for $\rho$ and $\tau$,
\begin{equation}\label{eq:haml-eqs}
\dot{\rho} = -\frac{\partial {\cal Z}^{(3,5)}}{\partial \tau} 
\qquad \dot{\tau}
= \frac{\partial {\cal Z}^{(3,5)}}{\partial \rho}~,
\end{equation}
which provide the flow induced by the normal form.  We numerically
integrate the normalized initial conditions, up to collecting about
2000 points, and keeping the relative energy error smaller than
$10^{-12}$. The resulting orbits lie on the level curves of the
integrable normal form ${\cal Z}^{(3,5)}(\rho,\Gamma,\tau)$
corresponding to the values $\Gamma=(\xi^2+\eta^2)/2$, given by
$\xi^{(3,5)}$, $\eta^{(3,5)}$ of each initial condition of ${\cal
  S}_{gb}$.  We complement for every point the values
$\xi=\sqrt{2\Gamma}$ and $\eta=0$ (equivalent to $M=0$).  Let us note
here that the use of numerically integrated orbits in order to obtain
the level curves of the normal form is only done for numerical
convenience. In principle, these curves are possible to obtain just
from the algebraic solution of the level curves equation ${\cal
  Z}^{(R_1,R_2)}=C$.  
Also, the condition $M=0$ in the normalized coordinates does not
correspond exactly to the surface of section $M=0$ in the original
variables. However, since ${\cal C}^{(R_1,R_2)}$ is by definition a
near-to-identity canonical transformation, we assume that the two
conditions do not differ too much and avoid computing the exact section
which involves quite cumbersome formul\ae. Finally, via ${\cal C}^{(3,5)}$,
we back-transform all the points of a level curve to the original
variables.

\subsubsection{Examples and results}
\begin{figure}
  \includegraphics[width=.99\textwidth]{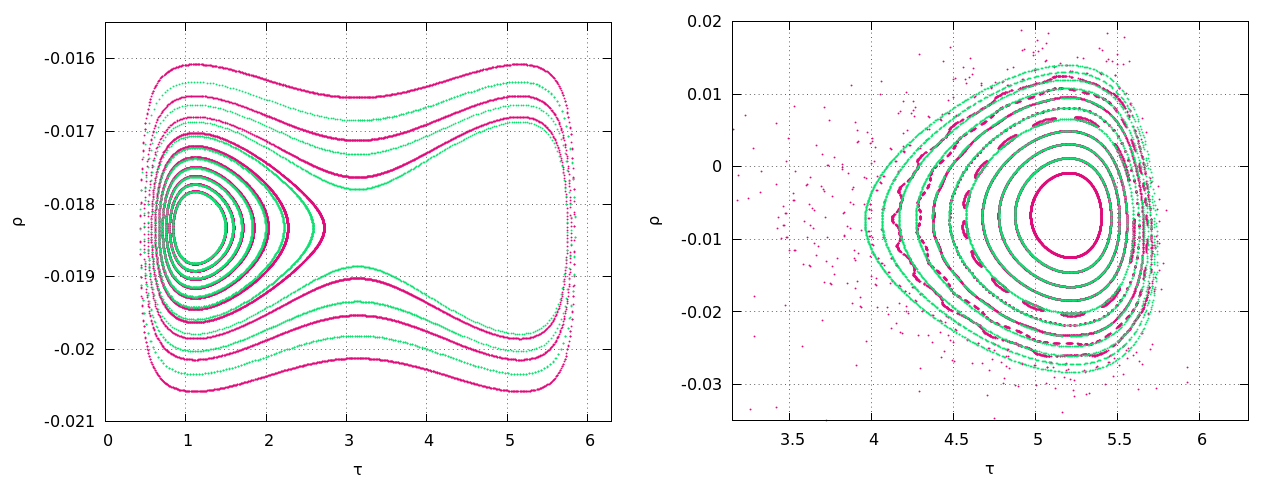}
  \caption[Comparison between level curves of the normal form and
    surfaces of section of the pCR3BP]{Comparison between the level
    curves produced by the averaged Hamiltonian in green and the
    points of the surface of section for the complete problem in pink,
    for the Sun-Earth problem (left panel) and Sun-Jupiter problem
    (right panel). In the Sun-Earth case, the generating body is the
    Earth Trojan 2010 TK$_7$. In the Sun-Jupiter case, the generating
    body is Trojan asteroid 1872 Helenos.}
  \label{fig:com_surf}
\end{figure}
\noindent
We choose two systems with very different values of the mass parameter
for a better contrast in the results.  The first case corresponds to
the Sun-Earth system, with $\mu=0.30003 \times 10^{-5}$.  The
generating body is the Earth's Trojan asteroid 2010 TK$_7$. We obtain its
coordinates from Jet Propulsion Laboratories JPL Ephemerides
Service\footnote{{\tt http://ssd.jpl.nasa.gov/?ephemerides}}, at epoch
2456987.5 JD (2014-Nov-26), that in Poincar\'e variables read
$\rho_{TK7} = -1.8401447 \times 10^{-2}$, $\tau_{TK7}
  = 3.5736334$, $\eta_{TK7} = 0.1152511$ and $\xi_{TK7} =
-0.1530054\,$.  The second case corresponds to the Sun-Jupiter system,
with $\mu=0.953855\times10^{-3}$. The generating body is the Trojan
asteroid 1872 Helenos, that librates around $L_5$.  The initial
conditions for Helenos are taken from Bowell Catalogue\footnote{ {\tt
    http://www.naic.edu/$\sim$nolan/astorb.html}}, at 2452600.5~JD
(2002-Oct-22), that in Poincar\'e variables read $\rho_{1872} =
-0.3836735\times10^{-2}$, $\tau_{1872} = 5.6716748$, $\eta_{1872} =
-0.0154266$ and $\xi_{1872} = -0.1104177\,$.

Figure~\ref{fig:com_surf} shows the comparison between the surface of
section and the level curves computed for the Sun-Earth system (left
panel), and Sun-Jupiter system (right panel). In both cases, the
points of the surface of section are shown in pink, while the level
curves of the normal form are shown in green.  In the case of the
Sun-Earth system, the agreement between the two representations is
very accurate. The averaged Hamiltonian reproduces accurately the
large variations of $\tau$ and distinguishes between tadpole and
horseshoe orbits. On the other hand, in the case of the Sun-Jupiter
system, due to a larger value of $\mu$, the system is substantially
more chaotic, a fact that the normal form cannot reproduce.
Nevertheless, the normal form is able to simulate well tadpole orbits
provided by the pCR3BP, as far as such orbits are not trapped in a
secondary resonance between the fast frequency and the synodic
libration frequency (see Sect.~\ref{sec:1.4.works_on_Trojans}
and~\ref{sec:3.X-secres}).

\subsection{Computation of quasi-actions}\label{sbs:comp_act}

As an additional test, we analyze some orbits that were not accurately
represented by normal forms in the past literature.
In~\cite{GabJorLoc-05}, the stability of some observed Jupiter's
Trojan asteroids was studied by constructing KAM tori solutions for
these asteroids.  Of the 34 initial conditions used
in~\cite{GabJorLoc-05}, the Kolmogorov normalization algorithm did not
work properly in seven cases. Here, we revisit the latter 7
initial conditions (1868~Thersites, 1872~Helenos,
2146~Stentor, 2207~Antenor, 2363~Cebriones, 2674~Pandarus and
2759~Idomeneus). These orbits either lie very close to the border of
stability or show an anomalous behavior with respect to the expected
tadpole orbit, so they represent an interesting test for our
new normal form.

The normal form ${\cal Z}^{(R_1,R_2)}$ contains two different actions
or integrals of motion. One, obtained by construction, is
$\Gamma$. The other, not explicitly obtained, is due to the fact that
the normal form is a 1 d.o.f. system.  This second constant of motion
is associated with the area enclosed by the level curves of $\tau$ and
$\rho$, and it should be reproduced by the normal form.  For the
comparison of the areas in the two Hamiltonian, we produce the
corresponding level curve as well as the numerical invariant curve for
each initial condition.  From the curves, we extract the maximum and
minimum values for $\rho$ and $\tau$, and we obtain the position of
the center for the curve provided by the averaged Hamiltonian
\begin{SCfigure}
  \centering
  \caption[Estimatives for the computation of quasi-actions]{Location
    of the maximum and minimum values for the coordinates $\rho$ and
    $\tau$ in the integration of the fictitious initial condition
    associated to 1872 Helenos, around L5. Green points correspond to
    the averaged level curve and pink points to the numerical surface
    of section.
    \vspace{1.5cm}}
  \includegraphics[width=0.6\textwidth]{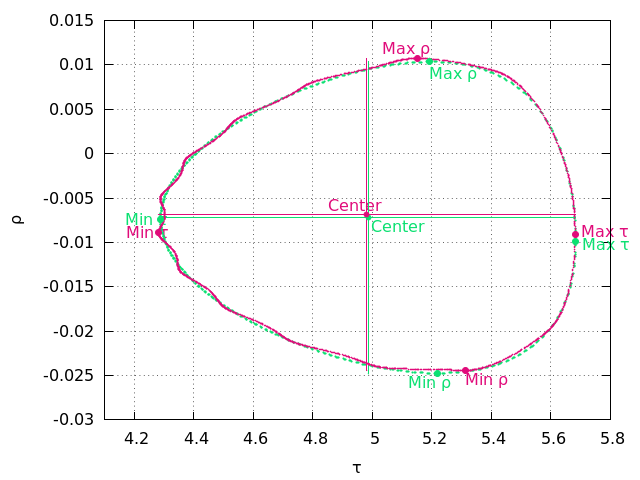}
  \label{fig:strg_guy}
\end{SCfigure}
\begin{equation}\label{eq:center-avrg}
\mathrm{C}_{\mathrm{avrg}} = \left(
\mathrm{C}_{(\tau,\mathrm{avrg})},
\mathrm{C}_{(\rho,\mathrm{avrg})} \right)= \left(
\frac{\left(\mathrm{Max}\,\tau_{\mathrm{avrg}} -
  \mathrm{Min}\,\tau_{\mathrm{avrg}} \right)}{2}, \frac{ \left(
  \mathrm{Max}\,\rho_{\mathrm{avrg}} -
  \mathrm{Min}\,\rho_{\mathrm{avrg}} \right) }{2} \right) ~~,
\end{equation}
and the center for the numerical orbit
\begin{equation}\label{eq:center-num}
\mathrm{C}_{\mathrm{num}} = \left(
\mathrm{C}_{(\lambda,\mathrm{num})}, \mathrm{C}_{(\rho,\mathrm{num})}
\right)= \left( \frac{\left(\mathrm{Max}\,\tau_{\mathrm{num}} -
  \mathrm{Min}\,\tau_{\mathrm{num}} \right)}{2}, \frac{ \left(
  \mathrm{Max}\,\rho_{\mathrm{num}} -
  \mathrm{Min}\,\rho_{\mathrm{num}} \right) }{2} \right) ~~.
\end{equation}
Fig.~\ref{fig:strg_guy} shows the curves, the extreme values
and the centers obtained in one of these cases. The discrepancy between
the two centers, which shows how much displacement there is between the
orbits, is computed by
\begin{equation}\label{eq:disp}
\delta \mathrm{C} = \frac{\left(\mathrm{C}_{(\rho,\mathrm{num})} -
  \mathrm{C}_{(\rho,\mathrm{avrg})}
  \right)}{\mathrm{C}_{(\rho,\mathrm{num})}} +
\frac{\left(\mathrm{C}_{(\tau,\mathrm{num})} -
  \mathrm{C}_{(\tau,\mathrm{avrg})}
  \right)}{\mathrm{C}_{(\tau,\mathrm{num})}} ~~.
\end{equation}

We re-express the points of each curve with respect to its center, by
introducing the quantities $\delta\tau=\tau-C_{\tau}$ and
$\delta\rho=\rho-C_{\rho}\,$. We obtain the distance to the center
$\mathrm{d} = \sqrt{\delta \rho^2 + \delta \tau^2}$ and the angle
$\theta$ with respect to the horizontal line ($ \theta = {\rm arctan}
(\delta\rho\,,\,\delta \tau)$). Re-ordering the points from small to
large values of $\theta$, we compute the triangular area generated by
two consecutive points and the center. The sum along all the triangles
represents the contained area within each curve, $A_{\mathrm{num}}$
for the complete system and $A_{\mathrm{avrg}}$ for the averaged
Hamiltonian flow\footnote{An alternative analytical way to estimate the
area is given by Green's formula}. For a further comparison, we compute also the
relative difference between the areas $\delta
A/A_{\mathrm{num}}=|A_{\mathrm{num}}-A_{\mathrm{avrg}}|/A_{\mathrm{num}}$,
and the displacement of the centers with respect to the position of
the corresponding equilateral Lagrangian point ($\rho = 0$, $\tau_{L4}=\pi/3$
for $L_4$ and $\tau_{L5}=5\pi/3$ for $L_5$).
\begin{table}
\centering
\begin{tabular}{|cccccc|}
\hline
Asteroid & $A_{\mathrm{num}}$ & $\delta A/A_{\mathrm{num}}$ &
$\delta \mathrm{C}$ & $\mathrm{C}_{(\rho,\mathrm{num})}$ &
$\mathrm{C}_{(\tau,\mathrm{num})}-\tau_{L4,L5}$ \\
\hline
\hline
1868 & $2.03\times 10^{-2}$ & $\, 3.21\times 10^{-3}$& $\, 6.95\times
10^{-3}$& $-1.08\times 10^{-2}$ & $-0.163$ \\
\hline
1872 & $3.75\times 10^{-2}$ & $\, 1.39\times 10^{-3}$& $\, 5.14\times
10^{-2}$ & $-6.86\times 10^{-3}$ & $-0.235$ \\
\hline
2146 & $1.67\times 10^{-2}$ & $\, 1.25\times 10^{-1}$& $\, 3.71\times
10^{-2}$ & $-1.94\times 10^{-1}$ & $-0.530$ \\
\hline
2207 & $2.31\times 10^{-2}$ & $\, 6.59\times 10^{-3}$& $\, 7.50\times
10^{-3}$& $-1.31\times 10^{-2}$ & $-0.196$ \\
\hline
2674 & $3.56\times 10^{-3}$ & $\, 1.51\times 10^{-2}$& $\, 3.61\times
10^{-3}$& $-1.43\times 10^{-2}$ & $-0.077$ \\
\hline
2759 & $2.67\times 10^{-2}$ & $\, 1.29\times 10^{-2}$& $\, 1.04\times
10^{-2}$& $-1.63\times 10^{-2}$ & $-0.232$ \\
\hline
\end{tabular}
\caption{Summary of the results for the quantities defining each
  averaged and numerical orbit}
\label{tab:table_data}
\end{table}

For six of the seven cases mentioned above, we obtain averaged orbits
that reflect the behavior of the numerical integrations. In
Table~\ref{tab:table_data} we report the results of the computations
described before.  The averaged areas clearly match their associated
numerical areas, with a relative error smaller than 2\%, except in one
case (asteroid 2146 Stentor), for which the error is about 13\%. This
may be due to the large displacement that its center has with respect
to the equilateral point. For the rest of the asteroids, the orbits on
the surface of section and the corresponding analytical level curves
agree. On the other hand, in Table 1 we do not present data for the
highly inclined ($39^{\circ}$) asteroid 2363~Cebriones. For this
asteroid, our normal form failed to provide an accurate orbit, using
the initial conditions provided in~\cite{GabJorLoc-05}. However, we
find that the numerical orbit generated by 2363~Cebriones presents a
very peculiar angular excursion (in $\tau$) with respect to the
Lagrangian point. Furthermore, the failure of the normal form could be
due to an imprecise evaluation of the initial condition, caused by a
non-consistent rotation to the plane of the primaries in the original
work~\cite{GabJorLoc-05}.

\chapter{The elliptic Trojan problem}\label{sec:3-ERTBP}

In the previous chapter we provided a new normalizing scheme that
allows to study the synodic libration of the Trojan orbits, in the
framework of the pCR3BP. Notwithstanding the degree of approximation
of this study, it is known that several aspects of the Trojan
orbits are ruled by the fact that the orbit of the primary is not
circular but elliptic.

In the present chapter, we revisit the main features of the planar ER3BP, by
means of a convenient Hamiltonian formalism where the three
d.o.f. appear well differentiated. This construction allows to
explicitely obtain the main frequencies of the motion, which interact
through different kinds of resonances.

By means of numerical experiments, we depict the resonance web in the
ER3BP in terms of stability maps and phase portraits. Our results from
a statistical study of escapes show that there exists a correlation
between escapes, sticky regions of the phase space and resonant
dynamics.


\section{Expansion in terms of modified-Delaunay variables}\label{sec:3.X-expan}

We start the construction of the Hamiltonian function
from~\eqref{eq:Ham_hel_RTBP_2} in Sect.~\ref{sec:1.3.3-ham_r3bp}: 
\begin{equation}
\begin{aligned}\label{eq:ini_ham_ertbp}
H_{ell} =&\, \frac{\|\mathbf{p}\|^2}{2} -\frac{{\cal
G}(m_0+m')}{\|\mathbf{r}\|} +\frac{{\cal G} m'}{\|\mathbf{r}\|} -
{\cal G} m' \left( \frac{1}{\Delta}
- \frac{\mathbf{r} \cdot \mathbf{r}'}{\|\mathbf{r}'\|^3} \right)~\\
=& \,\frac{\|\mathbf{p}\|^2}{2} -\frac{1}{\|\mathbf{r}\|} - {\cal G}
m' \left( \frac{1}{\Delta} - \frac{1}{\|\mathbf{r}\|}
-\frac{\mathbf{r} \cdot \mathbf{r}'}{\|\mathbf{r}'\|^3} \right)~,
\end{aligned}
\end{equation}
where $\Delta=\|\mathbf{r}-\mathbf{r}'\|$, $\mathbf{r}'$ and $\mathbf{r}$
are the heliocentric position vectors for the planet and for the massless body,
$\mathbf{p} = \dot{\mathbf{r}}$ and $\|\mathbf{r}\|$ is given
in~\eqref{eq:simply}.  However, since we now consider the elliptic
approximation, there holds that
\begin{equation}\label{eq:simply_3}
\|\mathbf{r'}\| = r'~,\quad \mathbf{r'}= (r \cos \theta', r \sin \theta')~,
\end{equation}
where $r'$ is the distance to the star and $\theta$, the
polar angle measured from the $x$-axis, accomplishes
$\theta = \varpi' + f'$, where $f'$ and $\varpi'$ are the
true anomaly and the longitude of the pericenter of the primary. 
With no other bodies
perturbing the motion of the planet, its ellipse is fixed in the space and
$\varpi'$ is constant in time. We expand each term of the disturbing
function as in the circular case,
\begin{equation}\label{eq:rdotrp_ellip}
\mathbf{r}\cdot \mathbf{r'}  = r r' \cos (\theta
- \theta') = r r' \cos \vartheta~,
\end{equation}
\begin{equation}\label{eq:delta_sol_ellip}
\Delta = \|\mathbf{r}-\mathbf{r}'\| = \sqrt{r^2 + r'^2 - 2r
r'\cos (\theta -\theta')} = \sqrt{r^2 +r'^2 - 2r r' \cos \vartheta}~. 
\end{equation}
where in both expressions $\vartheta = \theta - \theta'$. Substituting
Eqs.~\eqref{eq:simply}, \eqref{eq:simply_3}, \eqref{eq:rdotrp_ellip}
and~\eqref{eq:delta_sol_ellip} in Eq.~\eqref{eq:ini_ham_ertbp}, we obtain
\begin{equation}\label{eq:Ham_hel_RTBP_ellip}
H_{ell} = \frac{p^2}{2} - \frac{1}{r} - \mu \left( \frac{1}{\sqrt{r^2+r'^2-2\,r'\, r \cos\vartheta}} - \frac{1}{r} - \frac{r}{(r')^2} \cos \vartheta \right)~,
\end{equation}
where now also $r'$ is a function depending on the time.

We introduce, as first step, modified Delaunay
variables $(x,y,\lambda,\varpi)$, independent of the mass parameter
$\mu$ and given by
\begin{equation}
\begin{aligned}\label{eq:mod_delau_var1}
x =&\sqrt{a}-1~, &\lambda~, \qquad \\
y =&\sqrt{a}\left(\sqrt{1-e^2}-1\right)~, &\varpi~, \qquad
\end{aligned}
\end{equation}
where $\lambda$, $\varpi$, $a$ and $e$ are the mean longitude, the
longitude of the pericenter, major semiaxis and eccentricity of the
orbit of the Trojan body. The primed symbols correspond to the orbital
elements of the planet.  Considering 
$M = \lambda - \varpi$, 
$M' = \lambda' - \varpi'$ (see Sec.~\ref{sec:1.2.2-orb-elem}) and 
\begin{equation}\label{e_func_xandy}
e = \sqrt{1 - \left(1 + \frac{y}{x+1} \right)^2}~
\end{equation}
(Eq.~\ref{eq:mod_delau_var1}), the expansions for the true anomaly are
given by
\begin{equation}
\begin{aligned}\label{eq:cosv-series-expan-ellip}
\cos f &= -e +
\frac{2\big(1-e^2\big)}{e}\sum_{n=1}^{\infty}\big[J_n(ne)\cos(nM)\big] =
 \mathfrak{f}_{c}(x,y,\lambda,\varpi)~,\\
\sin f &= 2\sqrt{1-e^2}\sum_{n=1}^{\infty}\big[J_{n}^{\prime}(ne)\sin(nM)\big] = 
\mathfrak{f}_{s}(x,y,\lambda,\varpi)~,
\end{aligned}
\end{equation}
where $J_n$, $J_{n}^{\prime}\,$ are Bessel functions of the first kind
and their derivatives, given in
Eq.~\eqref{eq:def-Bessel-Jn-and-derivative-Chap2}.  Similarly, we
obtain $\cos f'= \mathfrak{f}'_{c}(x',y',\lambda',\varpi')$ and $\sin
f'= \mathfrak{f}'_{s}(x',y',\lambda',\varpi')$.  According to the
units of measure defined in Sect.~\ref{sec:1.3.2-rot_fram_Lag},
$x'=0$, while $\varpi'$ and $y'$ (through $e'$) act as
parameters. Thus, the expansions for the true anomaly of the primary
are
\begin{align}
\cos f'=  \mathfrak{f}'_{c}(\lambda';e',\varpi')~, \label{eq:cosvprim_series_ellip}\\
\sin f'=  \mathfrak{f}'_{s}(\lambda';e',\varpi')~. \label{eq:sinvprim_series_ellip}
\end{align}
On the other hand, by the definition of the actions
 in Eq.~\eqref{eq:mod_delau_var1}, we have
\begin{equation}
r = \frac{a(1 -e^2)}{1+e \cos f} = \frac{(x+y)^2}{1+e \cos f} 
\end{equation}
whereby, after replacing Eqs.~\eqref{eq:cosv-series-expan-ellip},
 we obtain
\begin{equation}\label{eq:r_ellip_series}
r = \mathfrak{f}_{r}(x,y,\lambda,\varpi)~,
\end{equation}
and
\begin{equation}\label{eq:r_prim_ellip_series}
r' = \mathfrak{f}'_{r}(\lambda';e',\varpi')~.
\end{equation}
Considering Eq.~\eqref{eq:r_ellip_series}, 
the series for $1/r$ in
terms of ($x,y,\lambda,\varpi$) is obtained as in
Chapter~\ref{sec:2-CRTBP}. 

Regarding $\cos \vartheta$, we have
\begin{equation}
\vartheta = \theta - \theta' = \varpi+ f - \varpi' - f'~,
\end{equation}
hence,
\begin{equation}
\begin{aligned}\label{eq:cos_varth_long}
\cos \vartheta = &\, \cos\varpi \cos f \cos \varpi' \cos f' 
- \cos\varpi \cos f \sin \varpi' \sin f' \\ 
& - \, \sin\varpi \sin f \cos\varpi' \cos f' + \sin\varpi \sin
f \sin \varpi' \sin f'\\ 
& + \, \sin\varpi \cos f \sin \varpi' \cos f' + \sin\varpi \cos f 
\cos \varpi' \sin f'\\ 
& + \, \cos\varpi \sin f \sin \varpi' \cos f' + \cos\varpi \sin f 
\cos \varpi' \sin f'~. 
\end{aligned}
\end{equation}
We then replace $\mathfrak{f}_c$, $\mathfrak{f}_s$, $\mathfrak{f}'_c$,
$\mathfrak{f}'_s$ into Eq.~\eqref{eq:cos_varth_long} and
obtain
\begin{equation}\label{eq:cos-varthet-ellip-series}
\cos \vartheta = \,
\mathfrak{f}_{\vartheta}(x,y,\lambda,\varpi,\lambda';e',\varpi')~.
\end{equation}

Gathering the previous expressions, we construct the disturbing
function in terms of modified Delaunay variables. The expansions up to
the second order in the eccentricities are presented in
Appendix~\ref{sec_app:expan_del_ellip}.  The
Hamiltonian~\eqref{eq:ini_ham_ertbp} in the new variables reads
\begin{equation}\label{ham3}
H_{ell} = -{1\over 2(1+x)^2} - \mu R(x,y,\lambda,\varpi,\lambda';e',\varpi')~,
\end{equation}
where $\lambda'=n t$. As shown below, in computing proper elements there turns
to be crucial to remove the dependence of the Hamiltonian on time by
introducing a `dummy' action variable $I$ conjugate to $\lambda'$,
namely
\begin{equation}\label{ham4}
H_{ell}=-{1\over 2(1+x)^2} + I - \mu R(x,y,\lambda,\varpi,\lambda';\varpi',e')~~.
\end{equation}
The present expression of the Hamiltonian corresponds to an autonomous
system of 3 d.o.f, while \eqref{ham3} is a $2+\frac{1}{2}$ d.o.f. system.

For the study of the Trojan dynamics, we define two new
angles, namely $\tau=\lambda-\lambda'$ and $\delta\varpi
= \varpi-\varpi'$.  The angle $\tau$ is the resonant angle
corresponding to the 1:1 MMR resonance, with value $\tau=\pi/3$ at the
Lagrangian point $L_4$. The angle $\delta\varpi$
expresses the relative position of the pericenter of the Trojan body from
the pericenter of the planet.  We introduce these new angles 
through a generating function ${\cal S}_2$
depending on the old angles ($\lambda$, $\lambda'$, $\varpi$) and the
new actions ($X_1$, $X_2$, $X_3$),
\begin{equation}\label{gensec}
{\cal S}_2=(\lambda-\lambda')X_1+\lambda'X_2+(\varpi-\varpi')X_3~,
\end{equation}
yielding the following transformation rules 
\begin{equation}
\begin{aligned}
 &\tau = \frac{\partial {\cal S}_2}{\partial X_1} = \lambda
 - \lambda'~,~~ &\tau_2& = \frac{\partial {\cal S}_2}{\partial X_2}
 = \lambda'~,~~ &\delta\varpi = \frac{\partial {\cal S}_2}{\partial
 X_3} = \varpi - \varpi'~,~~ \\ & x = \frac{\partial {\cal
 S}_2}{\partial \lambda} = X_1~,~~ & I & = \frac{\partial {\cal
 S}_2}{\partial \lambda'} = X_2 - X_1~,~~ & y = \frac{\partial {\cal
 S}_2}{\partial \varpi} = X_3~. \quad \qquad
\end{aligned}
\end{equation}
Note that preserving the canonical character of the variables requires
some modification of the dummy action variables as well.  We
keep the old notation for all variables involved in a identity
transformation ($X_1=x$, $\tau_2=\lambda'$, $X_3=y$). The Hamiltonian
then reads:
\begin{equation}\label{hamsec2}
H_{ell} = -{1\over 2(1+x)^2} -x + X_2  - \mu R(x,y,\tau,\delta\varpi,
\lambda';e',\varpi')~.
\end{equation}
This expression can be recast under the form:
\begin{equation}\label{hamsec3}
H_{ell} = \langle H \rangle + H_1
\end{equation}
where
\begin{equation}\label{eq:Haverg}
\langle H \rangle =-{1\over 2(1+x)^2} -x + X_2  -\mu \langle R \rangle(\tau,\delta\varpi,x,y;e',\varpi')
\end{equation}
and 
\begin{equation*}
H_1=  -\mu \tilde{R}(\tau,\delta\varpi,x,y,\lambda';e',\varpi')~,
\end{equation*}
with 
\begin{equation*}
\langle R \rangle ={1\over 2\pi}\int_{0}^{2\pi}R \df \lambda',~~~~\tilde{R}=R- \langle R \rangle~~.
\end{equation*}
The action $X_2$ is an integral of motion under the Hamiltonian flow
of $\langle H \rangle$. Thus, the Hamiltonian $\langle H \rangle$
represents a system of two d.o.f.  We call \emph{position of the forced equilibrium} $(\tau_0,\delta\varpi_0,x_0,y_0)$ the solution of the system
of equations
\begin{equation}\label{eq:force_equi_eq}
\dot{\tau}={\partial \langle H \rangle \over\partial x}=0,~~~
\dot{\delta\varpi}={\partial \langle H \rangle\over\partial y}=0,~~~
\dot{x}=-{\partial \langle H \rangle\over\partial\tau}=0,~~~
\dot{y}=-{\partial \langle H \rangle\over\partial \delta\varpi}=0~~~.
\end{equation}
We find
\begin{equation}\label{forced}
(\tau_0,\delta\varpi_0,x_0,y_0)= \big( \pi/3,\pi/3,0,\sqrt{1-e'^2}-1 \big)~~.
\end{equation}
From~\eqref{forced}, we can deduce that the equilibrium 
point is not given by a fixed point in the synodic frame of reference,
as it happens in the circular case. 
In particular, since $y_0 \neq 0$, a Trojan body with elements deduced
from~\eqref{forced} describes a fixed ellipse of eccentricity $e=e'$ in the
inertial frame. Thus, the body describes a short-period epicyclic loop
around $L_4$ in the synodic frame.

\section{The motion around the forced equilibrium and the three temporal scales}\label{sec:3.X-forced_equil}

We now introduce local action-angle variables around the point of
forced equilibrium. The purpose is to characterize the motion by two
approximate constants, one of which appears as an action variable
($J_s$) on the plane $(x,\tau)$ around the value $(x_0,\tau_0)$, while
the other appears as an action variable ($Y_p$) on the plane
$(y,\delta\varpi)$ around the value $(y_0,\delta\varpi_0)$.  To this
end, we introduce the `shift of center' canonical transformation given
by:
\begin{equation}\label{poincvar}
v=x-x_0,~~u=\tau-\tau_0,~~Y=-(W^2+V^2)/2,~~\phi=\arctan(V,W) 
\end{equation}
where
\begin{equation*}
V=\sqrt{-2y}\sin\delta\varpi-\sqrt{-2y_0}\sin\delta\varpi_0,~~~~
W=\sqrt{-2y}\cos\delta\varpi-\sqrt{-2y_0}\cos\delta\varpi_0~~,
\end{equation*}
where $Y$ is defined negative so as to keep the canonical structure
with respect to $\phi$.
Re-organising terms, the Hamiltonian \eqref{hamsec2} takes the form:
{\small
\begin{equation}\label{hamsec4}
H_{ell} = -{1\over 2(1+v)^2} -v + X_2 - \mu\left({\cal F}^{(0)}(u,\lambda'-\phi,v,Y;e',\varpi')+
{\cal F}^{(1)}(u,\phi,\lambda',v,Y;e',\varpi') \right)
\end{equation}}
where ${\cal F}^{(0)}$ contains terms depending on the angles
$\lambda'$ and $\phi$ only through the difference $\lambda'-\phi$, and
${\cal F}^{(1)}$ contains terms dependent on non-zero powers of
$e'$. This splitting suggests to perform one more canonical
transformation, through which the
part of the Hamiltonian corresponding to ${\cal F}^{(0)}$ can be formally
reduced to a system of 2 d.o.f.:
\begin{equation}\label{gencirc}
{\cal S}_3(u,\lambda',\phi,Y_u,Y_s,Y_p) = u Y_u + (\lambda'-\phi) Y_f
+ \phi Y_p
\end{equation}
yielding
\begin{equation}
\begin{aligned}
&\phi_u= \frac{\partial {\cal S}_3}{\partial Y_u} = u~,~~ 
&\phi_f&=\frac{\partial {\cal S}_3}{\partial Y_f} = \lambda'-\phi~,~~ 
&\phi_p= \frac{\partial {\cal S}_3}{\partial Y_p} = \phi~,~~\\
& v= \frac{\partial {\cal S}_3}{\partial u} = Y_u~,~~ 
& X_2 &= \frac{\partial {\cal S}_3}{\partial \lambda'} = Y_f~,~~ 
& Y=\frac{\partial {\cal S}_3}{\partial\phi}=Y_p-Y_f~.
\end{aligned}
\end{equation}
The subscripts `f' and `p' stand for `fast' and `proper' respectively,
for reasons we explain later. As before, we keep the old notation for
the variables transforming by the identities $\phi_u=u,\phi_p=\phi$,
and $Y_u=v$. However, it turns to be convenient to retain the new
notation for the action $Y_f\equiv X_2$.  The
Hamiltonian~\eqref{hamsec4} in the new canonical variables reads
{\small
\begin{equation}\label{eq:hamYfYsYp}
H_{ell} = -\frac{1}{2(1+v)^2} -v + Y_f -\mu{\cal
F}^{(0)}(v,Y_p-Y_f,u,\phi_f;e',\varpi') -\mu{\cal
F}^{(1)}(v,Y_p-Y_f,u,\phi_f,\phi;e',\varpi')~~~.
\end{equation}}
Collecting terms
linear in $(Y_p-Y_f)$, we find:
\begin{equation}\label{freqs}
\omega_f\equiv\dot{\phi_f} = \frac{\partial H_{ell}}{\partial Y_f} = 
1-{27\mu/8}+ {\cal O}(\mu^2)...,~~~
g\equiv\dot{\phi}  = \frac{\partial H_{ell}}{\partial Y_p} = 
{27\mu/8}+ {\cal O}(\mu^2)...~~~
\end{equation}
We identify $\omega_f$ and $g$ as the short-period and secular
frequencies, respectively, of the Trojan body. Comparing Eq.~\eqref{freqs}
with Eq.~\eqref{eq:lambdas_approx}, we see that the fast
(short-period) frequency recovers the same value as in 
the CRTBP.  Therefore, the set of variables constructed in~\eqref{gencirc}
allows to separate the 
three time-scales of the 3 d.o.f. The main contributions to the
frequencies come from the Keplerian part and ${\cal F}^{(0)}$, because
$\mu \,{\cal F}^{(1)}$ provides terms of at least first order in
$\mu\,e'$, where both the mass parameter $\mu$ and the eccentricity $e'$
are small parameters.

The above decomposition of the Hamiltonian allows to
consider various `levels' of perturbation. We call \emph{basic
model} the one of Hamiltonian
\begin{equation}\label{hambasic}
H_b=-{1\over 2(1+v)^2} -v + Y_f -\mu{\cal  F}^{(0)}(v,Y_p-Y_f,u,\phi_f;e',\varpi')~~~.
\end{equation}
The total Hamiltonian takes the form $H_{ell}=H_b+H_{sec}$, where
\begin{equation}\label{eq:Hsec_ER3BP}
H_{sec}= -\mu{\cal F}^{(1)}(v,Y_p-Y_f,u,\phi_f,\phi;e',\varpi')
\end{equation}
contains terms of at least order ${\cal O}(e'\,\mu)$. Since $\phi$ is
ignorable, $Y_p$ is an integral of the motion for the flow induced by
the Hamiltonian (\ref{hambasic}). The physical importance of $Y_p$ can
be understood as follows: the action variable $Y$ measures the radial
distance from the point of forced equilibrium in the plane $(V,W)$, in
which the forced equilibrium is located at the origin. In a first
approximation, the quasi-integral of the proper eccentricity can be
defined as
\begin{equation}\label{epr}
e_{p,0}=\sqrt{V^2+W^2}=\sqrt{-2Y}~~.
\end{equation}
\begin{figure}
\begin{center}
\includegraphics[width=1.01\textwidth,angle=0]{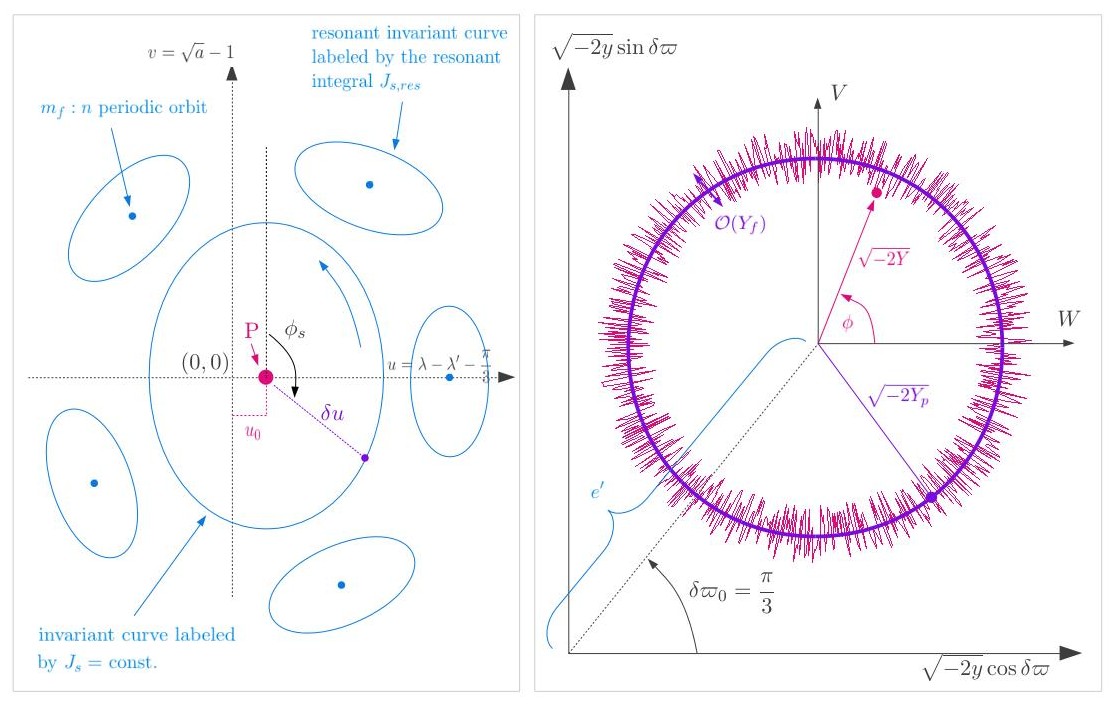}
\end{center}
\caption[Schematic representation of the physical meaning of the
  action-angle variables introduced for the $H_b$]{{\small Schematic
    representation of the physical meaning of the action-angle
    variables introduced in Sec~\ref{sec:3.X-forced_equil}.  The plane
    $(u,v)$ corresponds to the `synodic' motion of the Trojan
    body. Under the Hamiltonian $H_b$, the phase portrait can be
    represented by a Poincar\'{e} surface of section corresponding,
    e.g., to every time when the angle $\phi_f$ accomplishes a full
    cycle. The left panel shows schematically the form of the
    projection of this section on the plane $(u,v)$. The central point
    P represents a stable fixed point corresponding to the
    short-period periodic orbit around L4. The orbit has frequency
    $\omega_f$, while its amplitude increases monotonically with
    $Y_f$. The forced equilibrium corresponds to $u_0=0$, $Y_f=0$. The
    point P, however, has in general a shift to positive values
    $u_0>0$ for proper eccentricities larger than zero (see
    below). Far from resonances, the invariant curves around P are
    labeled by a constant action variable $J_s$, and its associated
    angle (phase of the oscillation) $\phi_s$. Resonances, and their
    island chains correspond to rational relations between the
    frequencies $\omega_f$ and $\omega_s$. Within the resonant
    islands, $J_s$ is no longer preserved, but we have, instead, the
    preservation of a resonant integral $J_{s,res}$. This integral
    will be computed in Chapter 5. On the other hand, the plane
    $(W,V)$ (right panel) depicts the evolution of the Trojan body's
    eccentricity vector under the Hamiltonian $H_b$. This panel is a
    more precise version of the right panel of
    Fig.~\ref{fig:paradoxeccvec.png}. The motion of the endpoint of
    the eccentricity vector can be decomposed to a circulation around
    the forced equilibrium, with angular frequency $g$, and a fast (of
    frequency $\omega_f$) `in-and-out' oscillation with respect to a
    circle of radius $e_p$, of amplitude which is of order ${\cal
      O}(Y_f)$. Under $H_b$ alone, the quantities $Y_p,J_s$, or
    $Y_p,J_{s,res}$ are quasi-integrals for all the regular
    orbits. Those quasi-integral allow to define 'proper elements' as
    in Sect.~\ref{sec:1.4.works_on_Trojans}. Furthermore, all extra
    terms with respect to $H_b$ in the Hamiltonian (\ref{hamsec2})
    depend on the slow angles $\phi$. Thus, all these terms can only
    slowly modulate the dynamics under $H_b$, and this modulation can
    produce a long-term drift of the values of $(Y_p,J_s)$, or
    $Y_p,J_{s,res}$ (see numerical experiments below). The drift can
    lead to large long term variations of the actions, and eventually
    to the escape of a Trojan body. In Chapter 4, we show that $H_b$
    is formally identical in the ERTBP and in a more general model
    called the Restricted Multi-Planet Problem. We conclude that the
    basic features of dynamics induced by $H_b$ apply in the same way
    with or without additional planets.}}
\label{fig:figure1.jpg}
\end{figure}
However, the above definition neglects the fact that $Y$ is subject 
to fast variations due to its dependence on $Y_f$. In fact, by 
Hamilton's equations we readily find that $\dot{Y}_f= 
{\cal O}(\mu)$. The time variation of $Y_f$ is associated 
to a fast frequency ${\dot{\phi_f}=1-g}$. In fact, by their 
definition we can see that the variables $(\phi_f,Y_f)$ describe 
epicyclic oscillations of the Trojan body, i.e. $\phi_f$ accomplishes 
one cycle every time when the Trojan body passes through a local 
pericenter. The time 
variations of $Y_f$ become particularly important when one of the 
following two conditions holds: i) $e'<\mu$, or ii) the orbit of 
the Trojan body is subject to a low-order resonance. On the other 
hand, $Y_p$ remains an exact 
integral of the Hamiltonian~\eqref{hambasic} even in the cases 
(i) or (ii). We thus adopt the following 
definition of the proper eccentricity:
\begin{equation}\label{eprnew}
e_p = \sqrt{-2Y_p}~~.
\end{equation}
In the Hamiltonian~\eqref{hambasic}, the integral $Y_p$ (or $e_p$)
becomes a label of a system of two degrees of freedom corresponding to
the canonical pairs $(u,v)$ and $(Y_f,\phi_f)$.  Since the function
${\cal F}^{(1)}$ contains terms of at least first order in $e'$,
$\dot{Y_p}={\cal O}(e')$ under the full
Hamiltonian~\eqref{eq:hamYfYsYp}. This implies that $Y_p$ (or $e_p$)
remains a good quasi-integral for not very high values of the
primary's eccentricity. On the other hand, a more accurate (${\cal
O}(e'^2)$) quasi-integral can be computed by a first order adiabatic
theory \cite{Pucacco-96}.  Figure~\ref{fig:figure1.jpg}
summarizes the physical meaning of the various action-angle variables
$(\phi_f,u,\phi,Y_f,v,Y_p)$. We emphasize that in
numerical computations one always stays with the original (Cartesian)
co-ordinates of the various bodies. Then, translation of the results
to action-angle variables and vice-versa is straightforward, passing
first to Delaunay elements, and then using the
transformations~\eqref{gensec}, \eqref{poincvar}, and~\eqref{gencirc}.
The functions ${\cal F}^{(0)}$ and ${\cal F}^{(1)}$, with an error
${\cal O}(x) \approx {\cal O}(\mu^{1/2})$, are given in the 
Appendix~\ref{appex:theHb}.

A second averaging over the fast angle
$\phi_f$ yields the Hamiltonian
\begin{equation}\label{hambasic3}
\overline{H_b}(u,v;Y_f,Y_p,e',\varpi')=- \frac{1}{2(1+v)^2}-v + Y_f 
-\mu\overline{{\cal F}^{(0)}}(u,v,Y_p-Y_f;e',\varpi')~~
\end{equation}
with
\begin{equation*}
\overline{{\cal F}^{(0)}}=\frac{1}{ 2\pi}\int_{0}^{2\pi}{\cal F}^{(0)} \df \phi_f~~.
\end{equation*}
In Chapter 4, we compute a more precise averaged model
$\overline{H_b}$ by means of a normalizing scheme similar to the one
of Chapter 2, but implemented in the ER3BP. For the moment, we just
focus on some basic properties of $\overline{H_b}$. The Hamiltonian
$\overline{H_b}(u,v;Y_f,Y_p,e',\varpi)$ represents a system of one
degree of freedom, all three quantities $Y_f,Y_p,e'$ serving now as
parameters, i.e. constants of motion under the dynamics of
$\overline{H_b}$. The Hamiltonian $\overline{H_b}$ describes the
synodic (guiding-center) motions of the Trojan body, with the
additional point that, since it depends on $e'$, it {\it does not}
correspond to the averaged (over fast angles) Hamiltonian of the
circular RTBP. From a physical point of view, this expresses the
possibility to find an integrable approximation to synodic motions
even when $e'\neq 0$.

The equilibrium point $(u_0,v_0)$ given by
\begin{equation*}
{\partial \overline{{\cal F}^{(0)}}\over\partial u}=
{\partial \overline{{\cal F}^{(0)}}\over\partial v}=0
\end{equation*}
corresponds to a short-period periodic orbit of the Hamiltonian $H_b$ 
around the forced equilibrium point. We define the action variable 
\begin{equation}\label{synodic}
J_s={1\over 2\pi}\int_C (v-v_0) \df (u-u_0)
\end{equation}
where the integration is over a closed invariant curve $C$ around
$(u_0,v_0)$ and `s' stands for `synodic' (see
Fig.~\ref{fig:figure1.jpg}). The angular variable $\phi_s$, conjugate
to $J_s$, evolves in time according to the synodic frequency
$\omega_s$ given by (see Eq.~\ref{omesyn})
\begin{equation}
\omega_s=\dot{\phi}_s= - \sqrt{27\mu\over 4}+... 
\end{equation}

Some manipulation of Eq.~\eqref{hambasic3} 
allows to find a first order approximation to the values of the
frequencies $\omega_s$ and $g$.
We deduce the shift in position, with respect to
L4, of the fixed point of $\overline{{\cal F}^{(0)}}$, corresponding
to the short-period orbit around L4 \cite{NamMurr-00}. 
The shift is given by $u_0=\tau_0-\pi/3$, where $\tau_0$ is the 
solution of ${\partial\overline{{\cal F}^{(0)}}/\partial\tau}=0$. 
We find:
\begin{equation}\label{u0}
u_0={29\sqrt{3}\over 24}e_{p,0}^2+... 
\end{equation}
where the error is of order 4 in the eccentricities $e_{p,0}$, $e'$.

We introduce the following canonical transformation, to analyze the motion
around the position of the periodic orbit given by $u_0,v_0$
\begin{equation}\label{eq:genedeltau}
{\cal S}_4(\phi_f,u,\phi,V,J_f,J_p) = (u - u_0) V + \phi_f \, J_f + \phi J_p~,
\end{equation}
where, in terms of the new actions, we have 
$u_0= \frac{29 \sqrt{3}}{12}\,(J_f-J_p)$, yielding
\begin{equation}
\begin{aligned}
& v= \frac{\partial {\cal S}_4}{\partial u} = V~,~~ 
&Y_f&=\frac{\partial {\cal S}_4}{\partial \phi_f} = J_f~,~~ 
&Y_p= \frac{\partial {\cal S}_4}{\partial \phi_p} = J_p~,~~\\
& \delta u = \frac{\partial {\cal S}_4}{\partial V} = u - u_0~,~~ 
& q_f &= \frac{\partial {\cal S}_4}{\partial J_f} = \phi_f - \frac{29 \sqrt{3}}{12}\,V ~,~~ 
& q_p=\frac{\partial {\cal S}_4}{\partial J_p}= \phi + \frac{29 \sqrt{3}}{12} \,
V~.
\end{aligned}
\end{equation}
In the new variables, the angles $q_p$ and $q_f$ contain a
trigonometric dependence on terms oscillating with the synodic
frequency, due to $V$.  Since $V \sim {\cal O}(\mu^{1/2})$, we have
$q_p = \phi + {\cal O}(\mu^{1/2})$, $q_f = \phi_f + {\cal
  O}(\mu^{1/2})$. However, since $\overline{H}_{b}$ in
Eq.~\eqref{hambasic3} does not explicitely depend on these angles, the
conjugated actions $J_p = Y_p$ and $J_f = Y_f$ remain integrals of
motion. As in the previous cases, we keep the notation for those
variables that were transformed by the identity ($Y_f$, $Y_p$,
$v$). Taylor-expanding $\overline{H}_{b}$, around $u_0$ up to terms of
order ${\cal O}(\delta u^2)$, we find (up to terms of first order in
$\mu$ and second order in the eccentricities):
\begin{equation}
\begin{aligned}\label{hbaveel}
\overline{H}_{b,ell} &= -\frac{1}{2} + Y_f -\mu\left({27\over 8}+... \right) 
\frac{e_{p,0}^2}{2}\\
&~ -{3\over 2}x^2+...
-\mu\left({9\over 8}+{63e'^2\over 16}+{129e_{p,0}^2\over 64}+...\right)
\delta u^2+...
\end{aligned}
\end{equation}
where $\frac{e_{p,0}^2}{2} = Y_f - Y_p$.  
Since $Y_f$ is ${\cal O}(\mu)$, up to terms linear in $\mu$ the 
part
\begin{equation}\label{harmsyn}
H_{syn}=-{3\over 2}v^2-\mu\left({9\over 8}+{63e'^2\over 16}
+{129e_{p}^2\over 64}+...\right)\delta u^2
\end{equation}
defines a harmonic oscillator for the synodic degree of freedom. 
The corresponding synodic frequency is
\begin{equation}\label{omesyn}
\omega_s=- \sqrt{
6\mu\left({9\over 8}+{63e'^2\over 16}+{129e_p^2\over 64}+...\right)
}~~.
\end{equation}
On the other hand, the secular frequency is given by
$g={\partial\overline{H}_{b}/\partial Y_p}$. Assuming a harmonic 
solution $\delta u =\delta u_0\cos(\omega_s t+\phi_{0s})$, and 
averaging over the synodic period $\langle \delta u^2 \rangle=\delta u_0^2/2$, 
we find
\begin{equation}\label{gell}
g= \mu\left({27\over 8} + {129\over 64}\delta u_0^2 +...\right)~~,
\end{equation}
completing the estimation of the frequencies. We remark here that
Eq.~\eqref{gell} applies for orbits in the neighborhood of the short
period orbit and it is in agreement with the results of 
\'Erdi~\cite{Erdi-81},~\cite{Erdi-88}. 

\section{Secondary resonances in the ER3BP}\label{sec:3.X-secres}

The Trojan domain describes itself a resonant regime, defined by the 1:1
commensurability of the mean motions of the Trojan body and the planet.
In addition, within this domain, we can find \emph{secondary resonances}
of the form
\begin{equation}\label{resgen}
m_f\omega_f+m_s\omega_s+m g = 0
\end{equation}
with $m_f,m_s,m$ integers. The most important of all these resonances
are those involving low order conmensurabilities between $\omega_f$ and
$\omega_s$. These resonances exist in the complete spectrum of
possible problems, from the pCR3BP ($e'=0$) up to the complete
Restricted Multi-Planet Problem RMPP (see Chapter 4). 
They are of the form
\begin{equation}\label{resba}
\omega_f+n\omega_s=0
\end{equation}
with $n=m_s$.  We briefly refer to a resonance of the
form \eqref{resba} as the '$1$:$n$' resonance, and to higher order
resonances as the $m_f$:$n$ resonances. For $m_f=1$ and $\mu$ in the
range $0.001\leq\mu\leq 0.01$, $n$ is in the range $4\leq n\leq
12$. In the frequency space $(\omega_f,\omega_s,g)$, the
relations \eqref{resba} represent planes normal to the plane
$(\omega_f,\omega_s)$ which intersect each other along the
$g$--axis. In the same space, all other resonances with $m \neq 0$
intersect transversally one or more planes of the main resonances. We
refer to such resonances as `transverse'.  In the numerical examples
below, we use the notation $(m_f,m_s,m)$, for the integers of the
resonant condition \eqref{resgen}. In Sect.~\ref{sec:3.X-chaoticdiff},
by means of numerical experiments, we show the effect of these
resonances in the rate of chaotic diffusion.

Under the Hamiltonian flow of $H_b$, a $m_f$:$n$ resonant periodic
orbit forms $n$ fixed points on a surface of section $(u,v)$, for
$mod(\phi_f,2\pi)=const$. Around the fixed points of a stable resonant
periodic orbit there are formed islands of stability (see
Fig.~\ref{fig:figure1.jpg}), surrounded by separatrix-like thin
chaotic layers passing through the unstable fixed
points. This kind of resonance bifurcates from the short-period orbit
at $\delta u=0$ (equivalent to $u=u_0$, see Eq.~\ref{eq:genedeltau})
provided that
\begin{equation}\label{bifres}
m_f\left( 1- {27\mu \over 8} +...\right)=
n\sqrt{
6\mu\left({9\over 8}+{63e'^2\over 16}+{129e_p^2\over 64}+...\right)
}~~,
\end{equation}
where we can identify the fast and synodic frequencies. Such orbits
appear in pairs, one stable and one unstable, and it is known that
they form 'bridges' connecting the short period family with the long
period family~\cite{DepRabe-69} (cf. discussion in
Sect.~\ref{sec:1.4.works_on_Trojans}). Under the full Hamiltonian
dynamics of $H_{ell}$, the bifurcation generates a 2D-torus, which is
the product of the above orbit times a circle on the plane $W,V$ with
frequency $g \approx 27\mu/8$.

Beyond the bifurcation point, as $e_p$ increases, the fixed points
move outwards, i.e., at larger distances from the central fixed point
$(x,u)=(0,u_0)$, while the resonant islands of stability grow in
size. The growth is faster for lower-order resonant periodic orbits
(i.e. for smaller $n$). This growth, however, stops when the islands
of stability enter in the main chaotic sea around the tadpole domain
of stability. Numerically computed examples of this behavior are given
in Sec.~\ref{sec:3.X.1-surfsec}.

\section{Numerical experiments}\label{sec:3.X-numerical}

In this section we present a parametric survey of the resonant
structures appearing in the space of proper elements ($J_s$,$Y_p$). We
focus in the main secondary resonances, of the form~\eqref{resba} and
its associated multiplets. As we show in the numerical experiments,
this kind of resonance dominates the phase space, affecting in particular
the domain of stable orbits. This study is based on two parts: we
first present a survey of phase portraits, illustrating the phase space
structure in the circular case. In this case, the phase
portraits can be visualized by a 2D surface of section, while in the
elliptic case the corresponding section is 4-dimensional (this issue
is discussed in detail in Chapter 4). On the other hand, based on
the phase portraits of the circular case we construct sets of initial
conditions which can be used in the elliptic problem as well. With
these initial conditions, we construct an atlas of stability maps
applying to the ER3BP, computed by means of a suitable chaotic
indicator.

\subsection{Parametric study of surfaces of section}\label{sec:3.X.1-surfsec}

As introduced in Sect.~\ref{2.3.X-suf.vs.levelc}, Poincar\'e
surfaces of section supply a good visualization of the dynamics of 2D
orbits.  Of all possible surfaces, it turns practical to consider
apsidal sections in which the orbits pass through consecutive local
pericentric or apocentric positions. Here we adopt the pericenter
crossing condition, $\dot{r} = 0$ and $\ddot{r} > 0$, where $\dot{r}$
is the radial velocity in the heliocentric frame. Note that this
is the same condition as for the phase portraits of Chapter 2, $M=0$
(Fig.~\ref{fig:cirsurfsec.png}).

The pericentric surface of section is two-dimensional if $e'=0$, and
four-dimensional if $e'>0$. In the circular case, $\mathcal{F}^{(1)}$
in the Hamiltonian~\eqref{eq:hamYfYsYp} becomes equal to zero by
identity. The exact invariance of $Y_p$ is
equivalent to the exact invariance of the Jacobi constant $C_J$ in the
barycentric rotating frame. In practice, it is more convenient to
construct surfaces of section of constant values of $C_J$ rather than
$Y_p$.  Yet, we label these surfaces of section using a corresponding
value of $e_p$. This correspondance is established in the following
way: to a given value of $Y_p$ corresponds a short-period orbit
crossing the chosen surface at a fixed point with coordinate $u_0$
(Eq.~\ref{u0}), with $e_p = \sqrt{-2Y_p}$.  Noticing that, for $e'=0$
the angles $\phi$ and $\varpi$ coincide ($\varpi'$ can be defined
without loss of generality equal to zero),
i.e. $\phi_f=\lambda'-\varpi$, the remaining initial conditions of the
fixed point are given by
\begin{equation}\label{init}
v_0 = 0, \qquad \phi_{f,0} = \lambda'-\varpi_0 = -u_0 - 
\frac{\pi}{3}, \qquad Y_f = 0~~.
\end{equation}
The condition on $\phi_f$ is the pericenter crossing condition.
Setting the Delaunay action $y_0$ as $y=Y_p-Y_f=Y_p$, and the angle
$\lambda_0=\lambda'+\pi/3+u_0$, with $\lambda'=0$ at $t=0$, one then
has all four values of the Delaunay variables
$(\lambda_0,\varpi_0,x_0,y_0)$, whereby cartesian position and
velocity vectors can be computed.  This allows to compute the Jacobi
constant $C_{J0}$ corresponding to the short-period orbit of given
$e_p$. We refer to the whole surface of section with $C_J=C_{J0}$ as
the section corresponding to a `given value of $e_p$' (referred to as
'the proper eccentricity'), although, for fixed $C_J$, $e_p$ actually
changes somewhat as we move on the section away from the point
$(u_0,x_0)$ (see Chapter 4). Now, for any other point $(u,x)$ on the
surface of section, the pericentric condition yields $\varpi=u+\pi/3$,
while $y$ (and hence the precise value of $Y_p$) can be computed by
solving numerically the Jacobi-constant equation $C_J=C_{J0}$. We
produce surface of section plots taking 35 equispaced initial
pericentric conditions along a fixed line of the form $v=B(u-u_0)$, up
to $u = 1.0$, and solving always the equation $C_J=C_{J0}$.  The
inclination $B$ is determined according to a rule explained below.
For each initial condition, we integrate the orbits and collect 1000
successive points on the surface of section, plotted in the plane
$(u,v)$.

\begin{figure}
\centering
\includegraphics[width=0.90\textwidth,angle=0]{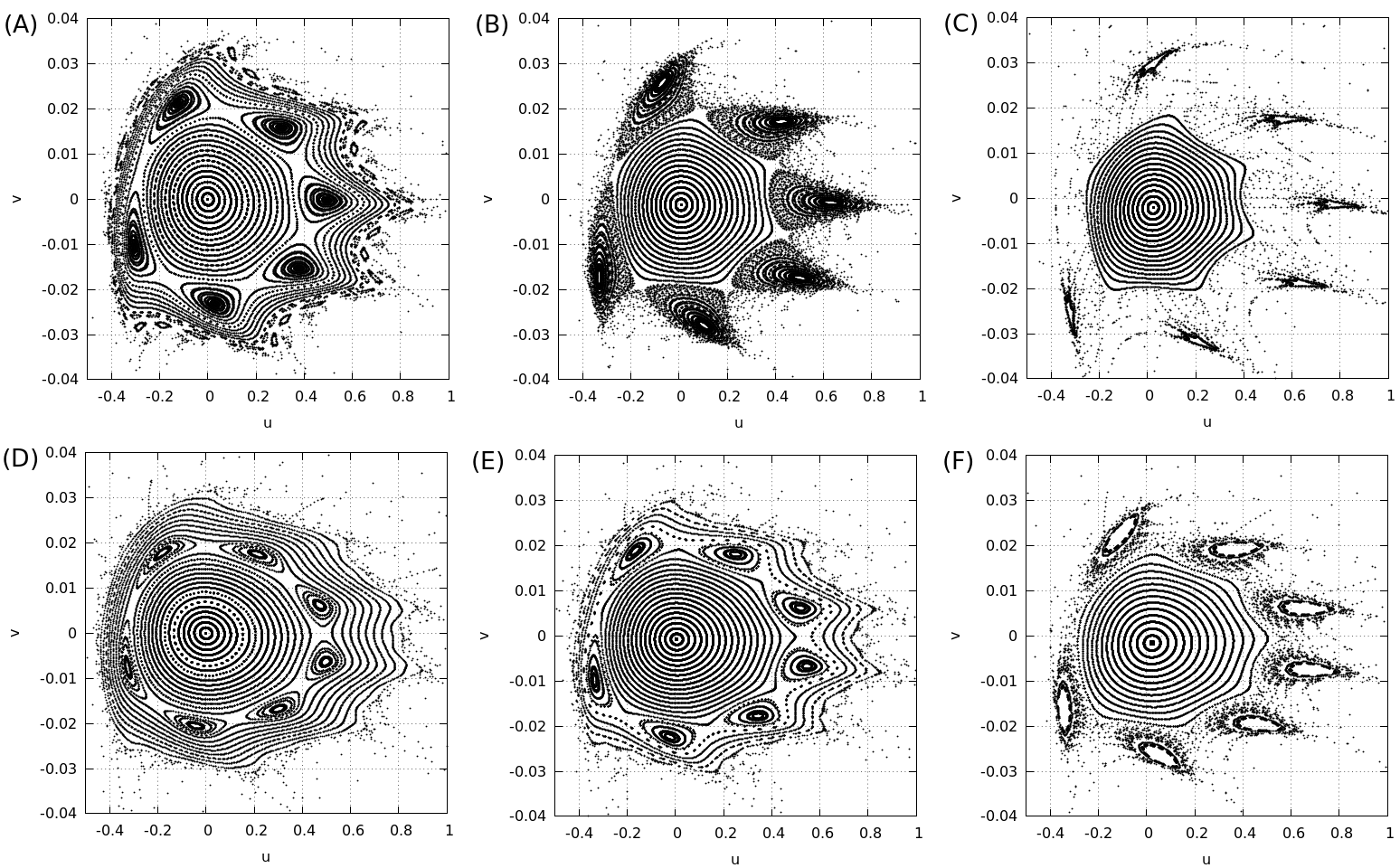}
\caption[Surfaces of section of the pCR3BP for $\mu=0.0041$ and
  $\mu=0.0031$]{Surface of section in the circular case ($e'=0$) for
  $\mu\,=\,0.0041$ (upper plots) and $\mu\,=\,0.0031$ (lower
  plots). The values of $e_p$ are, in each case: $e_p\,=\,0.0001$ (a),
  $e_p\,=\,0.06$ (b), $e_p\,=\,0.1$ (c), $e_p\,=\,0.0001$ (d),
  $e_p\,=\,0.05$ (e) and $e_p\,=\,0.1$ (f).}
\label{fig:pss67}
\end{figure}

We repeat this process for different values of $\mu$, from $0.001$ to
$0.06$, with an interval of $\Delta \mu = 0.001$, and with
$e_p=0$. This range of values contains the resonances of the form 1:n,
with $5\leq n\leq 12$. Higher order
resonances of the form $2$:$(2n+1)$, $3$:$(3n+1)$ or $3$:$(3n+2)$,
etc., are distinguishable by visual inspection.

In Figure \ref{fig:pss67}, we show some examples of this computation.
The plots for $\mu = 0.0041$ (upper row) and $\mu=0.0031$ (lower row)
correspond to the pericentric Poincar\'e surfaces of section in two
cases where the resonances $1$:$6$ and $1$:$7$, respectively, are
conspicuous in phase space. We note that one of the fixed points of
the stable periodic orbit corresponding to the 1:6 resonance lies on
the horizontal line $v=0$. This is so for all even resonances
(i.e. $1$:$n$ with $n$ even). On the other hand, for odd resonances
($1$:$n$ with $n$ odd) all stable fixed points lie on lines of the
form $v=B(u-u_0)$ with $B\neq 0$. In our stability maps, we use the
slopes $B$ given in Table~\ref{Tab:beta}. The line of initial
conditions in each case crosses the border of the stability islands
close to its mots widely separated points, a fact allowing a better
visualization of the resonance.
 
\begin{table}[h]
\begin{center}
\begin{tabular}{l c r}
Resonance & $\mu$ & slope $B$\\
\hline
$1$:$11$ & 0.0014 & 0.04 \\
$1$:$9$ & 0.0021 & 0.025 \\
$1$:$7$ & 0.0031 & 0.015 \\
$1$:$5$ & 0.0056 & 0.03 \\
\end{tabular}
\end{center}
\caption{The slopes $B$ used for the definition of the initial conditions in
the FLI maps, in the cases of odd secondary resonances (see text).}
\label{Tab:beta}
\end{table}

A resonant periodic orbit 1:n bifurcates from the central short-period 
orbit at pairs of values $(\mu,e_p)$ satisfying Eq.~(\ref{bifres}). 
As shown in Fig.~\ref{fig:pss67}, for fixed $\mu$, the resonant orbits 
move outwards as $e_p$ increases, while their corresponding island 
chains grow in size. The three panels in each row of Fig.~\ref{fig:pss67} 
correspond to three different values of $e_p$ (see caption), in increasing 
value from left to right. For small values of $e_p$, the stability islands
in both cases are surrounded by invariant tori. The stability domain
around $(u_0,v_0)$ extends from $u \simeq -0.4$ to $u \simeq 0.8$, for
$v=0$. Some small higher order resonances are visible at the border of
the stability domain. However, as $e_p$ increases, the resonant
islands grow in size, while most of the external invariant tori
are destroyed. For a critical value of $e_p$, the last KAM torus
surrounding the resonant island chain is destroyed. We find that this
value satisfies $e_{p,crit}<0.1$ in all studied cases. For
$e_p>e_{p,crit}$, the resonant islands are surrounded by the outer chaotic
sea, which penetrates the stability domain closer and closer to the
center. Thus, for $e_p=0.08$ the right boundary of the stability 
domain shrinks to $u=0.4$ or less.

Similar phenomena appear if $e_p$ is kept fixed while varying $\mu$.
As $\mu$ increases beyond the bifurcation value, the stability islands
of the resonance move outwards and increase in size. Reaching a
certain critical value of $\mu$, the last invariant torus at the
border of stability surrounding the islands is destroyed. This
mechanism also shrinks the stability region, although by abrupt
steps. On the other hand, different values of $\mu$ give rise to
different resonances.  Thus, the size of the domain of stability
undergoes abrupt variations connected to the bifurcations of new
resonances (see~\cite{Erdietal-07} for a quantitative study of this
effect in the case $e_p=0$ as well as Fig.~\ref{fig:mufli} below).

\subsection{FLI stability maps}\label{sec:3.X.2-flimaps}

As mentioned before, if $e'>0$, the pericentric surface of section
becomes 4-dimensional and, due to projection effects on the plane
$(u,v)$, a detailed visualization of the resonant structures becomes
unclear (see discussion in Sec.~\ref{sec:3.X-moddiff} and Chapter
4). Nevertheless, a convenient visualization is possible in the space
of the actions $(J_s,Y_p)$. In practice, we demonstrate all results in
a space of proper elements $\Delta u$ (libration angle) and $e_p$
(proper eccentricity), which are in one to one relation with the
action variables $(J_s,Y_p)$. The quantity $\Delta u$ is defined as
follows: for given $e_p$, we first determine $u_0$ via
Eq.~\eqref{u0}. Then, we consider all invariant curves around the
equilbrium point $(u=u_0,v=0)$ of the one degree of freedom
Hamiltonian $\overline{H}_b$~in Eq.~\eqref{hambasic3}, as well as a
line of initial conditions $v=B(u-u_0)$, where, in all examples below,
$B=0$ for even resonances, or as indicated in the Table~\ref{Tab:beta}
for odd resonances.  We call $u_p$ the point where the invariant curve
corresponding to the action value $J_s$ intersects the above line of
initial conditions. Finally, we set $\Delta u=u_p-u_0$. Using the
harmonic oscillator approximation of Eq.~\eqref{harmsyn}, the action
$J_s$ can be approximated as $J_s=E_s/\omega_s$, where $\omega_s$ is
given by Eq.~\eqref{omesyn}, while $E_s$ is the oscillator energy
$E_s=-H_{syn}$ found by substituting the initial conditions to
Eq.~\eqref{harmsyn}. Then, up to quadratic terms in $\Delta u$, one
has
\begin{equation}\label{jsdu}
J_s = \frac{3B^2/2 + \mu \left( 9/8 + 63e'^2/16 + 129 e_p^2/64 \right)}
{\left[ 6\mu \left( 9/8 + 63 e'^2/16 + 129 e_p^2/64 \right) \right]^{1/2}} 
\Delta u^2 + {\cal O}(\Delta u^4)
\end{equation}
Note that for odd resonances, $\Delta u$ is {\it not} equal to the 
half-width $D_p$ of the oscillation of the variable $u$ along the invariant 
curve of $\overline{H}_b$ corresponding to the action variable $J_s$, which 
is used as a standard definition of the proper libration angle. Instead, 
locating the point where the ellipse defined by $E_s=-H_{syn}$ 
intersects the axis $x=0$, we find
\begin{equation}\label{dudp}
D_p =  \left[ \frac{3B^2/2 + \mu \left( 9/8 + 63e'^2/16 
+ 129 e_p^2/64 \right)}{\mu \left( 9/8 + 63 e'^2/16 
+ 129 e_p^2/64 \right)}  \right]^{1/2} \Delta u 
+ {\cal O}(\Delta u^2)~.
\end{equation}

In the numerical simulations, after fixing $\mu$ and $e'$, we chose a 
$400\times 400$ grid of initial conditions in the square 
$0\leq\Delta u\leq 1$, $0\leq e_p\leq 0.1$, setting also 
$v=B\Delta u$ and $y=Y_p+Y_f$ with $Y_f=0$, $\phi_f=-\pi/3$, 
$\phi=\pi/3$. This completely specifies all Delaunay variables 
for one orbit, and hence its initial cartesian position and 
velocity vectors. We also set the value of the dummy action $I= Y_f + x$ 
in the Hamiltonian (\ref{ham4}). Finally, we express (\ref{ham4}) 
in the original Cartesian form
\begin{equation}\label{hamnum}
E=H\equiv {p^2\over 2} +I -{1\over r} - \mu \left({1\over\Delta}
-{\mathbf{r}\cdot\mathbf{r}'\over r'^3}-{1\over r}\right)~~
\end{equation}
and keep track of the constancy of the numerical value of the energy 
$E$ as a probe of the accuracy of numerical integrations.  

\begin{figure}
\centering
\includegraphics[width=0.9\textwidth]{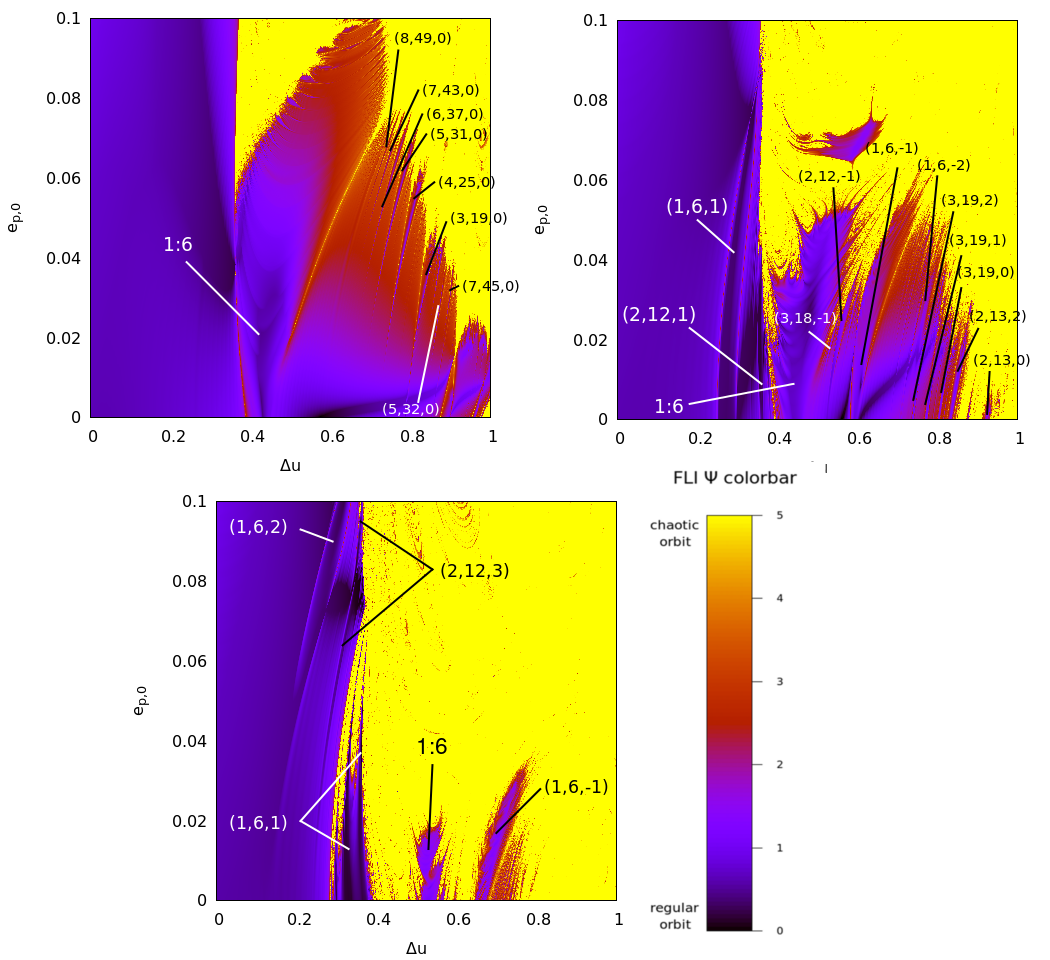}
\caption[Resonance identification in the FLI map for $\mu=0.0041$]{FLI
  maps with details of the resonances for the system with mass
  parameter $\mu=0.0041$, for $e^{\prime}\,=\,0$ (left upper panel),
  $e^{\prime}\,=\,0.02$ (rigth upper panel) and $e^{\prime}\,=\,0.06$
  (lower panel).}
\label{fig:freqanal}
\end{figure}
Stability maps are computed over the above grid of initial conditions 
by means of color-scaled plots of the value of a suitable chaotic indicator. 
Here we employ the Fast Lyapunov Indicator (FLI, see~\cite{Froeschetal-00}) given by 
\begin{displaymath}
\Psi(t) = \sup_{t} \, \log_{10}( \| {\mathbf \xi} \|) \,, 
\end{displaymath}
where ${\mathbf \xi}$ is the variational (deviation) vector computed
by solving the variational equations of motion along with the orbital
equations of motion and $\|\cdot\|$ denotes the $L_2$ norm. For
computing FLIs we implement a $7$th-$8$th order Runge-Kutta method,
with a timestep equal to $1/300$ of the period of the primary ($=2\pi$
in our units). We stop integrating orbits that have clearly reached
escape so as to avoid numerical overflows. The escapes are identified
by a sudden jump of the energy error to levels beyond $10^{-4}$, while
for non-escaping orbits the energy error at the maximum time of
integration $T = 10^3$ periods is less than $10^{-9}$. Also, we stop
integrating the variational equations of orbits reaching FLI values
larger than $50$.

Figure \ref{fig:freqanal} shows an example of computed stability maps for 
$\mu=0.0041$ and three values of $e'$, namely $e'=0$, $e'=0.02$, and 
$e'=0.06$. In each plot, the value of $\Psi$ is given in a color scale 
for all $400\times 400$ grid initial conditions in the plane of proper 
elements $(\Delta u,e_p)$. The color scale was set in the range 
$0\leq\Psi\leq 5$. Regular orbits correspond to darker colors (black) 
representing low values of $\Psi$, while the most chaotic orbits correspond 
to light colors (yellow). Orbits with $\Psi>5$ are shown also in yellow.  

A more detailed resonance identification is made by means of Frequency 
Analysis~\cite{Laskar-04}. The most conspicuous resonances are explicitly 
indicated in all three plots. For $e'=0$ (top left panel), the resonance 
$1$:$6$ dominates the stability map. Besides, several resonances of 
the type $(m,6m-1,0)$ produce strips penetrating the stability domain. 
In agreement with what was shown in the surface of section plots of 
Fig.~\ref{fig:pss67}, the width of the $1$:$6$ resonance increases, 
initially, as $e_p$ increases from zero up to a value $e_p\sim0.06$. 
Also, the chaotic separatrix-like layers around the resonance remain 
thin. However, for $e_p>0.06$ the resonance is detached from the main 
stability domain. Then, its corresponding islands of stability are 
embedded in a chaotic sea corresponding to orbits with a fast escape. 
For still larger $e_p$ (around 0.1) the central periodic orbit 
becomes unstable and the corresponding islands disappear. In general, 
this, as well as all higher order resonances, move outwards (towards 
higher values of $\Delta u$) as $e_p$ increases. Thus, all resonant 
strips have a small positive slope in Fig.~\ref{fig:freqanal}. 

On the other hand, increasing the value of $e'$ causes new {\it
transverse} resonances to appear. For $e'=0.02$ (top right panel),
resonances of the form $(n,6n,m)$, or $(n,6n+1,m)$, with $n=1,2,3$
and $m=-2$ up to $m=2$ are distinguishable. We note that in general
the angle formed between the transverse resonances ($m\neq 0$) and the
secondary resonances of the circular problem ($m=0$) is small (of
order $g/\omega_s$), where $g$ and $\omega_s$ are the respective
secular and synodic frequencies. This small transversality implies
that the intersection point of most mutually transverse resonances
lies in the chaotic zone, i.e. far from the domain of stability. In
particular, the transverse resonances with $m>0$ have no intersection
with the main resonance $1$:$6$ inside the stability domain. In fact,
these transverse resonances penetrate the domain of stability in
isolated, single-resonance strips, along which the orbits undergo
weakly chaotic diffusion that bears many features of Arnold diffusion.
On the other hand, the resonances
that are beyond the border of the inner stability domain form
multiplets, in which the chaotic diffusion has features of
modulational diffusion. This gives a mechanism of efficient escape
for chaotic orbits (see Section~\ref{sec:3.X-chaoticdiff}).

For still higher values of $e'$ (Fig.~\ref{fig:freqanal}, down panel for
$e'=0.06$), new transverse resonances appear. In fact, as $e'$
increases all transverse resonances move in the direction from top
left to bottom right, until reaching large values of $\Delta u$, after
which they enter into the main chaotic sea surrounding the domain of
stability. Then, they become less significant.

\begin{figure}
\centering
\includegraphics[width=1.0\textwidth]{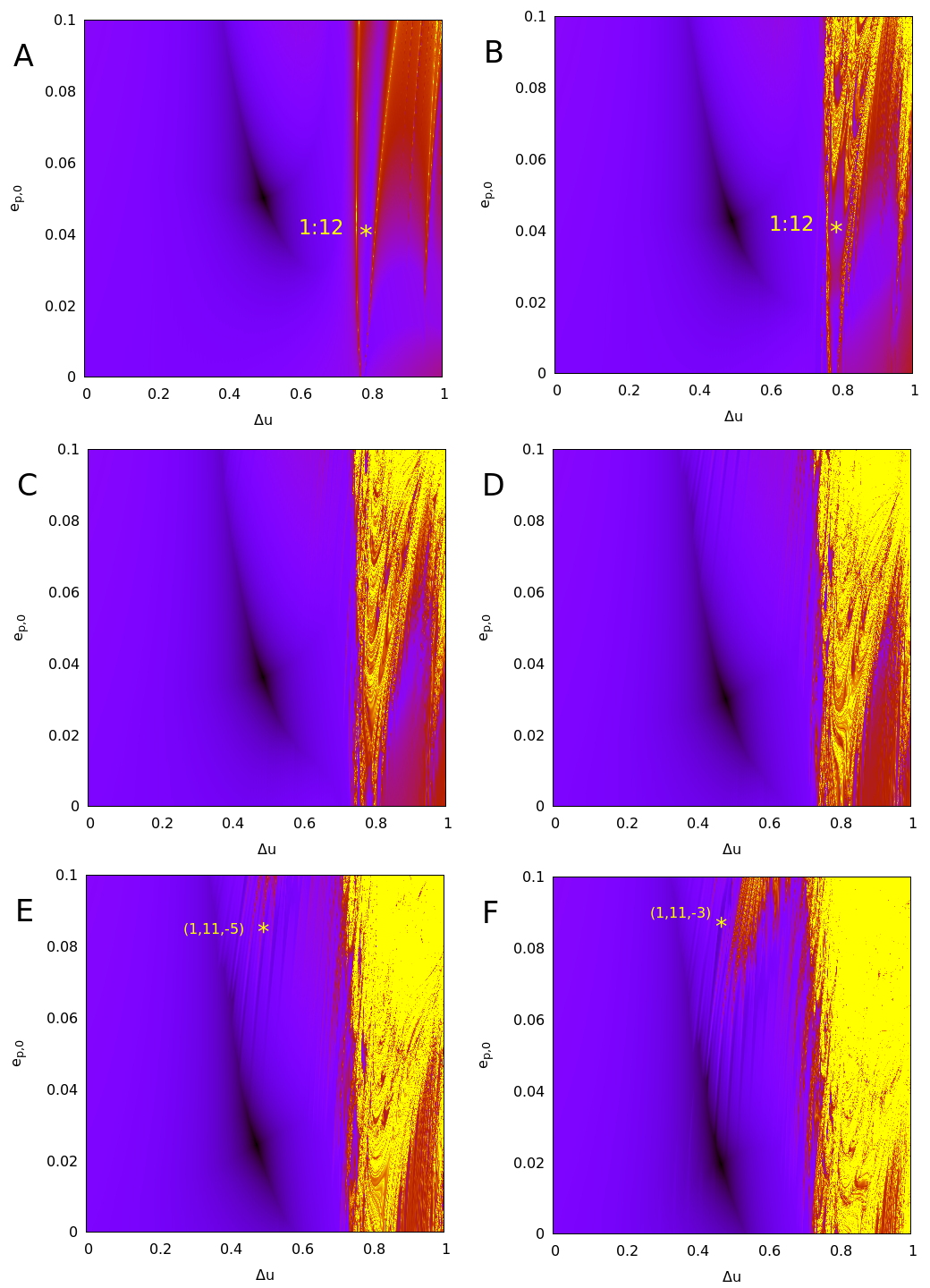}
\caption[FLI maps for the resonance 1:12, with $\mu=0.0012$]{FLI maps
  for the resonance 1:12, $\mu=0.0012$, for the values $e^{\prime}=0$
  (A), $e^{\prime}=0.02$ (B), $e^{\prime}=0.04$ (C), $e^{\prime}=0.06$
  (D), $e^{\prime}=0.08$ (E) and $e^{\prime}=0.1$ (F).}
\label{fig:fli1to12}
\end{figure}

\begin{figure}
\centering
\includegraphics[width=1.0\textwidth]{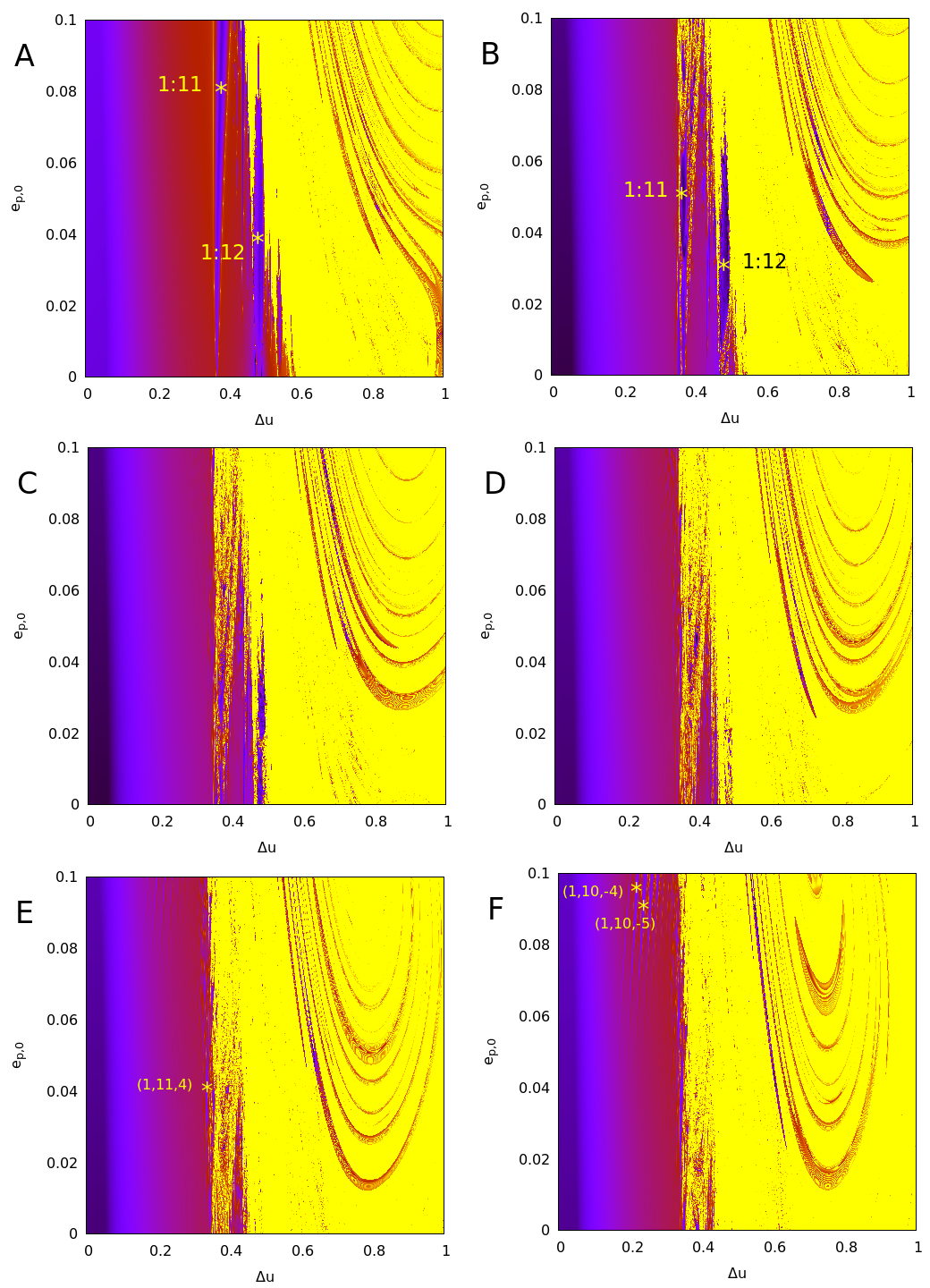}
\caption[FLI maps for the resonance 1:11, with $\mu=0.0014$]{FLI maps
  for the resonance 1:11, $\mu=0.0014$, for the values $e^{\prime}=0$
  (A), $e^{\prime}=0.02$ (B), $e^{\prime}=0.04$ (C), $e^{\prime}=0.06$
  (D), $e^{\prime}=0.08$ (E) and $e^{\prime}=0.1$ (F).}
\label{fig:fli1to11}
\end{figure}

\begin{figure}
\centering
\includegraphics[width=1.0\textwidth]{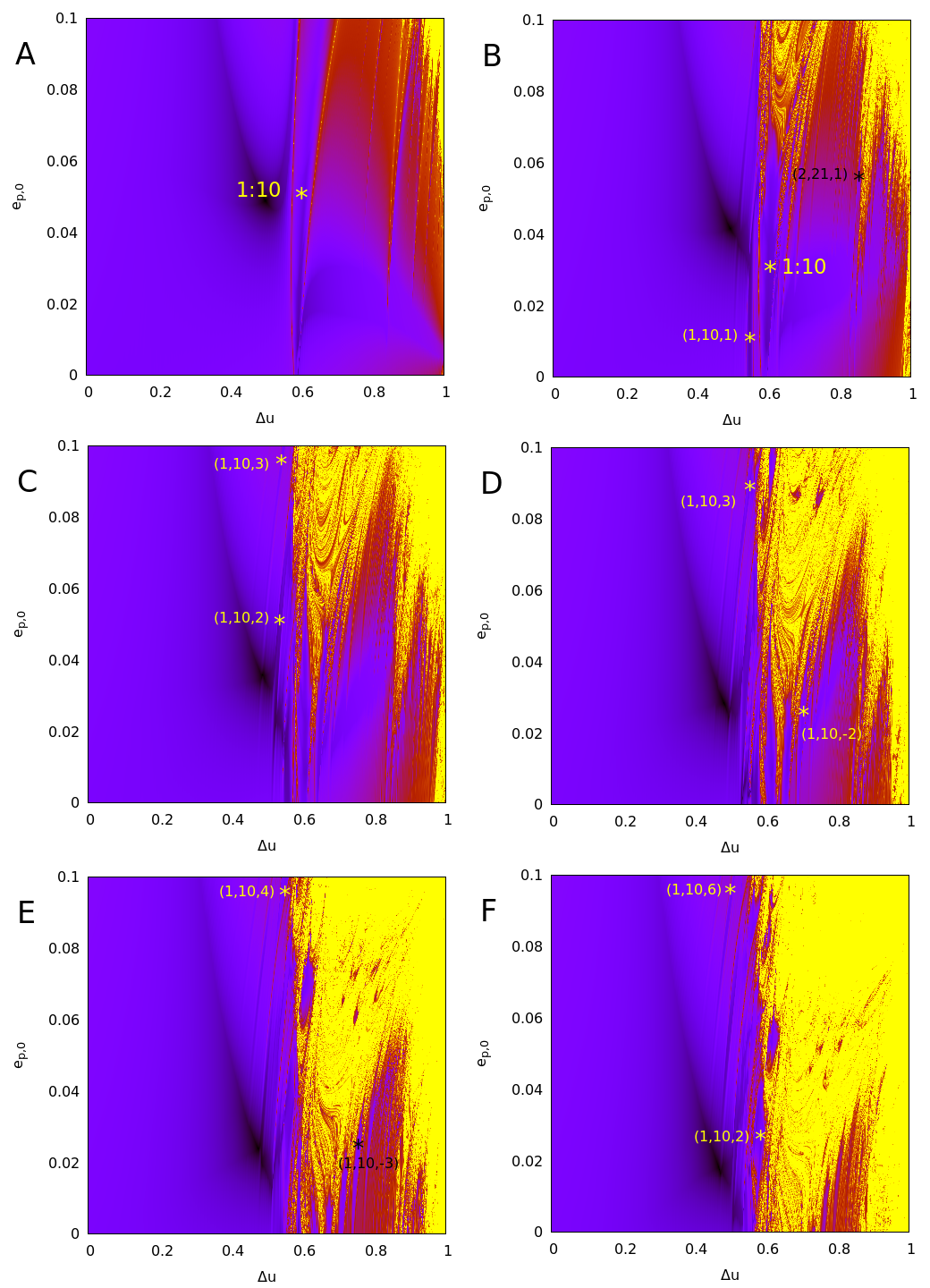}
\caption[FLI maps for the resonance 1:10, with $\mu=0.0016$]{FLI maps
  for the resonance 1:10, $\mu=0.0016$, for the values $e^{\prime}=0$
  (A), $e^{\prime}=0.02$ (B), $e^{\prime}=0.04$ (C), $e^{\prime}=0.06$
  (D), $e^{\prime}=0.08$ (E) and $e^{\prime}=0.1$ (F).}
\label{fig:fli1to10}
\end{figure}

\begin{figure}
\centering
\includegraphics[width=1.0\textwidth]{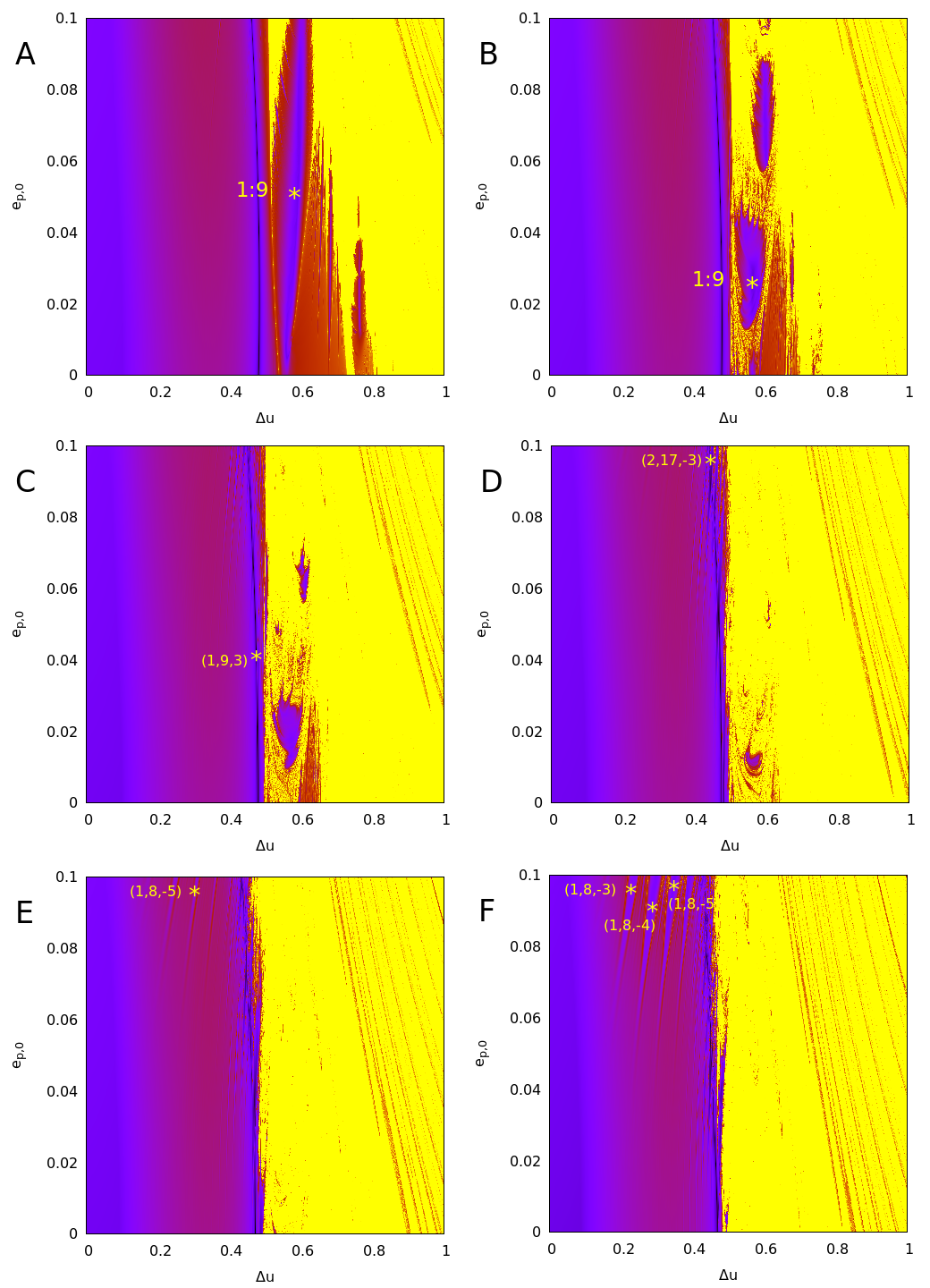}
\caption[FLI maps for the resonance 1:9, with $\mu=0.0021$]{FLI maps
  for the resonance 1:9, $\mu=0.0021$, for the values $e^{\prime}=0$
  (A), $e^{\prime}=0.02$ (B), $e^{\prime}=0.04$ (C), $e^{\prime}=0.06$
  (D), $e^{\prime}=0.08$ (E) and $e^{\prime}=0.1$ (F).}
\label{fig:fli1to9}
\end{figure}

\begin{figure}
\centering
\includegraphics[width=1.0\textwidth]{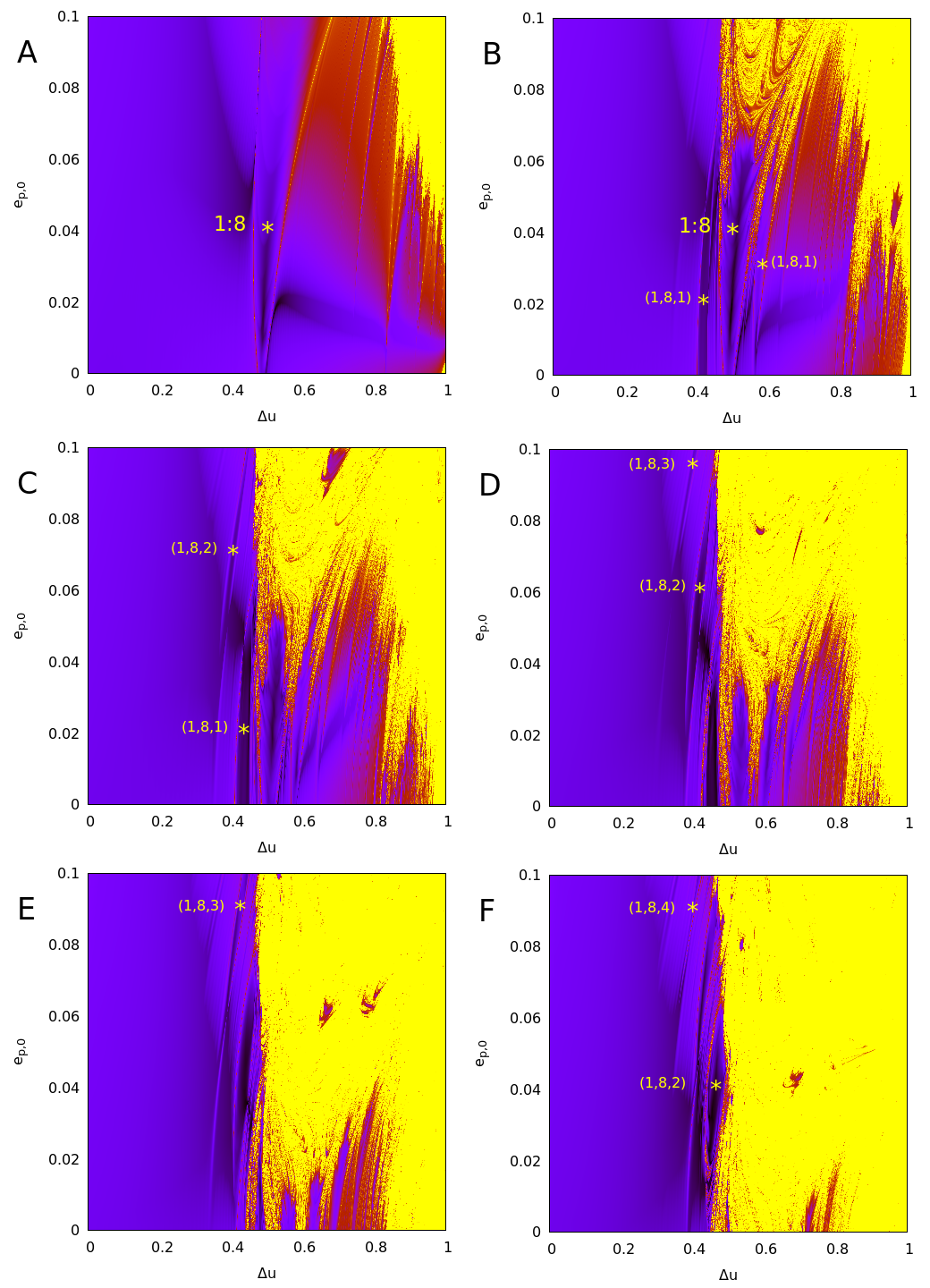}
\caption[FLI maps for the resonance 1:8, with $\mu=0.0024$]{FLI maps
  for the resonance 1:8, $\mu=0.0024$, for the values $e^{\prime}=0$
  (A), $e^{\prime}=0.02$ (B), $e^{\prime}=0.04$ (C), $e^{\prime}=0.06$
  (D), $e^{\prime}=0.08$ (E) and $e^{\prime}=0.1$ (F).}
\label{fig:fli1to8}
\end{figure}

\begin{figure}
\centering
\includegraphics[width=1.0\textwidth]{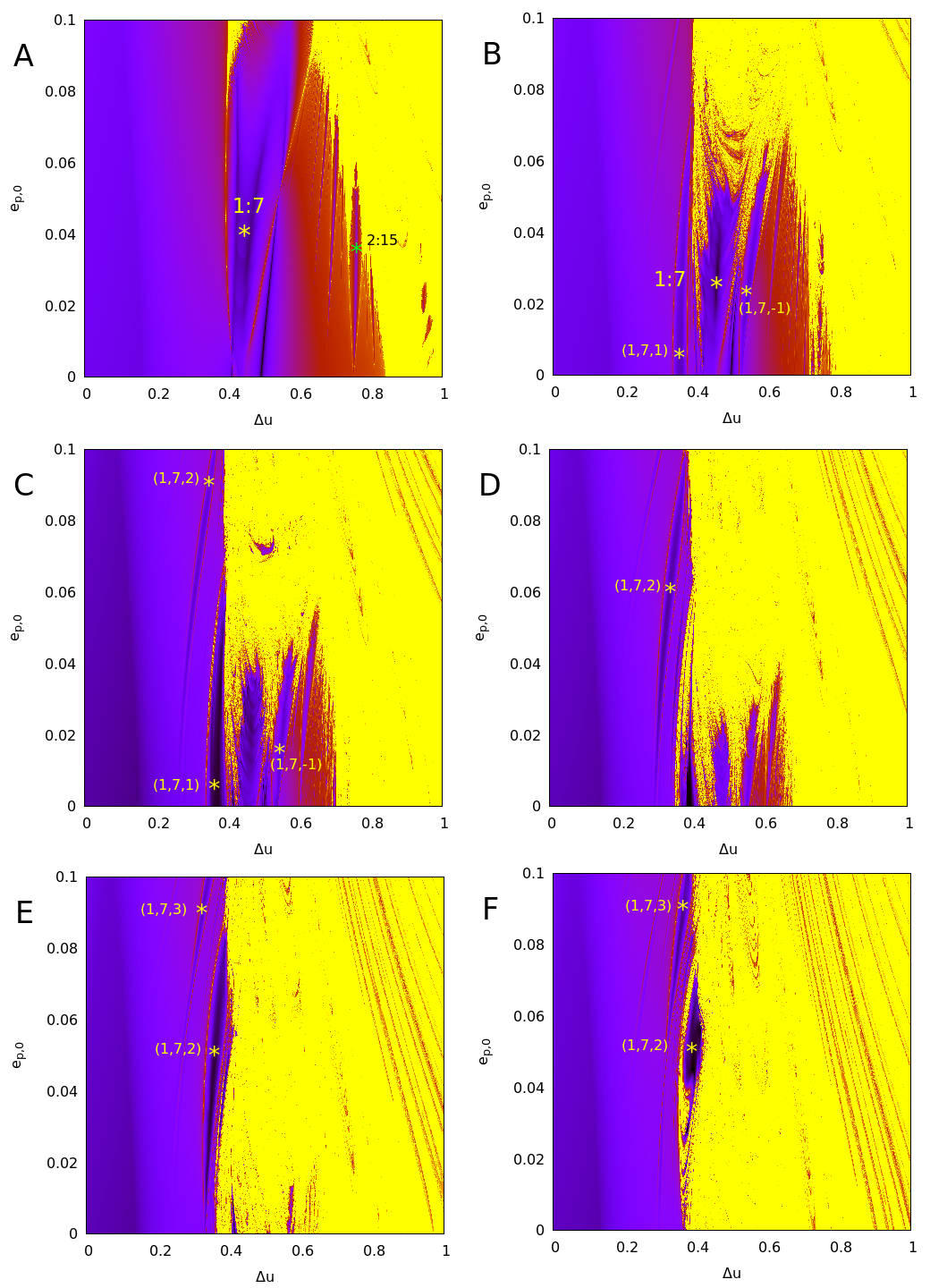}
\caption[FLI maps for the resonance 1:7, with $\mu=0.0031$]{FLI maps
  for the resonance 1:7, $\mu=0.0031$, for the values $e^{\prime}=0$
  (A), $e^{\prime}=0.02$ (B), $e^{\prime}=0.04$ (C), $e^{\prime}=0.06$
  (D), $e^{\prime}=0.08$ (E) and $e^{\prime}=0.1$ (F).}
\label{fig:fli1to7}
\end{figure}

\begin{figure}
\centering
\includegraphics[width=1.0\textwidth]{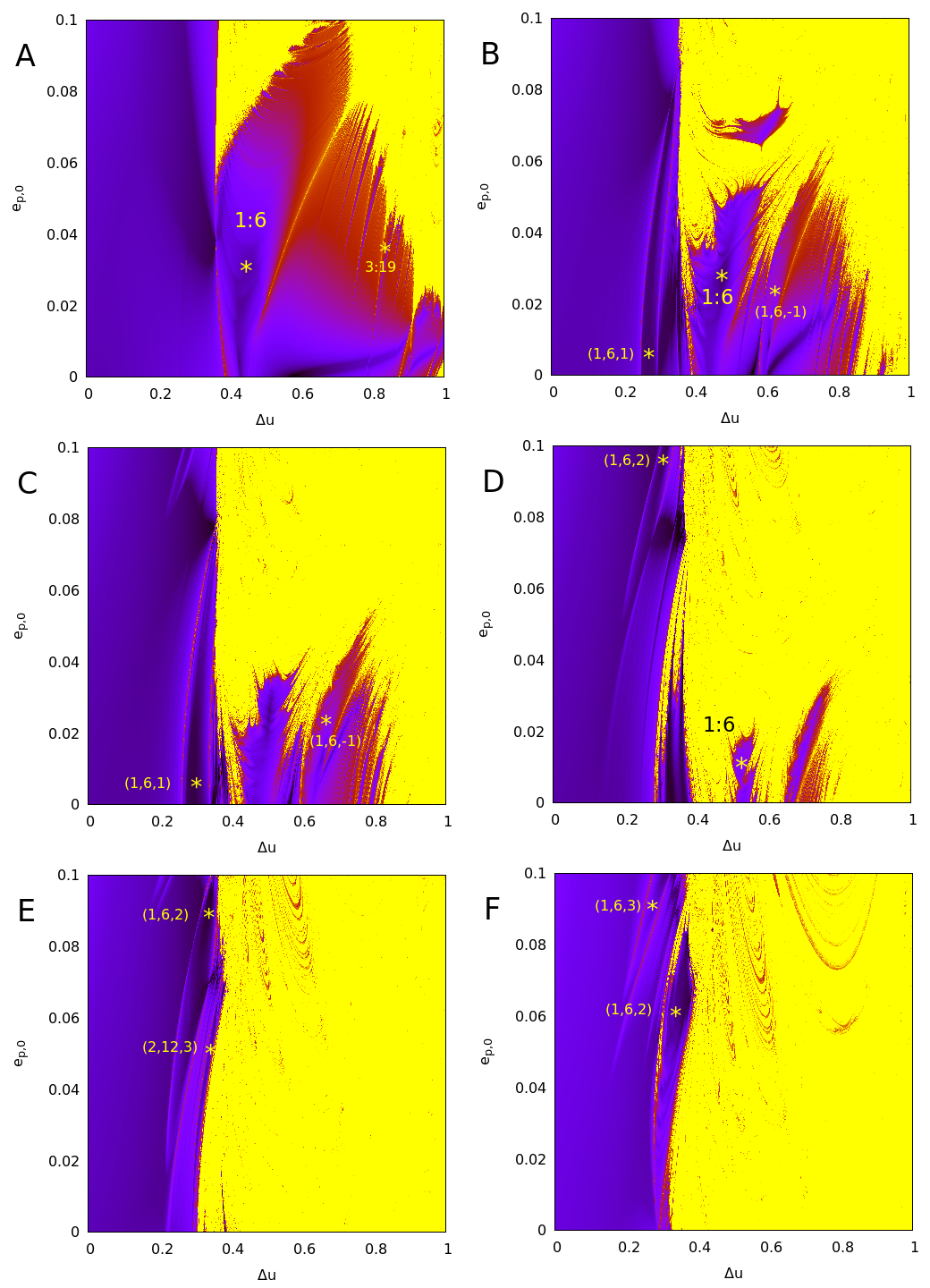}
\caption[FLI maps for the resonance 1:6, with $\mu=0.0041$]{FLI maps
  for the resonance 1:6, $\mu=0.0041$, for the values $e^{\prime}=0$
  (A), $e^{\prime}=0.02$ (B), $e^{\prime}=0.04$ (C), $e^{\prime}=0.06$
  (D), $e^{\prime}=0.08$ (E) and $e^{\prime}=0.1$ (F).}
\label{fig:fli1to6}
\end{figure}

\begin{figure}
\centering
\includegraphics[width=1.0\textwidth]{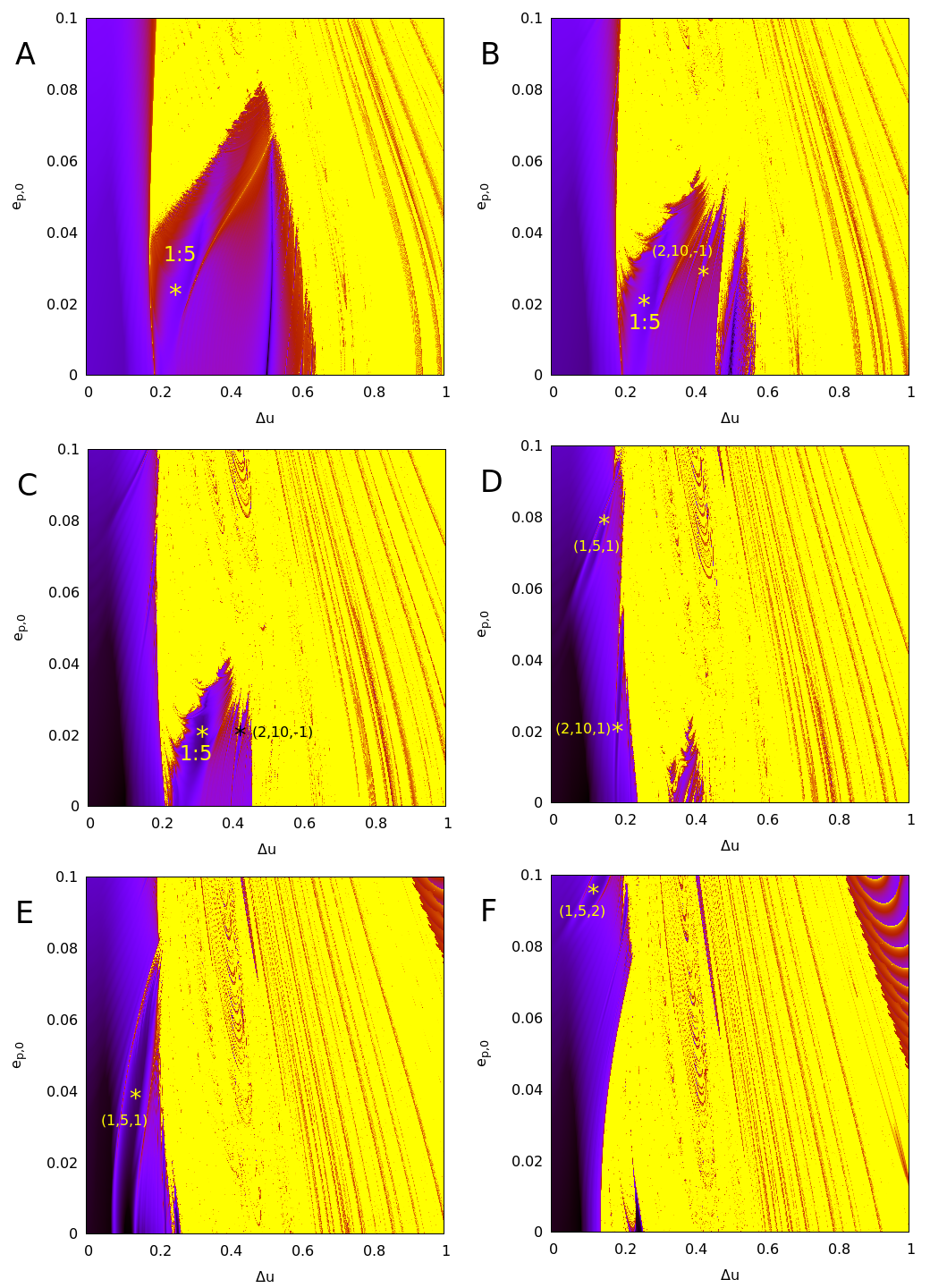}
\caption[FLI maps for the resonance 1:5, with $\mu=0.0056$]{FLI maps
  for the resonance 1:5, $\mu=0.0056$, for the values $e^{\prime}=0$
  (A), $e^{\prime}=0.02$ (B), $e^{\prime}=0.04$ (C), $e^{\prime}=0.06$
  (D), $e^{\prime}=0.08$ (E) and $e^{\prime}=0.1$ (F).}
\label{fig:fli1to5}
\end{figure}

\begin{figure}
\hspace{-1.cm}
\includegraphics[width=1.1\textwidth,angle=0]{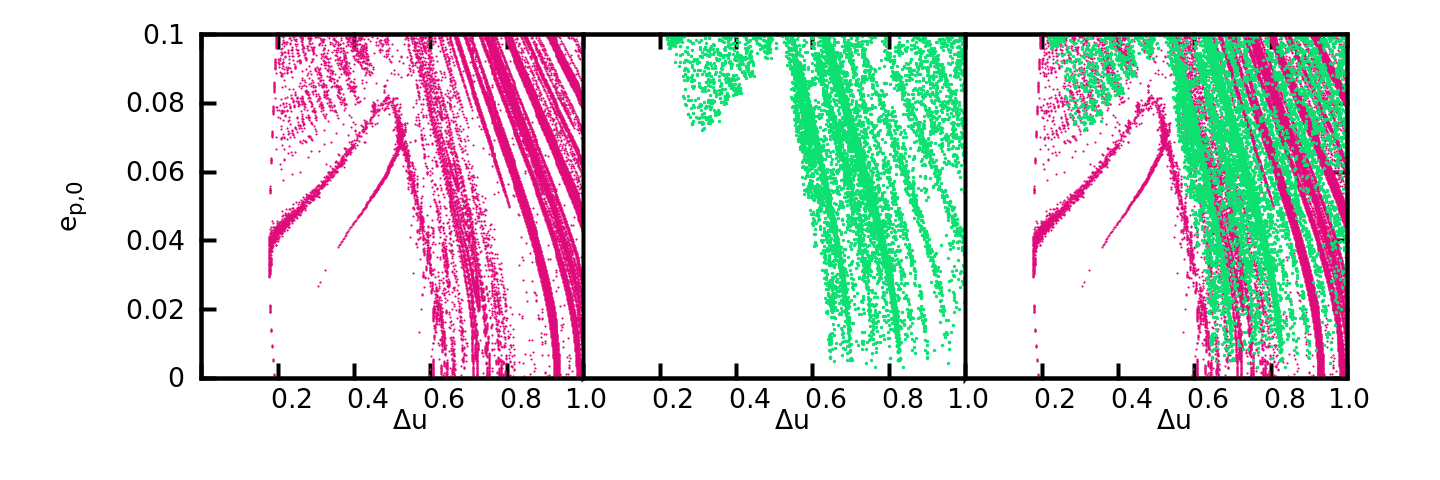}
\caption[Correlation between structures in the FLI maps and stable
  invariant manifolds]{Left panel: All initial conditions in the
  400$\times$400 grid of Fig.~\ref{fig:fli1to5}A for which the FLI
  value at the end of the integration is in the range $3.5\leq\Psi\leq
  6$. Middle panel: the points of intersection with the same section
  as for the FLI maps (see text) of the stable invariant manifolds of
  the family of short-period horizontal Lyapunov orbit around L3,
  computed for 400 different values of the Jacobi constant as
  indicated in the text. Right panel: superposition of the left and
  middle panels.}
\label{fig:flivsman}
\end{figure}

Figures \ref{fig:fli1to12} to \ref{fig:fli1to5} show the atlas of FLI
maps for differents cases of $e'$, and for the mass parameters
$\mu=0.0012$, $0.0014$, $0.0016$, $0.0021$, $0.0024$, $0.0031$,
$0.0041$ and $0.0056$. These are representative of the multiplets
formed around the resonances 1:12, 1:11, 1:10, 1:9, 1:8, 1:7, 1:6 and
1:5, respectively. In each plot, the values for $e^{\prime}$ are $0$,
$0.02$, $0.04$, $0.06$, $0.08$ and $0.1$, from top to bottom and left
to right.  Inspecting these plots, we emphasize some features.\\

\noindent
i) The size of the non-resonant domain does not change much with
variations of $e^{\prime}$. In fact, in all these plots we observe
that, despite the fact that, as $e^{\prime}$ evolves, new transversal
resonances appear, the non-resonant domain keeps its limits nearly
constant (at about $0.7$ in Fig \ref{fig:fli1to12}, $0.35$ in Fig
\ref{fig:fli1to11}, $0.6$ in Fig \ref{fig:fli1to10}, 
$0.5$ in Fig \ref{fig:fli1to9}, $0.45$ in Fig \ref{fig:fli1to8}, $0.4$ in
Fig \ref{fig:fli1to7}, $0.35$ in Fig \ref{fig:fli1to6} and $0.2$ in
Fig \ref{fig:fli1to5}).  On the other hand, the resonant domain, which
for low (but non-zero) values of $e^{\prime}$ is filled with small
transverse resonances, gradually shrinks within the chaotic sea, and
for values of $e^{\prime}$ around $0.1$, it completely
disappears. This gives a natural limit for the values of $e^{\prime}$
to consider, since no important resonances survive for $e'>0.1$.\\

\noindent
ii) Around a main resonance, we identify new emerging transverse
resonances, of the kind $(1,n,m)$, with $m$ a small integer.  Since
they involve a commensurability with $g$, they are not present for
$e^{\prime}=0$, but for greater values they become evident, especially
some isolated ones which penetrate inside the non-resonant region. As
$e^{\prime}$ increases, the whole structure moves outwards (towards
increasing values of $\Delta u$) but their upper limit also moves
downwards (towards smaller values of $e_p$). For bigger values of
$e^{\prime}$, the main resonances 1:n generally disappear or they are
small, leaving space for transverse resonances to dominate in action
space.\\

\noindent
iii) In Figures \ref{fig:fli1to11}, \ref{fig:fli1to7},
\ref{fig:fli1to6} and \ref{fig:fli1to5}, we can see traces, appearing
as thin darker lines in the chaotic domain (right part of the plot),
of the stable invariant manifolds emanating from lower-dimensional
invariant objects around $L_3$, such as the short-period planar
Lyapunov orbits in the case $e'=0$, or their associated 2D-tori, for
$e'\neq0$~\cite{AstFarr-04}.  Figure \ref{fig:flivsman} shows an
example of comparison of the structures found in the FLI maps with the
exact computation of the stable invariant manifolds of the Lyapunov
orbit around $L_3$ in the case $\mu=0.0056$, $e'=0$, corresponding to
panel A of Fig.~\ref{fig:fli1to5}. The left panel shows the same
structures in greater detail, plotting in pink all points for which
the FLI is in the limit $3.5\leq \Psi\leq 6$. These limits exclude all
points corresponding to regular orbits, as well as all escaping
orbits, for which the FLI evaluation quickly saturates to a high value
$\Psi\geq 50$. On the other hand, for the middle panel, we consider
the interval of Jacobi constant values $C_{min}\leq C\leq C_{max}$,
where $C_{min}=2.984$ and $C_{max}=3.00385$ represent the minimum and
maximum value of the Jacobi constant encountered in the whole
400$\times$400 grid of initial conditions of the FLI map of
Fig.~\ref{fig:fli1to5}A. Splitting this interval in 400 values of $C$,
for each value we compute numerically the corresponding horizontal
Lyapunov orbit around $L_3$ and its stable manifold and we collect all
the points in which the latter intersects the section of the FLI map
(given by the pericentric condition $\lambda'-\varpi-\pi/3=0$ as well
as $v-0.03\Delta u=0$). Numerically, we introduce some tolerence
$2\times 10^{-4}$ in the section determination, in order to collect a
sufficiently large number of points necessary for visualization of the
results. Plotting in the same co-ordinates as for the FLI map the
collected points for all 400 stable invariant manifolds corresponding
to the 400 different values of C yields the middle plot of
Fig.~\ref{fig:flivsman}. The relative loss of sharpness in the picture
of the manifolds is due to the numerical tolerance used in their
section's determination.  Despite this effect, we see clearly that the
structures formed by the invariant manifolds follow in parallel those
indicated by the corresponding FLI map. The possibility to use the FLI
(with a small number of iterations), in order to visualize invariant
manifolds was pointed out in~\cite{GuzzLeg-13}.  Here, this effect can
be considered as a manifestation of the so-called `Sprinkler'
algorithm (see~\cite{KovErdi-09} for a review). Namely, in a system of
fast escapes, plotting the initial conditions of the orbits which have
relatively large (forward or backward) stickiness times (and, thus,
relatively lower FLI values with respect to the escaping orbits)
allows to vizualize the (stable or unstable) manifolds of nearby
periodic orbits. In fact, the sticky chaotic orbits in the forward
sense of time are those trapped {\it within the lobes} defined by the
stable invariant manifolds (see, for example, figure 19
of~\cite{Efthyetal-97}). This effect is clearly shown in our example
by combining the left and middle panels of
Fig.~\ref{fig:flivsman}. The right panel of Fig.~\ref{fig:flivsman}
clearly shows that the points of greater stickiness in the forward
sense of time, as revealed by their relatively low (with respect to
fast-escaping orbits) FLI values, are located precisely between the
limits of the structures indicated by the stable invariant manifolds
of the family of the planar Lyapunov orbits around $L_3$. We note,
finally that when $e'>0$, instead of the foliation of all the
manifolds of the Lyapunov family, one has to consider the invariant
manifolds of a 2D unstable invariant torus around $L_3$. This
computation is numerically hardly tractable. Nevertheless, simple
inspection of all panels of Fig.~\ref{fig:fli1to5} clearly shows that
the structures found for $e'=0$ essentially continue to exist, in a
quite similar geometry, in the case $e'\neq 0$ as well.

\begin{SCfigure}
\hspace{-1cm}
\includegraphics[width=0.60\textwidth,angle=0]{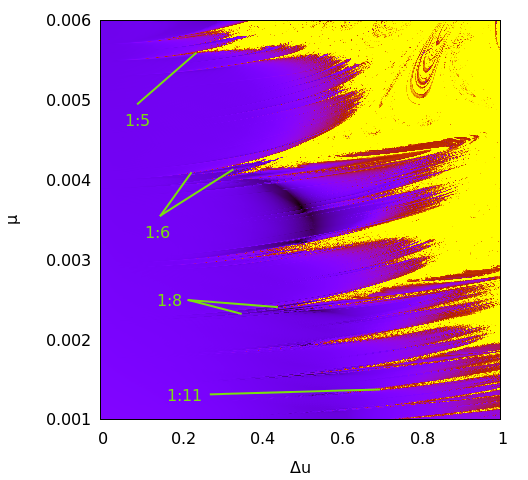}
\caption[Resonances bifurcation in the plane ($\Delta u,\mu$)]{FLI map
  for a grid of $400\times400$ initial conditions in the plane
  ($\mu$,$\Delta u$), for $e'=e_p=0.02$ (see text).  Dark colors
  (black) correspond to $\Psi=0$ (regular orbits) and light color
  (yellow) to $\Psi=5$ (chaotic orbits). The most important secondary
  resonances are labeled in the plot. Some compact bundles formed by
  transverse resonances near the main ones are also visible in the
  plot.
\vspace{1cm}}
\label{fig:mufli}
\end{SCfigure}

Returning to the discussion of the resonant structure, the overall 
effect of resonances on the size of the stability domain
is resumed in Fig.~\ref{fig:mufli}. For a set of initial values of
$\mu$, from $\mu = 0.001$ to $\mu = 0.006$, with step $\Delta \mu =
1.25\times 10^{-5}$, and fixed values for $e^{\prime} = 0.02$ and $e_p
= 0.02$, the figure shows the FLI map produced by integrations of the
set of initial conditions given by $v=0$, $\phi_f=-u-\frac{\pi}{3}$,
$Y_f=0$ and $\Delta u = u_0 + u$ varying from $0$ to $1$, with step
$0.0025$.  Darker (black) colors correspond to regular orbits, and
light (yellow) colors to chaotic orbits. The main resonances appear as
long yellow tongues that contain single or double thin chaotic layers
associated to the separatrix (depending on whether the resonance is
odd or even). In a similar way, a large number of smaller transverse
resonances fills the space between the main ones. By the fact that the
tongues are nearly horizontal, we can infer that the presence of
particular resonances is highly localized with respect to the value of
$\mu$, e.g. the resonance $1$:$6$ is important at $\mu=0.004$, it
completely disappears in the chaotic domain at $\mu=0.0045$.  Also, as
$\mu$ increases, a bifurcation of new secondary resonances
happens less frequently. Nevertheless, since they are of decreasing order,
their width and relative influence increases.

\section{Modulational diffusion}\label{sec:3.X-moddiff}

The resonant periodic orbits arising under the flow of $H_b$
correspond to resonant 2D tube tori under the flow of the full model
$H_{ell}$.  Respectively, the resonant fixed points correspond to
one-dimensional tori on a surface of section $mod(\phi_f,2\pi)=const$.
Projected on the plane $(u,v)$, these tori appear as thick curves (see
Fig.~\ref{fig:diffmodel} below). In the same projection, the islands
of stability and their delimiting separatrix-like chaotic layers are
observed to undergo `pulsations', i.e. some periodic shift in the
plane $(u,v)$ modulated at the frequency $g$. This pulsation phenomenon
is further described in Chapter 4 (see
Sect.~\ref{sec:feat_Hb}). Such pulsation is induced by the presence (in
$H_{ell}$ but not in $H_b$) of terms trigonometric in the angle $\phi$
and its multiples.

The modulation of all resonant motions by slow trigonometric terms
results in a long-term chaotic diffusion taking place in the space of
the action variables $(J_s,Y_p)$. In fact, based on the pulsation
width of the separatrices, we encounter the following two diffusion
regimes:\\

\noindent
i) {\it Non-overlapping resonances}: for small pulsation widths, the
separatrices of one resonance do not enter to the pulsation domain of
the separatrices of nearby resonances. In such cases the rate of the
chaotic diffusion is quite small, and the diffusion becomes
practically undetectable.  Also, the geometry of resonances in the
action space is closer to the paradigm of Arnold diffusion~\cite{Arnold-64}. 
An example of chaotic orbit in such regime is given in 
Fig.~\ref{fig:flimoddiff} (green orbit).\\

\noindent
ii) {\it Partially-overlapping resonances}: for large pulsation
widths, the pulsation domains of more than one separatrices of nearby
resonances partially overlap. In this case the rate of chaotic
diffusion increases dramatically. As shown in
Sec.~\ref{sec:3.X-chaoticdiff}, the chaotic orbits in the most
prominent chaotic layers exhibit a diffusion timescale of the order or
1Myr.  The diffusion leads finally to an escape from the resonant
domain and eventually from the overall tadpole domain. However, there
is also a weakly chaotic population exhibiting long times of
stickiness, with a power-law distribution of the stickiness times
characteristic of long-term correlated chaotic motions
(see~\cite{Meiss-92} p. 843, and references therein).  At any rate, in
most cases we find that the overlapping of resonances is not complete,
as is, for example, the case of resonant multiplets for Jupiter's
Trojan asteroids (see~\cite{RobGab-06}).  As a result, the overall
diffusion process in our experiments is closer to the paradigm of {\it
  modulational} diffusion~\cite{Chirikov-85}. Finally, there are
regular resonant orbits that never escape the system.\\
\begin{figure}[t]
\centering
\includegraphics[width=0.90\textwidth,angle=0]{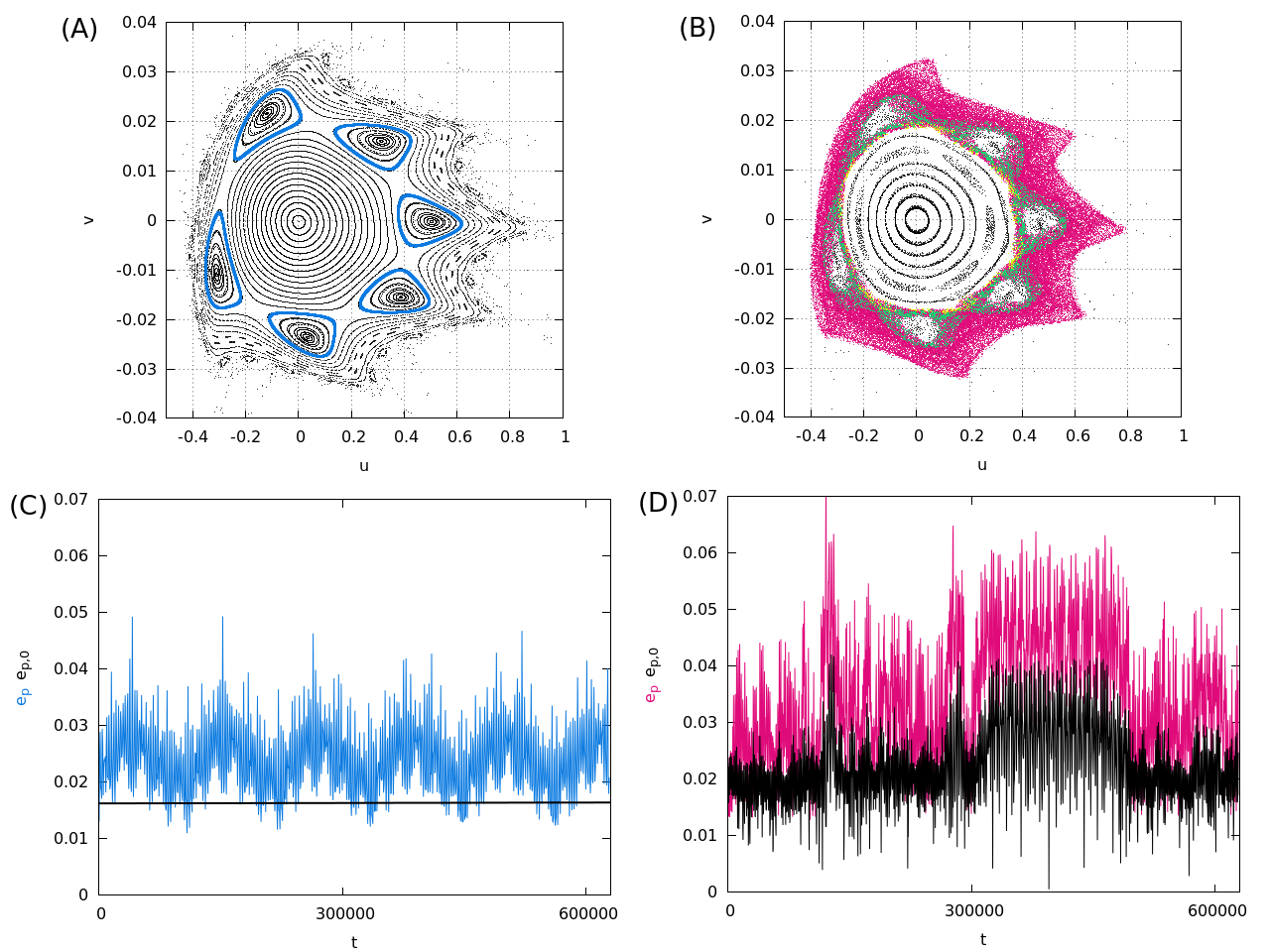}
\caption[Modulational diffusion mechanism and differences in the
  behavior of $e_p$ and $e_{p,0}$]{(A) - Phase portrait (pericenter
  surface of section) in the variables $u$, $v$, when $\mu=0.0041$,
  $e_p=0.01675$, $e'=0$ (circular case). An orbit moving in the thin
  separatrix layer of the $1$:$6$ resonance is shown in blue (with
  initial condition $v=0$, $u=0.376$).  (B) - Same as in (A) but now
  in the elliptic case $e'=0.02$. The chaotic orbit moves in the
  separatrix layer of the $1$:$6$ resonance up to a time $10^4$
  (green), but later it expands towards the chaotic layers of other
  adjacent resonances (pink). (C) - Time evolution of $e_p$
  (Eq. (\ref{eprnew}), black curve) and $e_{p,0}$ (Eq. (\ref{epr}),
  blue curve), for the blue orbit of (A). (D) - Time evolution of
  $e_p$ (black) and $e_{p,0}$ (pink) for the coloured chaotic orbit of
  (B).}
\label{fig:diffmodel}
\end{figure}

Figure \ref{fig:diffmodel} gives a typical example of the modulational
diffusion regime. The panel (A) shows an apsidal surface of
section $(u,v)$ (see Sec.~\ref{sec:3.X.1-surfsec}) which
depicts the structure of the phase space in the circular model
($e'=0$ for the dynamics under the Hamiltonian $H_b$), 
when $\mu=0.0041$, $e_{p}=0.01675$. For these
parameters, the phase portrait is dominated by the islands of the $1$:$6$
resonance. The separatrix-like chaotic layers surrounding the
resonance are very thin, while the resonant islands are delimited by
both inner and outer librational KAM curves. Thus, all orbits inside this
resonance cannot communicate with orbits of nearby resonances embedded
either in the remaining part of the stability domain or in the chaotic
sea surrounding the stability domain. An orbit near the
separatrix layer of the $1$:$6$ resonance is shown in blue in
Fig.~\ref{fig:diffmodel}(A). Note that $e_p=\sqrt{-2Y_p}$ is an exact
integral of motion of the flow under $H_b$, as confirmed by a
numerical computation of $e_p$ (Fig.~\ref{fig:diffmodel}(C), black curve). On
the contrary, the distance $(W^2+V^2)^{1/2}$ from the forced
equilibrium which coincides with the osculating value of the
eccentricity $e_{p,0}$ (Fig.~\ref{fig:diffmodel}(C), blue curve) undergoes
substantial short-period oscillations (of order $O(\mu)$). Thus, $e_p$
as defined via Eq.~\eqref{eprnew} is a much better measure of the
proper eccentricity than the usual definition $e_{p,0}$. 

Figure \ref{fig:diffmodel}(B), now, shows the projection on the plane
$(u,v)$ of a 4D pericentric surface of section in the elliptic case,
when $e'=0.02$ (black points), on which we superpose in colors the
points of {\it one} chaotic orbit undergoing modulated diffusion. The
$1$:$6$ resonance is still clearly visible on the $(u,v)$ projection,
giving rise to six islands, one of which intersects the line $v=0$ at
values around $u\approx 0.4$.  Another smaller 6-island chain,
intersecting the line $v=0$ at about $u\approx 0.3$ is also
distinguishable. As shown in Fig.~\ref{fig:freqanal}, the latter
corresponds to the transverse resonance $(1,6,1)$, whose extent,
however is limited and produces no substantial overlapping with other
low order resonances. On the other hand, the pulsation of the
separatrix of the $1$:$6$ resonance does result in a substantial
overlapping of this resonance with other outer resonances surrounding
the origin. As a result, an orbit started in the separatrix layer of
the $1$:$6$ resonance later communicates with the separatrix layers of
the outer resonances.  In the example of Fig.~\ref{fig:diffmodel}(B), the
orbit with initial conditions indicated in the caption remains
confined in the neighborhood of the $1$:$6$ resonance up to a time
$\sim 2\pi\times 10^4$ (green points), while at later times (up to
$\sim 2\pi\times 10^5$) the same orbit expands to embed several higher
order resonances of the form $m_f$:$n$ as well as some transverse
resonances of the elliptic problem (pink points). A careful inspection
shows that the orbit undergoes several outward and inward motions in
the whole domain from the $1$:$6$ resonance up to the outer resonances,
while the orbit eventually escapes from the system at still larger
times (of order $10^6$). The various outward or inward transitions are
abrupt, and they are marked by corresponding transitions in the value
of $e_p$, which is now only an approximate adiabatic invariant. Such
transitions are shown in Fig.~\ref{fig:diffmodel}(D) (black curve). Here,
an overall comparison with the time evolution of the quantity
$e_{p,0}$ (pink curve), shows that the definition of $e_p$ via
the action variable $Y_p$ still yields a useful measure of the proper
eccentricity, while $e_{p,0}$ presents wild variations even in
short timescales. In fact, the time behavior of $e_p$ presents jumps
at all outward or inward transitions of the corresponding orbit of
Fig.~\ref{fig:diffmodel}(B). A further analysis of how the diffusion
progresses in the space of action variables $(J_s,Y_p)$ is given in
the following section.

\section{Escapes statistics and chaotic diffusion}\label{sec:3.X-chaoticdiff}

The co-existence of different types of resonances renders non-trivial
the question in which domains of the phase space the chaotic
diffusion, due to the interaction of resonances, provides a more
efficient transport mechanism for orbits, thus affecting long term
stability. As explained in Sec.~\ref{sec:3.X-moddiff}, two main
regimes of chaotic transport exist. For isolated resonances located
inside the boundary of the main stability domain (like the transverse
resonances $(1,6,1)$ and $(1,6,2)$ of Fig.~\ref{fig:freqanal}), the
orbits in the stochastic layer have the possibility of slow diffusion
that bears features of Arnold diffusion. In any case, we find that the
diffusion rate is extremely small, thus it is practically
undetectable.  On the other hand, for resonances located beyond the
boundary of the main stability domain, the diffusion process is best
described by the paradigm of modulational diffusion. In particular,
the amplitude of pulsation of the separatrix-like chaotic layers at
the borders of the resonances is large enough to allow for
communication of the resonances, causing the orbits to undergo abrupt
jumps from one resonance to another, and eventually to escape.

\begin{SCfigure}
\centering
\includegraphics[width=0.55\textwidth]{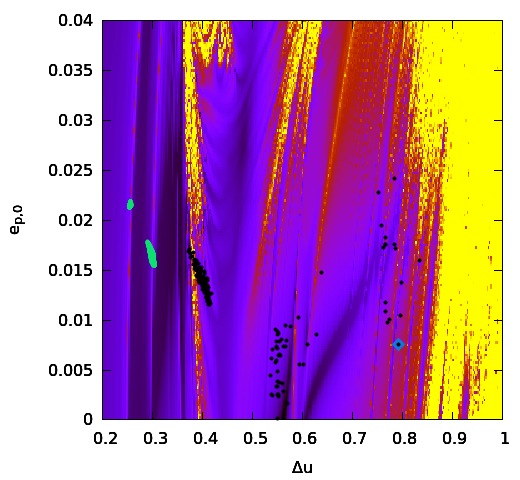}
\caption[Two different diffusion processes]{Different diffusion
  processes for two orbits with parameters $e^{\prime}\,=\,0.02$,
  $\mu\,=\,0.0041$, $e_p\,=\,0.01625$ and initial conditions
  $v\,=\,0$, $\phi\,=\,\pi/3$, $Y_f\,=\,0$ and $\Delta u\,=\,0.299$
  for the green orbit and $\Delta u\,=\,0.376$ for the black orbit.
\vspace{0.8cm}}
\label{fig:flimoddiff}
\end{SCfigure}
Figure \ref{fig:flimoddiff} provides evidence of the processes
mentioned above. Two orbits are shown superposed to the FLI map for
$\mu\,=\,0.0041$, $e'=0.02$. The initial conditions for both
correspond to $e_p=0.01625$, but different $\Delta u$ ($\Delta
u=0.299$, for the orbit in green, and $\Delta u=0.376$, for the orbit
in black). We plot the intersection points of both orbits with the
plane $(\Delta u,e_p)$ when $x=0$, $\phi_f=0$, with a tolerance
$0.001$ and $0.06$ respectively. According to these data, the first
orbit resides in the chaotic layer of the resonance $(1,6,1)$, while
the second is initially in the chaotic layer of the resonance
$1$:$6$. However, the first orbit is restricted to move essentially
only along the stochastic layer of the initial resonance, as no
resonance overlapping exists with any low-order adjacent resonance. As
a result, the orbit's diffusion is practically unobservable. By
contrast, the second orbit suffers a significant change of topology
over a timescale of only $10^5$ periods. The orbit visits many other
resonances besides the starting one, jumping stochastically between
the chaotic layers of the resonances 1:6. $(1,6,-1)$, $(1,6,-2)$,
$(3,19,2)$, and $(3,19,1)$, and possibly other ones of higher
order. The proper eccentricity $e_p$ also exhibits abrupt jumps in the
interval [$0.001$,$0.025$].

\begin{figure}
\vspace{0cm}
\centering
\includegraphics[width=0.7\textwidth]{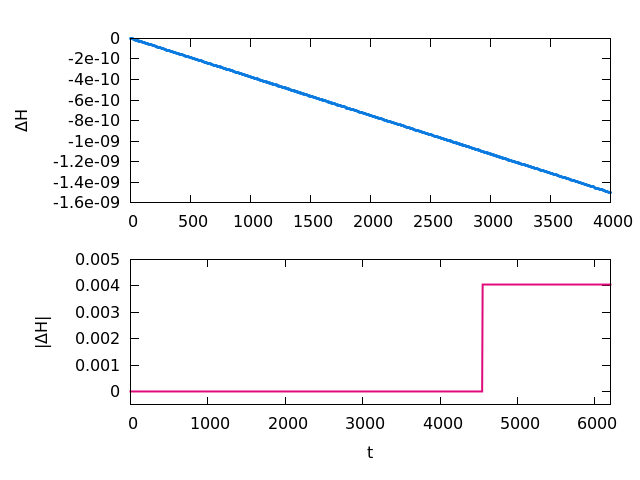}
\caption[Example of the time evolution of the round-off cumulative
  error $\Delta H$]{Time evolution of the value of the round-off
  cumulative energy error $\Delta H$ of an orbit that escapes at
  $t\sim 4600$.}
\label{fig:enererr}
\end{figure}

The long-term behavior of orbits in the modulational diffusion regime 
can be characterized by a statistical study. To this end, we consider 
an ensemble of orbits in a rectangle of initial conditions. As an 
example, setting, as before, $\mu=0.0041$, $e'=0.02$, we consider a 
$60\times 60$ grid of initial conditions in the interval $0.33\leq 
\Delta u\leq 0.93$, $0\leq e_p\leq 0.06$, with the remaining initial 
conditions defined as for the FLI maps above. The ensembles are 
processed at 5 different snapshots, corresponding to the integration 
times of $T=10^3$, $10^4$, $10^5$, $10^6$ and $10^7$ periods. In every 
snapshot (of final time $T$), the orbits are classified in three distinct 
groups:\\
\\
\noindent
{\it Regular orbits}: these are orbits whose value of the 
FLI satisfies the condition
\begin{equation}\label{flilim}
\Psi(T)<\log_{10}(\frac{N}{10}) = \log_{10} N -1
\end{equation}
where $N$ is the total number of periods for the integration. Since
for regular orbits the FLI grows linearly with $N$, the threshold of
Eq.~\eqref{flilim} allows to identify orbits which can be clearly
characterized as regular.  These orbits are exempt from further
integration.\\
\\
\noindent
{\it Escaping orbits:} an orbit is considered as escaping if the orbit
undergoes a sudden jump in the numerical energy error $\Delta H$
greater than $10^{-3}$. This threshold is determined by the
requirement that the jump surpasses by about two orders of magnitude
the worst possible accumulation of round-off energy errors at the end
of the integration time (i.e. after $10^7$ periods). We tested the
cumulative energy error as a function of time for different initial
conditions. Figure~\ref{fig:enererr} shows the evolution of $\Delta H$ for
one example of escaping orbit.  The first panel shows the increment of
$\Delta H$ up to a time $t=4600$.  The absolute cumulative error
grows linearly in time at a rate $\sim 4\times 10^{-13}$~per
period. This rate is characteristic of the orbits in the thin chaotic
layers between the resonances.  However, $\Delta H$ exhibits an abrupt
variation $\Delta H=4\times 10^{-3}$ at the moment of escape. Up to
the maximum integration time $10^7$, the cumulative energy error for
non-escaping orbits is smaller than $4\times 10^{-6}$. Thus, we set a
safe threshold value for escape identifications as $\Delta
H_{esc}=10^{-3}$.\\ \\
\noindent
{\it Transition orbits}: we characterize as transition orbits those whose 
FLI value violates condition~\eqref{flilim}, but which do not escape during 
the integration up to the time $T$. As we will see, part of these orbits 
remain at low FLI values up to the end of the integration, yielding 
a growth $\Psi\sim\log(T)$. Thus, the orbits exhibit a regular 
behavior up to at least $10^7$ periods. However, a second sub-group is 
formed among the transition orbits, containing truly sticky orbits 
with positive Lyapunov exponents and FLI values growing asymptotically 
linearly with $T$. \\
\\
With the results at the five different snapshots, a statistical 
study of the escaping times is constructed as follows: in the end of every 
snapshot, 
i) we count the number of orbits belonging to each of the three groups, 
ii) we compute the histogram of FLI values (from $0$ up to $50$) for the 
transition orbits, and iii) we store the values of $\Delta H$ and $\Psi$ 
for both the escaping and the transition orbits. The results of (i) are 
summarized in the following table:\\
\begin{center}
\begin{tabular}{ccccc}
\hline
Snapshot &(N. of periods) &Regular &Transition &Escaping\\
\hline
\hline
1 &$10^3$ &1220 ($33.8$\%) &2027 ($56.3$\%) &353 ($9.9$\%) \\
\hline
2 &$10^4$ &1263 ($35$\%) &1388 ($38.5$\%) &949 ($26.5$\%) \\
\hline
3 &$10^5$ &1296 ($36$\%) &966 ($26.8$\%) &1338 ($37.2$\%) \\
\hline
4 &$10^6$ &1299 ($36.1$\%) &699 ($19.4$\%) &1602 ($44.5$\%) \\
\hline
5 &$10^7$ &1309 ($36.3$\%) &603 ($16.8$\%) &1688 ($46.9$\%) \\
\hline
~&~&~&~&~\\
\end{tabular}
\end{center}

\begin{figure}
\centering
\includegraphics[width=0.9\textwidth]{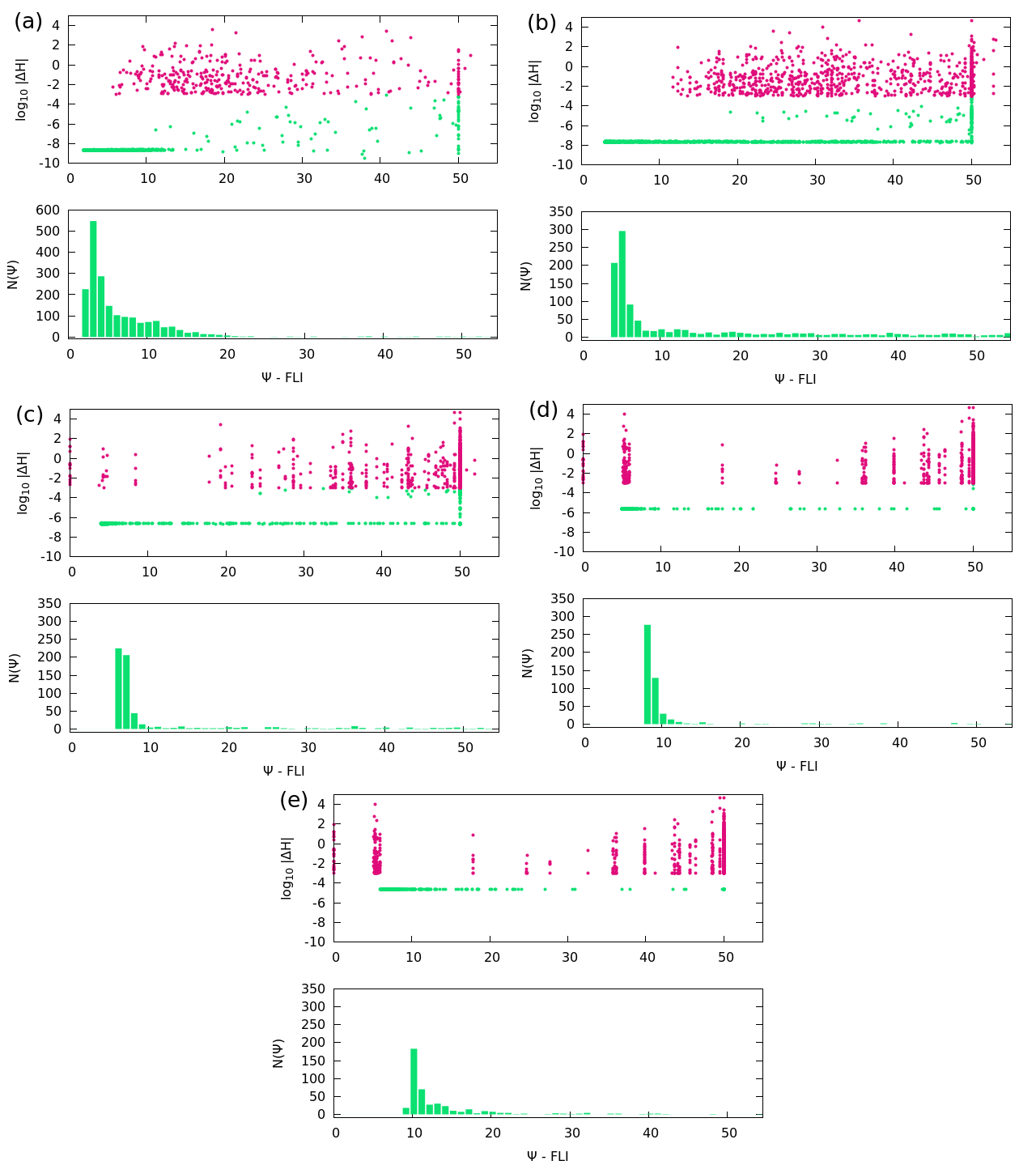}
\caption[Histograms of transition orbits and $\Delta H$
  distributions]{Cumulative energy round-off error $\Delta H$ vs FLI
  value for the groups of transitions (green) and escaping (pink)
  orbits, as well as the distribution of the FLI values for the
  transition group. The different panels refer to the time snapshots
  $T\,=\,10^3$ (a), $T\,=\,10^4$ (b), $T\,=\,10^5$ (c), $T\,=\,10^6$
  (d) and $T\,=\,10^7$ (e) periods of the primaries.}
\label{fig:histo}
\end{figure}
Focusing, now, on the groups of transition and escaping orbits, the
upper panels in Fig.~\ref{fig:histo}a-e show the distribution of $\Delta
H$ and FLI ($\Psi$) values for the transition orbits (green) and the
escaping orbits (pink) respectively. In Fig.~\ref{fig:histo}a (upper panel),
for $T=10^3$ periods, most of the transition orbits are found to keep
a relatively low value of the FLI, $\Psi<10$, and a cumulative energy
error $\Delta H$ of about $10^{-9}$. In fact, {\it all} the transition
orbits with larger $\Delta H$ become escaping orbits shortly after
$T=10^3$ periods. A second group of transition orbits, however, starts
being formed, with FLI larger than or equal to 50. The lower panel
shows the distribution of FLI values for all the transition
orbits. The left concentration represents regular or very sticky
orbits, while the more chaotic orbits are spread over larger values of
the FLI, with a small secondary peak formed in the right part of the
histogram at $\Psi=50$.  However, as the integration time increases, a
`stream' is formed that transports members of the left group towards
the right group. As a result, at the last snapshot, ($T=10^7$
periods), the right group contains about 30\% of the transition orbits
and 6\% of the total orbits considered. In fact, as visually clear in
all upper panels of Fig.~\ref{fig:histo}, most escapes occur at
intermediate values of the FLI, while the right group is nearly
completely detached from the left group of the transition orbits, the
latter moving to the right at a speed logarithmic in $T$, i.e. as
expected for regular orbits. Finally, the Lyapunov characteristic
times of the orbits in the right group are all substantially smaller
than $T_L=10^5$, while the orbits remain sticky for times
$T_{stickiness}> 2\pi 10^7$.  This behaviour is reminiscent of
\emph{stable chaos}~\cite{MilNob-92}.

In addition, we notice that the escaping orbits (pink) seem to form bands
of preferential values of the FLI. We do not not fully identify the
origin of these bands. Nevertheless, they could be connected to the
fact that the escape can occur only via the thin chaotic layers
between the resonances, so that the concentration to particular FLI
values could reflect the local FLI value for orbits residing for long
time within each one of such layers.

\begin{SCfigure}
\centering
\includegraphics[width=0.75\textwidth]{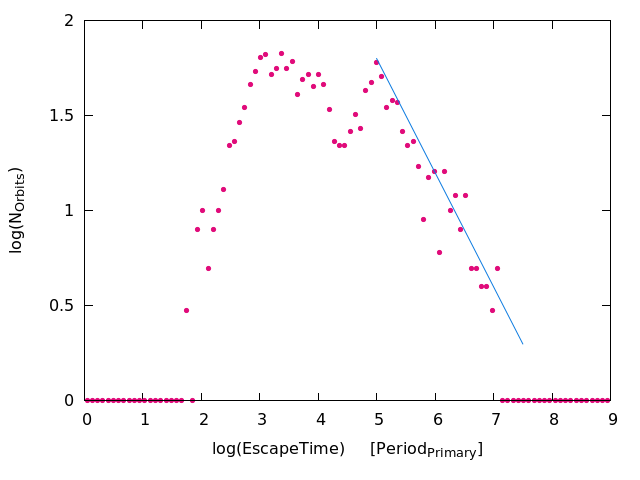}
\caption[Histogram of escaping times for the escaping
  orbits]{Histogram of escaping times for the escaping orbits.
        \vspace{1.5cm}}
\label{fig:histoesc}
\end{SCfigure}
Figure~\ref{fig:histoesc} shows the histogram of escaping times of all the
escaping orbits. It becomes evident that in the process of escaping
two distinct timescales can be distinguished, corresponding to two
peaks of the histogram. The first peak, at about $10^3$ periods,
corresponds to fast escapes, while the second, at about $10^5$
periods, corresponds to slow escapes. As shown below, the large
majority of fast escaping orbits are with initial conditions within
the chaotic sea surrounding the resonances, while slowly escaping
orbits are those with initial conditions in the thin chaotic layers
delimiting the resonances. In the latter case, we find that beyond a
time $t\approx 10^5$ periods, the distribution of the escape times 
shows a {\it power-law} tail. The straight line in Fig.\ref{fig:histoesc}
represents a power-law fit
\begin{equation}
P(t_{esc})\propto t_{esc}^{-\alpha},~~~~\alpha\approx 0.8~~~.
\end{equation}
We note in this respect that power-law statistics of the escape times
are a characteristic feature of stickiness and long-term chaotic
correlations of chaotic orbits~\cite{Meiss-92}.

\begin{figure}
\hspace{-0.7cm}
\includegraphics[width=1.05\textwidth]{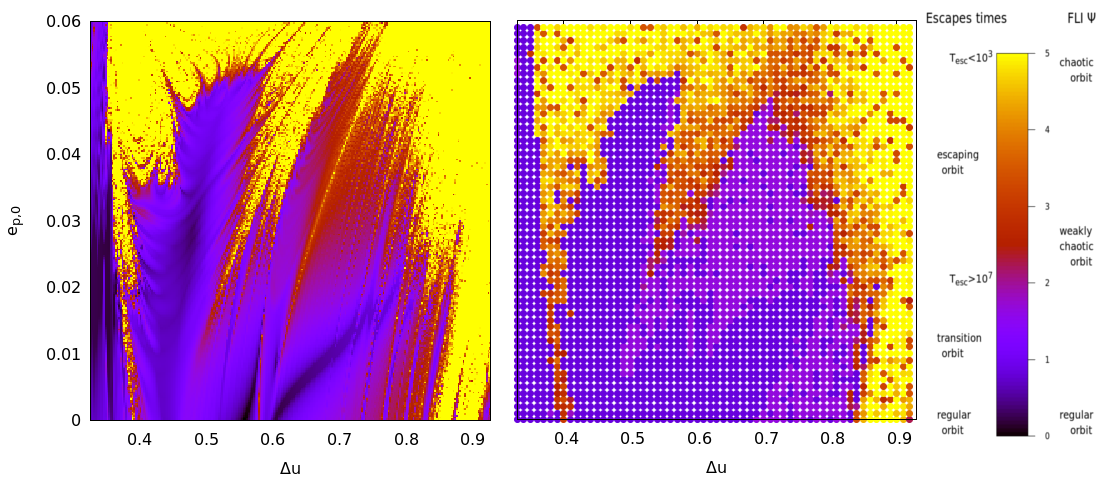}
\caption[FLI stability map and escaping times statistics]{FLI map for
  the grid of initial conditions $[0.33,0.93]\,
  \mathsf{x}\,[0.00,0.06]$ in the plane $(\Delta u,e_p)$ (left panel)
  and color map of the escaping times for the same grid (right panel),
  in both cases with parameters $\mu=0.0041$ and $e'=0.02$. See text
  for more details.}
\label{fig:fliesc}
\end{figure}
Finally, Fig.~\ref{fig:fliesc} shows a comparison between FLI values and
the residence (or escape) times for all the orbits of the integrated
ensemble.  The left panel shows the FLI map for the square of initial
conditions $[0.33,0.93]\times [0.00,0.06]$. The right panel, now,
shows in color scale the residence, or escaping times, for all $3600$
initial conditions of the ensemble. Yellow colors represent the faster
escaping times ($t_{esc}<10^3$ periods), while red the slower (between
$10^6$ and $10^7$ periods). In light purple are depicted the orbits
that remain in the group of transition orbits up to the end of the
integration ($10^7$ periods), while deep purple represents the orbits
belonging to the regular group. The first observation is that the
distribution of escaping times reproduces to some extent the main
features of the resonant structure found by the stability map. In
particular, the chaotic layer of the resonance $1$:$6$ appears
clearly marked by long escape times (larger than $10^5$ periods).
The stickiness is in general enhanced at the borders of all
resonances. On the other hand, most of the orbits with $\Psi=5$
or greater, that are qualified as chaotic in the stability map, belong
to the population of fast escapes, and we find that most need less
than $10^3$ periods to escape. But the most interesting feature is
that all the thin chaotic layers around the resonances $(2,12,-1)$,
$(1,6,-1)$, $(1,6,-2)$, $(3,19,2)$, $(3,19,1)$, $3$:$19$ and
$(2,13,2)$, contain orbits that appear \emph{not escaping} at least up to
$10^7$ periods. This, despite the fact that some of these orbits are
relatively strongly chaotic, i.e with $\Psi$ close to 5. Thus, we can
conclude that the stickiness phenomena in the thin chaotic layers
formed in the resonant domains of the action space can prolong the
stability of hypothetical Trojan bodies up to times comparable to the
age of the hosting system. 

\newpage

\chapter{The basic Hamiltonian $H_b$}\label{sec:4-HB}

In the previous chapter, we introduced the so-called 'basic
Hamiltonian' (Eq.~\ref{hambasic}). This model is a 2 d.o.f. system
representing the short period and synodic components of the
Trojan motion. Its expression comes naturally as a splitting of the
complete Hamiltonian in two parts: $H_b$ and a secular term $H_{sec}$
that gathers all the terms depending on the slow secular angle
$\phi$. By averaging the $H_b$ over its fast angle, we can formally
define action-angle variables for the synodic degree of freedom.
Since the $\overline{H_b}$ includes the 
eccentricity of the primary, it allows to 
find an integrable aproximation to synodic motions even when $e'\neq
0$. On the other hand, it makes a suitable starting point for
non-linear stability studies, from a more general point of view than
the circular problem (Section~\ref{sec:1.4.X.non_stab}).

In the present chapter we start by showing an important property of
the $H_b$: by means of some adjustments in the definitions of
variables and canonical transformations considered, it is shown that
the Hamiltonian $H_b$ found in the ER3BP is formally identical to a
corresponding 'basic model' $H_b$ found in a more complex model,
namely the 'Restricted Multi-Planet Problem' (RMPP), where we consider
the influence of more than one planets on the Trojan body.

We first construct the $H_b$ derived from the RMPP, comparing to the
construction of this Hamiltonian in the ER3BP (Chapter 3).  Having
established the formal correspondence between the two models, we then
return to the ER3BP and perform a more detailed analysis of the
properties of $H_b$, using both numerical and analytical
approaches. In particular, we numerically investigate up to what
extent the decomposition $H = H_{b} + H_{sec}$ provides a meaningful
model. Then, we apply the normalization scheme introduced in
Sec.~\ref{2.2-normalization}, which allows to average the Hamiltonian
over the fast angle circumventing in the ER3BP as well the convergence
problem of large radii.  Finally, using this averaged model, we
compute analytically the position of the most important secondary
resonances and compare the results with those found in the
numerical stability maps computed in the previous chapter.

\section{The Restricted Multi-Planet Problem}\label{sec:4.X-RMPP}

The decomposition of the Hamiltonian of the pER3BP that 
leads to the definition of the
$H_b$ (Sect.~\ref{sec:3.X-expan} and~\ref{sec:3.X-forced_equil}) 
can be generalized in a more representative
problem called planar \emph{Restricted Multi-Planet Problem} (RMPP). This
model is based on the pER3BP, but it includes also the secular effects of
additional planets on the Trojan body.
 
The RMPP Hamiltonian is derived as follows.  We assume that all the
planets are far from mean motion resonances. Their motion is thus
described by a set of secular frequencies: $g'$ for the primary and
$g_1,\, g_2,\,\ldots g_S$ for $S$ additional planets. These
frequencies are possible to compute either by a linear (Laplace)
theory, by a non-linear analytical extension (\cite{LibSans-13}), or
by purely numerical methods (e.g. frequency
analysis,~\cite{Laskar-04}). In any case, we consider that the
frequencies $g'$, $g_j$ are the frequencies of the leading terms in
the quasi-periodic representantion of the oscillations of the
planets' eccentricity vectors. Therefore, we can express their time
evolution as
\begin{equation}
\begin{aligned}\label{excvecprime}
e'\exp{i\varpi'}&=e_0'\exp{i(\varpi_0'+g't)}+\sum_{k=1}^{S} A_k
\exp{i(\varpi_{k0}'+g_kt)}~,\\
e_j\exp{i\varpi_j}&=B_{0j}\exp{i(\varpi_{0j}+g't)}+\sum_{k=1}^{S}
B_{kj} \exp{i(\varpi_{kj}'+g_kt)}~~,
\end{aligned}
\end{equation}
where $e'$, $\varpi'$ are the eccentricity and longitude of the
pericenter of the primary, and $e_j$, $\varpi_j$ the corresponding to
the $j-$th additional planet \cite{Morbi-11}. Without loss of
generality, the constant $\varpi_{0}'$ can be set equal to zero. The
positive quantities $A_{k}$, $B_{kj}$, with $k=1,...,S$, and $B_0$,
are referred to below as \emph{the amplitudes of oscillation of the
planetary eccentricities}. Also, we assume that $e_0' > \sum_{k=1}^{S}
A_k$. Then, at least the primary exhibits a constant, on the average,
precession of its eccentricity vector, by the frequency $g'$, i.e., we
can write $e' = e_{0}' + F$, $\varpi' = g' t + G$, where the functions
$F$ and $G$ depend trigonometrically on the angles $\phi'=g't$,
$\phi_{j}=g_{j} t$, $j=1,...,S$, while their size is of the order of
the amplitudes $A_k$, $k=1,...,S$. All these latter assumptions are
justified if the primary is the most massive planet in the system.

For each secular frequency we introduce now a pair of action-angle
variables, i.e. the `dummy' actions $I',I_j$ and the angles
$\phi'=g't$, $\phi_j=g_j t$, $j=1,2,...,S$. The primary's elements are
given by 
\begin{equation}\label{eq:sec_chang_prim}
e'=e_0'+F(\phi',\phi_1,\ldots,\phi_S)~,\qquad
\varpi'=\phi'+G(\phi',\phi_1,\ldots,\phi_S)~.
\end{equation}
Then, the Hamiltonian
of the RMPP is found by the following steps:\\

\noindent
i. Add to the Hamiltonian of the ER3BP (\ref{ham4}) the direct terms
for $S$ planets. The direct term for the $j$-th planet is:
\begin{equation*}
R_{j,direct}=-\mu_j \left({1\over\Delta_j}
-{\mathbf{r}\cdot\mathbf{r_j}\over r_j^3}\right)
\end{equation*}
where $\mu_j=m_j/(M+m')$ with $m_j$ equal to the mass of the $j$-th
planet, and $r_j=|\mathbf{r}_j|$, $\Delta_j = \vert \mathbf{r}
- \mathbf{r}_j \vert$, with $\mathbf{r}_j$ the heliocentric position
vector of the $j$-th planet. Transcribed to elements, $R_{j,direct}$
depends on $(\lambda,\varpi,\lambda_j,\varpi_j)$.  Assuming that the
$j$-th planet is far from a mean motion resonance with the primary
(and hence with the Trojan body), we compute
\begin{equation*}
{\cal R}_{j}\equiv \langle R_{j,direct} \rangle ={1
\over 4\pi^2}\int_{0}^{2\pi}\int_{0}^{2\pi}R_{j,direct}\df \lambda \df \lambda_j~~.
\end{equation*} 
By rotational symmetry, $\langle R_{j,direct} \rangle$ depends only on the
difference $\varpi-\varpi_j$, and hence, (see Eq.~\ref{excvec}), only
on the angles $\varpi$, $\phi_j$ and $\phi_j$, $j=1,2,...,S$. By
d'Alembert rules, this implies that it is also of first or higher
degree in the eccentricity $e_j$, i.e., it is of first or higher
degree in the amplitudes of oscillation of the planetary
eccentricities.\\

\noindent
ii. Consider now the indirect effects of the $S$ planets on the Trojan 
body. Far from mean motion resonances the primary's 
major semi-axis remains constant. 
Then, the indirect effects are 
accounted for by rendering the parameters $e',\varpi'$ in the 
expression~\eqref{ham4} time-dependent rather than constants. Replace now
Eq.~\eqref{eq:sec_chang_prim} in Eq.~\eqref{ham4}
and Taylor-expand,  around $e_0'$ and $\phi'$, 
assuming $F$ and $G$ small quantities. This leads to:
\begin{equation*}
R(\lambda,\varpi,x,y,\lambda',\varpi',e')=
R(\lambda,\varpi,x,y,\lambda',\phi';e_0') + R_2
\end{equation*}
where $R_2$ is of degree one or higher in the quantities $F,G$, 
and hence, of degree one or higher in the mass parameters $\mu_j$, 
$j=1,\ldots,S$.\\ 

\noindent
iii. Adding also terms for all dummy actions $I'$, $I_j$, the final 
Hamiltonian now reads
\begin{equation}\label{hamsec_rmpp}
H_{mp}=-{1\over 2(1+x)^2} + I + g'I'+ \sum_{j=1}^S g_j I_j - 
\mu R(\lambda,\varpi,x,y,\lambda',\phi';e_0') 
- \mu R_2 - \sum_{j=1}^S\mu_j{\cal R}_{j}~.
\end{equation}

Two important remarks are in order: i) The function $R$ in
Eq. (\ref{hamsec_rmpp}) is formally identical to the function $R$ in
(\ref{ham4}), apart from replacing $e'$ with $e_{0}'$ and $\varpi'$
with $\phi'$. ii) The functions $R_2$ and ${\cal R}_j$, $j=1,\ldots S$
are of first or higher degree in the amplitudes of oscillation of the
planetary eccentricity vectors. We note that the
case where mean motion resonances between the planets are present
necessitates a separate treatment, since then the domain of co-orbital
motion can be crossed by resonances of the type of the `great
inequality' (see~\cite{RobGab-06}).

From this point on, the procedure for deriving the final
Hamiltonian decomposition $H=H_b+H_{sec}$ follows as in
Chapter~\ref{sec:3-ERTBP}.  We only give a short sketch,
emphasizing the points where differences hold between the two
models. Unless explicitely differentiated, 
the conclusions of
Sec.~\ref{sec:3.X-expan} and~\ref{sec:3.X-forced_equil} hold also
here, with similar arguments.

The first canonical transformation ${\cal S}_2$ 
introduces the two resonant angles of the planar Trojan problem:
$\tau = \lambda -\lambda'$ and $\beta=\varpi-\phi'$, instead of
$\delta\varpi= \varpi-\varpi'$,
for representing the 
relative position of the pericenter of the Trojan body from the pericenter
of the planet (since one has $\beta = \varpi - g't +{\cal O}(\mu_j)$).
Altogether, the transformation reads
\begin{equation}\label{gensec_rmpp}
{\cal S}_2=(\lambda-\lambda')X_1+\lambda'X_2+(\varpi-\phi')X_3
+\phi' P' + \sum_{j=1}^{S} \phi_j P_k ~,
\end{equation}
leading to
{\small
\begin{equation}
\begin{aligned}
&\tau  = \lambda - \lambda'~,~~ 
&\tau_2&  = \lambda'~,~~ 
&\beta&  = \varpi - \phi'~,~~ 
&\phi'_{new}& = \phi'~,~~ 
&\phi_{j,new}& = \phi_j~,\\ 
& x = X_1~,~~ 
& I & = X_2 - X_1~,~~ 
& y & = X_3~,~~ 
& I' & = P'- X_3~,~~ 
& I_j& = P_j~,~~ \\
\end{aligned}
\end{equation}}
where $j=1,\ldots,S$.
As in Chapter 3, here also we keep the same notation for variables involved
in an identity transformation. After applying the transformation,
the Hamiltonian~\eqref{hamsec_rmpp} takes the following expression
\begin{equation}
\begin{aligned}\label{hamsec2_rmpp}
H_{mp} = &-{1\over 2(1+x)^2} -x + X_2
 - g'y + g'P'+ \sum_{j=1}^S g_j I_j - \mu R(\tau,\beta,x,y,
\lambda',\phi';e_0')\\
& - \sum_{j=1}^S\mu_j {\cal
R}_j(x,y,\beta,\phi',\phi_1,...,\phi_{s}) - \mu
R_2(x,y,\tau,\beta,\phi',\phi_1,...,\phi_s)~~.
\end{aligned}
\end{equation}

The latter Hamiltonian can be split into two terms $\langle
H_{mp} \rangle$ and $H_{1,mp}$. While $\langle H_{mp} \rangle$ has the
same form as $ \langle H \rangle$ in~\eqref{eq:Haverg} (with the
addition of the term $-g'y$), $H_{mp,1}$ reads
\begin{equation}
\begin{aligned}\label{eq:H1_rtpp}
H_{mp,1} = &\, g'P' + \sum_{j=1}^{S} g_j I_j - \mu  
\tilde{R}(\tau,\beta,x,y,\lambda',\phi';e'_0) \\
&-\sum_{j=1}^{S} \mu_j {\cal R}_j
(x,y,\beta,\phi',\phi_1,\ldots,\phi_S) - \mu R_{2}
(x,y,\tau,\beta,\phi',\phi_1,\ldots,\phi_S)~.
\end{aligned}
\end{equation}
The computation of the forced equilibrium position is done
 as in Chapter~\ref{sec:3-ERTBP}, 
\begin{equation*}
\dot{\tau}={\partial \langle H_{mp} \rangle \over\partial x}=0,~~~
\dot{\beta}={\partial \langle H_{mp} \rangle \over\partial y}=0,~~~
\dot{x}=-{\partial \langle H_{mp} \rangle \over\partial\tau}=0,~~~
\dot{y}=-{\partial \langle H_{mp} \rangle \over\partial \beta}=0~~~
\end{equation*}
resulting
\begin{equation}\label{forced_rmpp}
(\tau_0,\beta_0,x_0,y_0)= \Big( \pi/3,\pi/3,0,\sqrt{1-e_0'^2}-1 \Big)
+ {\cal O}(g')~~.
\end{equation}
For the expansion around the forced equilibrium, 
we consider the shift of center
\begin{equation}\label{poincvar_rmpp}
v=x-x_0,~~u=\tau-\tau_0,~~Y=-(W^2+V^2)/2,~~\phi=\arctan(V,W) 
\end{equation}
where
\begin{equation*}
V=\sqrt{-2y}\sin\beta-\sqrt{-2y_0}\sin\beta_0,~~~~
W=\sqrt{-2y}\cos\beta-\sqrt{-2y_0}\cos\beta_0~~.
\end{equation*}
We thus construct the synodic action variables $(v,u)$ in the plane $(x,\tau)$
around the value $(x_0,\tau_0)$, while the secular action variable
($Y_p$) measures the distance from the forced center $(y_0,\beta_0)$ in
the plane $(y,\beta)$.

Re-organising terms, the Hamiltonian~\eqref{hamsec2_rmpp} takes the form:
\begin{equation}
\begin{aligned}\label{hamsec4_rmpp}
H_{mp}=&\,-{1\over 2(1+v)^2} -v + X_2  - g'Y\\ 
-&\,\mu\left(
{\cal F}^{(0)}(u,\lambda'-\phi,v,Y;e_0')+
{\cal F}^{(1)}(u,\phi,\lambda',v,Y;e_0')
\right)\\
+&\, g'P'- \mu {\cal F}^{(2)}(u,\phi,\lambda',v,Y,\phi';e_0')\\
+&\, \sum_{j=1}^S g_j I_j - \sum_{j=1}^S\mu_j 
{\cal F}_j(u,\phi,v,Y,\phi,\phi',\phi_j,\omega_{0j},e_0',e_{0j})
\end{aligned}
\end{equation}
where (i) ${\cal F}^{(0)}$ contains terms depending on the angles
$\lambda'$, $\phi$ only through the difference $\lambda'-\phi$, (ii)
${\cal F}^{(1)}$ contains terms dependent only on non-zero powers of
$e_0'$ and independent of $\phi'$, and (iii) ${\cal F}^{(2)}$
contains terms dependent on $\phi'$ and also on non-zero powers
of either $e_0'$ or the oscillation amplitudes of the planetary
eccentricities. In terms of the current variables, we are ready
to introduce the basic Hamiltonian $H_b$ as follows
\begin{equation}\label{hambasic_rmpp}
H_b=-{1\over 2(1+v)^2} -v + X_2 - g'Y -\mu{\cal
  F}^{(0)}(u,\lambda'-\phi,v,Y;e_0')~~~.
\end{equation}
The total Hamiltonian takes the form $H=H_b+H_{sec}$, where
\begin{equation}
\begin{aligned}
H_{sec} &= 
-\mu{\cal F}^{(1)}(u,\phi,\lambda',v,Y;e_0')
+ g'P'- \mu {\cal F}^{(2)}(u,\phi,\lambda',v,Y,\phi';e_0') \label{eq:Hsec_RMPP}\\
&+ \sum_{j=1}^S g_j I_j - \sum_{j=1}^S\mu_j 
{\cal F}_j(u,\phi,v,Y,\phi,\phi',\phi_j,\omega_{0j},e_0',e_{0j})~~.
\end{aligned}
\end{equation}

Again, the fact that in~\eqref{hambasic_rmpp} the angles
$\lambda',\phi$ appear only under the combination $\lambda'-\phi$
implies that $H_b$ can be reduced to a system of two degrees of
freedom. The reduction is realized by the canonical transformation:
\begin{equation}\label{gencirc_rmmp}
{\cal S}_3(u,\lambda',\phi,Y_u,Y_s,Y_p) = u Y_u + (\lambda'-\phi) Y_f
+ \phi Y_p
\end{equation}
yielding
\begin{equation}
\begin{aligned}
\phi_u &\,={\partial {\cal S}_2 \over\partial Y_u}=u,~~~ \phi_f &\,={\partial
{\cal S}_2\over\partial Y_f}=\lambda'-\phi,~~~ \phi_p &\,={\partial
{\cal S}_2 \over\partial Y_p}=\phi,\\
v &\,={\partial {\cal S}_2 \over\partial u}=Y_u,~~~ J_3 &\,={\partial
{\cal S}_2 \over\partial \lambda'}=Y_f,~~~ Y &\,={\partial {\cal S}_2 \over\partial
\phi}=Y_p-Y_f~~.
\end{aligned}
\end{equation}
Keeping the old notation 
for $\phi_u=u,\phi_p=\phi$, $Y_u=v$, but, however, retaining the new 
notation for the action $Y_f\equiv X_2$, the Hamiltonian $H_b$ in the 
new canonical variables reads
\begin{equation}\label{hambasic2_rmpp}
H_b=-{1\over 2(1+v)^2}-v + (1+g')Y_f -g'Y_p
-\mu{\cal F}^{(0)}(u,\phi_f,v,Y_p-Y_f;e_0')~~~.
\end{equation}

We can now see that this form of the Hamiltonian $H_b$ is formally
identical in the RMPP and in the ER3BP. From
Eq.~\eqref{hambasic2_rmpp}, with the substitution $e_0'\rightarrow e'$
and setting $g'=0$, it is straightforward to re-obtain
Eq.~\eqref{hambasic}.  Hence, the basic features induced by $H_b$
apply in the same way with or without additional planets. Furthermore,
all extra terms with respect to $H_b$ in the
Hamiltonian~\eqref{eq:hamYfYsYp} depend on the slow angle $\phi$,
while in the case of the RMPP in~\eqref{hamsec4_rmpp}, they depend
also on the slow angles $\phi'$, $\phi_j$, $j=1,\ldots,S$, whose
corresponding frequencies are all secular. Thus, these terms can only
slowly modulate the dynamics under $H_b$. In the case of the ER3BP,
the modulation can produce a long-term drift of the values of
$(Y_p,J_s)$, or $Y_p,J_{s,res}$, as discussed in
Chapter~\ref{sec:3-ERTBP}, that may induce large long term variations
of the actions, and eventually lead to an escape of the Trojan body. A
similar phenomenon is expected in the RMPP. In the latter case, we have
additionally that the position of the forced equilibrium oscillates
quasi-periodically around the origin of the $(W,V)$ system of
axes. The amplitude of oscillation is of order ${\cal O}(\max(\mu_j
e_j))$, while the frequency is of first order in the planetary
masses. Considering that the expression of $H_b$ holds in the two
models, we re-introduce our definition of proper eccentricities,
\begin{equation}\label{epr}
e_{p,0}=\sqrt{V^2+W^2}=\sqrt{-2Y}~~,
\end{equation}
and
\begin{equation}\label{eprnew}
e_p = \sqrt{-2Y_p}~~.
\end{equation}
holding the same properties as in Sec.~\ref{sec:3.X-forced_equil}.

The last remark about the generalization to the RMMP regards the
definition of resonances. The most general form of planar secondary
resonances now is given by
\begin{equation}\label{resgen_rmpp}
m_f\omega_f+m_s\omega_s+m g + m'g'+m_1g_1+\ldots+m_Sg_S=0~,
\end{equation}
where we must take into account also the additional frequencies
induced by the precession of the primary ellipse ($g'$) and the
secular evolution of the extra planets $g_i$.  From the expression of
$H_{mp}$ in~\eqref{hamsec4_rmpp}, we can deduce that the dynamical
role played by transverse resonances involving the proper frequency
$g$~\eqref{resgen}, is quite similar to the one played by resonances
involving the secular planetary frequencies $g'$ and $g_j$,
$j=1,...,S$. Thus, we do not introduce any further diversification
between {\it transverse} resonances arising in the ER3BP and those due
to planetary secular dynamics. This suggests that most numerical
results found before regarding the chaotic diffusion at resonances in
the case of the ER3BP apply to the transverse resonances of more
general models involving more than one disturbing planets.

\section{Limits of applicability of the basic model 
$H_b$}\label{sec:feat_Hb}

The basic model $H_b$ represents a reduction of the number of degrees
of freedom with respect to the original problem. Thus, we expect that
its usefulness in approximating the full problem (ER3BP or RMPP) holds
to some extent only.  The following numerical examples aim to compare
the dynamical behavior of the orbits under the $H_b$ and the full
Hamiltonian. To this end, we compute and compare various phase
portraits (surfaces of section) arising under the two Hamiltonians. We
restrict ourselves to the comparison between $H_b$ and the full
Hamiltonian of the ER3BP only.  Then, as pointed out in the previous
section, all secular perturbations are accounted for by only one
additional degree of freedom with respect to $H_b$, represented by the
canonical pair $(\phi,Y_p)$. Integrating numerically the RMPP instead
of the ER3BP is considerably more demanding. Still, it is arguable
that the effect of the secular perturbations should remain
qualitatively similar by adding more degrees of freedom consisting of
slow action-angle pairs only.

Our numerical integrations of the full Hamiltonian model (ER3BP) are
performed in heliocentric Cartesian variables, in which the equations
of motion are straightforward to express. Whenever needed, translation
from Cartesian to the canonical variables appearing
in~\eqref{eq:hamYfYsYp} and vice versa is done following the sequence
of canonical transformations defined in Sec.~\ref{sec:3.X-expan}
and~\ref{sec:3.X-forced_equil}.

On the other hand, for the basic Hamiltonian $H_b$ in~\eqref{hambasic}
we have an explicit expression only in the latter variables.  However,
one can readily see that, for fixed $(u,v,\phi_f)$, all the initial
conditions of fixed difference $Y_f-Y_p$ lead to the same orbit,
independently of the individual values of $Y_f$ or $Y_p$. If we set
$Y_f=Y_{f,ref}=0$ and $Y_p=Y_{p,ref}=-e_{p,0}^2/2$ for one particular
orbit chosen in advance, denoted as 'reference orbit', this allows to
specify a certain appropriate value of the energy $E=E_{ref}=H_b$ for
that orbit. The proper eccentricity of the reference orbit satisfies
the condition $e_{p,ref}=e_{p,0}$, i.e., it becomes equal to the
modulus of the initial vector $\mathbf{e}-\mathbf{e}_{forced}$, where
$\mathbf{e}=(e\cos\varpi,e\sin\varpi)$ (the so-called 'eccentricity
vector', see Fig.~\ref{fig:paradoxeccvec.png},
Fig.~\ref{fig:figure1.jpg}), and $\mathbf{e}_{forced} =
(e'/2,e'\sqrt{3}/2)$. Now, keeping {\it both} $Y_p=Y_{p,ref}$ and
$E=E_{ref}$ fixed, but altering $(u,v,\phi_f)$, allows to solve the
equation $E_{ref}=H_b$ for $Y_f$ and specify new initial conditions
for more orbits at the same energy as the reference orbit. However, in
general this implies that the initial value of $Y_f$ for any of these
new orbits satisfies $Y_f\neq 0$. In terms of the initial eccentricity
vector, this implies that $e_{p_0}\neq e_{p,ref}$.  However, one
realizes that the so found orbit is precisely the same as one in which
we had set differently the initial value of $Y_p$, i.e. we set
$Y_p=-e_{p,0}^2/2\neq Y_{p,ref}$, and adjust, instead, $Y_f$ to the
value $Y_f=0$ so as to keep the initial difference $Y_f-Y_p$
constant. This means that while, for convenience, we formally proceed
with the former process (keeping $E=E_{ref}$ and $Y_p=Y_{p,ref}$ fixed
and adjusting $Y_f$ for different initial conditions), the correct
level value of the proper eccentricity for each of these initial
conditions is specified by the initial value $e_{p,0}$. Thus, we label
all plots by $e_{p,0}$ instead of $e_p$ in the FLI stability maps
presented before. Note also that in the case of the CR3BP, one readily
finds that $H_b$ becomes the exact full Hamiltonian, and furthermore
one has (apart from a numerical constant) $E=-C_j/2-e_p^2/2$, where $C_j$
is the Jacobi constant (see Sec.~\ref{sec:3.X.1-surfsec}). Then,
keeping $e_p=const=e_{p,ref}$ for all the initial conditions makes our
'isoenergetic' surfaces of section equivalent to surfaces of section
of a constant value of $C_j$. However, it is well known that, in the
circular case, while the value of $C_j$ fixes the overall level of
eccentricities of the trojan orbits, the eccentricity varies
nevertheless a little across different sets of initial conditions for
the same value of $C_j$.

Returning to our numerical computations, in order to choose a
reference orbit we select the one that corresponds to the short period
family around $L_4$. We then set for the
reference orbit $u=v=\phi_f=Y_f=0$, and consider different choices of
value for $Y_p=Y_{p,ref}$. Physically, this means to choose different
energy levels $E=E_{ref}$ at which the central short period orbit has
different proper eccentricity. Let us note that the property of the
central object being a periodic orbit is itself due to the use of the
basic model $H_b$; adding more degrees of freedom renders, instead,
the central object an invariant torus of dimension larger than one and
smaller than the full number of degrees of freedom.

Having selected $E_{ref}$ and $Y_{p,ref}$, we compute initial conditions for 
more orbits. More precisely, in each of the figures which follow, we define a 
set of $19$ initial conditions given by $u_j=0.05\,j$, $v_j=0$, 
$\phi_{f,j}=0$, for $j=0,\ldots,18$, and $Y_{f,j}$ computed as described 
above. With the above initial conditions, we numerically integrate the 
orbits, under the equations of motion of $H_b$, up to collecting, for each 
orbit, 500 points on the surface of section given by the condition $\phi_f=0$. 

Now, the same set of initial conditions is integrated under the equations 
of motion of the full ER3BP, for a time equivalent to 1000 revolutions 
of the primary, collecting about 990 points in the same surface of 
section. In the ER3BP, the surface of section is four-dimensional, 
but a two-dimensional projection on the plane $(u,v)$ allows to 
compare with the corresponding section of the basic model $H_b$. 

As an additional comparison, we also compute the surface of section 
provided by an intermediate model between the $H_b$ and the pER3BP.
We construct a 3 d.o.f Hamiltonian in the following way
\begin{equation}\label{eq:hbsec}
H_{b,sec} = H_b(Y_f,\phi_f,u,v,Y_p;\mu,e',e_{p,0}) + \langle {\cal F}^{(1)} \rangle
(Y_f,u,v,Y_p,\phi;\mu,e',e_{p,0})~~,
\end{equation}
where 
\begin{equation}\label{eq:f1ave}
\langle {\cal F}^{(1)} \rangle ={1\over 2\pi}\int_0^{2\pi}H_{sec}\df \phi_f~~,
\end{equation}
where $H_{sec}$ is given in Eq.~\eqref{eq:Hsec_ER3BP}.  Explicit
formul\ae~for $\langle {\cal F}^{(1)} \rangle$ can be found in the
Appendix~\ref{appex:theHb}. Such terms include a certain dependence on
the slow angle $\phi$, but are independent of the fast angle $\phi_f$.
Hence, $H_{b,sec}$ contains some but not all secular terms of the
disturbing function of the pER3BP. Up to first order in the mass
parameter $\mu$, the averaging \eqref{eq:f1ave} yields the same
Hamiltonian as the one produced by a canonical transformation
eliminating all secular terms depending in the fast angle $\phi_f$.
Thus, the model $H_{b,sec}$ captures the main effect of the secular
terms, as discussed in Sect.~\ref{sec:3.X-moddiff}: this is
a \emph{pulsation}, with frequency $g$, of the separatrices of all the
secondary resonances induced by $H_b$. Since the modulation due to
these secular terms is slow, far from secondary resonances we expect
that an adiabatic invariant holds for initial conditions close to the
invariant tori of $H_b$, thus yielding stable regular orbits. On the
other hand, as already discussed, close to secondary resonances the
pulsation provokes a weak chaotic diffusion best described by the
paradigm of modulational diffusion~\cite{Chirikov-85}.

\begin{figure}[h]
  \centering 
\includegraphics[width=0.80\textwidth]{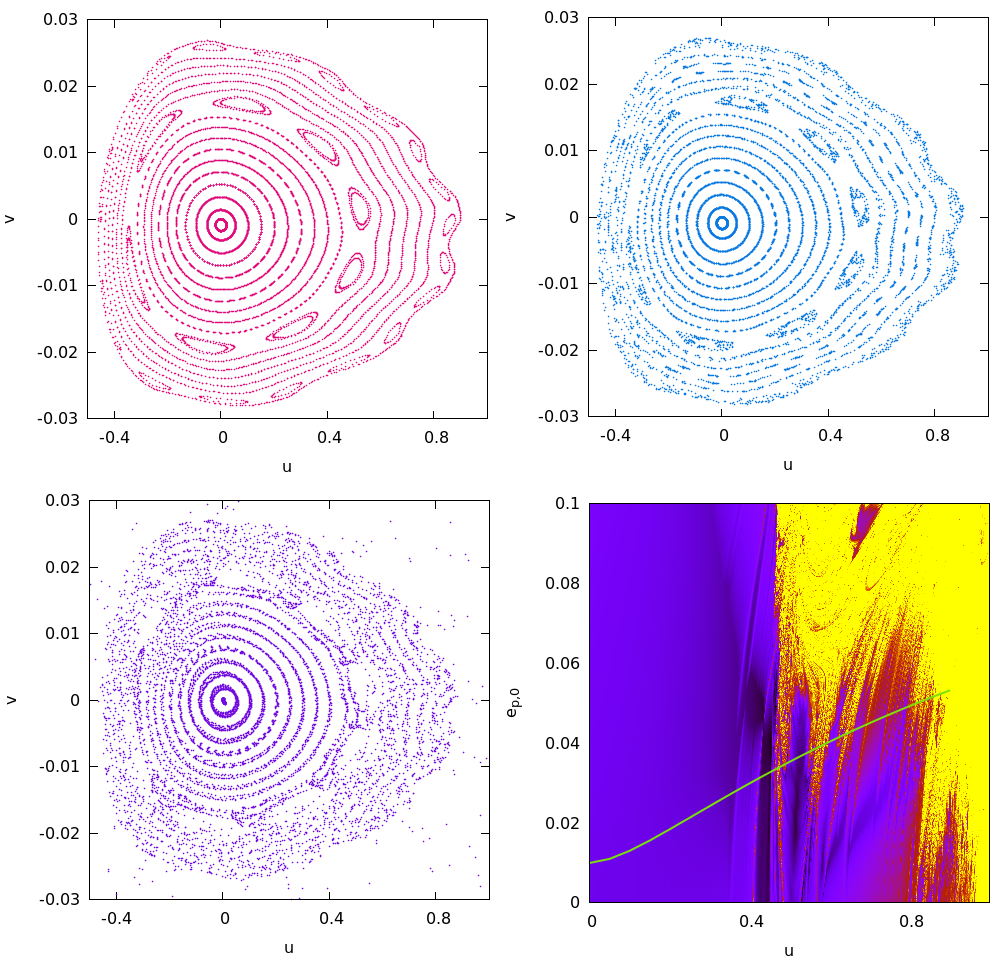} 
\caption[Comparison of surfaces of section for the $H_b$, the $H_{b,sec}$
and the ER3BP - 1]{Comparison of surfaces of section (section
  condition $\phi_f=0$) provided by different models.  The considered
  parameters are $\mu=0.0024$, $e'=0.04$ and $e_{p,ref}=0.01$. In pink
  points (upper left), we show the surface of section provided by
  $H_b$. In blue points (upper right), the one corresponding to
  $H_{b,sec}$. In purple points (lower left), the one corresponding to
  pERTBP. In lower right panel, we reproduce the FLImap of
  Sec.~\ref{sec:3.X.2-flimaps} corresponding to the physical
  parameters $\mu$ and $e'$ considered. The green line on the FLImap
  indicates the isoenergetic curve where the initial conditions are
  located.}  \label{fig:plot-epr01}
\end{figure}

\begin{figure}[t]
  \centering 
  \includegraphics[width=0.80\textwidth]{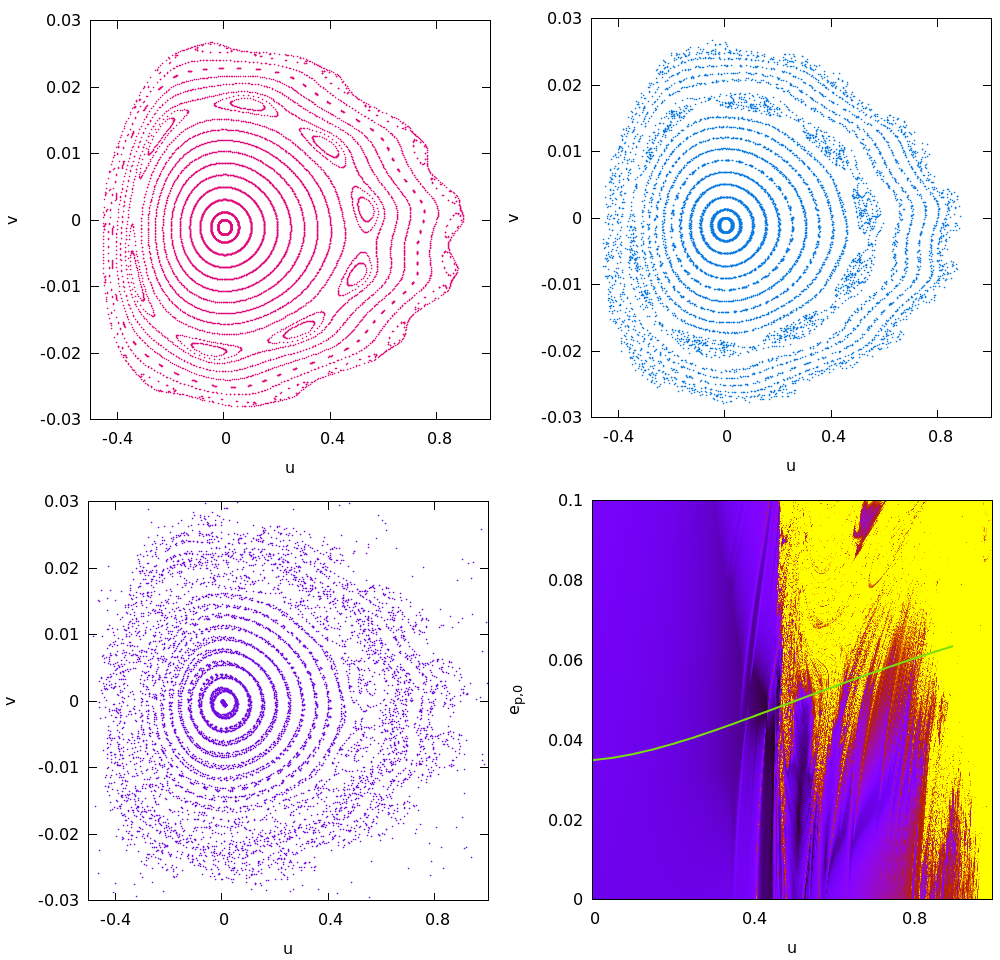} 
  \caption[Comparison of surfaces of section for the $H_b$, the $H_{b,sec}$ 
  and the ER3BP - 2]{Same as in Fig.\ref{fig:plot-epr01}, but for a higher 
  parameter value $e_{p,ref}=0.035$.}  
  \label{fig:plot-epr035}
\end{figure}

\begin{figure}[h]
  \centering
  \includegraphics[width=0.80\textwidth]{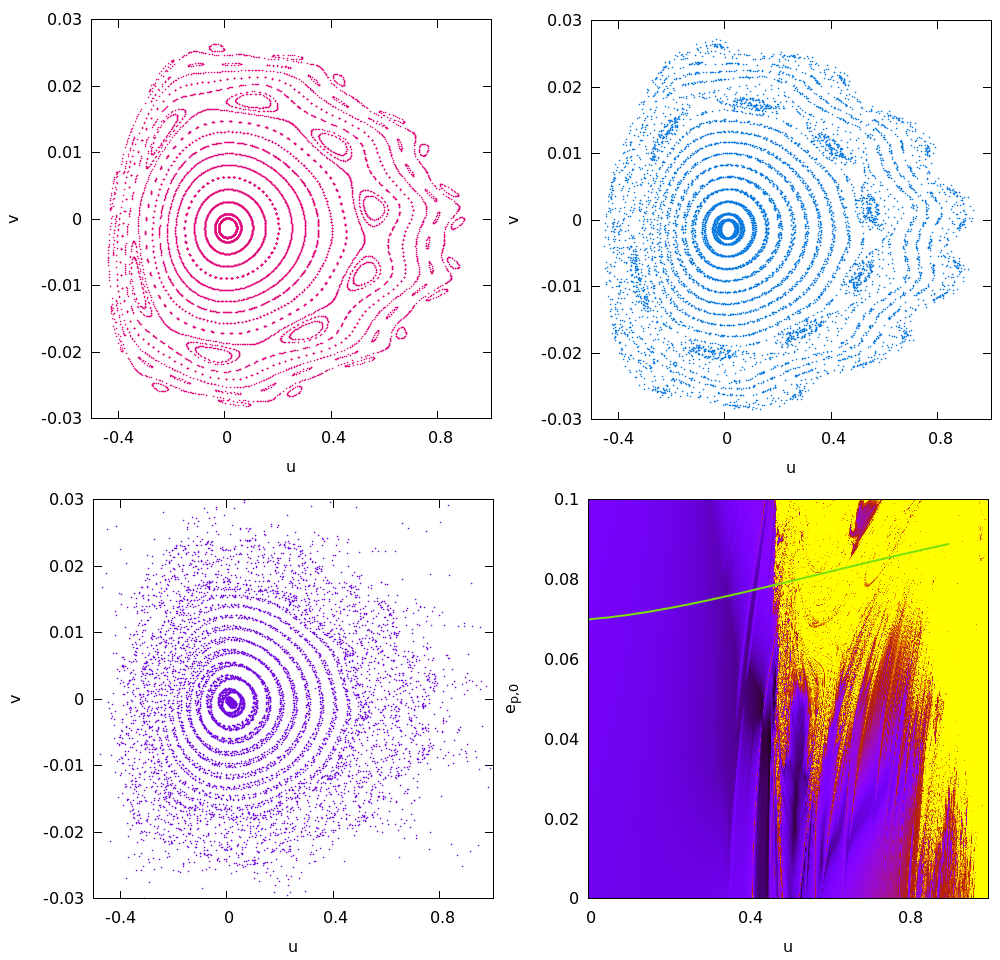}
  \caption[Comparison of surfaces of section for the $H_b$, the $H_{b,sec}$
  and the ER3BP - 3]{Same as in Fig.\ref{fig:plot-epr01}, but for a 
  still higher parameter value $e_{p,ref}=0.07$.}
  \label{fig:plot-epr07}
\end{figure}

Figures~\ref{fig:plot-epr01},~\ref{fig:plot-epr035}
and~\ref{fig:plot-epr07} show some examples of the comparison between the
three surfaces of section mentioned above. The physical parameters
chosen for these plots are $\mu=0.0024$ (which depicts clearly the 1:8
main secondary resonance) and $e'=0.04$. Figure~\ref{fig:plot-epr01}
shows the surface of section corresponding to $e_{p,ref}=0.01$,
Fig.~\ref{fig:plot-epr035} to $e_{p,ref}=0.035$ and
Fig.~\ref{fig:plot-epr07} to $e_{p,ref}=0.07$.  In each figure, the
upper left plot (pink points) corresponds to the surface of section
produced by the flow under the basic model $H_b$, the upper right plot
(blue points) to the flow under $H_{b,sec}$ and the lower left plot
(purple points) to the flow under the full Hamiltonian of the pER3BP.
As an additional information, we provide the FLI stability map
corresponding to the same parameters $\mu$ and $e'$, which was
computed in Sect.~\ref{sec:3.X.2-flimaps}.  On top of the FLI map, in
green we show the locus of initial conditions $(u,e_{p,0})$ on the
surface of section whose orbits have constant energy $E=E_{ref}$.

In Fig.~\ref{fig:plot-epr01}, in the approximation based on the model $H_b$, 
the absence of any dependence of the dynamics on the slow angle $\phi$ renders 
possible a clear display of the short period and synodic dynamics by means of 
the surface of section, which, for $H_b$, is two-dimensional. In fact, for 
more complex models like $H_{b,sec}$ or the full pER3BP, the corresponding 
surface of section is 4-dimensional and its 2D projection on the $(u,v)$ 
plane becomes blurred (top right and bottom left panels respectively). 
The blurring can be due partly to projection effects. However, we argue below 
that an important effect is caused also by the influence of the secular terms, 
absent in $H_b$, to the dynamics. 

Returning to the phase portrait of $H_b$, this allows to extract relevant
information such as: i) the position of the central fixed point, 
corresponding to the crossing of the section by the short period orbit, 
ii) several secondary resonances and the corresponding resonant islands 
of stability, and iii) the overall size of the libration domain of 
\emph{effective} stability. Also, this phase portrait allows to understand 
the structure of the stability map. In the phase portrait, as we move
from left to right along the line $v=0$, we encounter non-resonant
tori, interrupted by thin chaotic layers and the islands of some
secondary resonances, namely the resonances $1$:$8$ and $2$:$17$. The same
resonances are crossed as we move from left to right along the green
curve in the FLI stability map. Note, however, that no transverse
secondary resonances can be seen in the $H_b$ portrait, since these
resonances correspond, in general, to a non-resonant frequency ratio
of the fast and synodic frequencies $\omega_f$ and $\omega_s$; except
at resonance junctions, the exact resonance condition $m_f\omega_f +
m_s\omega_s + m_g\omega_g=0$ for some non-zero $m_s$, $m_f$, $m_g$
implies, in general, non-commensurable values of $\omega_f$ and
$\omega_s$. Since $g\ll \omega_s\ll \omega_f$, most transverse
resonances can only accumulate close to the main secondary resonances
forming resonant multiplets, as confirmed by visual inspection of the
stability maps presented in Sec.~\ref{sec:3.X.2-flimaps}. However, some
isolated transverse resonances may be embedded in the main domain of
stability whose border is marked by the most conspicuous secondary
resonance. In Fig.~\ref{fig:plot-epr01}, this domain extends up to
about $u\approx 0.5$. In the stability map of
Fig.~\ref{fig:plot-epr01}, the transverse resonances $[1,8,k]$, with
$k=-2,-1,1,2$, form a multiplet together with the conspicuous
resonance $1$:$8$. Two of these transverse resonances ($k=2$ and $k=1$)
are embedded in the main domain of stability. However, none of the
transverse resonances is visible in the phase portrait of the basic
model $H_b$.

As discussed in
Sec.~\ref{sec:3.X-forced_equil}, the amplitude of the secular terms
depends on the values of $e'$ and $e_{p,0}$.  For fixed $e'\neq 0$,
the amplitude of the pulsation generated by such terms increases with
$e_{p,0}$. For values of $e_{p,0}$ large enough, the pulsation
modifies the whole behavior in phase-space. Since, along the line
$v=0$, $e_{p,0}$ increases with $u$ (green curve in last panel of
Fig.~\ref{fig:plot-epr01}), the amplitude of the pulsation increases
as we move from the central fixed point outwards in the associated
phase portraits.

In regions where the resonant web is dense enough, this pulsation causes 
all narrow transverse resonances in a multiplet to overlap, increasing
the size of the chaotic domain and facilitating escaping mechanisms.  
In the set of parameters of Fig.~\ref{fig:plot-epr01}, we see from the 
corresponding FLI map that this happens for values of $e_{p,0}$ greater than 
about $0.06$.  Beyond this value, the effect induced by $H_{sec}$ implies 
that the blurring observed in the phase portraits (apart from the one of 
$H_b$) is not due just to projection effects but it has a dynamical origin, 
the nature of the orbits changes as they are converted from regular to 
chaotic. 

This latter effect is more conspicuous in Fig.~\ref{fig:plot-epr035} 
and~\ref{fig:plot-epr07}, in which, choosing a higher $e_{p,ref}$, we increase 
the level of proper eccentricities of all the orbits. 
In Fig.~\ref{fig:plot-epr035} the FLI stability map shows large domains 
of chaos which are not observed in the phase portrait of $H_b$, but they 
appear in the phase portrait of the full model. The separatrix pulsation 
of the $1$:$8$ resonance is not, however, large enough so as to completely 
wash out this resonance, which is therefore seen in all four panels of 
the plot. On the other hand, increasing still more the level of proper 
eccentricities (Fig.~\ref{fig:plot-epr07}) makes this pulsation large 
enough so as to completely introduce chaos in the position of the $1$:$8$ 
resonance. This limit of eccentricity levels marks the overall validity 
of the approximation based on $H_b$ regarding the position of secondary 
resonances. Beyond this value, $H_b$ still represents fairly well the 
dynamics only inside the main librational domain of stability. We note 
also that the elimination of the main secondary resonance $1$:$8$ by the 
separatrix pulsation is already present in the model $H_{b,sec}$ 
(compare the corresponding phase portraits in  
Fig.~\ref{fig:plot-epr01}, \ref{fig:plot-epr035}, \ref{fig:plot-epr07}). 

In conclusion, the pulsation mechanism induced by the secular terms in
the Hamiltonian affects essentially those regions of the phase space
where resonances accumulate in the form of multiplets. For libration
orbits, these are the regions beyond the main secondary resonance
$1$:$n$, which always dominates the phase-space.  The regions inner to
that resonance are not influenced considerably and the representation
of the dynamics via the basic model $H_b$ remains accurate there, even
for high values of the proper eccentricity. The value of the latter at
which the separatrix pulsation of the $1$:$n$ resonance completely
washes this resonance marks the overall limit of approximation of the
basic model. On the other hand, most orbits beyond that limit turn to
be chaotic and fast-escaping the libration domain, thus of lesser
interest in applications related to Trojan astronomical objects.

\section{Normalization of the $H_b$}\label{sec:norm_Hb}

We devote the rest of the chapter to the application of the
normalizing scheme introduced in Sec.~\ref{2.2-normalization} to the
Hamiltonian model $H_b$.  As shown in the previous section, this model
allows to efficiently separate the secular part of the Hamiltonian
from the part representing the dynamics in the fast and synodic
degree of freedom.  In Chapter 2, this computation was done for the
pCR3BP. Here, we show how it can be extended, via the Hamiltonian
$H_b$, to problems of higher complexity, like the ER3BP or the RMPP.
As a practical example, we show a method for the computation of the
position of several resonances, based on normal form computations,
applicable to in a wide spectrum of physical parameters.

\subsection{Preparation of the Hamiltonian $H_b$}\label{sec:preparation}

We start by first expressing the basic model $H_b$ in variables 
appropriate for introducing the normalization scheme of Chapter 2. 
To this end, the synodic degree of freedom is re-expressed
by the variables
\begin{equation}\label{eq:shift_cent}
x = v + x_0, \quad \tau = u +\tau_0~~.
\end{equation}
The constants $x_0$ and $\tau_0$ give the position of the forced
equilibrium (Eq.~\ref{forced_rmpp}).  In the expression of the
$H_b$~\eqref{hambasic_rmpp}, it turns convenient to introduce new
canonical pairs, though the transformation
\begin{equation}\label{gensec_rmpp_4}
{\cal S}_4= (Y_f - Y_p) \theta_1 + Y_p \theta_2 + x \theta_3~,
\end{equation}
yielding
\begin{equation}
\begin{aligned}
&Y_1  = Y_f - Y_p ~,~~ 
&\phantom{\theta_2  - } Y_p =\,& Y_2 ~,~~ 
& x   = Y_3~, \\ 
& \theta_1 = \phi_f ~,~~ 
 &\theta_2  -\theta_1 =\,& \phi ~,~~ 
& \theta_3  = \tau~.\\
\end{aligned}
\end{equation}
We keep, as before, the same notation for variables
transformed by the identity. In addition, since 
only one action ($Y_1$) and angle ($\theta_2$) variable are introduced
by the transformation, we refer to them as ${\cal Y}$ and $\theta$.
After these preliminary transformations,
the basic model $H_b$ reads
\begin{equation}\label{eq:hbasic_xtau}
H_b = -\frac{1}{2(1+x)^2}-x + {\cal Y}+Y_p
- \mu {\cal F}^{(0)}(x,{\cal Y},\tau,\phi_f;e')~~.
\end{equation}
In terms of these variables, the dependence of $H_b$ on $\tau$ is of
the form $\frac{\cos^{k_1} \tau} {(2-2\cos\tau)^{j/2}}$ or
$\frac{\sin^{k_2} \tau}{(2-2\cos\tau)^{j/2}}$, $j=2n-1$ with $k_1$,
$k_2$ and $n$ integers.

In order to initialize the normalization procedure, we write and
expand the Hamiltonian in \eqref{eq:hbasic_xtau}, by introducing
modified Poincar\'{e} variables, as in Eq.~\eqref{eq:Garfinkel-coord},
\begin{equation}
\begin{aligned}\label{eq:delau-garf-coord}
& x~, &\tau~,\phantom{=\,\sqrt{2{\cal Y}}\sin\phi_f\,} \\
&\xi =\,\sqrt{2{\cal Y}}\cos\phi_f\,, &\eta=\,\sqrt{2{\cal Y}}\sin\phi_f\,~~.\\
\end{aligned}
\end{equation}
The new expression for the Hamiltonian reads
\begin{equation}\label{eq:hbasic_xtauxieta}
H_b(\tau,x,\xi,\eta,Y_p) = -\frac{1}{2(1+x)^2}-x + Y_p +
\frac{\xi^2+\eta^2}{2} - \mu {\cal F}^{(0)}
(\tau,x,\xi,\eta;Y_p,e')~~.
\end{equation}
Finally, we expand the Hamiltonian in terms of every variable except
$\tau$, obtaining
\begin{equation}
\begin{aligned}\label{eq:hb_xtauxieta_exp}
H_b (x,\tau,\xi,\eta,Y_p) &= -x
+ \sum_{i=0}^{\infty} \,(-1)^{i-1}(i+1)\,
\frac{x^i}{2}\, +\, \frac{\xi^2+\eta^2}{2} \,+ \,Y_p \\ +&
\,\mu \sum_{\substack{ m_1,m_2,m_3\\ k_1,k_2,k_3,j}}
a_{m_1,m_2,m_3,k_1,k_2,j}\, e'^{k_3} x^{m_1} \, \xi^{m_2}\, \eta^{m_3}
\, \cos^{k_1}(\tau) \, \sin^{k_2}(\tau) \,
\beta^j(\tau)~~,
\end{aligned}
\end{equation}
where the $a_{m_1,m_2,m_3,k_1,k_2,j}$ are constant coefficients and
$\beta(\tau)=\frac{1}{\sqrt{2-2\cos\tau}}$. The Hamiltonian $H_b$ 
in~\eqref{eq:hb_xtauxieta_exp} corresponds to the
'zero-th' step in the normalizing scheme, i.e., before any normalization. 
This we denote as $H^{(1,0)}$.

\subsection{Normalization scheme}\label{sec:norm_sche}

As in Sect.~\ref{2.2-normalization}, the normalizing algorithm defines
a sequence of Hamiltonians by an iterative procedure. Since the
idea behind the scheme remains the same as before, just for the sake
of completeness, we introduce all the necessary formul\ae~for 
the normalization in terms of the current set of variables.

The main formal difference with respect to the scheme presented 
at Sect.~\ref{2.2-normalization} lies on the corresponding definition
of the class of functions ${\cal P}_{l,s}$, that must include also
the contribution of the planet's eccentricity. Thus these functions are now
of the form
{\small
\begin{equation}\label{eq:classPwitheprime}
\sum_{2m_1+m_2+m_3=l}\, \, \sum_{\substack{k_1+k_2\leq l+4s-3\\ j\leq 2l+7s-6}} 
a_{m_1,m_2,m_3,s,k_1,k_2,j}\, \mu^s e'^{k_3} x^{m_1} \, 
\xi^{m_2}\, \eta^{m_3}
\, \cos^{k_1}(\tau) \, \sin^{k_2}(\tau) \,\beta^j(\tau)~,
\end{equation}}
where, despite the fact that $e'$ appears separated from the
coefficient $a_{m_1,m_2,m_3,s,k_1,k_2,j}$, it plays no role in the
normalization scheme and it is carried on (along with its powers) in the
normalization as a parameter.

At a generic normalizing step ($r_1$,$r_2$), the expansion of the
Hamiltonian is given by
{\small
\begin{equation}
\begin{aligned}\label{eq:hr1r2-1}
H^{(r_1,r_2)}(x,\xi,\tau,\eta,Y_p) = &\, Y_p+\frac{\xi^2 + \eta^2}{2} +
\sum_{l\geq4}Z_{l}^{(0)} \left(x,(\xi^2+\eta^2)/2 \right)\\
 + &\, \sum_{s=1}^{r_1-1} \left( \sum_{l=0}^{R_2} \mu^s
 Z_{l}^{(s)} \left( x,(\xi^2+\eta^2)/2,\tau \right) 
+ \sum_{l>R_2}\mu^{r_1} f_l^{(r_1,r_2-1;s)} (x,\xi,\eta,\tau) \right)\\
 + &\, \sum_{l=0}^{r_2} \mu^{r_1} Z_{l}^{(r_1)} (x,(\xi^2+\eta^2)/2,\tau) 
 + \sum_{l\geq r_2+1} \mu^{r_1}
 f_{l}^{(r_1,r_2-1;r_1)} (x,\tau,\xi,\eta) \\ 
+ &\, \sum_{s>r_1} \sum_{l\geq 0} \mu^{s}
 f_{l}^{(r_1,r_2-1;s)}(x,\tau,\xi,\eta)~~. 
\end{aligned}
\end{equation}}
All the terms $Z_{l}^{(s)}$ and $f_{l}^{(r_1,r_2;s)}$ appearing
in~\eqref{eq:hr1r2-1} are made by expansions including a \emph{finite}
number of monomials of the type given by the class ${\cal
  P}_{l,s}$. More specifically $Z_l^{(0)}\in {\cal P}_{l,0}$
$\forall\ l\ge 4$, $Z_l^{(s)}\in {\cal P}_{l,s}$ $\forall\ 0\le l\le
R_2\,,\ 1\le s<r_1\,$, $Z_l^{(r_1)}\in {\cal P}_{l,r_1}$
$\forall\ 0\le l<r_2\,$, $f_l^{(r_1,r_2-1;r_1)}\in {\cal P}_{l,r_1}$
$\forall\ l\ge r_2\,$, $f_l^{(r_1,r_2-1;s)}\in {\cal P}_{l,s}$
$\forall\ l>R_2\,,\ 1\leq s<r_1\,$ and $\forall\ l\ge 0,\ s>r_1\,$.

As before, we can distinguish the terms in normal form ${\cal Z}$
(i.e. the terms depending on $\xi$ and $\eta$ exclusively through
$(\xi^2+\eta^2)/2$), from those that still keep a generic dependence
on these variables.

The $(r_1,r_2)$--th step of the
algorithm formally defines the latter Hamiltonian $H^{(r_1,r_2)}$ by
\begin{equation}\label{eq:hr1r2Lie}
H^{(r_1,r_2)} = \exp \left({\cal L}_{\mu^{r_1}\chi_{r_2}^{(r_1)}}\right) 
H^{(r_1,r_2-1)}~~,
\end{equation}
where the Lie series operator 
$\exp {\cal L}_{\mu^{r_1}\chi_{r_2}^{(r_1)}}$  is given in~\eqref{eq:Lie-exp-oper}.
The generating function $\mu^{r_1}\chi_{r_2}^{(r_1)}$ is
determined by solving the following homological equation with respect
to the unknown
$\chi_{r_2}^{(r_1)}=\chi_{r_2}^{(r_1)}(x,\xi,\tau,\eta)$:
\begin{equation}\label{eq:homol_eq}
{\cal L}_{\mu^{r_1}\chi_{r_2}^{(r_1)}} Z_2^{(0)} +
f_{r_2}^{(r_1,r_2-1;r_1)} = Z_{r_2}^{(r_1)}~~,
\end{equation}
where $Z_{r_2}^{(r_1)}$ is the new term in the normal form, and
$Z_2^{(0)}$ represents the kernel of the homological equation. By
construction, the Hamiltonian produced at ever step inherits the
structure presented in~\eqref{eq:hr1r2-1}.  From the latter, we point
out that the splitting of the Hamiltonian in sub-functions of the form
${\cal P}_{l,s}$, organizes the terms in groups with the same order of
magnitude $\mu^s$ and total degree $l/2$ (possibly semi-odd) in the
variables $x$ and ${\cal Y}=\frac{\xi^2+\eta^2}{2}$. It is easy to
check that the Hamiltonian $H_b$ in~\eqref{eq:hb_xtauxieta_exp} is in
suitable form for the first normalizing step, as Hamiltonian $H^{(1,0)}$,
according to~\eqref{eq:hr1r2-1}.

Let $R_1$ and $R_2$ be the maximum orders considered for the 
normalization scheme, thus 
the algorithm requires $R_1\cdot R_2$ normalization steps, 
constructing the finite sequence of Hamiltonians 
$H^{(1,0)} = H_b,H^{(1,1)},\,\ldots,H^{(R_1,R_2)}$.
We remark here that $H^{(r_1+1,0)} = H^{(r_1,R_2)}\, \forall \, 
1 \leq r_1 \leq R_1$.
Hence, the final Hamiltonian, reads
\begin{equation}\label{eq:HR1R2}
H^{(R_1,R_2)}(x,\xi,\tau,\eta,Y_p) = {\cal Z}^{(R_1,R_2)} 
\left(x,\frac{(\xi^2+\eta^2)}{2}, \tau,Y_p \right) 
+ {\cal R}^{(R_1,R_2)} (x,\xi,\tau,\eta)~~,
\end{equation}
where we distinguish the normal form ${\cal Z}^{(R_1,R_2)}$ from the
remainder ${\cal R}^{(R_1,R_2)}$.
While the dependence of ${\cal Z}^{(R_1,R_2)}$ on $x$ and $\tau$
remains generic, it depends on $\xi$ and $\eta$ \emph{only} though the
form $\frac{\xi^2+\eta^2}{2}$.  Thus, we have
\begin{equation}\label{eq:HR1R2renamed}
H^{(R_1,R_2)}(x,\tau,{\cal Y},\phi_f,Y_p) = 
{\cal Z}^{(R_1,R_2)}\left(x,\tau,{\cal Y},Y_p\right) 
+ {\cal R}^{(R_1,R_2)} (x,\tau,{\cal Y},\phi_f)~~. 
\end{equation}
The key remark is that $\phi_f$ becomes ignorable in the normal form
and, therefore, ${\cal Y}$ becomes an integral of motion of ${\cal
Z}^{(R_1,R_2)}$). Then, the normal form can be viewed as a Hamiltonian
of one d.o.f. depending on two constant actions ${\cal Y}$ and $Y_p$,
i.e. ${\cal Z}^{(R_1,R_2)}$ represents now a formally integrable
dynamical system. Formally speaking, the normalization over the fast
angle $\phi_f$ corresponds to the canonical method for reducing $H_b$
to its averaged version. Of course, since the true system is not
integrable, it is natural to expect that the normalization procedure
diverges in the limit of $R_1,R_2\rightarrow \infty$.  The divergence
corresponds to the fact that the size of the remainder function ${\cal
R}^{(R_1,R_2)}$ cannot be reduced to zero as the normalization order
tends to infinity. Then, the {\it optimal} normal form approximation
is obtained by choosing the values of both integer parameters $R_1$
and $R_2$ so as to reduce the size of the remainder ${\cal
R}^{(R_1,R_2)}$ as much as possible. In practice, there are
computational limits that compromise the choice of values of $R_1$ and
$R_2$. In all subsequent computations, the values are $R_1=2$ and
$R_2=4$, corresponding to a second order expansion and truncation on
the mass parameter $\mu$ and fourth order for the polynomial degree of
$\xi$ and $\eta$ (second order expansion in the eccentricity $e$; note
also that the expansion is of second order as well in the the
primary's eccentricity $e'$). In the following, these normalization
orders are shown to be sufficient for the normal form to provide a
good representation of the original Hamiltonian in the domain of
regular motions.  In the next section, we employ this possibility in
order to compute the positions of different resonances, based on the
integrable approximation provided by the normal form.

\section{Application: normal form determination of the location of resonances}\label{sec:location}

The obtention of a normal form by averaging the basic Hamiltonian
allows to extract information of the resonant structure by pure
analytical means. In this section, we focus on the use of the normal
form approximation $Z^{(R_1,R_2)}$ in \eqref{eq:HR1R2renamed} for the
computation of the values of the three main frequencies of motion.
With these values, it is possible to locate the position of the most
important resonances for a certain combination of physical parameters.

Consider an orbit with initial conditions as specified in terms of the
two parameters $u=\tau-\tau_0$ and $e_{p,0}$ as detailed in the
previous section, referring to stability maps like the ones of
figures \ref{fig:plot-epr01},
\ref{fig:plot-epr035}, \ref{fig:plot-epr07}.
The computation proceeds by the following steps.

1) We first evaluate the synodic frequency $\omega_s$, i.e., the 
frequency of libration of the synodic variables $\tau$ and $x$. 
The normal form $Z^{(R_1,R_2)}$ leads to Hamilton's equations:
\begin{equation}\label{eq:xdot}
\frac{\textrm{d}x}{\textrm{d}t} = f (x,\tau;{\cal Y}) =
-\frac{\partial Z^{(R_1,R_2)}}{\partial \tau}~~
\end{equation}
and 
\begin{equation}\label{eq:taudot}
\frac{\textrm{d}\tau}{\textrm{d}t} = g (x,\tau;{\cal Y}) =
\frac{\partial Z^{(R_1,R_2)}}{\partial x}~~.
\end{equation}
For every orbit we can define the constant energy 
\begin{equation}\label{eq:Zhamilt}
Z^{(R_1,R_2)}(x,\tau;{\cal Y},Y_p)-Y_p \equiv 
\zeta^{(R_1,R_2)}(x,\tau;{\cal Y})= {\cal E}~~.
\end{equation}
Note that since $Y_p$ appears only as an additive constant in 
$Z^{(R_1,R_2)}$, the function $\zeta^{(R_1,R_2)}$ does not depend 
on $Y_p$. Also, according to \eqref{epr} and 
\eqref{eq:delau-garf-coord}, we have ${\cal Y}=\frac{e_{p,0}^2}{2}$. 
Then, for a fixed value of ${\cal E}$, if
$\frac{\partial \zeta^{(R_1,R_2)}}{\partial \tau} \neq 0$, we can
express $\tau$ as an explicit function of $x$,
\begin{equation}\label{eq:taufunct}
\zeta^{(R_1,R_2)}(x,\tau;{\cal Y}) = {\cal E}  \quad \Longrightarrow 
\quad \tau = \tau({\cal E},x;{\cal Y})~~. 
\end{equation}
Thus, replacing in~\eqref{eq:xdot}, 
\begin{equation}\label{eq:xdot2}
\frac{\textrm{d}x}{\textrm{d}t} = f (x,\tau({\cal E},x;{\cal Y})
;{\cal Y}) \quad \Longrightarrow \quad \text{d}t = 
\frac{\text{d}x}{f (x,\tau({\cal E},x;{\cal Y}) ;{\cal Y})}~~,
\end{equation}
by which we can derive an expression for the synodic period $T_{syn}$
\begin{equation}\label{eq:synper_int}
T_{syn} = \oint \frac{\text{d}x}
{f (x,\tau({\cal E},x;{\cal Y}) ;{\cal Y})}~~,
\end{equation}
and thus the synodic frequency
\begin{equation}\label{eq:synfreq}
\omega_{s} = \frac{2\pi}{T_{syn}}~~.
\end{equation}
In practice, \eqref{eq:taufunct} is hard to invert analytically, 
and likewise, the integral (\ref{eq:synper_int}) cannot be explicitly 
computed. We thus compute both expressions numerically on grids of 
points of the associated invariant curves on the plane $(\tau,x)$, 
or by integrating numerically \eqref{eq:xdot2} as a first order 
differential equation (we found that the latter method is more 
precise than the former).

2) We now compute the fast and secular frequencies $\omega_f$, 
$g$. Since 
$Z^{(R_1,R_2)}(x,\tau;{\cal Y},Y_p) = Y_p + \zeta^{(R_1,R_2)}(x,\tau;{\cal Y})$, 
we find 
$\dot{\theta} =\frac{\partial {\cal Z}^{(R_1,R_2)}}{\partial Y_p} =1$ 
implying $g=1-\omega_f$. To compute now $\omega_f$, we use the
equation
\begin{equation}\label{eq:fastfreq1}
\omega_{f} = \, \frac{1}{T_{syn}} \int_{0}^{T_{syn}} 
\frac{\textrm{d}\phi_f}{\textrm{d}t} \, \textrm{d}t = 
\, \frac{1}{T_{syn}} \int_{0}^{T_{syn}} 
\frac{\partial Z^{(R_1,R_2)} (x,\tau;{\cal Y})}{ \partial {\cal Y}} \, 
\textrm{d}t~~.
\end{equation}
Replacing \eqref{eq:xdot2} in \eqref{eq:fastfreq1}, we generate an
explicit formula for the fast frequency
\begin{equation}\label{eq:fastfreq3}
\omega_{f} = \frac{1}{T_{syn}} \oint
\frac{1}{f (x,\tau({\cal E},x;{\cal Y}) ;{\cal Y})} \,
\frac{\partial Z^{(R_1,R_2)} (x,\tau({\cal E},x;{\cal Y});{\cal Y})}
{ \partial {\cal Y}}\, \textrm{d}x~~.
\end{equation}

Both frequencies $\omega_f$ and $\omega_s$ are functions of the labels
${\cal E}$ and ${\cal Y}$, which, in the integrable normal form
approximation, label the proper libration and the proper eccentricity
of the orbits. In the normal form approach one has
$e_{p,0}=e_p=$~const, implying ${\cal Y}=e_p^2/2$. If, as for the FLI
maps in Sect.~\ref{sec:3.X.2-flimaps}, we fix a scanning line of
initial conditions $x_{in} = B\, u_{in}= B\,(\tau_{in}-\tau_0)$, with $B$ a
constant, the energy ${\cal E}$, for fixed $e_p$, becomes a function
of the initial condition $u_{in}$ only. Thus, $u_{in}$ represents an
alternative label of the proper libration~\cite{Erdi-78}.
With these conventions, all three frequencies become functions of the
labels $(u_{in},e_p)$. A generic resonance condition then reads
\begin{equation}\label{eq:rescond_functPhi}
\Phi_{m_f,m_s,m} = m_f \omega_f(e_p,u_{in}) 
+ m_s \omega_s(e_p,u_{in})+ m g(e_p,u_{in})=0~~.
\end{equation}
For fixed resonance vector $(m_f,m_s,m)$, \eqref{eq:rescond_functPhi}
can be solved by root-finding, thus specifying the position of the
resonance in the plane of the proper elements $(u_{in},e_p)$.  

\begin{SCfigure}
  \includegraphics[width=.55\textwidth]{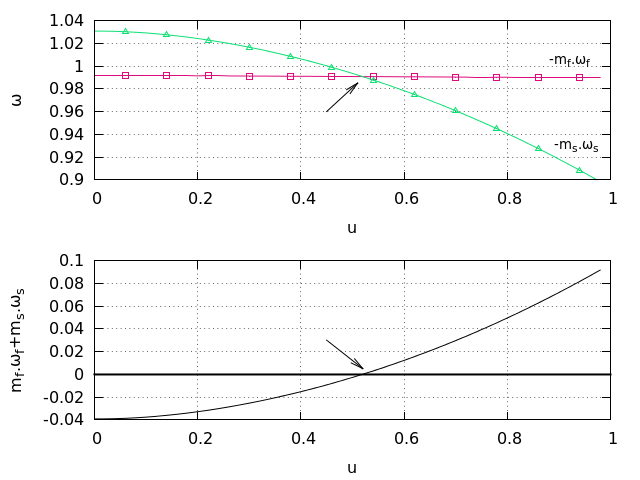} 
  \caption[Representation
  of the evolution of the frequencies in the normal form ${\cal
  Z}^{(R_1,R_2)}$ as function of $u$]{Representation of the evolution
  of the frequencies as function of $u$.  In the upper panel,
  $m_f \omega_f$ (pink line, square points) and $-m_s\omega_s$ (green
  line, triangle points).  In the lower panel, the evolution of the
  function $m_f\omega_f+m_s\omega_s$ (black line). The arrows denote
  the point where the frequencies accomplish the resonant condition
  $m_f\omega_f+m_s\omega_s=0$, giving the position of the resonance in
  terms of $u$. For this example, we choose the resonance $1:8$,
  corresponding to $m_f=1$, $m_s=8$, $\mu=0.0024$, $e'=0.04$ and a
  representative value for
  $e_{p,0}=0.05$.  \vspace{0.5cm}} \label{fig:freq_fig}
\end{SCfigure}

As an example, Fig.~\ref{fig:freq_fig}, shows $\omega_f$ and
$\omega_s$, as well as the function $\Phi_{1,-8,0}(e_p,u_{in})$, as
functions of $u_{in}$ for the parameters $\mu=0.0024$, $e'=0.04$ and a
fixed value of $e_p=0.05$. The arrow in the lower panel marks the
position of the resonance. Changing the value of $e_p$ in the same
range as the one considered in our numerical FLI stability maps
($0<e_{p,0}<0.1$), we specify $u_{in}$ all along the locus of the
resonance projected in the stability map. Repeating this computation
for several transverse resonances $(m_f,m_s,m)$ we are able to trace
the location of each of them.

\begin{figure}[h]
  \centering \includegraphics[width=.95\textwidth]{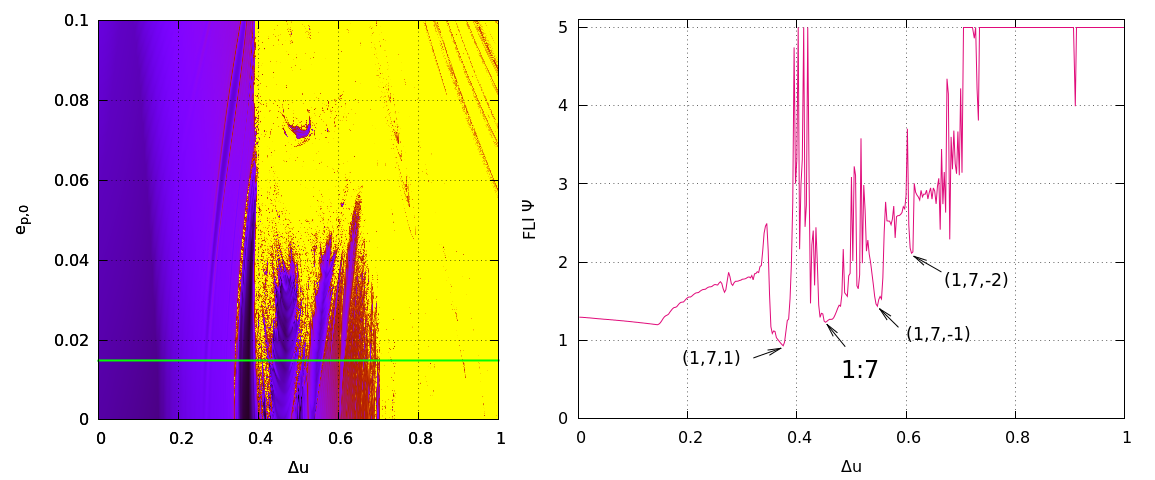} 
  \caption[Local minima of the FLI as tracers of the resonance center]{FLI
  $\Psi$ as function of the libration amplitude $\Delta u$, for fixed
  parameters $\mu~=~0.0031$, $e'~=~0.04$ and $e_{p,0}~=~0.015$ (right
  panel). The local minima give a good approximation of the position
  of the centers of each resonance. The orbits whose corresponding FLI
  values are plotted in the left panel lie on the green line on top of
  the FLI map (right panel). The confirmation of each canditate is 
  done by frequency analysis.}  \label{fig:linesFLI.png}
\end{figure}

In order to test the accuracy of the above method, we compare the
results of the analytical estimation with the position of the
resonances extracted from the FLI maps computed in
Sect.~\ref{sec:3.X.2-flimaps}. Under the assumption that the local
minimum of the FLI in the vicinity of a resonance gives a good
approximation of the resonance center, we study the curves of the FLI
$\Psi$ as a function of the libration amplitude $\Delta u$, for a
fixed value of $e_{p,0}$.  Figure~\ref{fig:linesFLI.png} exemplifies
the case for $\mu=0.0031$, $e'=0.04$, $e_{p,0} = 0.015$, where we
choose four candidates as centers of the resonances $(1,7,1)$,
$1$:$7$, $(1,7,-1)$ and $(1,7,-2)$. The confirmation of the resonant
nature of the candidate orbits is done by means of Frequency
Analysis~\cite{Laskar-04}. By changing the value of $e_{p,0}$ along
the interval $[0,0.1]$, we can depict the centers of the resonances on
top of the FLI maps.

\begin{SCfigure}
  \centering \includegraphics[width=.60\textwidth]{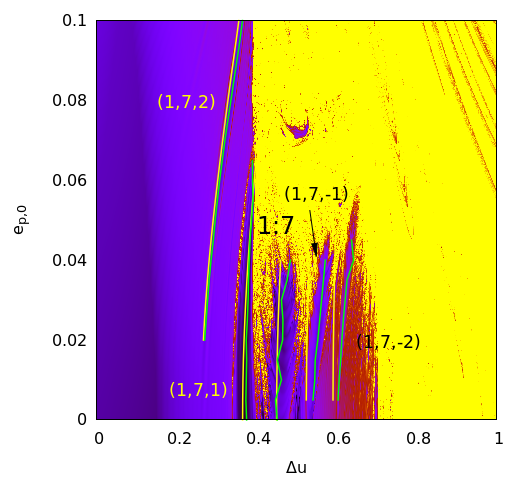} 
  \caption[Main
  and transverse secondary resonances located by $Z^{(R_1,R_2)}$
  and by FLI $\Psi$ minima - 1]{Main
  and transverse secondary resonances located by $Z^{(R_1,R_2)}$
  (yellow) and the estimation of FLI $\Psi$ minima (green).  In this
  example, $\mu=0.0031$, $e'=0.04$, $m_f=1$, $m_s=7$, $m=0,\pm 1,\pm
  2$.  Labels indicate the corresponding resonance in each
  case. \vspace{0.8cm}} 
  \label{fig:transverses1.png}
\end{SCfigure}

\begin{SCfigure}
  \centering 
  \includegraphics[width=.60\textwidth]{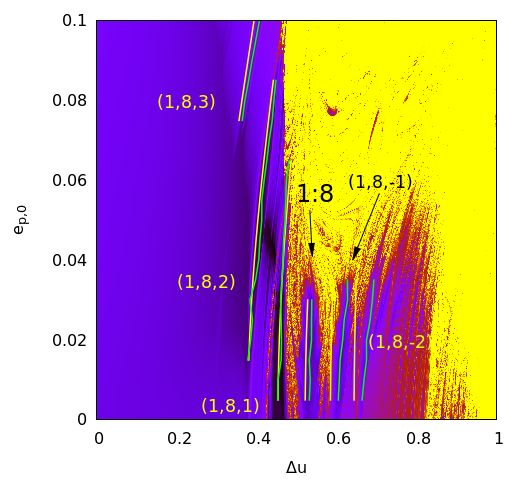} 
  \caption[Main
  and transverse secondary resonances located by $Z^{(R_1,R_2)}$
  and by FLI $\Psi$ minima - 2]{Same
  as Fig.~\ref{fig:transverses1.png}, for $\mu=0.0024$, $e'=0.06$, and
  $m_f=1$, $m_s=8$, $m=0,\pm 1,\pm 2,3$.  
  \vspace{0.8cm}} 
  \label{fig:transverses2.png}
\end{SCfigure}

\begin{SCfigure}
  \centering
  \includegraphics[width=.60\textwidth]{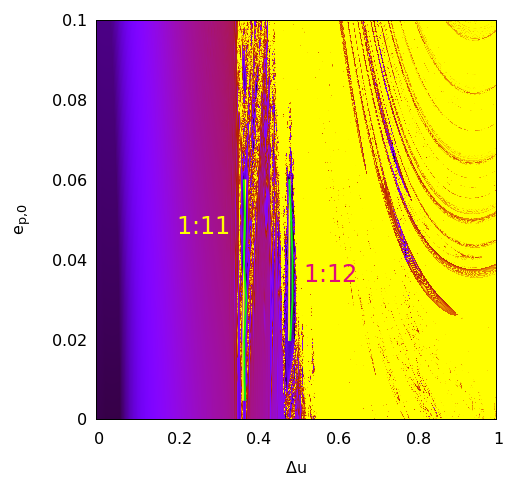}
  \caption[Main
  and transverse secondary resonances located by $Z^{(R_1,R_2)}$
  and by FLI $\Psi$ minima - 3]{Same
  as Fig.~\ref{fig:transverses1.png}, for $\mu=0.0014$, $e'=0.02$,
  and  $m_f=1$, $m_s=11,12$, $m=0$.
  \vspace{0.8cm}} 
  \label{fig:transverses3.png}
\end{SCfigure}

\begin{table}[h]
\begin{center}
\begin{tabular}{|c c c c c|}
\hline
Resonance & $\mu$, $e'$ & $\overline{u_{{\cal Z}}}$ & $\overline{u_{\Psi}}$
& $\overline{\delta u_{in}}$ \\
\hline
$1$:$7$   & $0.0031$, $0.04$ & 0.453908 & 0.463308  &  2.129422$\snot[-2]$ \\
$(1,7,1)$ &      $''$       & 0.377456 & 0.380947  &  1.417910$\snot[-2]$ \\
$(1,7,2)$ &      $''$       & 0.306036 & 0.312011  &  1.880279$\snot[-2]$ \\
$(1,7,-1)$&      $''$       & 0.527218 & 0.554430  &  4.885329$\snot[-2]$ \\
$(1,7,-2)$&      $''$       & 0.593373 & 0.618057  &  3.964370$\snot[-2]$ \\
$1$:$8$   & $0.0024$, $0.06$ & 0.524485 & 0.535153  &  1.993063$\snot[-2]$ \\
$(1,8,1)$ &      $''$       & 0.465475 & 0.464924  &  6.377401$\snot[-3]$ \\
$(1,8,2)$ &      $''$       & 0.406439 & 0.412246  &  1.605145$\snot[-2]$ \\
$(1,8,3)$ &      $''$       & 0.374879 & 0.385020  &  2.617987$\snot[-2]$ \\
$(1,8,-1)$&      $''$       & 0.587834 & 0.616093  &  4.572688$\snot[-2]$ \\
$(1,8,-2)$&      $''$       & 0.646464 & 0.679154  &  4.796435$\snot[-2]$ \\
$1$:$11$  & $0.0014$, $0.02$ & 0.367663 & 0.370842  &  9.264243$\snot[-3]$ \\
$1$:$12$  &      $''$       & 0.482117 & 0.486631  &  1.021940$\snot[-2]$ \\
\hline
\end{tabular}
\end{center}
\caption{Averaged values of $u_{{\cal Z}}$, $u_{{\Psi}}$ and $\delta u_{in}$ for the resonances in Figures~\ref{fig:transverses1.png}, 
\ref{fig:transverses2.png} and~\ref{fig:transverses3.png}}
\label{tab:errors}
\end{table}

Figures~\ref{fig:transverses1.png},~\ref{fig:transverses2.png}
and~\ref{fig:transverses3.png} show examples of these computations,
for the parameters $\mu=0.0031$ and $e'=0.04$, $\mu=0.0024$ and $e'=0.06$, 
$\mu=0.0014$ and $e'=0.02$, respectively. The normal form predictions are
superposed as yellow lines upon the underlying FLI stability maps and
the resonant candidates extracted from the FLI maps denote the green
curves. Due to the noise in the FLI curves, it is not possible to
clearly extract the position of the resonance centers for all values
of $e_{p,0}$, while an analytic estimation (with varying levels of
accuracy) is always possible.  At any rate, in 
Fig.~\ref{fig:transverses1.png}-\ref{fig:transverses3.png},
we plot the values of the centers only
in the cases when both methods provide clear
results. Table~\ref{tab:errors} summarizes the results for the
location of the centers ($u_{{\cal Z}}$, $u_{\Psi}$) and the relative
errors ($\delta u_{in} = \frac{|u_{{\cal Z}} - u_{\Psi}|}{u_{\Psi}}$),
on average, for the resonances shown in the figures.

Regarding the overall performance of the analytic estimation, we can
note that the level of approximation is very good for relatively low
values of $\mu$, $e_p$ and $u_{in}$, while the error in the predicted
position of the resonance increases to a few percent for greater
values of those parameters, with an upper (worst) value $6\%$ (see
Table~\ref{tab:errors}). This is the expected behavior for a normal
form method, whose approximation becomes worse with higher values of
the method's small parameter(s).  Nevertheless, we demonstrate the
overall efficiency of the normal form approach in order to
analytically determine the locus of resonances in the space of proper
elements. This confirms that the basic Hamiltonian is able to well
approximate the fast and synodic dynamics of the ER3BP. Additionally,
the fact that we do not consider expansions in terms of $\tau$ allows
to retain accurate information about higher order harmonics.  We also
showed that by the use of the relation between the fast action $Y_f$
and the secular action $Y_p$, it is possible to estimate, via $H_b$,
the value of the secular frequency, and to determine the position of
transverse resonances, even when though these resonances are not
represented in the dynamics under the $H_b$.



\newpage

\chapter{Asymmetric expansions and resonant normal form for $H_b$}\label{sec:asym-expand-res-nf}

As already discussed in Chapters 3 and 4, one of the most interesting
features in the libration domain of the Trojan motion is the existence
of secondary resonances. For some combinations of physical parameters,
these resonances occupy a large fraction of the domain of stability.
In the previous chapter, we provided an analytical method for locating
the centers of these resonances, based on reducing the system to an
integrable model of 1 d.o.f.  Nevertheless, this approach cannot
estimate analytically the \emph{size} of the secondary resonance
given, for example, by the width of the separatrix-like thin chaotic
layer which typically surrounds the libration domain of the resonance.
These features can only be estimated if we transform the system into
a \emph{pendulum-like} Hamiltonian, that represents the motion in the
resonant domain. To this end, in the present chapter we provide the
construction of a \emph{resonant} normal form for the basic model
$H_b$ introduced in previous chapters. The application of this
algorithm requires a complete Fourier decomposition of the Hamiltonian
in terms of the angles involved in the resonance. In practice, this
means that we can no longer keep terms depending on the quantity
$\beta(\tau) = \frac{1}{\sqrt{2 -2 \cos \tau}}$, as in the approach described in
Chapters 2 and 4. As discussed in Chapter 2, this could imply a loss
of the good convergence properties of the series expansions. However,
in the sequel, we provide a novel expansion allowing to partly remedy this
problem.  This allows to compute a resonant normal form for secondary
resonances. We provide tests of the latter's accuracy, by locating
resonances and comparing with the results of the previous chapters.

\section{Asymmetric expansion}

The resonant normal form computed below provides a model for studying
the dynamics involved in the domain of a secondary resonance of the
form:
\begin{equation}\label{eq:main_sec_res}
m_f\omega_f+m_s\omega_s = 0~,
\end{equation}
where $\omega_f$ and $\omega_s$ are the frequencies associated to the
fast and synodic angles $\phi_f$, $\phi_s$ (related to $x$ and
$\tau$), and $m_f$ and $m_s$ are small integers. In 
Sect.~\ref{sec:norm_sche}, we introduced a version
of the $H_b$ that was only partially expanded, since we retained
powers of the quantity $\beta(\tau)=\frac{1}{\sqrt{2-2\cos\tau}}$.
We recall here the expression of this
Hamiltonian,
{\small
\begin{equation}
\begin{aligned}\label{eq:hb_with_beta_xi_eta}
H_b (&\tau,x,\xi,\eta,Y_p) = -x + \sum_{i=0}^{\infty} \,(-1)^{i-1}(i+1)\,
\frac{x^i}{2}\, +\, \frac{\xi^2+\eta^2}{2} \,+ \,Y_p \\ +&
\,\mu \sum_{\substack{ m_1,m_2,m_3\\ k_1,k_2,k_3,j}}
a_{m_1,m_2,m_3,k_1,k_2,j}\, e'^{k_3} x^{m_1} \, \xi^{m_2}\, \eta^{m_3}
\, \cos^{k_1}(\tau) \, \sin^{k_2}(\tau) \,
\beta^j(\tau)~~, 
\end{aligned}
\end{equation}}
where $a_{m_1,m_2,m_3,k_1,k_2,j}$ are rational numbers. Setting 
$\xi = \sqrt{2 {\cal Y}} \cos \phi_f$ and 
$\eta= \sqrt{2 {\cal Y}} \sin \phi_f$, 
the Hamiltonian reads
{\small
\begin{equation}
\begin{aligned}\label{eq:hb_with_beta_calY} 
H_b (&x,{\cal Y},\tau,\phi_f,Y_p) = -x
+ \sum_{i=0}^{\infty} \,(-1)^{i-1}(i+1)\,
\frac{x^i}{2}\, +\, {\cal Y} \,+ \,Y_p \\ +&
\,\mu \sum_{\substack{ m_1,m_2,m_3\\ k_1,k_2,k_3,j}}
a_{m_1,m_2,m_3,k_1,k_2,j}\, e'^{k_3} x^{m_1} \, \cos^{k_1}(\tau) \, 
\sin^{k_2}(\tau) {\cal Y}^{m_4} \,
\cos^{m_2} \phi_f \, \sin^{m_3} \phi_f \,
\beta^j(\tau)~~,
\end{aligned}
\end{equation}}
where $m_4 = (m_2+m_3)/2$. The librations in $\tau$ are represented in
terms of a synodic angle $\phi_s$, which represents the phase of the
synodic libration (see Fig.~\ref{fig:figure1.jpg} and
Eq.~\ref{eq:ys_phis} below). The computation of a resonant normal form
requires to explicitly Fourier expand the terms in \emph{both} angles
$\phi_f$ and $\phi_s$. Although the
Hamiltonian~\eqref{eq:hb_with_beta_calY} represents a Fourier
expansion for the fast d.o.f. (angle $\phi_f$), there still remain the
powers of $\beta$ that must be expanded in order to obtain a complete
Fourier expansion in the angle $\phi_s$ as well.

\begin{figure}[t]
\includegraphics[width=0.99\textwidth]{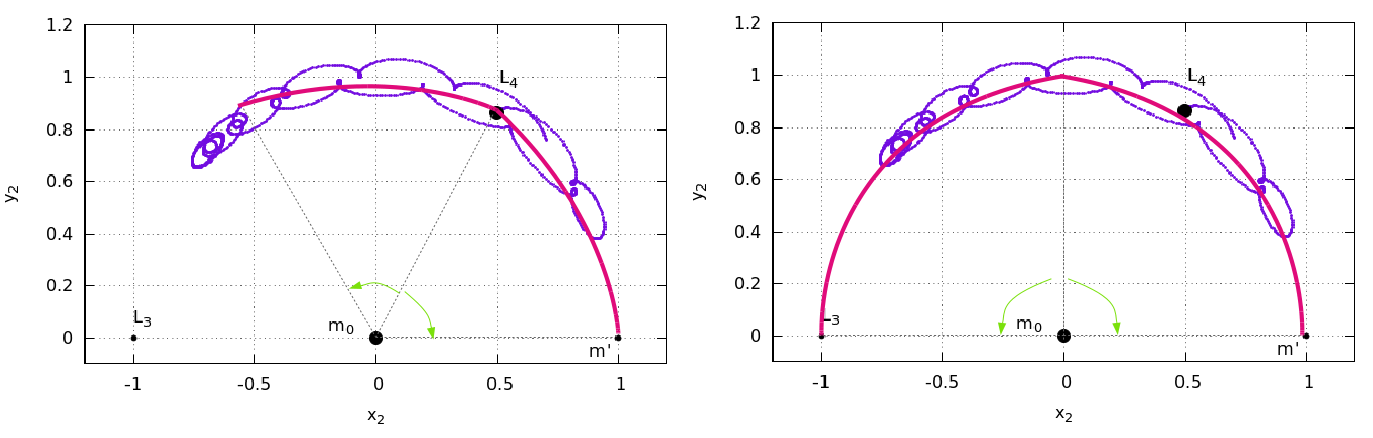} 
\caption[Schematic comparison of the convergence domain for symmetric 
and asymmetric expansions]{Schematic comparison of the convergence
domain when the expansion of functions of type $\beta(\tau)^N$ is
around $\tau_0 = \pi/3$ (left) or $\tau_0 = \pi/2$ (right). The latter
turns to be much more suitable to represent Trojan orbits with large
angular excursion on $\tau$.}
\label{fig:convergdomain.png}
\end{figure}

The functions $ \beta(\tau)^N = \frac{1}{\left( 2
-2\cos\tau \right)^{N/2}}$, with $N\in \mathbb{N} $, present a
singularity at $\tau=0$. As already discussed in
Sect.~\ref{2.1-expansion}, this implies that any Taylor expansion of
these functions around a certain $\tau_0$ is convergent only in the
domain $\mathscr{D}_{\tau_0,\delta}$ centered at $\tau_0$ and of
radius $\delta=\tau_0-0$.  The most common approach consists of Taylor
expansions around the equilibrium point, located at
$\tau_0=\frac{\pi}{3}$ (or equivalently, at
$\tau_0=\frac{5\pi}{3}$). The radius of convergence of such a series
is equal to $\frac{\pi}{3}$. Besides the necessity to introduce many terms 
in the expansion in order to represent well the Hamiltonian close to 
the limit of the convergence domain, we stressed already that many Trojan
orbits cross this domain (Fig.~\ref{fig:convergdomain.png}).

We will argue now that this problem
 can be addressed by considering an expansion around the
\emph{non-equilibrium point} $\tau_0 = \frac{\pi}{2}$. In this case,
we obtain a polynomial expansion of the Hamiltonian in powers of the
quantity $(\tau-\pi/2)$. This can be re-ordered as a polynomial
expression in powers of $u = \tau-\pi/3$. It is immediate to see
that any finite truncation of this expression yields a different
polynomial than the one obtained by a finite truncation of the direct
Taylor expansion around $\tau=\pi/3$. However, the new expression
yields a better approximation in a domain widely extended up to
$\tau \sim \pi$, a fact that brings many benefits for the
representation of extended tadpole orbits. We call the expansion around
$\tau_0 = \frac{\pi}{2}$ \emph{asymmetric}, while the one around $\tau_0
= \frac{\pi}{3}$ \emph{symmetric}.
Figure~\ref{fig:convergdomain.png} shows a schematic comparison of the
convergence domains in the two cases.

\begin{figure}[h]
\includegraphics[width=0.99\textwidth]{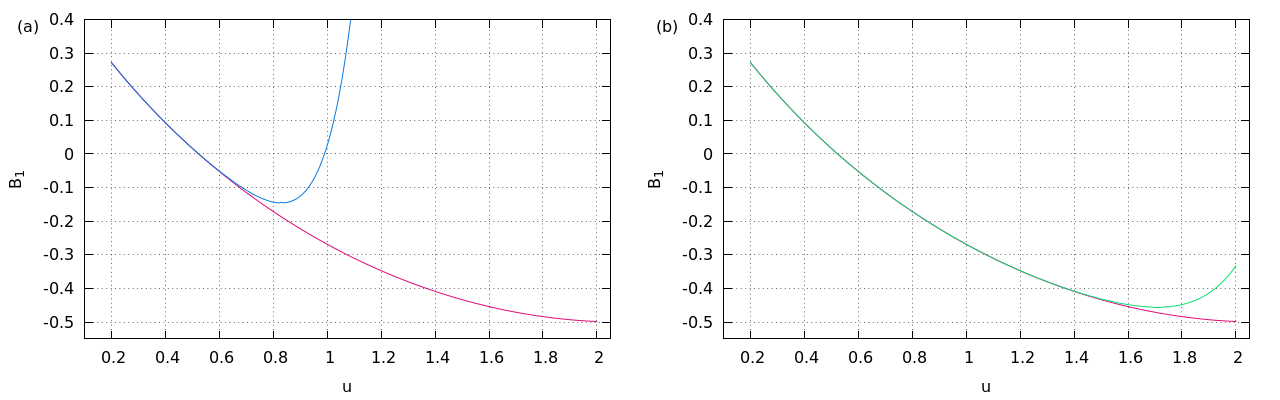}
\caption[Comparison of the functions $B_{1}$, $B_{1,\pi/3}$ and 
$B_{1,\pi/2}$]{Evalution of the functions $B_{1}$ (pink) and
$B_{1,\pi/3}$ (blue) in the left panel, and $B_{1}$ (pink) and
$B_{1,\pi/2}$ (green) in the right panel, for $u \in [0.2,2]$.}
\label{fig:b1.png}
\end{figure}

\begin{figure}[h]
\includegraphics[width=0.99\textwidth]{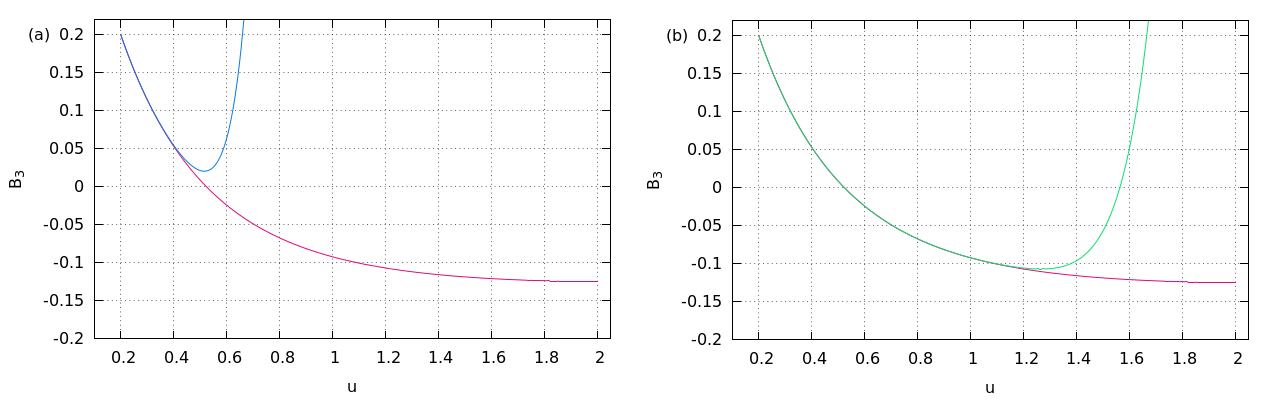}
\caption[Comparison of the functions $B_{3}$, $B_{3,\pi/3}$ and 
$B_{3,\pi/2}$]{Evalution of the functions $B_{3}$ (pink) and $B_{3,\pi/3}$ 
(blue) in the left panel, and $B_{3}$ (pink) and $B_{3,\pi/2}$ (green)
in the right panel, for $u \in [0.2,2]$.}
\label{fig:b3.png}
\end{figure}

\begin{figure}[h]
\includegraphics[width=0.99\textwidth]{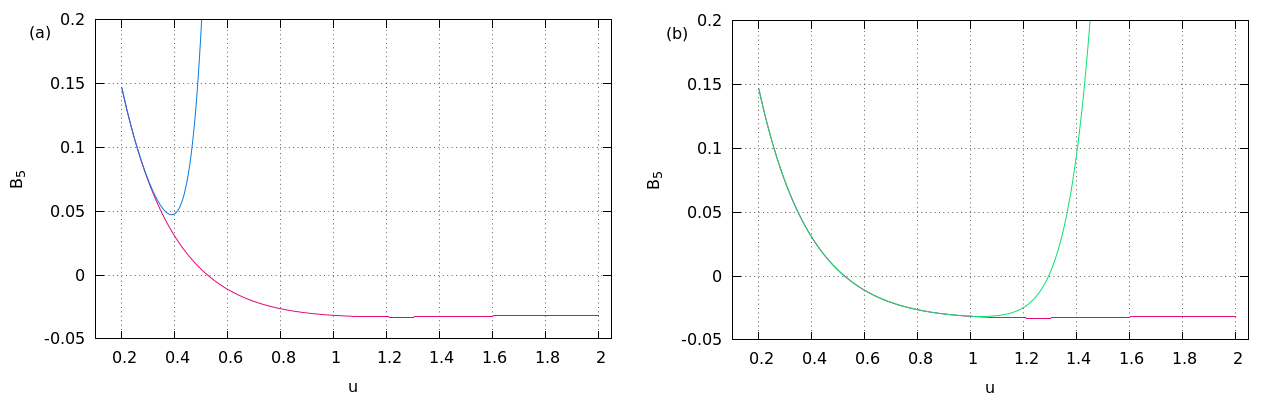}
\caption[Comparison of the functions $B_{5}$, $B_{5,\pi/3}$ and 
$B_{5,\pi/2}$]{Evalution of the functions $B_{5}$ (pink) and $B_{5,\pi/3}$ 
(blue) in the left panel, and $B_{5}$ (pink) and $B_{5,\pi/2}$ (green)
in the right panel, for $u \in [0.2,2]$.}
\label{fig:b5.png}
\end{figure}

\begin{figure}[h]
\includegraphics[width=0.99\textwidth]{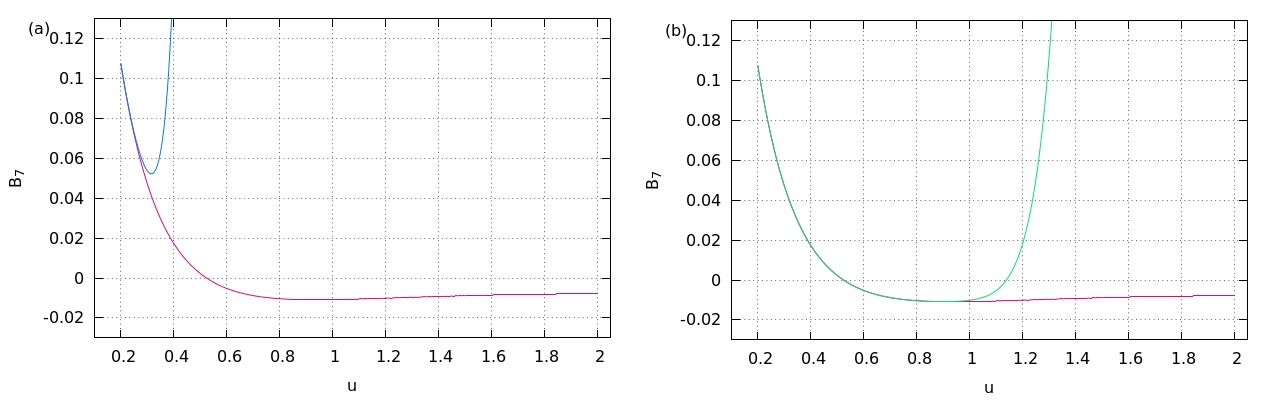}
\caption[Comparison of the functions $B_{7}$, $B_{7,\pi/3}$ and 
$B_{7,\pi/2}$]{Evalution of the functions $B_{7}$ (pink) and $B_{7,\pi/3}$ 
(blue) in the left panel, and $B_{7}$ (pink) and $B_{7,\pi/2}$ (green)
in the right panel, for $u \in [0.2,2]$.}
\label{fig:b7.png}
\end{figure}

A more formal comparison of the asymmetric and symmetric expansions
is presented in Figures~\ref{fig:b1.png}, \ref{fig:b3.png},
\ref{fig:b5.png} and \ref{fig:b7.png}. We consider the functions
\begin{equation}
\begin{aligned}\label{eq:the_betas}
B_{1}(\tau) & = \frac{\cos \tau}{\beta(\tau)} = \frac{\cos \tau}{(2-2\cos\tau)^{1/2}}~,
& B_{3}(\tau)  = \frac{\cos \tau}{\beta^3(\tau)} = \frac{\cos \tau}{(2-2\cos\tau)^{3/2}}~,\\
B_{5}(\tau) & =\frac{\cos \tau}{\beta^5(\tau)} = \frac{\cos \tau}{(2-2\cos\tau)^{5/2}}~,
& B_{7}(\tau)  = \frac{\cos \tau}{\beta^7(\tau)} = \frac{\cos \tau}{(2-2\cos\tau)^{7/2}}~,
\end{aligned}
\end{equation}
which represent the most common terms in powers of $\beta(\tau)$ appearing 
in Eq.~\eqref{eq:hb_with_beta_calY}. 
We consider the Taylor expansion of $B_{1}$, $B_{3}$, $B_{5}$ and
$B_{7}$ around $\tau_0= \pi/3$,
\begin{equation}\label{eq:the_betas_sym}
B_{M,\pi/3}  = B_{M}\left(\frac{\pi}{3}\right) +  B^{(1)}_{M}\left(\frac{\pi}{3}\right) \, u + 
\frac{1}{2} B^{(2)}_{M}\left(\frac{\pi}{3}\right) \, u^2 + 
\frac{1}{6} B^{(3)}_{M}\left(\frac{\pi}{3}\right) \, u^3 + \ldots~,
\end{equation}
where $B^{(n)}_{M}$ is the $n$-th derivative of the function $B_{M}$, 
$M = 1,3,5,7$  and
$u = \tau - \pi/3$.
On the other hand, we consider the Taylor expansions around $\tau_0 = \pi/2$,
{\small
\begin{equation}\label{eq:the_betas_asym}
B_{M,\pi/2}  = B_{M}\left(\frac{\pi}{2}\right) +  B^{(1)}_{M}\left(\frac{\pi}{2}\right) \, (u-\frac{\pi}{6}) + 
\frac{1}{2} B^{(2)}_{M}\left(\frac{\pi}{2}\right) \, (u-\frac{\pi}{6})^2 + 
\frac{1}{6} B^{(3)}_{M}\left(\frac{\pi}{2}\right) \, (u-\frac{\pi}{6})^3 + \ldots~.
\end{equation}}
\noindent
Figure~\ref{fig:b1.png} compares the function $B_{1}$ (pink, both
panels) with the two corresponding expansions $B_{1,\pi/3}$ (blue, left
panel) and $B_{1,\pi/3}$ (green, right panel) up to order 10 in $u$.
Figures~\ref{fig:b3.png},~\ref{fig:b5.png} and~\ref{fig:b7.png} show
the same comparison for $B_{3}$, $B_{5}$ and $B_{7}$, respectively.
From the figures, the difference in the
representation of functions $B_{M}$ given by the two expansions
becomes evident. In
the case of $B_1$, the asymmetric expansion $B_{1,\pi/2}$ reproduces
the correct behavior up to values of $u\sim 1.6$. Higher order
truncations are accurate nearly all the way to the position of
$L_3$, at $u \equiv 2.1$. On the other hand, the corresponding
symmetric expansion hardly reaches half of that domain. For increasing
order of $B$ ($B_{3}$, $B_{5}$, $B_{7}$, in
Fig.~\ref{fig:b3.png},~\ref{fig:b5.png} and~\ref{fig:b7.png}), both
expansions (at order 10) loose accuracy, this effect being always much
more notorious for the symmetric expansion. In the case of $B_{7}$,
the domain reproduced by the symmetric expansion at order 10 does not
reach even one third of the domain reproduced by the asymmetric
expansion. On the other hand, the symmetric expansion needs less terms
for accurately representing the functions $B_M$ for values $u \sim
0$. This fact is expected, since that expansion takes place exactly
around this value, while the asymmetric expansion is around the value
$u=\pi/6$.  Nevertheless, the loss of accuracy for small values of $u$
is remedied by including high order terms in the expansions. The
number of terms added in order to counteract this loss is radically smaller
than the number of terms required for an accurate
representation of values of $u \sim \frac{2\pi}{3}$ in the symmetric
expansion. As a matter of fact, as shown in the figures, by considering
expansions of order 10 in $u$, the difference between the two
representations in the domain of small values of $u$ is
negligible. Similar studies may be done with functions of the type
$ \frac{\sin \tau}{\beta(\tau)}$, yielding similar results.
\begin{SCfigure}
\includegraphics[width=0.60\textwidth]{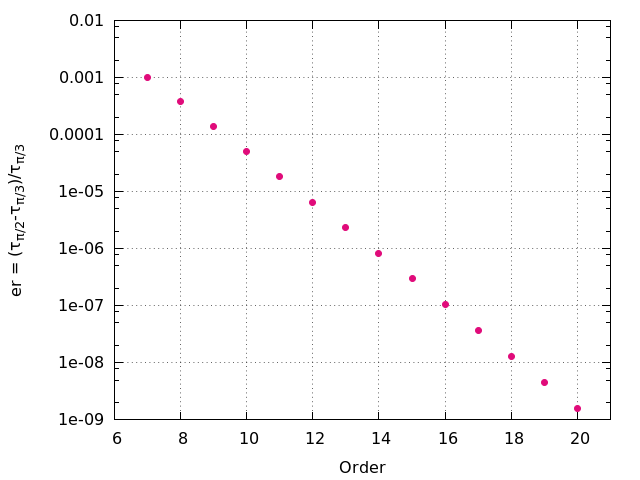}
\caption[Error in the equilibrium position $L_4$ for asymmetric expansions,
as function of the expansion order]{Computation of the error in the
estimation of the value of $u_{equil} = \frac{\df \phi}{\df \tau}$,
when $\phi$ is asymmetrically expanded, as function of the truncation
order of the expansion. The convergence to the true value of the
equilibrium point is exponential. At order of truncation equal 10, it
is possible to recover the position with 4 significant digits.
\vspace{0.8cm}}
\label{fig:center.png}
\end{SCfigure}

As an additional test, we compute the accuracy by which a finite
truncantion of the asymmetric expansion recovers the position of the
Lagrangian equilibrium point itself. This is given by
$\frac{\df \phi}{\df \tau}$, where $\phi = \cos \tau
- \frac{1}{\sqrt{2}\,\sqrt{1-\cos\tau}}$. The symmetric expansion
satisfies the above equation by identity at $u =0$ ($\tau
= \pi/3$). Figure~\ref{fig:center.png} shows the error in the value of
$u$ found by the asymmetric expansion. This is found to fall
exponentially fast with the truncation order, so that a truncation
order $r\sim 10$ recovers the correct equilibrium position with about 4
significant digits, while at $r \sim 20$ it recovers more than
8 significant digits.
\begin{figure}[h]
\includegraphics[width=1.05\textwidth]{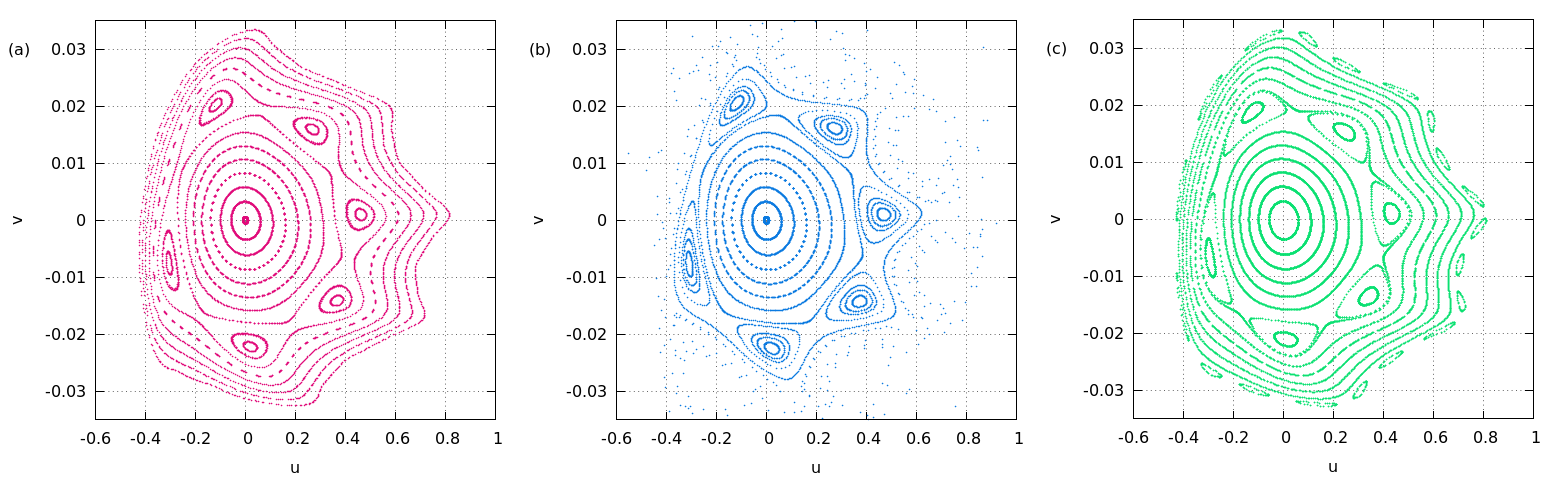}
\caption[Comparison of surfaces of section of $H_b$, the symmetric and
asymmetric expansions of the $H_b$, for $\mu=0.0041$.]{Comparison of
surfaces of section of the non expanded $H_b$ (left panel, pink
points), the symmetric expansion of $H_b$ (mid panel, blue points) and
the asymmetric expansion of $H_b$ (right panel, green points). The
corresponding mass parameter is $\mu = 0.0041$ and $e'=0$. For
information about the initial conditions, see text. }
\label{fig:3surfaces.png}
\end{figure}
\begin{SCfigure}
\includegraphics[width=0.60\textwidth]{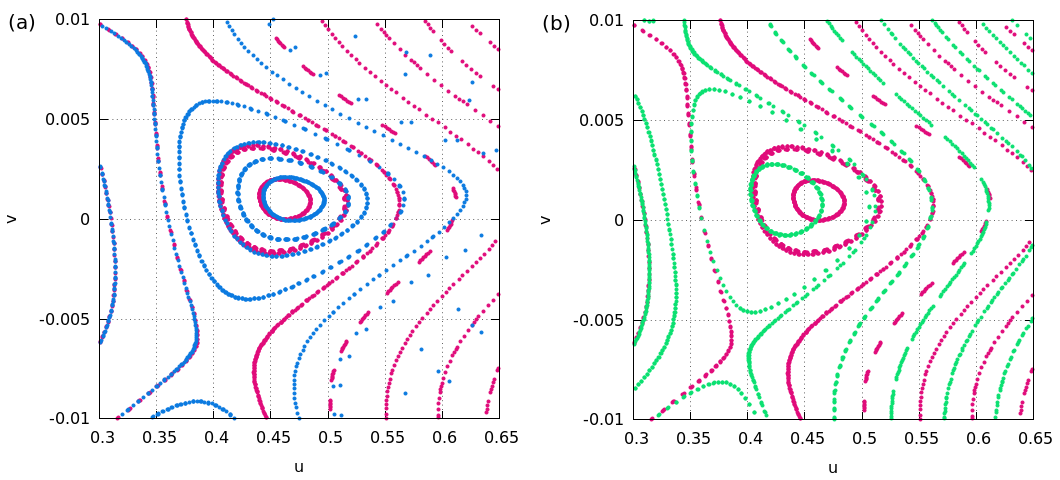}
\caption[Distortion in the stability islands of a secondary resonances
due to the symmetric and asymmetric expansions]{Detail of
Fig.~\ref{fig:3surfaces.png} showing the differences of the symmetric
expansion (left panel, blue points) and the asymmetric expansion
(right panel, green points) with respect to the non expanded $H_b$
(both, pink points) regarding the size of the $1$:$6$ stability
island.
\vspace{0.4cm}}
\label{fig:islands_big.png}
\end{SCfigure}
\begin{figure}[h]
\includegraphics[width=1.05\textwidth]{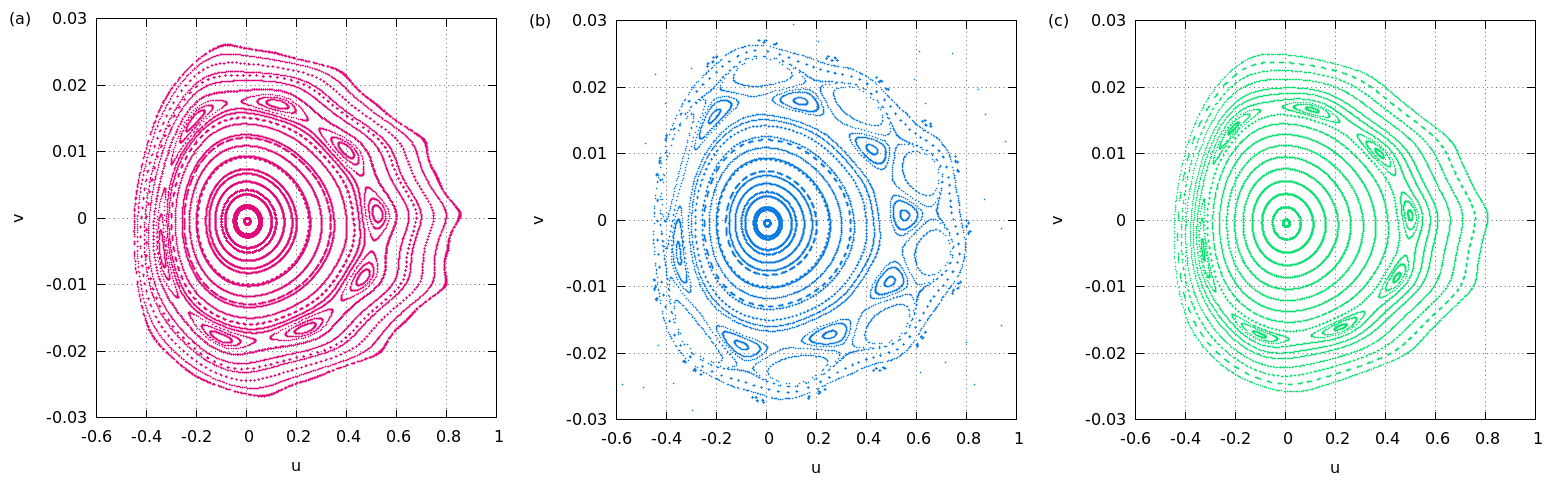}
\caption[Comparison of surfaces of section of $H_b$, the symmetric expansion
of the $H_b$ and the asymmetric expansion, for
$\mu=0.0024$.]{Comparison of surfaces of section of the non expanded $H_b$
(left panel, pink points), the symmetric expansion of $H_b$ (mid
panel, blue points) and the asymmetric expansion of $H_b$ (right
panel, green points). The corresponding mass parameter is $\mu =
0.0024$ and $e'=0.04$. The loss of convergence in the symmetric expansion
induces spureous dynamical structures in the form of meandric tori,
that are inexistent in the complete $H_b$.}
\label{fig:3surfaces-2.png}
\end{figure}

A last test of the benefits in the asymmetric expansion is
provided by computing phase portraits. Figure~\ref{fig:3surfaces.png}
shows three different surfaces of section, computed by the technique
introduced in Sect.~\ref{sec:feat_Hb} for 16 initial conditions
($\phi_f= -\pi/3$, $x=0$, $0.0 \leq u \leq 0.8$ and ${\cal Y}$
satisfying the isoenergetic condition defined by ${\cal Y}_0
= \frac{e_{p,0}^2}{2}$, for $\mu =0.0041$, $e'=0$, showing the $1$:$6$
main secondary resonance). In the left panel (pink), we present the
portrait of the orbits under the dynamics of $H_b$, without
introducing any expansion in its expression. In the middle panel 
(blue), we show the surface of section obtained by integrating
numerically the orbits under the $H_b$ symmetrically expanded (i.e.,
$H_b$ with the replacements of Eq.~\ref{eq:the_betas_sym}).  The
right panel (green) shows the surface corresponding to the $H_b$
asymmetrically expanded (using Eq.~\ref{eq:the_betas_asym}).  The
expansions in both cases are of order 10 in $u$, as in
Figures~\ref{fig:b1.png}--\ref{fig:b7.png}.  In the portraits, we can
distinguish two different features: the blue (middle panel) surface of
section, that corresponds to the symmetric expansion, represents with
good accuracy the orbits, up to $u\sim 0.5$. From that point on, the
loss of convergence of the expansions renders impossible to represent
any orbit, regular or chaotic. For some values of the mass parameter,
the main secondary resonances, which are always the most conspicuous
resonances in phase space, lie outside that limit. On the other hand,
the asymmetric expansion is able to represent the orbits in the whole
domain. We may also note that both expansions slightly distort the
stability islands.  Figure~\ref{fig:islands_big.png} presents a small
region of Fig.~\ref{fig:3surfaces.png} focusing on one of the
stability islands.  We can see that the asymmetric expansion
underestimates the size of the island, while the symmetric expansion
overestimates it.

Finally, Fig.~\ref{fig:3surfaces-2.png} shows a surface of
section as in Fig.\ref{fig:3surfaces.png}, for $\mu = 0.0024$, $e'= 0.04$
and $e_{p,0} = 0.04$. We see that, in the middle panel, the loss of
accuracy due to the symmetric expansion induces the creation of
structures (as meandric tori,~\cite{Carvalho-13}) that are not present
in the original model.

Overall, the use of the asymmetric expansion helps to improve the accuracy.
We will now employ this fact  in order to compute a
resonant normal form based on the asymmetric expansion of 
the Hamiltonian $H_b$.

\section{Preparation of the $H_b$}

The construction of the asymmetrically expanded $H_b$ is done by
considering the asymmetric expansions~\eqref{eq:the_betas_asym} when
replacing the functions of~\eqref{eq:the_betas} in
Eq.~\eqref{eq:hb_with_beta_calY}. In Appendix~\ref{app:asymm_exp}, we
give the analytical formul\ae~for the asymmetric expansions of
$\frac{\cos\tau}{\left(2-2\cos\tau\right)^{N/2}}$,
$\frac{\sin\tau}{\left(2-2\cos\tau\right)^{N/2}}$, $\cos^M\tau$ and
$\sin^M \tau$, in terms of $u = \tau-\pi/3$, with $N,\,
M\in \mathbb{N}$.  Regarding $x$, it is enough to consider the Taylor
expansion of the functions depending on this variable, around $x_0
=0$, i.e. in powers of $v=x-x_0$ (see Eq.~\ref{poincvar}). All the
expansions are carried out up to order 20 in the book-keeping
paramenter (see Rule~\ref{bkp:rule-1}). With these replacements and
applying certain trigonometric rules, the $H_b$ takes the form
\begin{equation}\label{eq:hb_asym_expand}
H_b (v,{\cal Y},u,\phi_f,Y_p) = Y_p + \sum_{\substack{m_1,m_2,\\m_3,m_4}}
\mathtt{a}_{(m_1,m_2,m_3,m_4)}\,  v^{m_1} \, u^{m_2}\, (\sqrt{{\cal Y}})^{m_3}\, \cos (m_4 \phi_f)~,
\end{equation}
where we gather all dependence of the series terms on the parameters
$\mu$ and $e'$ in the real coefficients $\mathtt{a}_{m_1,m_2,m_3,m_4}$.

The next step corresponds to a re-organization of the terms of the
Hamiltonian, according to a \emph{book-keeping} parameter
$\lambda$, as in Eq.~\eqref{eq:1stordHam}. To every term in the
Hamiltonian~\eqref{eq:hb_asym_expand}, we asign a power of $\lambda$ by the
following rule:
\begin{bookrule}\label{bkp:rule-1}
To every monomial of the type 
\begin{equation*}
\mathtt{a}_{(m_1,m_2,m_3,m_4)}\,  v^{m_1} \, u^{m_2}\, (\sqrt{{\cal Y}})^{m_3}\, 
{\textstyle {{\cos}\atop{\sin}}} (m_4 \phi_f)~,
\end{equation*}
there corresponds a book-keeping parameter of type $\lambda^{r(m_1,m_2,m_3,m_4)}$,
given by
\begin{equation*}
 r(m_1,m_2,m_3,m_4) =
   \begin{cases}
   \mathrm{Max}(0\,,\,m_1+m_2+m_3-2) & \text{if }\,m_4 = 0 \\
   \mathrm{Max}(0\,,\,m_1+m_2+m_3-2)+1 & \text{if }\,m_4 \neq 0 
   \end{cases}~. 
\end{equation*}
\end{bookrule}
This book-keeping choice effectively separates the terms according to
their smallness, while it ensures that the ${\cal O}(\lambda^0)$
terms that will appear below in the 
kernel of the homological equation are
exclusively terms up to second order in $u$ and $v$ and
linear in ${\cal Y}$. The addition of one power of $\lambda$ to those
monomials containing harmonics of $\phi_f$ aims to exclude combined
terms of the form ${\cal Y}{\textstyle {{\cos}\atop{\sin}}} (m_4\,\phi_f)
{{u}\atop{v}}$ from this kernel. Such terms have a very small size,
but without the book-keeping rule~\ref{bkp:rule-1} they would
necessarily enter into the diagonalization of the 
quadratic part of the Hamiltonian, a fact that
would complicate the computations.

After the application of the book-keeping Rule~\ref{bkp:rule-1}, the
Hamiltonian reads
{\small
\begin{equation}
\begin{aligned}\label{eq:hb_asym_expand_lambd_1}
H_b (v,{\cal Y},u,\phi_f,Y_p) = &\, Y_p  \, + \,
\mathtt{a}_{(1,0,0,0)}\, v \, + \, \mathtt{a}_{(0,1,0,0)} \, u \, + \,
\mathtt{a}_{(0,0,1,1)}\, \sqrt{{\cal Y}} \cos \phi_f \, + \, 
\mathtt{a}_{(0,0,1,1)}\, \sqrt{{\cal Y}} \sin \phi_f \\
+ & \, \mathtt{a}_{(2,0,0,0)} \, v^2 \, + \, \mathtt{a}_{(1,1,0,0)} \,v \,u\,+
\, \mathtt{a}_{(0,2,0,0)} u^2 + \mathtt{a}_{(0,0,2,0)} {\cal Y} \,\\
+ & \sum_{r=1}^{r_{max}}
\mathtt{a}_{(m_1,m_2,m_3,m_4)}\, \lambda^r v^{m_1} \, u^{m_2}\, 
(\sqrt{{\cal Y}})^{m_3}\, {\textstyle {{\cos}\atop{\sin}}} (m_4\,\phi_f)~.
\end{aligned}
\end{equation}}
Inspecting Eq.~\eqref{eq:hb_asym_expand_lambd_1}, at zero-th order of
$\lambda$ there appear some terms linear in $v$, $u$, and $\sqrt{{\cal
Y}} \cos \phi_f$ and $\sqrt{{\cal Y}} \sin \phi_f$, whose magnitude
turns to be less than $10^{-8}$. These terms correspond to the error
in the position of the Lagrangian equilibrium point with respect to
$L_4$ (see Fig.~\ref{fig:center.png}).  Since they are extremely small,
for convenience we simply neglect them.  Thus, the Hamiltonian is now
given by
\begin{equation}
\begin{aligned}\label{eq:hb_asym_expand_lambd_2}
H_b (v,{\cal Y},u,\phi_f,Y_p) = &\, Y_p  \, + 
\, \mathtt{a}_{(2,0,0,0)} \, v^2 \, + \, \mathtt{a}_{(1,1,0,0)} \,v \,u\,+
\, \mathtt{a}_{(0,2,0,0)} u^2 + \mathtt{a}_{(0,0,2,0)} {\cal Y} \,\\
+ & \sum_{r=1}^{r_{max}}
\mathtt{a}_{(m_1,m_2,m_3,m_4)}\, \lambda^r v^{m_1} \, u^{m_2}\, (\sqrt{{\cal Y}})^{m_3}\, 
{\textstyle {{\cos}\atop{\sin}}} (m_4\,\phi_f)~.
\end{aligned}
\end{equation}

The resonant normal form requires replacing the synodic variables
$(u,v)$ by corresponding action-angle variables $(Y_s,\phi_s)$.
However, in Eq.~\eqref{eq:hb_asym_expand_lambd_2} we observe that
there appears a term of the form $\mathtt{a}_{(1,1,0,0)} \,v \,u\,$
besides $\mathtt{a}_{(2,0,0,0)} \, v^2$ and $\mathtt{a}_{(0,2,0,0)} \,
u^2$. Thus, before introducing the transformation to action-angle
variables, we introduce a diagonalization of the synodic d.o.f.,
through an intermediate change of variables. Writing
$\mathtt{a}_{(2,0,0,0)}$ as $A$, $\mathtt{a}_{(0,2,0,0)}$ as $B$ and
$\mathtt{a}_{(1,1,0,0)}$ as $C$, the quadratic part $Av^2 + C u\, v\,
+ Bu^2$ in Eq.~\eqref{eq:hb_asym_expand_lambd_2} yields Hamilton's
equations of motion
\begin{equation}
\dot{\mathscr{Q}} = \mathbf{M} \, \mathscr{Q}= 
\begin{pmatrix}
C & 2A\\
-2B & -C \\
\end{pmatrix}
\mathscr{Q}
\end{equation}
where $\mathscr{Q} = (u,v)$. The corresponding diagonalized system in terms of
variables $(U,V)$ is represented by equations of motion of the form
\begin{equation}
\dot{\mathscr{W}} = 
\begin{pmatrix}
0 & \omega_s \\
-\omega_s & 0 \\
\end{pmatrix}
\mathscr{W}
\end{equation}
where $\mathscr{W} = (U,V)$, and $\omega_s$ is such that the
eigenvalues of the matrix $\mathbf{M}$ are $\lambda_{1,2}
= \pm \omega_s$. The canonical transformation connecting $\mathscr{W}$
with $\mathscr{Q}$ is given by
\begin{equation}\label{eq:fromWtoQ}
\mathscr{Q} = \frac{1}{\sqrt{\mathrm{Det}(\mathbf{E})}}\, \left( \mathbf{E} \cdot \mathbf{B} \right) \,
\mathscr{W} 
\end{equation}
where
\begin{equation}
\mathbf{B} = 
\begin{pmatrix}
\frac{1}{\sqrt{2}} & \frac{-\mathrm{i}}{\sqrt{2}}\\
\frac{-\mathrm{i}}{\sqrt{2}} & \frac{1}{\sqrt{2}}\\
\end{pmatrix}~,
\end{equation}
and $\mathbf{E}$ is the matrix that has as columns the eigenvectors
$e_{1,2}$ associated with the eigenvalues $\lambda_{1,2}$. After the
latter substitution, the transformed Hamiltonian reads
\begin{equation}
\begin{aligned}\label{eq:hb_asym_expand_lambd_3}
H_b (V,{\cal Y},U,\phi_f,Y_p) = &\, Y_p  \, + 
\, \mathtt{A}_{(2,0,0,0)} \, V^2 \, +\, \mathtt{A}_{(0,2,0,0)}\, U^2 + 
\mathtt{A}_{(0,0,2,0)}\, {\cal Y} \, \\
+ & \,\sum_{r=1}^{r_{max}}
\mathtt{A}_{(m_1,m_2,m_3,m_4)}\, \lambda^r V^{m_1} \, U^{m_2}\, (\sqrt{{\cal Y}})^{m_3}\, 
{\textstyle {{\cos}\atop{\sin}}} (m_4\,\phi_f)~,
\end{aligned}
\end{equation}
where $\mathtt{A}_{(2,0,0,0)} = \mathtt{A}_{(0,2,0,0)} = \omega_s$.
The last canonical transformation consists of introducing the
action-angle variables of the harmonic oscillator for the variables
$U$ and $V$. It is formally given by the transformation
\begin{equation}
\begin{aligned}\label{eq:ys_phis}
&{\cal Y}~, &\phi_f~, \phantom{\sqrt{2Y_s} \cos \phi_s} \\
&U = \sqrt{2Y_s} \sin \phi_s~, &V  = \sqrt{2Y_s} \cos \phi_s~.
\end{aligned}
\end{equation}
Finally, re-arranging the trigonometric terms depending on the two 
angles $\phi_f$ and $\phi_s$, the Hamiltonian takes the form:
\begin{equation}
\begin{aligned}\label{eq:hb_asym_expand_lambd_4}
H_b (Y_s,{\cal Y},\phi_s,\phi_f,Y_p) = &\, Y_p  \, + 
\, \omega_s \, Y_s \, +\, 
\omega_f \, {\cal Y} \,\\
+ & \,\sum_{r=1}^{r_{max}}
\mathtt{c}_{(k_1,k_2,k_3,k_4)}\, \lambda^r (\sqrt{Y_s})^{k_1} (\sqrt{{\cal Y}})^{k_2}\, 
{\textstyle {{\cos}\atop{\sin}}} (k_3 \phi_s + k_4\,\phi_f)~.
\end{aligned}
\end{equation}
In terms of this set of variables, the book-keeping 
Rule~\ref{bkp:rule-1} reads
\begin{bookrule}\label{bkp:rule-2}
To every monomial of the type 
\begin{equation*}
\mathtt{c}_{(k_1,k_2,k_3,k_4)}\, (\sqrt{Y_s})^{k_1} (\sqrt{{\cal Y}})^{k_2}\, 
{\textstyle {{\cos}\atop{\sin}}} (k_3 \phi_s + k_4\,\phi_f)~,
\end{equation*}
there corresponds a book-keeping parameter of type $\lambda^{r(k_1,k_2,k_4)}$,
given by
\begin{equation*}
 r(k_1,k_2,k_4) =
   \begin{cases}
   \mathrm{Max}(0\,,\,k_1+k_2-2) & \text{if }\,k_4 = 0 \\
   \mathrm{Max}(0\,,\,k_1+k_2-2)+1 & \text{if }\,k_4 \neq 0 
   \end{cases}~.
\end{equation*}
\end{bookrule}
From the canonical transformation in Eq.~\eqref{eq:ys_phis}, it is
straighforward to check that the harmonics of the angles $\phi_f$ and
$\phi_s$ have the same parity as the powers of the corresponding
functions in the variables $\sqrt{{\cal Y}}$ and $\sqrt{Y_s}$. This
property can be checked in Appendix~\ref{Appe:res-norm-form}, where we
present an example of the above series expansion for $\mu=0.0056$ and
$e'=0$.

\section{Resonant normal form}\label{sec:res_normal_form}

In Sections~\ref{2.2-normalization} and~\ref{sec:norm_Hb}, starting
from a Hamiltonian of 2 d.o.f. of freedom (CR3BP and $H_b$,
respectively), we normalized it to an integrable system 
with a constant of motion independent from the Hamiltonian.
Since in those cases we just focused on eliminating the
dependence of the Hamiltonian on just \emph{one} angle, the
normalizing scheme does not present any inconvenience:
Eq.~\eqref{eq:gral_gen_funct} in the $r$-th step is simply reduced to
the form
\begin{equation}\label{eq:gral_gen_funct_1dof}
\chi_{r} = \lambda^{r} \sum_{k} 
\frac{b^{r-1}(p^{(r-1)})}{{\rm i}(k\,\omega)}
 e^{{\rm i}(k q^{(r-1)})}~,
\end{equation}
where $q^{(r-1)}$ is the angle that we want to remove in the variables
of the $(r-1)$-th step, $\omega$ its frequency and $p^{(r-1)}$ its
conjugate action. The denominator is different from zero as long as
$k \neq 0$. On the other hand, if we proceed with a normalization of a
Hamiltonian in two pairs of action-angle variables, the generating
function constructed in each normalization step reads
\begin{equation}\label{eq:gral_gen_funct_2dof-A}
\chi_{r} = \lambda^{r} \sum_{k_1,k_2} 
\frac{b(p^{(r-1)}_{1},p^{(r-1)}_{2})}{{\rm i\,}(k_1\,\omega_1+k_2 \,\omega_2)}
 e^{{\rm i}(k_1 q^{(r-1)}_{1}+ k_2 q^{(r-1)}_{2})}~.
\end{equation}
As long as the orbits we want to represent guarantee that
the condition
\begin{equation}\label{eq:res_cond_2dof_gen}
k_1 \omega_1 + k_2 \omega_2 \neq 0
\end{equation}
holds for, at least, small $k_1$, $k_2 \:\in\: \mathbb{N}$, we can
proceed with a scheme as those already introduced. On the other hand,
in the vicinity of a secondary resonance, the condition of
Eq.~\eqref{eq:res_cond_2dof_gen} is \emph{not} accomplished, and
\emph{small divisors} appear in Eq.~\eqref{eq:gral_gen_funct_2dof-A}.
Let $m_1$, $m_2$ be two integers chosen a priori so that
$\frac{m_1}{m_2} \approx \frac{\omega_1}{\omega_2}$.  We define
the \emph{resonant module} ${\cal M}$ as the set of integer vectors
that accomplish the condition $k_1 m_1 + k_2 m_2 = 0$.  The generating
function is well defined if the Fourier terms
$e^{i(k_1\,q_1+k_2\,q_2)}$ chosen to be normalized are such that
$k_1,\,k_2\notin {\cal M}$. Then, 
\begin{equation}\label{eq:gral_gen_funct_2dof}
\chi_{r} = \lambda^{r} \sum_{k_1,k_2 \notin {\cal M}} 
\frac{b}{{\rm i}(k_1\omega_1+k_2 \omega_2)}
 e^{{\rm i}(k_1 q_1 + k_2 q_2)}~.
\end{equation}
The 1 d.o.f. cases developed so far are accounted for by choosing the
resonant module as
\begin{equation}\label{eq:res_mod_1dof}
{\cal M}_{1dof} = \{ k: |k| = 0\}~,
\end{equation}
while, in 2 d.o.f. systems, far from resonances, we can construct a 
non-resonant normal form setting 
\begin{equation}\label{eq:res_mod_2dof_nonsres}
{\cal M}_{nonres} = \{k_1,\,k_2: k_1 = k_2 = 0\}~.
\end{equation}

We introduce now the general recursive normalization algorithm for a
$n$ d.o.f. system. In the general case, the
resonant module corresponds to
\begin{equation}\label{eq:res_mod_gen}
{\cal M} = \{ \mathbf{k} = (k_1,k_2,...,k_n): k_1\, m_1 + k_2\, m_2 +
... + k_n\,m_n= 0 \}~,
\end{equation}
where $\sum_{i=1}^{n} |m_i| \neq 0$ (the application of resonant
normalization presented in Sect.~\ref{sec:location} corresponds to
$n=2$).  Let us assume that the Hamiltonian is in normal form up to
order $r$ in the book-keeping parameter, i.e.
\begin{equation}
{\cal H} = {\cal Z}_0 + \lambda {\cal Z}_1 + \ldots + \lambda^r {\cal Z}_r +
\lambda^{r+1} {\cal H}^{(r)}_{r+1} + \lambda^{r+2} {\cal H}^{(r)}_{r+2} + \ldots~.
\end{equation}

From the terms of order
$\lambda^{r+1}$, in the Fourier expansion,
\begin{equation}
{\cal H}^{(r)}_{r+1} = \sum_{\mathbf{k}} b (\mathbf{p}^{(r)}) \,
\mathrm{e}^{\,\mathrm{i}(\mathbf{k} \cdot \mathbf{q}^{(r)})}~,
\end{equation}
where $\mathbf{q}^{(r)} = (q^{(r)}_{1},\ldots,q^{(r)}_{n})$,
$\mathbf{p}^{(r)}=(p^{(r)}_{1},\ldots,p^{(r)}_{n})$ and
$\mathbf{k}=(k_1,\ldots,k_n)$, we isolate the terms that we want to
eliminate in the present step, denoted by
\begin{equation}\label{eq:phantomHrp1}
\phantom{{\cal H}}^{\ast}{\cal H}^{(r)}_{r+1} = 
\sum_{\mathbf{k} \notin {\cal M}} b (\mathbf{p}^{(r)}) \,
\mathrm{e}^{\,\mathrm{i}(\mathbf{k}\, \cdot\, \mathbf{q}^{(r)})}~.
\end{equation}
The homological equation
\begin{equation}\label{eq:homol_eq_res_nf}
\lambda^{r+1}\phantom{|}^{\ast}{\cal H}^{(r)}_{r+1} +
\{ {\cal Z}_0,\chi_{r+1} \} = 0
\end{equation}
has the solution
\begin{equation}\label{eq:gene_func_res_nf}
\chi_{r+1} = \lambda^{r+1}
\sum_{\mathbf{k} \notin {\cal M}} \frac{ b (\mathbf{p}^{(r)})}{\mathrm{i\,(\mathbf{k}\, \cdot \,\boldsymbol{\omega})}} \,
\mathrm{e}^{\,\mathrm{i}(\mathbf{k}\, \cdot\, \mathbf{q}^{(r)})}~,
\end{equation} 
with $\boldsymbol{\omega} = (\omega_1,\ldots,\omega_n)$. Having
the expression of the generating function, we compute the transformed
Hamiltonian 
\begin{equation}
{\cal H}^{(r+1)} = \mathrm{exp}({\cal L}_{\chi_{r+1}}) {\cal H} ^{(r)}~,
\end{equation}
which, by construction, is in normal form up to order $\lambda^{r+1}$, i.e.
\begin{equation}
{\cal H} = {\cal Z}_0 + \lambda {\cal Z}_1 + \ldots + \lambda^r {\cal Z}_r +
\lambda^{r+1} {\cal Z}_{r+1} + \lambda^{r+2} {\cal H}^{(r)}_{r+2} + 
\lambda^{r+3} {\cal H}^{(r)}_{r+3} + \ldots~.
\end{equation}

We can see now that, due to Eq~\eqref{eq:phantomHrp1}, the normal
 form, besides including terms depending just on the actions, also
 contains terms of the form
\begin{equation}\label{eq:struc-res-nf}
b (\mathbf{p}_{r}) \,
\mathrm{e}^{\,\mathrm{i}(\mathbf{k} \cdot \mathbf{q}_r)}~,
\end{equation}
with $\mathbf{k} \in {\cal M}$. Thus, the normal form includes
pendulum-like terms which allow to represent the selected
resonance.

\section{Location of the resonance and resonance widths by means of the resonant normal form}\label{sec:location}

Let us consider the function $H_b$ given in
Eq.~\eqref{eq:hb_asym_expand_lambd_4} as the starting Hamiltonian
$H_b^{(0)}$ of the normalizing scheme. We apply the normalizing scheme
presented in Sect.~\ref{sec:res_normal_form}, up to a maximum
normalization order $R$ in $\lambda$. In all the examples that follow,
we set $R=14$. An example of the first order normalization and the
corresponding computations, for a particular set of parameters $\mu$
and $e'$ is presented in Appendix~\ref{Appe:res-norm-form}.

Let $H_b^{(R)}$ be 
the final normalized Hamiltonian. According to Eq.~\eqref{eq:struc-res-nf},
the form of $H_b^{(R)}$ is given by
\begin{equation}\label{eq:nf-Hb-con-lambd}
H_b^{(R)} = \sum_{\substack{r=0\\(k_f,k_s) \in {\cal M}}}^{R} \lambda^r \mathtt{b} ({\cal
Y},Y_s) \, \mathrm{e}^{\,\mathrm{i}(k_f \phi_f + k_s \phi_s)}~.
\end{equation}
If we replace the book-keeping parameter $\lambda$ for its value equal to 1,
we recover the final normal form, depending on the actions and the
angles through the combination,
\begin{equation}\label{eq:nf-Hb-con-lambd}
H_b^{(R)} = \sum_{(k_f,k_s) \in {\cal M}} \mathtt{c}_{(d_f,d_s,k_f,k_s)}\, \sqrt{{\cal
Y}}^{\,d_f}\, \sqrt{Y_s}^{\,d_s} \, \mathrm{e}^{\,\mathrm{i}(k_f \phi_f + k_s \phi_s)}~,
\end{equation}
where the pairs $(d_f,k_f)$ and $(d_s,k_s)$ have the same parity, and
the values of the Fourier wavenumbers are bounded by $|k_f|\leq d_f$ and
$|k_s|\leq d_s$.  The integers $(d_f,d_s)$ are limited by the value of
$R$, through the book-keeping Rule~\ref{bkp:rule-2}. We define the
quantity $\Psi = m_1\,{\cal Y} + m_2 \, Y_s $ as a \emph{resonant
integral}, where $m_1$ and $m_2$ are the integers that define the
resonant module ${\cal M}$ in Eq.~\eqref{eq:res_mod_gen} with $n=2$. Let us
consider a single term of the Hamiltonian $H_b^{(R)}$, with generic
coefficients $(d_f,d_s,k_f,k_s)$, denoted by $\mathfrak{h}$. 
The Poisson bracket (Eq.~\ref{eq:poisson_brack_def}) of $\mathfrak{h}$
and $\Psi$ is
\begin{align*}\label{eq:poss-term-Hbnf}
\{\mathfrak{h},\Psi\} & =\, \left\{ \mathtt{c}_{(d_f,d_s,k_f,k_s)}\, 
\sqrt{{\cal Y}}^{\,d_f}\, \sqrt{Y_s}^{\,d_s} \, 
\mathrm{e}^{\,\mathrm{i}(k_f \phi_f + k_s \phi_s)}, 
\Psi \right\}~, \\
& = \, \frac{\partial \mathfrak{h}}{\partial {\cal Y}} 
\frac{\partial \Psi}{\partial \phi_f} - \frac{\partial \mathfrak{h}}{\partial
\phi_f} \frac{\partial \Psi}{\partial {\cal Y}}  +
\, \frac{\partial \mathfrak{h}}{\partial Y_s} 
\frac{\partial \Psi}{\partial \phi_s} - \frac{\partial \mathfrak{h}}{\partial
\phi_s} \frac{\partial \Psi}{\partial Y_s} ~,\\
& = \, \frac{d_f}{2} \frac{1}{{\cal Y}} \, \mathfrak{h} \,.\, 0 -
\mathrm{i}\,k_f\,  \mathfrak{h} \, m_1 + 
\, \frac{d_s}{2} \frac{1}{Y_s} \, \mathfrak{h} \,.\, 0 -
\mathrm{i}\,k_s\,  \mathfrak{h} \, m_2 ~, \\
& = \, -\mathrm{i}  \mathfrak{h} \left(k_f\,.\,m_1 + k_s\,.\,m_2 \right)~.
\end{align*}
Since $(k_f,k_s) \in {\cal M}$, we have
\begin{equation}\label{eq:poss-term-Hbnf_2}
\{\mathfrak{h},\Psi\} = 0~.
\end{equation}
Hence,
\begin{equation}\label{eq:res-intgral-cond}
{\cal L}_{H_b^{(R)}}\, \Psi = \{H_b^{(R)}, \Psi\} = 0~,
\end{equation}
i.e. $\Psi$ is a formal integral of $H_b^{(R)}$. 

We may remark here that all the above definitions refer to 
the variables after the last normalization step,
i.e. those induced by the sequential
application of the canonical transformations related to $\chi_r$,
$r=1,\ldots,R$. In proper notation (see Sect.~\ref{sec:1.1.2-pert_theory})
\begin{equation}\label{eq:norm-res-int}
\Psi = m_1\,{\cal Y^{(R)}} + m_2 \, Y_s^{(R)}
\end{equation}
where
\begin{equation}\label{eq:from-old-to-new}
{\cal Y}^{(0)} = \mathscr{C}^{(R)} \, {\cal Y}^{(R)}~,\qquad
Y^{(0)}_{s} = \mathscr{C}^{(R)} \, Y^{(R)}_s~,
\end{equation}
\begin{equation}\label{eq:transfC}
\mathscr{C}^{(R)} = \varphi^{(1)} \circ \varphi^{(2)} \circ \ldots \circ
\varphi^{(R-1)} \circ \varphi^{(R)}~,
\end{equation}
and the transformations $\varphi^{(r)}$ are given by
\begin{equation}\label{eq:varphi-r}
\varphi^{(r)} = \exp\left({\cal L}_{\chi_r} \right)({\cal Y}^{(r)},Y_2^{(r)},
\phi_f^{(r)}, \phi_s^{(r)})~.
\end{equation}

\begin{figure}[t]
\centering
\includegraphics[width=0.99\textwidth]{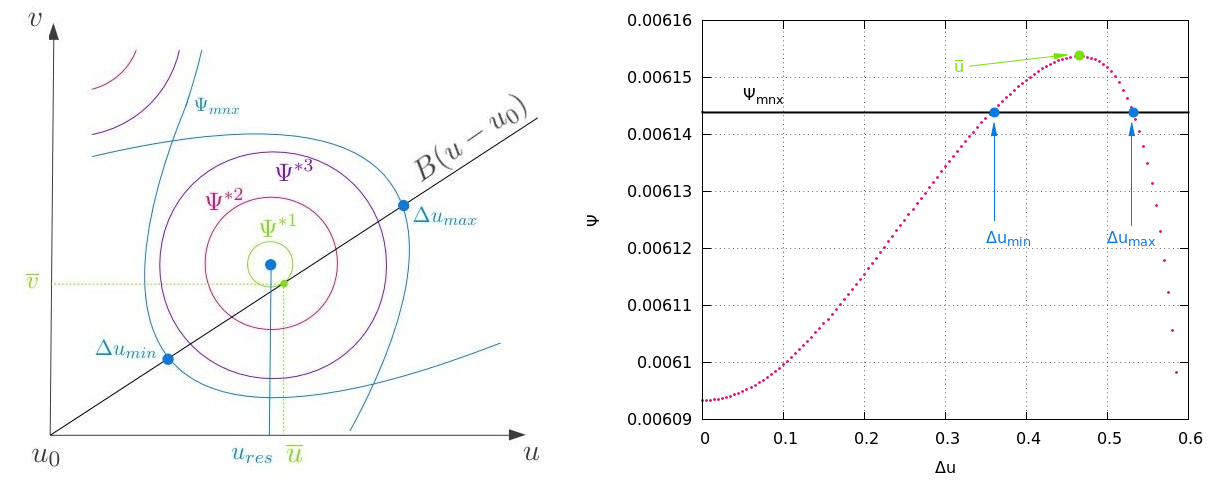}
\caption[Schematic representation of the level curves of the resonant
integral in the plane $(u,v)$. The resonant integral as function of
 $\Delta u$]{Left panel - Schematic representation of the plane
 ($u,\,v$) for a surface of section of the $H_{b}$. The central blue
 dot represents the location of a stable periodic orbit, whose co-ordinate
 in equal to $u=u_{res}$. At this point, the resonant integral $\Psi$
 presents a global extremum. Additional quasi-periodic orbits inside
 the island of stability are labeled with the corresponding values of
 $\Psi$, i.e. $\Psi^{\ast 1}$, $\Psi^{\ast 2}$, $\Psi^{\ast 3}$,
 $\Psi_{mnx}$, accomplishing $\Psi^{\ast 1} > \Psi^{\ast 2}
 > \Psi^{\ast 3} > \Psi_{mnx}$. The value $\Psi_{mnx}$ represents a
 theoretical separatrix of the resonance in the resonant integral
 approximation (in reality, instead of the separatrix we have a thin
 separatrix-like chaotic layer). For the initial conditions taken
 along the line $B(u-u_0)$, the orbit satisfying
 Eq.~\eqref{eq:deriv_psi} corresponds to a level curve tangent to the
 line, labeled $\Psi^{\ast 1}$. The initial condition for u along this
 line, $\overline{u}$, represents a good approximation to the exact
 resonant position $u_{res}$. The two values of $u$ on the line
 $B(u-u_0)$ satisfying $\Psi = \Psi_{mnx}$ correspond to the
 intersection of the separatrix with the line $B(u-u_0)$ ($\Delta
 u_{min}$ and $\Delta u_{min}$, in blue), and provide an estimation of
 the width of the resonance. Right panel - Values of the resonant
 integral $\Psi$ along the line $B(u-u_0)$. The position of the
 maximum of the function corresponds to $\overline{u}$ (green
 dot). The value of $\Psi_{mnx}$ (black line) defines the position of
 the two borders of the resonance $\Delta u_{min}$ and $\Delta
 u_{max}$ (blue dots).}
\label{fig:minresint.jpg}
\end{figure}

By considering the tranformation $\mathscr{C}^{(R)}$
we can represent the resonant integral in terms of the original
variables $({\cal Y}^{(0)},Y_s^{(0)},\phi_f^{(0)},\phi_s^{(0)})$
\begin{equation}\label{eq:non-norm-res-int}
\Psi({\cal Y}^{(0)},Y_s^{(0)},\phi_f^{(0)},\phi_s^{(0)}) = 
\Psi\left(\mathscr{C}^{(R)}({\cal Y}^{(R)},Y_s^{(R)},\phi_f^{(R)},\phi_s^{(R)}) \right)~.
\end{equation}
Applying the inverse transformations to those introduced in
Eqs.~\eqref{eq:ys_phis} and~\eqref{eq:fromWtoQ}, we are able to
express the resonant integral in \eqref{eq:non-norm-res-int} as
function of the variables used in Sect.~\ref{sec:3.X.2-flimaps} and
Sect.~\ref{sec:location},
\begin{equation}\label{eq:uvYphif-res-int}
\Psi = \Psi(v,{\cal Y},u,\phi_f)~.
\end{equation}
We now show that the form of $\Psi$ as in Eq.~\eqref{eq:uvYphif-res-int}
is appropriate so as to find the position and the size of the main
secondary resonances in the space of proper elements, in a way comparable
to the one used in Sect.~\ref{sec:location}. 

As first step, we choose a set of parameters $\mu$ and $e'$, that
refer to a particular case of the FLI maps presented in
Sect.~\ref{sec:3.X.2-flimaps}, characterized by the presence of a 
conspicuous main secondary
resonance. Additionally, we fix the value of the fast angle
$\phi_f = -\pi/3$, to coincide with the surfaces considered in the
stability maps, and we also replace the fast action ${\cal Y}$ by
the corresponding proper eccentricity, ${\cal Y}= \frac{e_{p,0}^2}{2}$.
This way, the resonant integral $\Psi$ becomes a function of $(e_{p,0},v,u)$.

Let us consider a generic surface of section of the $H_b$, computed as
described in Sect~\ref{sec:feat_Hb}. After fixing the value of
$e_{p,0}$, each orbit in the surface is generated by a pair of initial
conditions ($u^{\ast},v^{\ast}$). Since $\Psi$ is a first integral of
the basic Hamiltonian, independently of the variables used, we can
label each invariant curve within the resonant island of stability by
the associated value of $\Psi(e_{p,0}^{\ast},u^{\ast},v^{\ast})$ (left
panel of Fig.~\ref{fig:minresint.jpg}, schematic).  In other words,
for fixed $e_{p,0}$, the orbits in the surface of section lie on level
curves of the function $\Psi(u,v)$.  It is straightforward to prove
that $\Psi$ has a stationary point at the position of the periodic
orbit, which, for the stable orbit, is a maximum or
minimum. Thus, to find the position of the stable resonant periodic
orbit in the plane $(u,v)$ for fixed $e_{p,0} = e_{p,0}^{\ast}$ is
equivalent to locating the corresponding extrema of the function
$\Psi(u,v)$.

However, the choice of the parameter $B$ in the FLI maps determines
also the value of $v$ as function of $u$, via $v = B(u-u_0)$ (the
coefficients $B$ for various resonances are given in
Table~\ref{Tab:beta}). Therefore, when dealing with the initial
conditions of a certain FLI map, the resonant integral is reduced to
$\Psi = \Psi(e_{p,0},u)$. However, applying the same rules to the
constant energy condition, $E = H_b(e_{p,0},u)$, we can solve
for $e_{p,0}$, for fixed energy $E$ as function of $u$. Hence, $\Psi$
now depends on the value of $u$ only.  Let us
consider the function $\psi(u)=\Psi$ with the above replacements. The
point $(\overline{u},B(\overline{u}-u_0))$ accomplishing
\begin{equation}\label{eq:deriv_psi}
\frac{\df \psi(u)}{\df u}|_{\overline{u}} = 0
\end{equation}
corresponds to the point of contact of the line of initial conditions
$v = B(u-u_0)$ with the contour curve 
\emph{tangent} to such a line (see Fig.~\ref{fig:minresint.jpg}, left panel).
Thus, even if we choose a line of initial conditions that does not cross
exactly the position of the resonant orbit, the extrema of the
function $\psi(u)$ still provide a good approximation to the location
of the resonance.  

On the other hand, the resonant integral provides a method for
estimating the values of the borders of the resonance.  Let us
consider initial conditions along lines of the form $v = A (u-u_0)$,
with different slopes $A$. As discussed above, for the line crossing
exactly the position of the stable point, the resonant integral
acquires its maximum value at $u_{res}$. For any other slope, $\Psi$
has a maximum value along the line at the position that generates an
orbit tangent to the line itself. However, this maximum value of
$\Psi$ \emph{decreases} as the distance from the position of the
stable point increases. On the line crossing exactly
the \emph{unstable} point, the function $\Psi_{max}(A)$ (that gives
the maximum values of $\Psi$ as function of the slope $A$) has its own
minimum, namely $\Psi_{mnx}$. The value $\Psi = \Psi_{mnx}$ labels
the whole separatrix, which, besides going through the unstable point,
delimits the outermost border of the resonance. After computing 
the value of $\Psi_{mnx}$, returning to the
line of slope $B$, the orbits satisfying $\Psi = \Psi_{mnx}$
correspond to the interior and exterior borders of the resonance,
$\Delta u_{min}$ and $\Delta u_{max}$, as projected on the FLI map (see
Fig.~\ref{fig:minresint.jpg}, left panel).

For the estimation of $\overline{u}$,
$\Delta u_{min}$ and $\Delta u_{max}$, we proceed as follows. We first
consider a reference orbit given by
($u_{ref},\,v=0,\,\phi_f=-\pi/3,/,{\cal Y} = (e_{p,0}^{\ast})^2/2$),
that defines a reference energy $E^{\ast}$ and a minumum value of
$\Psi_{max}(A) =\Psi_{mnx}$ .  We produce several initial conditions
by varying $u$ and $v = B(u-u_0)$, with fixed $\phi_f=-\pi/3$.  The last
value for each initial condition (${\cal Y} = e_{p,0}^2/2$) is derived
from the value of the energy. In this set of isoenergetic points
($u,\,e_{p,0}$), we look for the extreme of the function
$\Psi(u,e_{p,0})$, that provides the value of $\overline{u}$, and the
two values for which $\Psi(u,e_{p,0}) = \Psi_{mnx}$, that provide
$\Delta u_{min}$ and $\Delta u_{max}$, and their associated $e_{p,0}$,
for the chosen energy (see Fig.~\ref{fig:minresint.jpg}, right panel).
We repeat this procedure for different values of $e_{p,0}^{\ast}$ for
fixed $\mu$ and $e'$, and we trace the whole distribution of resonant
positions and borders on the FLI map. With other pairs of physical
parameters, this whole scheme may be repeated by locating the position of
different main secondary resonances.

\begin{figure}[h]
\includegraphics[width=0.99\textwidth]{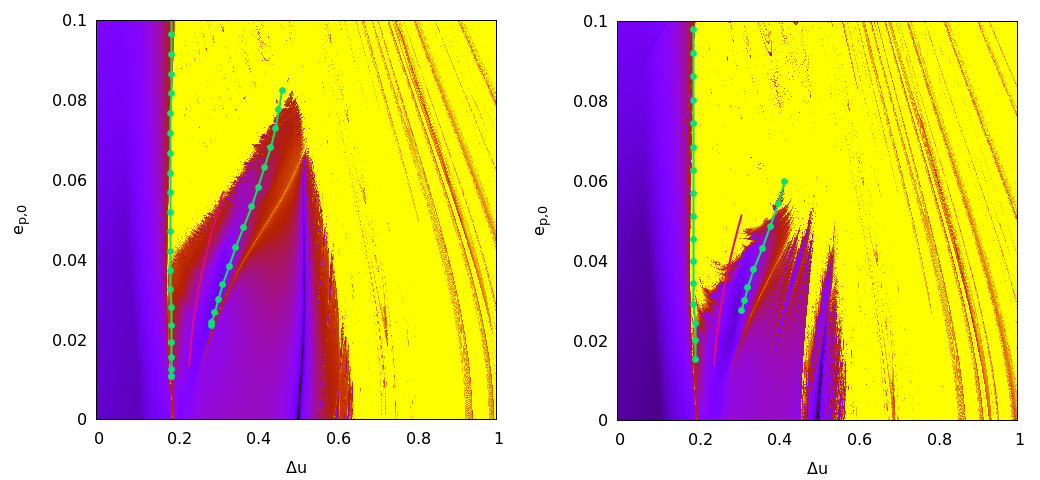}
\caption[Estimation of the center of the resonance $\overline{u}$ and the
borders $\Delta u_{min}$ and $\Delta u_{max}$ for
$\mu=0.0056$]{Estimation of the center of the resonance $\overline{u}$
(pink line in left panel, blue line in right panel) and the borders
$\Delta u_{min}$ and $\Delta u_{max}$ (green points), for different
values of the energy $e_{p,0}^{*}$ and parameters $\mu=0.0056$, $e'=0$
(left panel) and $e'=0.02$ (right panel). The estimation is plotted on
top of the corresponding FLI map for those parameters. The secondary
resonance is $1$:$5$.}
\label{fig:1to5resint.png}
\end{figure}
\begin{figure}
\includegraphics[width=0.99\textwidth]{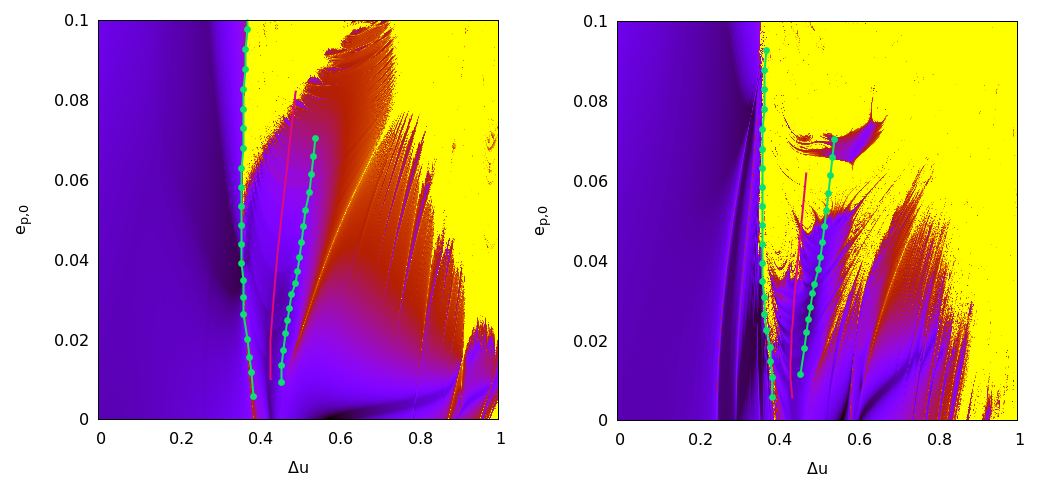}
\caption[Estimation of the center of the resonance $\overline{u}$ and the
borders $\Delta u_{min}$ and $\Delta u_{max}$ for
$\mu=0.0041$]{As in Fig.~\ref{fig:1to5resint.png}, for $\mu=0.0041$.
The secondary resonance is $1$:$6$.}
\label{fig:1to6resint.png}
\end{figure}
\begin{figure}
\includegraphics[width=0.99\textwidth]{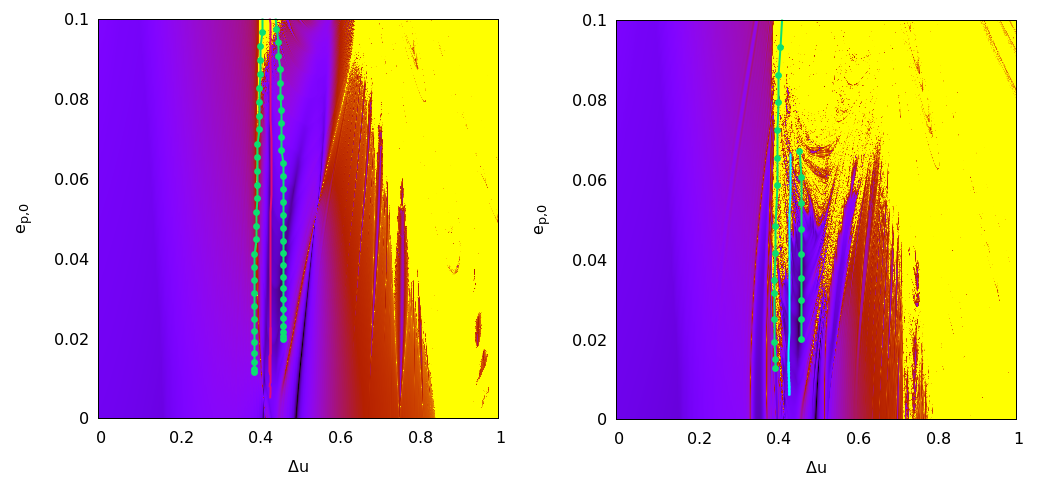}
\caption[Estimation of the center of the resonance $\overline{u}$ and the
borders $\Delta u_{min}$ and $\Delta u_{max}$ for
$\mu=0.0031$]{As in Fig.~\ref{fig:1to5resint.png}, for $\mu=0.0031$.
The secondary resonance is $1$:$7$.}
\label{fig:1to7resint.png}
\end{figure}

In Figs.~\ref{fig:1to5resint.png}, 
\ref{fig:1to6resint.png} and~\ref{fig:1to7resint.png}, we 
present a few examples of this method of location of resonances, applied to the
main secondary resonances $1$:$5$ ($\mu=0.0056$), $1$:$6$
($\mu=0.0041$) and $1$:$7$ ($\mu=0.0031$) respectively. In the figures
we show the position of the centers (pink or blue lines) and of the
borders (green dots). Left panels correspond to $e'=0$ and right
panels to $e'=0.02$.

The method turns to be particularly efficient for the estimation of
the centers for high values of the mass parameter $\mu$ (see
Fig.~\ref{fig:1to5resint.png}, Fig.~\ref{fig:1to6resint.png}),
associated to important secondary resonances, and of decreasing
accuracy for decreasing $\mu$ (Fig.~\ref{fig:1to7resint.png}).
Regarding the location of the borders, we see that in all cases
the determination of $\Delta u_{min}$ is very accurate, while
the location of the external border $\Delta u_{max}$ is understimated.

From the plots, we conclude that the performance of the method seems
to be strictly related with the normalization order used to the
computation of the resonant integral.  On one hand, resonances of the
type $1$:$n$ appear in the normalized Hamiltonian $H_{b}^{(R)}$ at
orders $R = n$ or greater. Thus, higher order resonances (associated
with smaller values of $\mu$) require the computation of many
normalizing steps for being well represented.  In the examples
provided in Fig.~\ref{fig:1to5resint.png}-\ref{fig:1to7resint.png}, we
consider an initial asymmetric expansions of the Hamiltonian up to
order ${\cal O}(\lambda^{20})$ as well as 14 normalization steps. From the
figures, it is clear that these limits (defined by the limitations of
the software and of the computational resources) are not sufficient
for a clear analytical representation of the resonance $1$:$7$, while
they are so for lower order resonances $1$:$5$ and $1$:$6$.
\begin{SCfigure}
\includegraphics[width=0.6\textwidth]{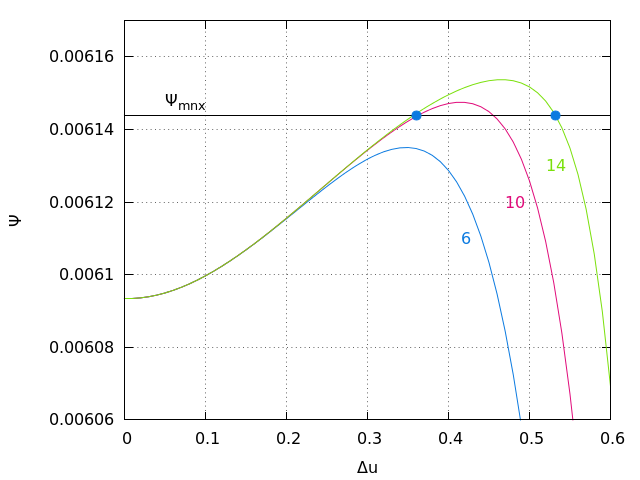}
\caption[Examples of the influence of the normalization order $R$ on the behavior of the resonant integral along an isoenergetic line $B(u-u_0)$]
{Examples of the influence of the normalization order $R$ on the
behavior of the resonant integral $\Psi$ along an isoenergetic line
$B(u-u_0)$.  The normalization orders considered are $R =6$ (blue), $R
=10$ (pink) and $R = 14$ (green). The black line denotes the value of
$\Psi_{mnx}$ for the resonance separatrix. The blue points
denote the intersections $\Psi = \Psi_{mnx}$ and the
green curve indicates the position of $\Delta u_{min}$ and $\Delta
u_{min}$, for the 14-th order approximation.
\vspace{0.8cm}}
\label{fig:res_int_orders.png}
\end{SCfigure}

On the other hand, our experiments show that the resonant integral
$\Psi$ as function of $u$, converges very slowly with the increment of
the normalization order (at least up to the optimal order, which in
all the examples considered seems to be well beyond $R=14$). In
Fig.~\ref{fig:res_int_orders.png} we present examples of the
computation of $\Psi$ along the line of initial conditions $x=0$ for
the $1$:$6$ resonance ($\mu =0.0041$, $e'=0$, $E^{\ast}=-0.496949$),
for different orders of normalization $R = 6$ (blue), $10$ (pink),
$14$ (green). From the image, it turns clear that from the order $10$
on, the position of $\Delta u_{min}$ is accurately obtained, and it
will not change with the addition of higher order terms. This does not
hold for the external border $\Delta u_{max}$, whose position greatly
depends on the order $R$. An accurate representation of $\Delta
u_{max}$ can only be achieved when long expansions and high order
normalizations are considered.

At any rate, it should be emphasized that, in both the ER3BP and the
RMPP, the inner domain (the region from $u = u_0$ to $u = u_0+ \Delta
u_{min}$) is characterized for containing nearly exclusively regular orbits or
resonances that do not overlap when pulsating.  In other words, the
diffusion mechanisms in the inner domain are very inefficient, and the
whole region remains stable for extremely long times.  Thus, an
accurate determination of its border, as the one provided by the
resonant integral obtained from the $H_b$, turns to be essential for
the determination of the size of the most important stability domain,
i.e., the domain before the appearence of the most conspicuous
secondary resonance (for the relevance of this inner border to Trojan
dynamics in exoplanetary systems, see \cite{Leleuetal-15}).

\chapter{Conclusions and perspectives}

In the present thesis we implemented the method of Hamiltonian normal
forms in the 'Trojan problem', i.e. the problem of the motion 
of a small body
in the
neighborhood of the equilateral Lagrangian points of a gravitational
system of two primary bodies, like a star-planet or planet-moon
system. Modern applications of this problem were reviewed in
Sect.~\ref{sec:1.4.works_on_Trojans}. In this thesis, we focused on
developing new methods of analytical study of the Trojan motions based
on tools provided by the canonical formalism and by the 
Hamiltonian perturbation theory. These new methods were implemented in models
of increasing complexity and the results were checked against
numerical simulations. In the sequel, we summarize the most relevant
results.

In Chapter 2, we presented a novel normal form approach for Trojan
orbits in the framework of the planar Circular Restricted 3-Body
Problem (pCR3BP). The main result is that the new method allows to
circumvent the problem of bad convergence for Trojan orbits with large
synodic amplitudes. We used a normalizing scheme, based on Lie series,
which allows to average the Hamiltonian without making expansions in
terms of the synodic d.o.f (Sect.~\ref{2.1-expansion}). The so-found
normal form is a 1 d.o.f. integrable Hamiltonian representing the
synodic component. We tested the normal form by comparing its
analytical predictions with numerical computations of orbits with
initial conditions derived from real objects (several Trojan asteroids
of Jupiter and the Earth's Trojan asteroid $2010$
$\mathrm{TK}_7$). The results of the tests show that the averaging
process keeps unaltered the main dynamics of the Trojan orbits in the
framework of the circular approximation
(Sect.~\ref{sbs:comp_act}). Also, the new method allows to define a
semi-analytical determination of quasi-actions for the synodic degree
of freedom.

In Chapter 3, we revisited some of the main features of the Trojan
problem in the framework of the planar Elliptic Restricted 3-Body
Problem (ER3BP). We introduced a new set of action-angle
variables (Fig.~\ref{fig:figure1.jpg}), allowing to characterize the
three d.o.f. associated with different time scales (fast, synodic and
secular). We decomposed the original Hamiltonian in terms of two
components: i) the basic Hamiltonian $H_b$, representing the fast and
synodic d.o.f., and ii) the secular Hamiltonian $H_{sec}$, gathering
the terms depending on the secular angle (Eq. \ref{hambasic}). The
discrimination of the three d.o.f. allowed to properly define the
associated frequencies and to make a categorization of the resonances
appearing in the problem (Sect.~\ref{sec:3.X-secres}). These
resonances exist for different combinations of the physical parameters
of the problem, namely the mass parameter $\mu$ and the primary's
eccentricity $e'$, and, in general, come by multiplets
(Fig.~\ref{fig:freqanal}, Fig.~\ref{fig:fli1to12}--\ref{fig:fli1to5}).
Numerical stability maps allowed to depict the resonance web and its
dependence on the parameters of the system
(Sect.~\ref{sec:3.X.2-flimaps})

The formulation based on $H_b$ motivated a new definition of the
quasi-integral of the 'proper eccentricity' $e_p$, that can be used instead of
the less accurate definition, $e_{p,0}$ (Eq.~\ref{epr}, \ref{eprnew}),
i.e. the distance from the point of the forced equilibrium to the
endpoint of the eccentricity vector. The quantity $e_p$ is an exact
integral of motion of the dynamics of the $H_b$, but it remains also a
quasi-integral of the complete problem, for not very large values of
the primary's eccentricity $e'$ (Fig.~\ref{fig:diffmodel}). In
addition, $e_p$ does not present the large variations displayed by
$e_{p,0}$ in cases when $e'< \mu$ or when the Trojan body is trapped
in a resonance.

The secular terms in $H_{sec}$ make all the resonances of the $H_b$ to
slowly pulsate. The effect of such a pulsation divides the space of
proper elements in two different regions: the inner domain ($0 <
\Delta u < \Delta u_{min}$, where $\Delta u_{min}$ corresponds to the
inner border of the most conspicuous secondary resonance in terms of
proper libration amplitudes), and the outer domain ($\Delta u_{min} <
\Delta u$). In the inner domain, where the resonances are isolated one
from the other, the pulsation does not induce essential changes in the
dynamics with respect to the dynamics under the $H_b$
(Sect.~\ref{sec:feat_Hb}). In the outer domain, that it is densely
filled with multiplets, the resonances may partially overlap due to
the pulsation and, thus, accelerate the rate of diffusion (modulational
diffusion), in comparison with that of the inner domain
(Sect.~\ref{sec:3.X-moddiff}, Sect.~\ref{sec:feat_Hb}). As a consequence, the
orbits in the outer domain turn to be much more chaotic, with shorter
characteristic Lyapunov times.  We tested the effect of modulational
diffusion on the distribution of the escape times, by making a
statistical numerical study of a set of orbits of the outer domain
(Sect.~\ref{sec:3.X-chaoticdiff}). The distribution of escape times
turns to be bimodal, with two well differentiated peaks
(Fig.~\ref{fig:histoesc}). The fast escapes are associated with highly
chaotic orbits. The slow escapes are associated with the orbits
belonging to the chaotic layers of the lower order resonances
(Fig.~\ref{fig:fliesc}).  Finally, the orbits laying in the chaotic
layers of high order resonances show features of the so-called 'stable
chaos', presenting rather short Lyapunov times, although not escaping
for the whole integration time (Fig.~\ref{fig:histo},
Fig.~\ref{fig:fliesc}).

In Chapter 4, we focused on a detailed study of the properties of
the 'basic' Hamiltonian model $H_b$. It is shown that the function
$H_b$ obtained from the ER3BP is formally the same as the one obtained
from a more general and complex model, called the Restricted
Multi-Planet Problem (Sect.~\ref{sec:4.X-RMPP}). The $H_b$, a 2
d.o.f. system, accurately represents the dynamics in the outer domain
of the complete problem as long as the proper eccentricity does not
surpass a critical value $e_{crit}$. This limit is reached when the
separatrix pulsation of the dominant secondary resonance reaches an
amplitude larger than the width of the
resonance itself. Beyond this critical value, the difference between the
$H_b$ and the complete model acquires a different dynamical nature
(Fig.~\ref{fig:plot-epr01}--\ref{fig:plot-epr07}). On the other hand,
the inner domain is well represented by the $H_b$, independently of
the value of $e_{p,0}$. Altogether, the illustration of the dynamics of the
$H_b$ using surfaces of section allows to obtain a first
numerical estimation of the position, size and importance of
the resonance web in the phase-space (Sect.~\ref{sec:feat_Hb}).

Finally, in the same chapter, we produced an averaged $H_b$ following the same
novel scheme as introduced in Chapter 2.  With the
estimation of the frequencies, we could locate the position of several
secondary and transverse resonances
(Fig.~\ref{fig:transverses1.png}--\ref{fig:transverses3.png}), with
errors of the order of a few percent (Table~\ref{tab:errors}). This
last result confirmed that the $H_b$ provides a good approximation
for all the three frequencies.

In Chapter 5, we introduced a resonant normal form construction for
some of the main secondary resonances described in previous chapters.
This construction requires an expansion in terms of the action-angle
variables involved in the secondary resonance under study. Since the
usual Taylor expansions exhibit severe convergence problems,
we applied a different kind of expansion called asymmetric,
that helps extending the series convergence domain
(Fig.~\ref{fig:b1.png}--\ref{fig:b7.png}). The dynamics of the $H_b$
are well represented by the asymmetrically expanded Hamiltonian, while the
symmetric expansion induces spurious behaviors
(Fig.~\ref{fig:3surfaces.png}, Fig.~\ref{fig:3surfaces-2.png}).

Through the normalization scheme, we constructed resonant integrals
that allow to analytically estimate the center $\overline{u}$ and the
width $(\Delta u_{max} - \Delta u_{min})$ of the most important
secondary resonances. Since higher order resonances appear in the
normalized Hamiltonian only at high order, a large number of 
normalizing steps are required in order to analytically represent
these resonances in the resonant normal form. Thus, the method is much
more efficient for low order (associated with large values of $\mu$)
resonances, that are also the most relevant cases. With an asymmetric
expansion of order 20
and normalization  of order 14, as considered in the computations, we
were able to reproduce the position of the $1$:$5$, $1$:$6$ and, up to
some extent, $1$:$7$ resonances
(Fig.~\ref{fig:1to5resint.png}--\ref{fig:1to7resint.png}).

The method based on the computation of resonant integrals turns to be
very slowly converging with the normalization order $R$.  This makes
the estimation of the outer border $\Delta u_{max}$ quite difficult
(Fig.~\ref{fig:res_int_orders.png}). However, for the inner border,
the convergence can be achieved rather fast, and the determination of
the position is accurate from the order $R = 10$ and beyond. Since the
position of $\Delta u_{min}$ represents the limit of the inner
(non-resonant) domain, we can accurately estimate the size of this
latter domain, which is the most important domain in the applications
where astronomical Trojan objects are expected to be found in stable
orbits.\\

While the above results emphasize the analytical framework of the
methods developed in Chapter 2--5, they also open new possibilities
for extension and/or applications in the context of concrete models
of astronomical interest. 

1) An obvious extension regards adding the third dimension, i.e, 
non-planar motions, in the formalism of Chapter 3. Analogously to
the definition of the proper eccentricity, this should lead to a 
convenient definition of the quasi-integral of the 'proper inclination'.
Also, the addition of one more degree of freedom adds new frequencies
to the problem, and thus is expected to modify the resonant structure
in the space of proper elements.

2) The asymptotic regime of the series developed in Chapter 2, 4 and 5
remains to be explored. The usual Birkhoff series exhibit an
exponentially small remainder at the optimal normalization order, a
fact allowing to obtain estimates of the domain of practical stability
in the spirit of the Nekhoroshev theorem
(Sect.\ref{sec:1.4.X.non_stab},
Sect.~\ref{sec:1.4.works_on_Trojans}). However, it is yet unclear how
to implement such estimates in series computed by the modified
algorithm of Chapter~2.

3) Applications of the above methods in the Solar System include the
case of asteroids in the tadpole domain of the 1:1 MMR with the
planets. A classical application (e.g~\cite{Efthy-13} and
references therein) refers to the Trojan asteroids of Jupiter, which,
however, have been so far examined from the analytical point of view
only in the framework of simplified models like the CR3BP or the
ER3BP. Since the numerical simulations
(e.g~\cite{RobGab-06},~\cite{MarzSch-07}) show that secular effects by,
at least, the outer planets of the Solar System are important in the
stability of Jupiter's Trojan asteroids, an analytical investigation
in more complex models like the RMPP of Chapter 4 would be desired. In
addition, analytical studies of the Trojan asteroids in other planets
of the Solar System, like Uranus, Neptune or the Earth, are sparse and
should be enriched.

4) The case of the equilateral points of the Sun-Earth system presents
interest also from the astrodynamical point of view, since these
points can serve for the station-keeping of man-made observatories.  A
preliminary application of the normal form method of Chapter 2 to the
problem of low-cost transfer at $L_4$ or $L_5$ in the Sun-Earth system
was presented in P\'aez and Locatelli (2015) (see page
\pageref{cha:prefacio}), in the framework of the CR3BP. However, a
realistic application would require considering more disturbing
bodies, like the Moon or the planets, as well as the non-circular
features of the Earth's orbit.

5) Finally, the prolific discovery of extrasolar planetary systems in
recent years leaves completely open the question of the existence and
stability of \emph{Trojan exoplanets}. No such planets have been identified
so far in exoplanet surveys. This may indicate that such planets are
rare, which case would necessitate a dynamical explanation, or that
there exist yet unsurpassed constraints in exo-Trojan detectability. It
has been proposed that the complexity of the orbits of Trojan bodies
may itself introduce intricacies in possible methods of
detection. This emphasizes the need to understand in detail the
orbital dynamics of the 1:1 mean motion commensurability, for a wide
variety of systems.

The above and other examples demonstrate that the classical problem of
Trojan dynamics remains nowadays a very vivid area of research. The
prospects opened by analytical methods and tools as those developed in
the present thesis would hopefully prove helpful in future 
dynamical investigations.

\appendix 
\chapter{Power series expansion of the Hamiltonian of the pCR3BP}
\label{sec_app:expansions}

In this Appendix we give the full construction of the
Hamiltonian~\eqref{eq:initial-Ham-rewritten},
starting from Eq.~\eqref{eq:Ham_hel_RTBP_4}. 
Each of the terms of the disturbing function,
\begin{equation*}
\frac{1}{r}\ ,
\qquad
r\cos\vartheta~,
\qquad{\rm and}\qquad
r^2~,
\end{equation*}
is treated independently. In practice, the most convenient approach
is based on some non trivial manipulation of these series.  We
preliminarly expand some expressions in terms of the orbital elements
$e$ and $M$. Then, we multiply these formul{\ae} by some factors
depending also on the other orbital elements and some mixed variables
which are not always canonical. This intermediate step allows us to
introduce the final expansions with respect to the canonical
coordinates $(\rho,\xi,\tau,\eta)\,$ and powers of $\beta(\tau)$.

\section{Preliminary expansions in terms of $e$ and $M$}
\label{sbs:strategy}

For the term $1/r$, we consider Eq.~\eqref{eq:r_solo},
\begin{equation}
r = \frac{G^2}{1+e\cos f}\,\ ,
\label{eq:initial_r}
\end{equation}
where $f$ is the true anomaly. 
Let $g^{(I)}$ be
\begin{equation}
g^{\rm(I)}(e,f)=G^2/r=1+e\cos f~.
\label{eq:def-gI}
\end{equation}
Let us introduce the expansions of $\cos f$ and $\sin f$, in terms of the
eccentricity and mean anomaly (\S2.5 of~\cite{MurrDerm-99})
\begin{equation}
\vcenter{\openup1\jot\halign{
 \hbox {\hfil $\displaystyle {#}$}
&\hbox {\hfil $\displaystyle {#}$\hfil}
&\hbox {$\displaystyle {#}$\hfil}\cr
\cos f &= &-e +
\frac{2\big(1-e^2\big)}{e}\sum_{n=1}^{\infty}\big[J_n(ne)\cos(nM)\big]\ \ ,
\cr
\sin f &=
&2\sqrt{1-e^2}\sum_{n=1}^{\infty}\big[J_{n}^{\prime}(ne)\sin(nM)\big]\ \ ,
\cr
}}
\label{eq:cosv-sinv-series-expan}
\end{equation}
where
\begin{equation}
J_{n}(x) =
\sum_{j=0}^{\infty}\frac{(-1)^j}{j!\,(j+n)!}\left(\frac{x}{2}\right)^{n+2j}
\ \quad{\rm and}\ \quad
J_{n}^{\prime}(x) =
\sum_{j=0}^{\infty}\frac{(-1)^j(2j+n)}{2\big(j!\,(j+n)!\big)}
\left(\frac{x}{2}\right)^{n+2j-1}
\label{eq:def-Bessel-Jn-and-derivative}
\end{equation}
are the Bessel functions of first kind $J_n$ and their derivatives
$J_{n}^{\prime}\,$, respectively. 
Replacing Eq.~\eqref{eq:cosv-sinv-series-expan} into Eq.~\eqref{eq:def-gI},
we obtain $h^{\rm(I)}(e,M)$, given by
\begin{equation}
h^{\rm(I)}(e,M) = g^{\rm(I)}\big(e,f(e,M)\big)=
\sum_{j=0}^{\infty}e^j {\cal P}^{\rm(I)}_j(M)\ .
\label{eq:def-fI}
\end{equation}
According to~\eqref{eq:def-Bessel-Jn-and-derivative}, each function
${\cal P}^{\rm(I)}_j(M)$ is a trigonometric polynomial of degree
$j\,$, given by a sum of cosines with Fourier harmonics having the
same parity of $j\,$.

For the term $r\cos\theta\,$, we have (see Fig.~\ref{fig:vartheta.jpg})
\begin{equation}
\vartheta = \lambda-\lambda'+f - M= \tau+f-M~,
\label{eq:initial_theta}
\end{equation}
where  $\tau=\lambda-\lambda'$ (Sec.~\ref{2.1-expansion}).
Using~(\ref{eq:initial_r})
and~(\ref{eq:initial_theta}), we obtain
\begin{equation}
\frac{r\cos\vartheta}{G^2}=
\frac{\cos\tau\cos(f-M)-\sin\tau\sin(f-M)}{1+e\cos f}\,\ .
\label{expr:rcostheta_over_G^2}
\end{equation}
For the expansion of Eq.~\eqref{expr:rcostheta_over_G^2}, it
is convenient to consider first the parts not depending on
$\tau$. Thus, we introduce
\begin{equation}\label{def:lot_of_g}
\vcenter{\openup1\jot\halign{
 \hbox {\hfil $\displaystyle {#}$}
&\hbox {\hfil $\displaystyle {#}$\hfil}
&\hbox {$\displaystyle {#}$\hfil}\cr
g^{\rm(II)}(e,f) &= &\frac{1}{g^{\rm(I)}(e,f)} = \frac{1}{1+e\cos f}\ ,
\cr
g^{\rm(III)}(e,f) &= &\frac{\cos f}{g^{\rm(I)}(e,f)} = \frac{\cos f}{1+e\cos f}\ ,
\cr
g^{\rm(IV)}(e,f) &= &\frac{\sin f}{g^{\rm(I)}(e,f)} = \frac{\sin f}{1+e\cos f}\ .
\cr
}}
\end{equation}

We expand $g^{\rm(II)}$ with respect to $e$ around $e=0$,
\begin{equation}\label{eq:gII-series}
g^{\rm(II)}(e,f)= \frac{1}{1+e \cos
  f}=\sum_{j=0}^{\infty}\big[(-1)^je^j\cos^j f\big]~.
\end{equation}
Using~\eqref{eq:cosv-sinv-series-expan}, we obtain
the functions
${\cal P}^{\rm(II)}_j(M)\,$,
${\cal P}^{\rm(III)}_j(M)$ and ${\cal P}^{\rm(IV)}_j(M)\,$, such that
\begin{equation}\label{def:lot_of_f}
\vcenter{\openup1\jot\halign{
 \hbox {\hfil $\displaystyle {#}$}
&\hbox {\hfil $\displaystyle {#}$\hfil}
&\hbox {$\displaystyle {#}$\hfil}\cr
h^{\rm(II)}(e,M) &=& g^{\rm(II)}(e,f(e,M))=
\sum_{j=0}^{\infty}e^j{\cal P}^{\rm(II)}_j(M)\ ,
\cr
h^{\rm(III)}(e,M) &=& g^{\rm(III)}(e,f(e,M))=
\sum_{j=0}^{\infty}e^j{\cal P}^{\rm(III)}_j(M)\ ,
\cr
h^{(IV)}(e,M) &=& g^{\rm(IV)}(e,f(e,M))=
\sum_{j=0}^{\infty}e^j{\cal P}^{\rm(IV)}_j(M)\ .
\cr
}}
\end{equation}
${\cal P}^{\rm(II)}_j(M)$ is a
trigonometric polynomial of degree $j\,$, while both
${\cal P}^{\rm(III)}_j(M)$ and ${\cal P}^{\rm(IV)}_j(M)$ are of degree
$j+1\,$. Furthermore, the Fourier expansions
${\cal P}^{\rm(II)}_j(M)$ and ${\cal P}^{\rm(III)}_j(M)$ are given by sums of
cosines, while that of ${\cal P}^{\rm(IV)}_j(M)$ is a sum of sines. All
the expansions of such trigonometric polynomials contain Fourier
harmonics having the same parity as the maximal degree.

We now combine Eqs.~\eqref{def:lot_of_g} and Eq.~\eqref{def:lot_of_f} 
so that
\begin{equation}
\vcenter{\openup1\jot\halign{ \hbox {\hfil $\displaystyle {#}$} &\hbox
    {\hfil $\displaystyle {#}$\hfil} &\hbox {$\displaystyle
      {#}$\hfil}\cr h^{\rm(V)}(e,M) &=
    &\frac{\cos(f(e,M)-M)}{1+e\cos(f(e,M))} = h^{\rm(III)} \cos
    M+h^{\rm(IV)} \sin M = \sum_{j=0}^{\infty}e^j{\cal
      P}^{\rm(V)}_j(M)\ , \cr h^{\rm(VI)}(e,M) &=
    &\frac{\sin(f(e,M)-M)}{1+e\cos(f(e,M))} = h^{\rm(IV)} \cos
    M-h^{\rm(III)} \sin M\ = \sum_{j=0}^{\infty}e^j{\cal
      P}^{\rm(VI)}_j(M) ,  \cr }}
\label{eq:def-r_over_Gsquared_cosv-l_sinv-l}
\end{equation}
where $\{{\cal P}^{\rm(V)}_j(M)\}_{j\ge 0}$ and $\{{\cal P}^{\rm(VI)}_j(M)\}_{j\ge
  0}\,$, can be computed from~(\ref{def:lot_of_f})

By the d'Alembert rules, the terms appearing in the trigonometric
polynomials of order ${\cal O}(e^j)$ must have the same parity of $j$
and being at most of degree $j$, for the function $R$ to be analytic
in the coordinates $\xi$ and $\eta$ (Eq.~\ref{eq:Garfinkel-coord}) in
a neighborhood of the origin.  Therefore, ${\cal P}^{\rm(V)}_j(M)$ and
${\cal P}^{\rm(VI)}_j(M)$ must be trigonometric polynomials of degree
at most $j\,$.

For the expansions of $r^2$, let
$h^{\rm(VII)}(e,M)=r^2/G^4$ be
\begin{equation}
h^{\rm(VII)}(e,M)=
\big[h^{\rm(II)}(e,M)\big]^2 =
\sum_{j=0}^{\infty}e^j{\cal P}^{\rm(VII)}_j(M)\ .
\label{eq:def-fVII}
\end{equation}
Thus, from Eqs.~(\ref{def:lot_of_f}) it is possible to calculate
explicitly the Fourier coefficients ${\cal P}^{\rm(VII)}_j(M)$.  This
expansion in terms of ${\cal P}^{\rm(VII)}_j(M)$ shares the same
properties as those of ${\cal P}^{\rm(I)}_j(M)$ and ${\cal
  P}^{\rm(V)}_j(M)\,$.

\section{Expansions in terms of canonical coordinates}
\label{sbs:to_exp_can_coord}

We now replace the orbital elements $e,\, M$ by
the Poincar\'e variables, given in Eqs.~\eqref{eq:Garfinkel-coord}.
The functions $h^{\rm(I)}$, $h^{\rm(V)}$ and $h^{\rm(VII)}$ are of
the form
\begin{equation}
h(e,M)=
\sum_{j=0}^{\infty}\sum_{k=0}^{\lfloor j/2\rfloor}
\big[c_{j,k}\,e^j\cos\big((j-2k)M\big)\big]
\ ,
\label{eq:def-generic-function-of-e,M}
\end{equation}
where $c_{j,k}\in\mathbb{R}\,$. The expansion of $h^{\rm(VI)}$ is
equivalent, but with sines instead of cosines. For replacing $e$ and
$M$ with the canonical coordinates ($G\,$, $\xi$, $\eta\,$), we first
introduce the variable
$\zeta=\sqrt{2\Gamma/G}\,$ which, according
Eqs.~\eqref{eq:Delaunay-coord_cR3BP},
\begin{equation}
\zeta=\frac{\sqrt{2}\sqrt{1-\sqrt{1-e^2}}}{\sqrt[4]{1-e^2}}\ .
\label{eq:def_gr_zeta}
\end{equation} 
Let us consider that $e\in[0,1)$. 
It is possible to invert Eq.~\eqref{eq:def_gr_zeta} as follows: we
consider $\zeta=\zeta(e)$ is a monotone map such that
$\zeta:[0,1)\mapsto[0,\infty)\,$.
Thus, we introduce $z=\sqrt[4]{1-e^2}$ ($z:[0,1)\mapsto(0,1]$) and
we solve the equation $\zeta^2\,z^2=2(1-z^2)$ in the unknown
$z$. Finally, from the definition of $z$, we express
$e$ as a function of $\zeta$
\begin{equation}
e=\sqrt{1-\left(\frac{2}{\zeta^2+2}\right)^2}\ .
\label{eq:inv_def_gr_zeta}
\end{equation}

Hence, we construct the Taylor-Fourier series
of~(\ref{eq:def-generic-function-of-e,M}) in terms of
$\big(\xi/\sqrt{G},\eta/\sqrt{G}\big)\,$:
\begin{enumerate}
\item{} Let $h_{{\rm tmp}}$ be one of $h^{\rm(I)}$, $h^{\rm(V)}$,
  $h^{\rm(VI)}$ or $h^{\rm(VII)}$, where we replace the
  expression~\eqref{eq:inv_def_gr_zeta} and expand. Thus,
\begin{equation}
h_{{\rm tmp}}(\zeta,M)= \sum_{j=0}^{\infty}\sum_{k=0}^{\lfloor
  j/2\rfloor} \left[{\bar c}_{j,k}\,\zeta^j \,{\scriptstyle
    {{\cos}\atop{\sin}}} \big((j-2k)M\big)\right] \ .
\label{eq:expan-temporary-f}
\end{equation}

\item{} 
According to Vi\`ete's formul{\ae} for multiple angles
\begin{equation*}
    \vcenter{\openup1\jot\halign{
        \hbox {\hfil $\displaystyle {#}$}
        &\hbox {\hfil $\displaystyle {#}$\hfil}
        &\hbox {$\displaystyle {#}$\hfil}\cr
    \cos(n\alpha) &=
    &\sum_{k=0}^{n}\left({{n}\atop{k}}\right)\cos^k\alpha\sin^{n-k}\alpha
    \cos\left(\frac{(n-k)\pi}{2}\right)\ ,
    \cr
    \sin(n\alpha) &=
    &\sum_{k=0}^{n}\left({{n}\atop{k}}\right)\cos^k\alpha\sin^{n-k}\alpha
    \sin\left(\frac{(n-k)\pi}{2}\right)\ ,
    \cr
    }}
\end{equation*}
where $n\in \mathbb{N}$ and $\theta\in \mathbb{R}$,
each term ${\cal O}(\zeta^j)$ 
in~(\ref{eq:expan-temporary-f}) generates monomials of the type
$\zeta^{2j_0}\big(\zeta\cos M\big)^{j_1}\big(\zeta\sin M\big)^{j_2}$,
where $j=2j_0+j_1+j_2\,$, $j_0$,$j_1$,$j_2 > 0$.
We apply the following substitution
\begin{equation*}
\vcenter{\openup1\jot\halign{
 \hbox {\hfil $\displaystyle {#}$}
&\hbox {$\displaystyle {#}$\hfil}\cr
\zeta^{2j_0}\big(\zeta\cos M\big)^{j_1}\big(\zeta\sin M\big)^{j_2} &=
\left(\frac{\xi^2+\eta^2}{G}\right)^{j_0}\left(\frac{\xi}{\sqrt{G}}\right)^{j_1}
\left(\frac{\eta}{\sqrt{G}}\right)^{j_2}
\cr
&= \frac{1}{G^{j/2}}\big(\xi^2+\eta^2\big)^{j_0}\xi^{j_1}\eta^{j_2}\ ,
\cr
}}
\end{equation*}
in each monomial.

\item{} We sum the coefficients corresponding to the 
same monomials on $(\xi,\eta)\,$, obtaining
\begin{equation}
{\cal F}\left(\frac{\xi}{\sqrt{G}},\frac{\eta}{\sqrt{G}}\right)=
\sum_{j=0}^{\infty}\sum_{{j_1\ge 0\,,\,j_2\ge 0}\atop{j_1+j_2=j}}
\left[d_{j_1,j_2}\,\left(\frac{\xi}{\sqrt{G}}\right)^{j_1}
\left(\frac{\eta}{\sqrt{G}}\right)^{j_2}\right]
\ ,
\label{eq:expan-final-F-of-entrycsieta}
\end{equation}
where $d_{j_1,j_2}\in \mathbb{R}$ $\forall$ $j_1\ge 0\,$, $j_2\ge 0\,$.
\end{enumerate}

\noindent The final function ${\cal F}\,$
inherits the parity properties of the functions $h\,$. Applying this
procedure to $h^{\rm(I)}$, $h^{\rm(V)}$, $h^{\rm(VI)}$ and $h^{\rm(VII)}$,
we obtain
\begin{equation}
\vcenter{\openup1\jot\halign{
   \hbox {\hfil $\displaystyle {#}$}
   &\hbox {\hfil $\displaystyle {#}$\hfil}
   &\hbox {$\displaystyle {#}$\hfil}\cr
{\cal F}^{\rm(I)}\left(\frac{\xi}{\sqrt{G}},\frac{\eta}{\sqrt{G}}\right) &=
&h^{\rm(I)}\left(\,e\left(\frac{\xi}{\sqrt{G}},\frac{\eta}{\sqrt{G}}\,\right)\,,
\,M\left(\frac{\xi}{\sqrt{G}},\frac{\eta}{\sqrt{G}}\,\right)\,\right)
\ ,
\cr
{\cal F}^{\rm(V)}\left(\frac{\xi}{\sqrt{G}},\frac{\eta}{\sqrt{G}}\right) &=
&h^{\rm(V)}\left(\,e\left(\frac{\xi}{\sqrt{G}},\frac{\eta}{\sqrt{G}}\,\right)\,,
\,M\left(\frac{\xi}{\sqrt{G}},\frac{\eta}{\sqrt{G}}\,\right)\,\right)
\ ,
\cr
{\cal F}^{\rm(VI)}\left(\frac{\xi}{\sqrt{G}},\frac{\eta}{\sqrt{G}}\right) &=
&h^{\rm(VI)}\left(\,e\left(\frac{\xi}{\sqrt{G}},\frac{\eta}{\sqrt{G}}\,\right)\,,
\,M\left(\frac{\xi}{\sqrt{G}},\frac{\eta}{\sqrt{G}}\,\right)\,\right)
\ ,
\cr
{\cal F}^{\rm(VII)}\left(\frac{\xi}{\sqrt{G}},\frac{\eta}{\sqrt{G}}\right) &=
&h^{\rm(VII)}\left(\,e\left(\frac{\xi}{\sqrt{G}},\frac{\eta}{\sqrt{G}}\,\right)\,,
\,M\left(\frac{\xi}{\sqrt{G}},\frac{\eta}{\sqrt{G}}\,\right)\,\right)
\ .
\cr
}}
\end{equation}

For introducing $\rho\,$, we consider
{\small
\begin{equation}
\vcenter{\openup1\jot\halign{
 \hbox {\hfil $\displaystyle {#}$}
&\hbox {\hfil $\displaystyle {#}$\hfil}
&\hbox {$\displaystyle {#}$\hfil}
&\hbox to 2 ex{\hfil$\displaystyle {#}$\hfil}
&\hbox {\hfil $\displaystyle {#}$}
&\hbox {\hfil $\displaystyle {#}$\hfil}
&\hbox {$\displaystyle {#}$\hfil}\cr
F^{\rm(I)}(G,\xi,\eta) &=
&\frac{1}{G^2}\cdot
{\cal F}^{\rm(I)}\left(\frac{\xi}{\sqrt{G}},\frac{\eta}{\sqrt{G}}\right)\ ,
& &F^{\rm(V)}(G,\xi,\eta) &=
&G^2\cdot {\cal F}^{\rm(V)}\left(\frac{\xi}{\sqrt{G}},\frac{\eta}{\sqrt{G}}\right)\ ,
\cr
F^{\rm(VI)}(G,\xi,\eta) &=
&G^2\cdot {\cal F}^{\rm(VI)}\left(\frac{\xi}{\sqrt{G}},\frac{\eta}{\sqrt{G}}\right)\ ,
& &F^{\rm(VII)}(G,\xi,\eta) &=
&G^4\cdot
{\cal F}^{\rm(VII)}\left(\frac{\xi}{\sqrt{G}},\frac{\eta}{\sqrt{G}}\right)\ ,
\cr
}}
\label{eq:def-Ffirst-Fseventh}
\end{equation}}
and we replace with the transformation $G = 1+\rho$ given in~\eqref{eq:Garfinkel-coord}
\begin{equation}
\vcenter{\openup1\jot\halign{
 \hbox {\hfil $\displaystyle {#}$}
&\hbox {\hfil $\displaystyle {#}$\hfil}
&\hbox {$\displaystyle {#}$\hfil}
&\hbox to 2 ex{\hfil$\displaystyle {#}$\hfil}
&\hbox {\hfil $\displaystyle {#}$}
&\hbox {\hfil $\displaystyle {#}$\hfil}
&\hbox {$\displaystyle {#}$\hfil}\cr
{\cal K}^{\rm(I)}(\rho,\xi,\eta) &= &F^{\rm(I)}(1+\rho,\xi,\eta)\ ,
& &{\cal K}^{\rm(V)}(\rho,\xi,\eta) &= &F^{\rm(V)}(1+\rho,\xi,\eta)\ ,
\cr
{\cal K}^{\rm(VI)}(\rho,\xi,\eta) &= &F^{\rm(VI)}(1+\rho,\xi,\eta)\ ,
& &{\cal K}^{\rm(VII)}(\rho,\xi,\eta) &= &F^{\rm(VII)}(1+\rho,\xi,\eta)\ .
\cr
}}
\label{eq:def-Kfirst-Kseventh}
\end{equation}
Finally, we 
construct the series for $1/r$, $r\cos\vartheta$ and $r^2$, 
\begin{equation}
\frac{1}{r}=\frac{1}{r(\rho,\xi,\eta)}={\cal K}^{\rm(I)}(\rho,\xi,\eta)\ ,
\label{eq:1/r-as-a-function-of-can-coord}
\end{equation}
\begin{equation}
r\cos\vartheta=
r(\rho,\xi,\tau,\eta)\cos\big(\vartheta(\rho,\xi,\tau,\eta)\big)=
{\cal K}^{\rm(V)}(\rho,\xi,\eta)\cos\tau-
{\cal K}^{\rm(VI)}(\rho,\xi,\eta)\sin\tau\ ,
\label{eq:rcostheta-as-a-function-of-can-coord}
\end{equation}
\begin{equation}
r^2 =\big(r(\rho,\xi,\eta)\big)^2={\cal K}^{\rm(VII)}(\rho,\xi,\eta)\ ,
\label{eq:rsquared-as-a-function-of-can-coord}
\end{equation}
that are polynomial series in $(\rho,\xi,\eta)\,$
but not in $\tau$.

Let us consider the term of the disturbing function
\begin{equation}\label{eq:damn_Delta_1}
\frac{1}{\Delta} = \frac{1}{\sqrt{1+r^2-2r \cos \vartheta}}~.
\end{equation}
We decompose
\begin{equation}\label{eq:damn_Delta_2}
\frac{1}{\Delta} = \frac{1}{\sqrt{ 2-2\cos \tau +
    \delta(\rho,\xi,\eta,\tau)}} = \frac{1}{\sqrt{2 -2\cos \tau} \,
  \sqrt{1+ \frac{\delta(\rho,\xi,\eta,\tau)}{(2-2 \cos\tau)}}} =
\frac{1}{\sqrt{2-2\cos \tau}} \, \frac{1}{\sqrt{1+\psi}}~,
\end{equation}
where $\delta(\rho,\xi,\eta,\tau) = (\Delta(\rho,\xi,\eta,\tau))^2 -
(2-2\cos \tau)$, and $\psi = \delta(\rho,\xi,\eta,\tau)/(2-2\cos
\tau)$. The term $\beta = 1/\sqrt{2-2\cos \tau}$ corresponds to the
first order approximation of $1/\Delta$. For the expansion of the
rest, we consider the Taylor series in terms of the small quantity
$\psi$
\begin{equation}\label{eq:exp_psi}
\frac{1}{\sqrt{1+\psi}} = 1+ \sum_{i=1}^{\infty}
\frac{(-1)^i}{i!\,2^i} \psi^i \prod_{j=1}^{i} (2j-1) =
\sum_{i=0}^{\infty} \mathsf{c}_{i} \psi^i~,
\end{equation}
where $\mathsf{c}_{i} \in \mathbb{R}$.
From Eqs.~\eqref{eq:rcostheta-as-a-function-of-can-coord}--~\eqref{eq:rsquared-as-a-function-of-can-coord}, we obtain the expansion of $\Delta$, from
which we derive also the expansion for $\delta$. Thus, the powers of 
$\psi$ in~\eqref{eq:exp_psi} have the form
\begin{equation}\label{eq:psi-s}
\psi^i =  \left(\frac{\delta(\rho,\xi,\tau,\eta)}{2-2 \cos \tau} \right)^i =  \frac{1}{(2-2\cos\tau)^i} \, \mathsf{d}_{k_1,k_2,k_3,k_4,k_5} \, \xi^{k_1} \, \rho^{k_2}\, \eta^{k_3}\, \cos^{k_4} \tau \, \sin^{k_5} \tau~,
\end{equation}
where $\mathsf{d}_{k_1,k_2,k_3,k_4,k_5} \in \mathbb{R} $ are
coefficients arising from the expansions in
Eqs.~\eqref{eq:rcostheta-as-a-function-of-can-coord}
and~\eqref{eq:rsquared-as-a-function-of-can-coord}, and $k_i \in
\mathbb{N}$.  Replacing with
Eqs.~\eqref{eq:exp_psi},~\eqref{eq:psi-s}, we obtain the final series
for the term
\begin{equation}\label{eq:1ovdeltaconB}
\frac{1}{\Delta} = \beta(\tau) +
\sum_{l=1}^{\infty}\,\sum_{{\scriptstyle{m_1+m_2}}\atop{\scriptstyle
        {+m_3=l}}} \ \sum_{{\scriptstyle{k_1+k_2\le
          l}}\atop{\scriptstyle {j\le 2l+1}}}
\mathsf{d}_{m_1,m_2,m_3,k_1,k_2,j} \, \xi^{m_2} \, \rho^{m_1}\, \eta^{m_3}\, \cos^{k_1} \tau \, \sin^{k_2} \tau \, \beta(\tau)^{j}~,
\end{equation}
where
$\beta(\tau) = 1/\sqrt{2-2\cos\tau}$. Gathering 
Eqs.~\eqref{eq:1/r-as-a-function-of-can-coord}--~\eqref{eq:rcostheta-as-a-function-of-can-coord}--~\eqref{eq:1ovdeltaconB}, we construct the complete
expansion of the disturbing function in Eq.~\eqref{eq:initial-Ham-rewritten}.

\chapter{The disturbing function of the ER3BP in terms of orbital elements}
\label{sec_app:expan_del_ellip}

Up to second order in the eccentricities , the expansions of the three
terms of the disturbing function of the ER3BP
(Eq.~\ref{eq:Ham_hel_RTBP_ellip}), in terms of the orbital elements
are given by:

\begin{align*}
- \frac{1}{r} =&  - \frac{1}{a} - \frac{e\,\cos\,(\lambda-\varpi)}{a} 
- \frac{e^2\,\cos\,(2\lambda-2\varpi)}{a}~,\phantom{\big)}\\
\frac{r}{r'}\,\cos \vartheta = & -\frac{a\,\cos\,(\lambda-\lambda')}{a'^2} 
-\frac{a\,e\,\cos\,(2\lambda-\lambda'-\varpi)}{2\,a'^2}+ \frac{3\,a\,e\,\cos(\lambda'-\varpi)}{2\,a'^2}
-\frac{2\,a\,e'\,\cos\,(\lambda-2\lambda'+\varpi')}{a'^2} \\ 
& + \frac{a\,e^2\,\cos\,(\lambda-\lambda')}{2\,a'^2}+ \frac{a\,e'^2\,\cos\,(\lambda-\lambda')}{2\,a'^2}
-\frac{3\,a\,e^2\,\cos\,(3\lambda-\lambda'-2\varpi)}{8\,a'^2} - 
\frac{a\,e^2\,\cos\,(\lambda+\lambda'-2\varpi)}{8\,a'^2} \\
& - \frac{a\,e'^2\,\cos\,(\lambda +\lambda'-2 \varpi')}{8\,a'^2} + 
\frac{3\,a\,e\,e'\,\cos\,(2 \lambda'-\varpi-\varpi')}{a'^2}
-\frac{a\,e\,e'\,\cos\,(2\lambda-2\lambda'-\varpi+\varpi')}{a'^2}\\
&-\frac{27\,a\,e'^2\,\cos\,(\lambda-3\lambda'+2\varpi')}{8\,a'^2}~,
\end{align*}

\vfill \eject

{\small
\begin{align*} 
& \frac{1}{\Delta} = 
   \frac{3\,a^4\,e^2\,\cos\,(2\lambda-2\varpi)}{4\left(a^2+a'^2-2\,a\,a'\,\cos\,(\lambda-\lambda')\right)^{5/2}} 
   +\frac{3\,a^4\,e^2}{4\left(a^2+a'^2-2\,a\,a'\,\cos\,(\lambda-\lambda')\right)^{5/2}}
   +\frac{3\,a^3\,a'\,e^2\,\cos\,(3\lambda-\lambda'-2\varpi)}{4\left(a^2+a'^2-2\,a\,a'\,\cos\,(\lambda-\lambda')\right)^{5/2}}\\
   &-\frac{9\,a^3\,a'\,e^2\,\cos\,(\lambda+\lambda'-2\varpi)}{4\left(a^2+a'^2-2\,a\,a'\,\cos\,(\lambda-\lambda')\right)^{5/2}}
   -\frac{9\,a^3\,a'\,e\,e'\,\cos\,(2\lambda-\varpi-\varpi')}{4\left(a^2+a'^2-2\,a\,a'\,\cos\,(\lambda-\lambda')\right)^{5/2}}
   +\frac{3\,a^3\,a'\,e\,e'\,\cos\,(2\lambda'-\varpi-\varpi')}{4\left(a^2+a'^2-2\,a\,a'\,\cos\,(\lambda-\lambda')\right)^{5/2}}\\
   &-\frac{9\,a^3\,a'\,e\,e'\,\cos\,(\varpi-\varpi')}{4\left(a^2+a'^2-2\,a\,a'\,\cos\,(\lambda-\lambda')\right)^{5/2}}
   +\frac{3\,a^3\,a'\,e\,e'\,\cos\,(2\lambda-2\lambda'-\varpi+\varpi') }{4\left(a^2+a'^2-2\,a\,a'\,\cos\,(\lambda-\lambda')\right)^{5/2}}
   -\frac{3\,a^3\,a'\,e^2\,\cos\,(\lambda-\lambda')}{2\left(a^2+a'^2-2\,a\,a'\,\cos\,(\lambda-\lambda')\right)^{5/2}}\\
   &+\frac{a^2\,e^2\cos\,(2\lambda-2\varpi)}{4\left(a^2+a'^2-2\,a\,a'\,\cos\,(\lambda-\lambda')\right)^{3/2}}
   -\frac{9\,a^2\,a'^2\,e^2\,\cos\,(2\lambda-2\varpi)}{8\left(a^2+a'^2-2\,a\,a'\,\cos\,(\lambda-\lambda')\right)^{5/2}}
   +\frac{3\,a^2\,a'^2\,e^2\,\cos\,(4\lambda-2\lambda'-2\varpi)}{16\left(a^2+a'^2-2\,a\,a'\,\cos\,(\lambda-\lambda')\right)^{5/2}}\\
   &+\frac{27\,a^2\,a'^2\,e^2\,\cos\,(2\lambda'-2\varpi)}{16\left(a^2+a'^2-2\,a\,a'\,\cos\,(\lambda-\lambda')\right)^{5/2}}
   +\frac{a^2\,e\,\cos\,(\lambda-\varpi)}{\left(a^2+a'^2-2\,a\,a'\,\cos\,(\lambda-\lambda')\right)^{3/2}}
   +\frac{27\,a^2\,a'^2\,e'^2\,\cos\,(2\lambda-2\varpi')}{16\left(a^2+a'^2-2\,a\,a'\,\cos\,(\lambda-\lambda')\right)^{5/2}}\\
   &-\frac{9\,a^2\,a'^2\,e'^2\,\cos\,(2\lambda'-2 \varpi')}{8\left(a^2+a'^2-2\,a\,a'\,\cos\,(\lambda-\lambda')\right)^{5/2}}
   -\frac{9\,a^2\,a'^2\,e\,e'\,\cos\,(3\lambda-\lambda'-\varpi-\varpi')}{8 \left(a^2+a'^2-2\,a\,a'\,\cos\,(\lambda -\lambda')\right)^{5/2}}
   +\frac{21\,a^2\,a'^2\,e\,e'\,\cos\,(\lambda+\lambda'-\varpi-\varpi')}{4\left(a^2+a'^2-2\,a\,a'\,\cos\,(\lambda-\lambda')\right)^{5/2}}\\
   &+\frac{27\,a^2\,a'^2\,e\,e'\,\cos\,(\lambda-\lambda'+\varpi-\varpi')}{8\left(a^2+a'^2-2\,a\,a'\,\cos\,(\lambda-\lambda')\right)^{5/2}}
   +\frac{3\,a^2\,a'^2\,e\,e'\,\cos\,(3\lambda-3\lambda-\varpi+\varpi')}{8\left(a^2+a'^2-2\,a\,a'\,\cos\,(\lambda-\lambda')\right)^{5/2}}
   -\frac{3\,a^2\,a'^2\,e\,e'\,\cos\,(\lambda-\lambda'-\varpi+\varpi')}{4\left(a^2+a'^2-2\,a\,a'\,\cos\,(\lambda-\lambda')\right)^{5/2}}\\
   &-\frac{9\,a^2\,a'^2\,e\,e'\,\cos\,(\lambda-3\lambda'+\varpi+\varpi') }{8\left(a^2+a'^2-2\,a\,a'\,\cos\,(\lambda-\lambda')\right)^{5/2}}
   +\frac{3\,a^2\,a'^2\,e'^2\,\cos\,(2\lambda-4 \lambda'+2\varpi')}{16\left(a^2+a'^2-2\,a\,a'\,\cos\,(\lambda-\lambda')\right)^{5/2}}
   -\frac{3\,a^2\,e^2}{4\left(a^2+a'^2-2\,a\,a'\,\cos\,(\lambda-\lambda')\right)^{3/2}}\\
   &+\frac{15\,a^2\,a'^2\,e^2}{8\left(a^2+a'^2-2\,a\,a'\,\cos\,(\lambda-\lambda')\right)^{5/2}}
   +\frac{15\,a^2\,a'^2\,e'^2 }{8\left(a^2+a'^2-2\,a\,a'\,\cos\,(\lambda-\lambda')\right)^{5/2}}
   -\frac{9\,a^2\,a'^2\,e^2\cos\,(2\lambda-2\lambda')}{8\left(a^2+a'^2-2\,a\,a'\,\cos\,(\lambda-\lambda')\right)^{5/2}}\\
   &-\frac{9\,a^2\,a'^2\,e'^2\,\cos\,(2\lambda-2\lambda')}{8\left(a^2+a'^2-2\,a\,a'\,\cos\,(\lambda-\lambda')\right)^{5/2}}
   +\frac{3\,a\,a'\,e^2\,\cos\,(3\lambda-\lambda'-2\varpi)}{8\left(a^2+a'^2-2\,a\,a'\,\cos\,(\lambda-\lambda')\right)^{3/2}}
   +\frac{a\,a'\,e^2\,\cos\,(\lambda+\lambda'-2\varpi)}{8\left(a^2+a'^2-2\,a\,a'\,\cos\,(\lambda-\lambda')\right)^{3/2}}\\
   &+\frac{a\,a'\,e\,\cos\,(2\lambda-\lambda'-\varpi)}{2\left(a^2+a'^2-2\,a\,a'\,\cos\,(\lambda-\lambda')\right)^{3/2}}
   -\frac{3\,a\,a'\,e\cos\,(\lambda'-\varpi)}{2\left(a^2+a'^2-2\,a\,a'\,\cos\,(\lambda-\lambda')\right)^{3/2}}
   +\frac{a\,a'\,e'^2\,\cos\,(\lambda+\lambda-2\varpi')}{8\left(a^2+a'^2-2\,a\,a'\,\cos\,(\lambda-\lambda')\right)^{3/2}}\\
   &-\frac{9\,a\,a'^3\,e'^2\,\cos\,(\lambda+\lambda'-2\varpi')}{4\left(a^2+a'^2-2\,a\,a'\,\cos\,(\lambda-\lambda')\right)^{5/2}}
   -\frac{3\,a\,a'\,e'\,\cos\,(\lambda-\varpi')}{2\left(a^2+a'^2-2\,a\,a'\,\cos\,(\lambda-\lambda')\right)^{3/2}}
   -\frac{3\,a\,a'\,e\,e'\,\cos\,(2\lambda-\varpi-\varpi')}{4\left(a^2+a'^2-2\,a\,a'\,\cos\,(\lambda-\lambda')\right)^{3/2}}\\
   &+\frac{3\,a\,a'^3\,e\,e'\,\cos\,(2\lambda-\varpi-\varpi')}{4\left(a^2+a'^2-2\,a\,a'\,\cos\,(\lambda-\lambda')\right)^{5/2}}
   -\frac{3\,a\,a'\,e\,e'\,\cos\,(2\lambda'-\varpi-\varpi')}{4\left(a^2+a'^2-2\,a\,a'\,\cos\,(\lambda-\lambda')\right)^{3/2}}
   -\frac{9\,a\,a'^3\,e\,e'\,\cos\,(2\lambda'-\varpi-\varpi')}{4\left(a^2+a'^2-2\,a\,a'\,\cos\,(\lambda-\lambda')\right)^{5/2}}\\
   &+\frac{9\,a\,a'\,e\,e'\,\cos\,(\varpi-\varpi')}{4\left(a^2+a'^2-2\,a\,a'\,\cos\,(\lambda-\lambda')\right)^{3/2}}
   -\frac{9\,a\,a'^3\,e\,e'\,\cos\,(\varpi-\varpi')}{4\left(a^2+a'^2-2\,a\,a'\,\cos\,(\lambda-\lambda')\right)^{5/2}}
   +\frac{a\,a'\,e'\,\cos\,(\lambda-2\lambda'+\varpi')}{2\left(a^2+a'^2-2\,a\,a'\,\cos\,(\lambda-\lambda')\right)^{3/2}}\\
   &+\frac{a\,a'\,e\,e'\,\cos\,(2\lambda-2\lambda'-\varpi+\varpi')}{4\left(a^2+a'^2-2\,a\,a'\,\cos\,(\lambda-\lambda')\right)^{3/2}}
   +\frac{3\,a\,a'^3\,e\,e'\,\cos\,(2\lambda-2\lambda'-\varpi+\varpi')}{4\left(a^2+a'^2-2\,a\,a'\,\cos\,(\lambda-\lambda')\right)^{5/2}}
   +\frac{3\,a\,a'\,e'^2\,\cos\,(\lambda-3\lambda'+2\varpi')}{8\left(a^2+a'^2-2\,a\,a'\,\cos\,(\lambda-\lambda')\right)^{3/2}}\\
   &+\frac{3\,a\,a'^3\,e'^2\,\cos\,(\lambda-3\lambda'+2\varpi')}{4\left(a^2+a'^2-2\,a\,a'\,\cos\,(\lambda-\lambda')\right)^{5/2}}
   -\frac{a\,a'\,e^2\,\cos\,(\lambda-\lambda')}{2\left(a^2+a'^2-2\,a\,a'\,\cos\,(\lambda-\lambda')\right)^{3/2}}
   -\frac{a\,a'\,e'^2\,\cos\,(\lambda-\lambda')}{2\left(a^2+a'^2-2\,a\,a'\,\cos\,(\lambda-\lambda')\right)^{3/2}}\\
   &-\frac{3\,a\,a'^3\,e'^2\,\cos\,(\lambda-\lambda')}{2\left(a^2+a'^2-2\,a\,a'\,\cos\,(\lambda-\lambda')\right)^{5/2}}
   +\frac{a'^2\,e'^2\,\cos\,(2\lambda'-2\varpi')}{4\left(a^2+a'^2-2\,a\,a'\,\cos\,(\lambda-\lambda')\right)^{3/2}}
   +\frac{3\,a'^4\,e'^2\,\cos\,(2\lambda'-2\varpi')}{4\left(a^2+a'^2-2\,a\,a'\,\cos\,(\lambda-\lambda')\right)^{5/2}}\\
   &+\frac{a'^2\,e'\,\cos\,(\lambda'-\varpi')}{\left(a^2+a'^2-2\,a\,a'\,\cos\,(\lambda-\lambda')\right)^{3/2}}
   +\frac{1}{\sqrt{a^2+a'^2-2\,a\,a'\,\cos\,(\lambda-\lambda')}}
   -\frac{3\,a'^2\,e'^2}{4\left(a^2+a'^2-2\,a\,a'\,\cos\,(\lambda-\lambda')\right)^{3/2}}\\
   &+\frac{3\,a'^4\,e'^2}{4\left(a^2+a'^2-2\,a\,a'\,\cos\,(\lambda-\lambda')\right)^{5/2}}
\end{align*}
}

\newpage

\chapter{Form of the function $H_b$}\label{appex:theHb}

Neglecting ${\cal O}(x)$ terms, and setting $b_0=2-2\cos\tau$,
$e_{p,0}=\sqrt{2(Y_f-Y_p)}$, the functions ${\cal F}^{(0)}$ and ${\cal
  F}^{(1)}$ of 
\begin{equation*}
H_b=-{1\over 2(1+v)^2} -v + Y_f -\mu{\cal
  F}^{(0)}(v,Y_p-Y_f,u,\phi_f;e',\varpi') -\mu{\cal F}^{(1)}(v,Y_p-Y_f,u,\phi_f,\phi;e',\varpi')
\end{equation*}
(Eq.~\ref{eq:hamYfYsYp}), up to second order in $e_{p,0}$ and $e'$,
are analyzed in trigonometric terms in the angles $\tau$, $\phi_f$,
and $\phi$, as follows:

\section{- $\langle{\cal F}^{(0)} \rangle$}
\begin{center}
\begin{tabular}{||l c ||}
\hline
Constant & $-1+{1\over b_0^{1/2}}+{1\over b_0^{3/2}}\left(-{3e'^2\over
  8}-{3e_{p,0}^2\over 4}\right)+{1\over
  b_0^{5/2}}\left(3e'^2+{21e_{p,0}^2\over 8}\right)$ \\ 
\hline
$\cos\tau$ & $-1+{e_{p,0}^2\over 2}+e'^2+{1\over b_0^{3/2}}\left(-e'^2-{e_{p,0}^2\over 2}\right)+{1\over
  b_0^{5/2}}\left(-{27e'^2\over 16}-{3e_{p,0}^2\over  2}\right)$ \\ 
\hline 
$\cos2\tau$ & $-{e'^2\over 2}+{1\over b_0^{3/2}}\left({e'^2\over 8}\right)
+{1\over b_0^{5/2}}\left(-{3e'^2\over 2}-{9e_{p,0}^2\over 8}\right)$ \\ 
\hline 
$\cos 3\tau$ & $+{1\over b_0^{5/2}}\left({3e'^2\over 16}\right)$ \\
\hline
$\sin \tau$ & ${1\over
  b_0^{5/2}}\left(-{33\sqrt{3}e'^2\over 16}\right)$\\ 
\hline
$\sin 2\tau$ & $\quad -{\sqrt{3}e'^2\over 2}+{1\over
  b_0^{3/2}}\left({\sqrt{3}e'^2\over 8}\right)+{1\over
  b_0^{5/2}}\left({3\sqrt{3}e'^2\over 4}\right)$\\ 
\hline
$\sin 3\tau$ & $+{1\over b_0^{5/2}}\left({3\sqrt{3}e'^2\over 16}\right)$\\ 
\hline
\end{tabular}\\
\end{center}

\vfill \eject

\section{- $\langle {\cal F}^{(1)} \rangle$}
\begin{center}
\begin{tabular}{||l c||}
\hline
$\cos(\phi)$ & $+{1\over b_0^{3/2}}\left({3e_{p,0} e'\over 2}\right)+{1\over b_0^{5/2}}\left({-15e_{p,0} e'\over 8}\right)$ \\ 
\hline
$\cos(\tau-\phi)$ & ${e_{p,0} e'\over 4}+{1\over b_0^{3/2}}\left({-e_{p,0} e'\over 4}\right)+{1\over b_0^{5/2}}\left({-3e_{p,0} e'\over 2}\right)$ \\
\hline
$\cos(\tau+\phi)$ & ${e_{p,0} e'\over 4}+{1\over b_0^{3/2}}\left({-e_{p,0} e'\over 4}\right)+{1\over b_0^{5/2}}\left({21e_{p,0} e'\over 8}\right)$ \\ 
\hline
$\cos(2\tau-\phi)$ & $-e_{p,0} e'+{1\over b_0^{3/2}}\left({e_{p,0} e'\over 4}\right)+{1\over b_0^{5/2}}\left({15e_{p,0} e'\over 16}\right)$ \\
\hline
$\cos(2\tau+\phi)$ & $+{1\over b_0^{5/2}}\left(-{9e_{p,0} e'\over 16}\right)$ \\ 
\hline
$\cos(3\tau-\phi)$ & $+{1\over b_0^{5/2}}\left({3e_{p,0} e'\over 8}\right)$ \\
\hline
$\sin(\phi)$ & $+{1\over b_0^{3/2}}\left(-{3\sqrt{3}e_{p,0} e'\over 4}\right)+{1\over b_0^{5/2}}\left({21\sqrt{3}e_{p,0} e'\over 8}\right)$ \\ 
\hline
$\sin(\tau-\phi)$ & $-{\sqrt{3} e_{p,0} e'\over 4}+{1\over b_0^{3/2}}\left({\sqrt{3}e_{p,0} e'\over 4}\right)+{1\over b_0^{5/2}}\left({3\sqrt{3}e_{p,0} e'\over 4}\right)$ \\
\hline
$\sin(\tau+\phi)$ & ${\sqrt{3} e_{p,0} e'\over 4}+{1\over b_0^{3/2}}\left(-{\sqrt{3}e_{p,0} e'\over 4}\right)+{1\over b_0^{5/2}}\left(-{3\sqrt{3}e_{p,0} e'\over 4}\right)$ \\ 
\hline
$\sin(2\tau-\phi)$ & $+{1\over b_0^{5/2}}\left({9\sqrt{3}e_{p,0} e'\over 16}\right)$\\
\hline
$\sin(2\tau+\phi)$ & $+{1\over b_0^{5/2}}\left(-{9\sqrt{3}e_{p,0} e'\over 16}\right)$ \\ 
\hline
\end{tabular}\\
\end{center}

\section{- $\tilde{{\cal F}}^{(0)} = {\cal F}^{(0)}-\langle{\cal F}^{(0)}\rangle$}
\begin{center}
\begin{tabular}{||l c ||}
\hline
$\cos(\phi_f)$ & ${3e_{p,0}\over 2}+{1\over b_0^{3/2}}\left(-{3e_{p,0}\over 2}\right)$ \\
\hline
$\cos(\phi_f+\tau)$ & $-e_{p,0}+{1\over b_0^{3/2}}\left(e_{p,0}\right)$ \\
\hline
$\cos(\phi_f+2\tau)$ & $-{e_{p,0}\over 2}+{1\over b_0^{3/2}}\left({e_{p,0}\over 2}\right)$ \\
\hline
$\cos(2\phi_f)$ & ${1\over b_0^{5/2}}\left({27e_{p,0}^2\over 16}\right)$ \\
\hline
$\cos(2\phi_f+\tau)$ & $-{e_{p,0}^2\over 8}+{1\over b_0^{3/2}}\left({e_{p,0}^2\over 8}\right)+{1\over b_0^{5/2}}\left(-{9e_{p,0}^2\over 4}\right)$ \\ 
\hline
$\cos(2\phi_f+2\tau)$ & $-e_{p,0}^2+{1\over b_0^{3/2}}\left({e_{p,0}^2\over 4}\right)+{1\over b_0^{5/2}}\left(-{3e_{p,0}^2\over 8}\right)$ \\
\hline
$\cos(2\phi_f+3\tau)$ & $-{3e_{p,0}^2\over 8}+{1\over b_0^{3/2}}\left({3e_{p,0}^2\over 8}\right)+{1\over b_0^{5/2}}\left({3e_{p,0}^2\over 4}\right)$ \\
\hline
$\cos(2\phi_f+4\tau)$ & ${1\over b_0^{5/2}}\left({3e_{p,0}^2\over 16}\right)$ \\
\hline
\end{tabular}\\
\end{center}

\vfill \eject

\section{- $\tilde{{\cal F}}^{(1)} = {\cal F}^{(1)}-\langle{\cal F}^{(1)}\rangle$}

\subsubsection{Terms with $\cos$}
\begin{center}
\begin{tabular}{||l c ||}
\hline $\cos(\phi_f-\tau+\phi)$ & $-2e'+{1\over
  b_0^{3/2}}\left({e'\over 2}\right)$ \\ \hline $\cos(\phi_f+\phi)$ &
${3e'\over 4}+{1\over b_0^{3/2}}\left({e'\over 4}\right)$ \\ \hline
$\cos(\phi_f+\tau+\phi)$ & $-{e'\over 2}+{1\over
  b_0^{3/2}}\left(-e'\right)$\\ \hline $\cos(\phi_f+2\tau+\phi)$ &
$-{e'\over 4}+{1\over b_0^{3/2}}\left({e'\over 4}\right)$ \\ \hline
$\cos(2\phi_f-2\tau+2\phi)$ & $+{1\over b_0^{5/2}}\left({3e'^2\over
  16}\right)$\\ \hline $\cos(2\phi_f-\tau+\phi)$ & $+{1\over
  b_0^{5/2}}\left(-{9e_{p,0} e'\over 8}\right)$ \\ \hline
$\cos(2\phi_f-\tau+2\phi)$ & $-{27e'^2\over 8}+{1\over
  b_0^{3/2}}\left({3e'^2\over 8}\right)+{1\over
  b_0^{5/2}}\left({3e'^2\over 16}\right)$ \\ \hline
$\cos(2\phi_f+\phi)$ & $3 e_{p,0} e'+{1\over
  b_0^{3/2}}\left(-{3e_{p,0} e'\over 4}\right)+{1\over
  b_0^{5/2}}\left({3e_{p,0} e'\over 16}\right)$ \\ \hline
$\cos(2\phi_f+2\phi)$ & ${3e'^2\over 2}+{1\over
  b_0^{3/2}}\left(-{e'^2\over 8}\right)+{1\over
  b_0^{5/2}}\left(-{63e'^2\over 32}\right)$ \\ \hline
$\cos(2\phi_f+\tau+\phi)$ & $-{e_{p,0} e'\over 8}+{1\over
  b_0^{3/2}}\left({e_{p,0} e'\over 8}\right)+{1\over b_0^{5/2}}\left(3
e_{p,0} e'\right)$ \\ \hline $\cos(2\phi_f+\tau+2\phi)$ & $-{e'^2\over
  16}+{1\over b_0^{3/2}}\left({e'^2\over 16}\right)+{1\over
  b_0^{5/2}}\left({3e'^2\over 2}\right)$ \\ \hline
$\cos(2\phi_f+2\tau+\phi)$ & $-e_{p,0} e'+{1\over
  b_0^{3/2}}\left(-{e_{p,0} e'\over 2}\right)+{1\over
  b_0^{5/2}}\left({-15e_{p,0} e'\over 8}\right)$ \\ \hline
$\cos(2\phi_f+2\tau+2\phi)$ & ${e'^2\over 2}+{1\over
  b_0^{3/2}}\left(-{e'^2\over 2}\right)+{1\over
  b_0^{5/2}}\left({9e'^2\over 8}\right)$ \\ \hline
$\cos(2\phi_f+3\tau+\phi)$ & $-{3e_{p,0} e'\over 8}+{1\over
  b_0^{3/2}}\left({3e_{p,0} e'\over 8}\right)+{1\over
  b_0^{5/2}}\left(-{3 e_{p,0} e'\over 8}\right)$ \\ \hline
$\cos(2\phi_f+3\tau+2\phi)$ & ${3e'^2\over 16}+{1\over
  b_0^{3/2}}\left(-{3e'^2\over 16}\right)+{1\over
  b_0^{5/2}}\left(-{15e'^2\over 16}\right)$ \\ \hline
$\cos(2\phi_f+4\tau+\phi)$ & $+{1\over b_0^{5/2}}\left({3 e_{p,0}
  e'\over 16}\right)$ \\ \hline $\cos(2\phi_f+4\tau+2\phi)$ &
$+{1\over b_0^{5/2}}\left(-{3e'^2\over 32}\right)$\\ \hline
\end{tabular}
\end{center}

\vfill \eject

\subsubsection{Terms with $\sin$}
\begin{center}
\begin{tabular}{||l c ||}
\hline
$\sin(\phi_f+\phi)$ & ${3\sqrt{3}e'\over 4}+{1\over b_0^{3/2}}\left(-{3\sqrt{3}e'\over 4}\right)$ \\
\hline
$\sin(\phi_f+\tau+\phi)$ & $-{\sqrt{3}e'\over 2}+{1\over b_0^{3/2}}\left({\sqrt{3}e'\over 2}\right)$ \\
\hline
 $\sin(\phi_f+2\tau+\phi)$ & $-{\sqrt{3}e'\over 4}+{1\over b_0^{3/2}}\left({\sqrt{3}e'\over 4}\right)$ \\
\hline
$\sin(2\phi_f-\tau+2\phi)$ & $+{1\over b_0^{5/2}}\left(-{9\sqrt{3}e'^2\over 16}\right)$ \\
\hline
 $\sin(2\phi_f+\phi)$ & $+{1\over b_0^{5/2}}\left({27\sqrt{3}e_{p,0} e'\over 16}\right)$ \\
\hline
$\sin(2\phi_f+2\phi)$ & ${3\sqrt{3}e'^2\over 2}+{1\over b_0^{3/2}}\left(-{3\sqrt{3}e'^2\over 8}\right)+{1\over b_0^{5/2}}\left({3\sqrt{3}e'^2\over 32}\right)$\\
\hline
$\sin(2\phi_f+\tau+\phi)$ & $-{\sqrt{3}e_{p,0} e'\over 8}+{1\over b_0^{3/2}}\left({\sqrt{3}e_{p,0} e'\over 8}\right)+{1\over b_0^{5/2}}\left(-{9\sqrt{3}e_{p,0} e'\over 4}\right)$ \\
\hline
$\sin(2\phi_f+\tau+2\phi)$ & $-{\sqrt{3}e'^2\over 16}+{1\over b_0^{3/2}}\left({\sqrt{3}e'^2\over 16}\right)+{1\over b_0^{5/2}}\left({3\sqrt{3}e'^2\over 2}\right)$\\
\hline
$\sin(2\phi_f+2\tau+\phi)$ & $-\sqrt{3}e_{p,0} e'+{1\over b_0^{3/2}}\left({\sqrt{3}e_{p,0} e'\over 4}\right)+{1\over b_0^{5/2}}\left(-{3\sqrt{3}e_{p,0} e'\over 8}\right)$\\
\hline
$\sin(2\phi_f+2\tau+2\phi)$ & $-{\sqrt{3}e'^2\over 2}+{1\over b_0^{3/2}}\left(-{\sqrt{3}e'^2\over 4}\right)+{1\over b_0^{5/2}}\left(-{15\sqrt{3}e'^2\over 16}\right)$\\
\hline
 $\sin(2\phi_f+3\tau+\phi)$ & $-{3\sqrt{3}e_{p,0} e'\over 8}+{1\over b_0^{3/2}}\left({3\sqrt{3}e_{p,0} e'\over 8}\right)+{1\over b_0^{5/2}}\left({3\sqrt{3}e_{p,0} e'\over 4}\right)$ \\
\hline
$\sin(2\phi_f+3\tau+2\phi)$ & $-{3\sqrt{3}e'^2\over 16}+{1\over b_0^{3/2}}\left({3\sqrt{3}e'^2\over 16}\right)+{1\over b_0^{5/2}}\left(-{3\sqrt{3}e'^2\over 16}\right)$ \\
\hline
 $\sin(2\phi_f+4\tau+\phi)$ & ${1\over b_0^{5/2}}\left({3\sqrt{3}e_{p,0} e'\over 16}\right)$ \\
\hline
$\sin(2\phi_f+4\tau+2\phi)$ & $ {1\over b_0^{5/2}}\left({3\sqrt{3}e'^2\over 32}\right)$ \\
\hline
\end{tabular}
\end{center}

\chapter{Analytical formul\ae~for the asymmetric expansions}\label{app:asymm_exp}

\section{Asymmetric expansions for $\frac{\cos\tau}{(2-2 \cos \tau)^{N/2}}$
and $\frac{\sin\tau}{(2-2 \cos \tau)^{N/2}}$}

The asymmetric expansion in terms of $u=\tau-\pi/3$, up to a generic order $K$
for the functions $\frac{\cos\tau}{(2-2 \cos \tau)^{N/2}}$,
$\frac{\sin\tau}{(2-2 \cos \tau)^{N/2}}$, $\cos^M \tau $ and 
$\sin^M \tau $, with $N \in \mathbb{N}$ and $M \in \cal{N}$ fixed
is given by 
\begin{displaymath}
\frac{\cos\tau}{(2-2 \cos \tau)^{N/2}} = \frac{1}{2^{N/2}} \, \sum_{k=0}^{K} {\cal M}_{1}(k) \, u^k + {\cal O}(u^K)~,\quad \mathrm{where} \quad
{\cal M}_{1}(k) =  \sum_{i=k}^{K} \frac{1}{i!} \,F^{(i)}(\pi/2)\,
{i \choose k}
 \left(-\frac{\pi}{6} \right)^{i-k}~~,
\end{displaymath}
\begin{displaymath}
\frac{\sin\tau}{(2-2 \cos \tau)^{N/2}} = \frac{1}{2^{N/2}} \, \sum_{k=0}^{K} {\cal M}_{2}(k) \, u^k + {\cal O}(u^K)~,\quad \mathrm{where} \quad 
{\cal M}_{2}(k) =  \sum_{i=k}^{K} \frac{1}{i!} \,G^{(i)}(\pi/2)\,
{i \choose k}
 \left( 
-\frac{\pi}{6} \right)^{i-k}~~,
\end{displaymath}
\begin{displaymath}
\cos^M\tau =  \sum_{k=0}^{K} {\cal M}_{3}(k) \, u^k~ + {\cal O}(u^K), \quad \mathrm{where} \quad
{\cal M}_{3}(k) = \sum_{i=k}^{K} \frac{1}{i!} \,B_{M,M}^{(i)}\,
{i \choose k}
 \left( -\frac{\pi}{6} \right)^{i-k}~~,
\end{displaymath}
\begin{displaymath}
\sin^M\tau =  \sum_{k=0}^{K} {\cal M}_{4}(k) \, u^k~ + {\cal O}(u^K), \quad \mathrm{where} \quad
{\cal M}_{4}(k) = \sum_{i=k}^{K} \frac{1}{i!} \,C_{M,M}^{(i)}\,
{i \choose k} 
 \left( -\frac{\pi}{6} \right)^{i-k}~~,
\end{displaymath}
and
\begin{align*}
F^{(n)}(\pi/2) &= \sum_{i=1}^{[\frac{n-1}{2}]} (n,2i-1)\, (-1)^{i} \,f^{(n-(2i-1))}(\pi/2)~~, \\
G^{(n)}(\pi/2) &= \sum_{i=0}^{[\frac{n}{2}]} (n,2i)\, (-1)^{i} \,f^{(n-2i)}(\pi/2)~~,
\end{align*}
with $[\frac{n-1}{2}]$ the integer part of $\frac{n-1}{2}$, and
$[\frac{n}{2}]$ the integer part of $\frac{n}{2}$; the derivatives
$f^{(n)}$ are given by
\begin{displaymath}
f^{(n)}\left( \pi/2 \right) = \sum_{m=1}^{n} A_{m,m}^{(n)} ~~;
\end{displaymath}
the coefficients $A_{m,m}^{(n)}$, $B_{M,M}^{(n)}$ and $C_{M,M}^{(n)}$ are given by
\begin{align*}
A_{m,m}^{(n)} &= -A_{m,m-1}^{(n-1)} - \left( \frac{2(m-1)+N}{2} \right) A_{m-1,m-1}^{(n-1)}~, 
 &A_{1,1}^{(1)} &= - \frac{N}{2}~~, \\
B_{M,M}^{(n)} &= - B_{M,M-1}^{(n-1)} + \left(M+1\right) B_{M,M+1}^{(n-1)},
 &B_{1,1}^{(1)} &= -M~~, \phantom{\Big( A \Big)} \\
C_{M,M}^{(n)} &= C_{M,M-1}^{(n-1)} - \left(M+1\right) C_{M,M+1}^{(n-1)},
 &C_{1,1}^{(1)} &= M~~. \phantom{\Big( A \Big)}
\end{align*}

\section{Proofs}
\subsection{Derivatives of $f$: general formula}\label{deriv_f_gen}

\begin{lemma}\label{lemma1}
Let the function $f(\tau)$ be 
\begin{equation}\label{function}
f(\tau) = \frac{1}{\left(1 -\cos \tau \right)^{N/M}} ~~,
\end{equation}
and $f^{(n)}$ its derivative of $n$ order. Then, a general formula for $f^{(n)}$ is given by
\begin{equation}\label{formula_f}
f^{(n)}\left( \tau \right) = \sum_{m=1}^{n} \sum_{l=0}^{m} A_{m,l}^{(n)} \, T_{m,l}^{(n)} = \sum_{m=1}^{n} \sum_{l=0}^{m} A_{m,l}^{(n)} \frac{\cos^{m-l} (\tau) \, \sin^{l} (\tau)}{\left( 1 - \cos \tau \right)^{\frac{M.m+N}{M}}} ~~,
\end{equation}
where the coefficients $A_{m,l}^{(n)}$ are given by
\begin{equation}\label{coefficient}
A_{m,l}^{(n)} = -\left( m - l +1 \right) A_{m,l-1}^{(n-1)} + \left(l+1\right) A_{m,l+1}^{(n-1)} - \left( \frac{M(m-1)+N}{M} \right) A_{m-1,l-1}^{(n-1)}~~,
\end{equation}
and
\begin{equation}\label{f1}
f^{(1)}(\tau) = A_{1,1}^{(1)} \,\frac{\sin \tau}{\left( 1 - \cos \tau \right)^{(M+N)/M}} \qquad \mathrm{with} \quad A_{1,1}^{(1)} = - \frac{N}{M}~~.
\end{equation}
\end{lemma}

\subsubsection{Proof}

The proof is based on an inductive argument. Consider first the 2nd
order derivative.  Considering the definition in (\ref{f1}) for
$f^{(1)}$, we derive with respect to $\tau$
\begin{equation}\label{f1deriv}
f^{(2)} = \frac{\mathrm{d} f^{(1)}}{\mathrm{d} \, \tau} = -\frac{N}{M} 
\frac{\cos(\tau)}{(1-\cos(\tau))^{(M+N)/M}} + \frac{N(M+N)}{M^2} \frac{\sin^2(\tau)}{(1-\cos(\tau))^{(2M+N)/M}} ~~.
\end{equation}
On the other hand, from (\ref{f1}), we obtain
\begin{equation}\label{coeffn1}
A^{(1)}_{1,0} = 0~~,\qquad A^{(1)}_{1,1} = -\frac{N}{M}~~
\end{equation}
Then, following formula (\ref{formula_f}), we can express $f^{(2)}$ as
\begin{equation}
\begin{aligned}\label{f2form1}
f^{(2)} (\tau) &= A^{(2)}_{1,0} \, \frac{\cos(\tau)}{(1-\cos(\tau))^{(M+N)/M}} +
A^{(2)}_{1,1} \, \frac{\sin(\tau)}{(1-\cos(\tau))^{(M+N)/M}}   \\
&+ A^{(2)}_{2,0} \, \frac{\cos^2(\tau)}{(1-\cos(\tau))^{(2M+N)/M}}
+ A^{(2)}_{2,1} \, \frac{\cos(\tau) \sin(\tau)}{(1-\cos(\tau))^{(2M+N)/M}} \\
&+ A^{(2)}_{2,2} \, \frac{\sin^2(\tau)}{(1-\cos(\tau))^{(2M+N)/N}} ~~, 
\end{aligned}
\end{equation}
where, according to (\ref{coefficient})\footnote{In agreement with
  Eq.~\eqref{formula_f}, if $l \notin [0,m]$, the corresponding
  coefficient $A_{m,l}^{(n)} = 0$.}
\begin{displaymath}
A^{(2)}_{1,0} = - (1-0+1) A^{(1)}_{1,-1} + (0+1) A^{(1)}_{1,1} - \frac{(M.0+N)}{M}A^{(1)}_{0,-1}~~,
\end{displaymath}

\begin{displaymath}
A^{(2)}_{1,1} = - (1-1+1) A^{(1)}_{1,0} + (1+1) A^{(1)}_{1,2} - \frac{(M.0+N)}{M}A^{(1)}_{0,0}~~,
\end{displaymath}

\begin{displaymath}
A^{(2)}_{2,0} = - (2-0+1) A^{(1)}_{2,-1} + (0+1) A^{(1)}_{2,1} - \frac{(M.0+N)}{M}A^{(1)}_{1,-1}~~,
\end{displaymath}

\begin{displaymath}
A^{(2)}_{2,1} = - (2-1+1) A^{(1)}_{2,0} + (1+1) A^{(1)}_{2,2} - \frac{(M.1+N)}{M}A^{(1)}_{1,0}~~,
\end{displaymath}

\begin{displaymath}
A^{(2)}_{2,2} = - (2-2+1) A^{(1)}_{2,1} + (2+1) A^{(1)}_{2,3} - \frac{(M.1+N)}{M}A^{(1)}_{1,1}~~.
\end{displaymath}
Considering (\ref{coeffn1}), we obtain
{\small
\begin{displaymath}
A^{(2)}_{1,0} = A^{(1)}_{1,1} = -\frac{N}{M},\quad A^{(2)}_{1,1} = 0~~,\quad
A^{(2)}_{2,0} = 0, \quad A^{(2)}_{2,1} = 0, \quad A^{(2)}_{2,2} = -\frac{M+N}{M}\,
 A^{(1)}_{1,1} = \frac{N(M+N)}{M^2}~~.
\end{displaymath}}
Then, 
\begin{equation}\label{f2form1}
f^{(2)} = -\frac{N}{M} \frac{\cos(\tau)}{(1-\cos(\tau))^{(M+N)/M}} + \frac{N(M+N)}{M^2} \frac{\sin^2(\tau)}{(1-\cos(\tau))^{(2M+N)/M}}~~,
\end{equation}
which coincides with (\ref{f1deriv}) and proves the proposition for $n=2$.
We now assume the proposition to be true for an arbitrary integer $n$, i.e.,
\begin{equation}\label{formulafn}
f^{(n)}\left( \tau \right) = \sum_{m=1}^{n} \sum_{l=1}^{m} A_{m,l}^{(n)} \, T_{m,l}^{(n)} = \sum_{m=1}^{n} \sum_{l=1}^{m} A_{m,l}^{(n)} \frac{\cos^{m-l} (\tau) \, \sin^{l} (\tau)}{\left( 1 - \cos \tau \right)^{\frac{(mM+N)}{n}}} ~~,
\end{equation}
with
\begin{equation}\label{coefficientfn}
A_{m,l}^{(n)} = -\left( m - l +1 \right) A_{m,l-1}^{(n-1)} + \left(l+1\right) A_{m,l+1}^{(n-1)} - \left( \frac{M(m-1)+N}{M} \right) A_{m-1,l-1}^{(n-1)}~~.
\end{equation} 

Let $A_{m,l}^{(n)}\, T^{(n)}_{m,l}$ be a generic term of $f^{(n)}$, where
\begin{equation}\label{tndefn}
T^{(n)}_{m,l} =  \, \frac{\cos^{m-l}(\tau) \sin^l(\tau)}{(1-\cos(\tau))^{(mM+N)/N}}~~.
\end{equation}
Computing the derivative of $A_{m,l}^{(n)}\, T^{(n)}_{m,l}$ with
respect to $\tau$, we obtain
{\small
\begin{equation}
\begin{aligned}\label{eq:altramente}
\frac{\mathrm{d} \, \left(A_{m,l}^{(n)} \,T^{(n)}_{m,l} \right)
}{\mathrm{d} \tau} = & \, \frac{A^{(n)}_{m,l}}{(1-\cos(\tau)^{2(mM+N)/M}}
\, \left[ -(m-l) \cos^{m-l-1}(\tau) \sin^{l+1}(\tau) \left(1-
\cos(\tau)\right)^{(mM+N)/M} \phantom{\frac{A}{B}} \right. \\ 
+ & \, l \cos^{m-l+1}(\tau) \sin^{l-1}(\tau) \left( 1 -
\cos(\tau)\right)^{(mM+N)/M} \phantom{\Big(A \Big)} \\ 
- & \, \left. \frac{mM+N}{M}
\cos^{m-l}(\tau) \sin^{l+1}(\tau) \left(1-\cos(\tau) \right)^{(mM+N)/M
  -1} \right]~~,\\
 = & \, A^{(n)}_{m,l} \, \left[ -(m-l) \frac{\cos^{m-l-1}(\tau)
   \sin^{l+1}(\tau)}{ (1- \cos(\tau))^{(mM+N)/M}} + l \,
 \frac{\cos^{m-l+1}(\tau) \sin^{l-1}(\tau)}{(1 -\cos(\tau))^{(mM+N)/M}} \right.\\ 
 - & \, \left. \frac{mM+N}{M} \frac{\cos^{m-l}(\tau)
   \sin^{l+1}(\tau)}{(1-\cos(\tau))^{(mM+M+N)/2}} \right]~\nonumber.
\end{aligned}
\end{equation}}
Rearranging the indices, we get
{\small
\begin{equation}
\begin{aligned}\label{tns}
\frac{\mathrm{d} \, \left( A_{m,l}^{(n)} \,T^{(n)}_{m,l} \right) }{\mathrm{d} \tau} = & A^{(n)}_{m,l}\left[ -(m-l)  \frac{\cos^{m-(l+1)}(\tau) \sin^{(l+1)}(\tau)}{ (1- \cos(\tau))^{(mM+N)/M}} + l \, \frac{\cos^{m-(l-1)}(\tau) \sin^{(l-1)}(\tau)}{( 1 - \cos(\tau))^{(mM+N)/M}} \right. \\
\quad & {} - \left. \frac{(mM+N)}{M} \frac{\cos^{(m+1)-(l+1)}(\tau) \sin^{(l+1)}(\tau)}{(1-\cos(\tau))^{(M(m+1)+N)/N}} \right]~,\\
 = & -(m-l) \,A_{m,l}^{(n)} \, T_{m,l+1}^{(n+1)} + l \, A_{m,l}^{(n)} \, T_{m,l-1}^{(n+1)} - \frac{(mM+N)}{M} \,A_{m,l}^{(n)} \, T_{m+1,l+1}^{(n+1)} ~~.
\end{aligned}
\end{equation}}

Then, we can see that terms of the form $T_{m,l}^{(n)}$ generate
derivatives of the same form. We remark here that it is not possible
to generate terms of the form $\frac{1}{\cos(\tau)}$. The latter
would correspond to the cases were $m=l$ or $l=0$. However, 
for $m=l$ the first coefficient vanishes, while for $l=0$ the second
coefficient vanishes. In conclusion
\begin{displaymath}
\frac{\mathrm{d} \left(A_{m,l}^{(n)} \, T_{m,l}^{(n)} \right)}{\mathrm{d} \tau } = \sum_{m',l'} A_{m',l'}^{(n+1)} \, T_{m',l'}^{(n+1)} ~~,
\end{displaymath}
and therefore, covering all possible indices
{\small
\begin{equation}
f^{(n+1)}  =\, \frac{\mathrm{d} \, f^{(n)}}{\mathrm{d} \tau} =
\sum_{m'=1}^{n+1} \sum_{l'=0}^{m'} A_{m',l'}^{(n+1)}
T_{m',l'}^{(n+1)}= \, \sum_{m'=1}^{n+1} \sum_{l'=0}^{m'}
A_{m',l'}^{(n+1)} \frac{\cos^{m'-l'}(\tau)
  \sin^{l'}(\tau)}{(1-\cos(\tau))^{(m'M+N)/M}} ~. \label{fnplus1}
\end{equation}}
For the expression of the coefficients $A_{m',l'}^{(n+1)}$, let us
consider a generic term of (\ref{fnplus1}),
\begin{displaymath}
A_{m',l'}^{(n+1)} T_{m',l'}^{(n+1)} = A_{m',l'}^{(n+1)} \frac{\cos^{m'-l'}(\tau) \sin^{l'}(\tau)}{(1-\cos(\tau))^{(m'M+N)/M}} ~~.
\end{displaymath}
According to (\ref{tns}), $\mathrm{d} \left( A_{m,l}^{(n)} \, T_{m,l}^{(n)} \right)/\mathrm{d} \tau$ contributes to (and only to) three terms, that are of the form $T_{m,l+1}^{(n+1)}$, $T_{m,l-1}^{(n+1)}$ and $T_{m+1,l+1}^{(n+1)}$. Hence, the
term $m',l'$ can only get contributions from three different terms
of $f^{(n)}$:
\begin{itemize}
\item[] $T^{(n)}_{m_1,l_1}$ where $m_1 = m'$ and $l_1 +1 =l'$
\item[] $T^{(n)}_{m_2,l_2}$ where $m_2 = m'$ and $l_2 -1 =l'$
\item[] $T^{(n)}_{m_3,l_3}$ where $m_3 +1 = m'$ and $l_3 +1 =l'$
\end{itemize}
Furthermore,
\begin{align}
A_{m',l'}^{(n+1)} \,T^{(n+1)}_{m',l'}  = &\, -(m_1-l_1)\, A^{(n)}_{m_1,l_1} \, \frac{\cos^{m_1-(l_1+1)}(\tau) \sin^{(l_1+1)}(\tau)}{ (1- \cos(\tau))^{(Mm_1+N)/M}} \nonumber \\
 + &\, \, l_2 \, A^{(n)}_{m_2,l_2} \, \frac{\cos^{m_2-(l_2-1)}(\tau) \sin^{(l_2-1)}(\tau)}{( 1 - \cos(\tau))^{(Mm_2+N)/M}} \label{eq:previousterm} \\
- & \frac{(Mm_3+N)}{N} \, A^{(n)}_{m_3,l_3} \frac{\cos^{(m_3+1)-(l_3+1)}(\tau) \sin^{(l_3+1)}(\tau)}{(1-\cos(\tau))^{(M(m_3+1)+N)/M}} ~~. \nonumber
\end{align}
Replacing with the rules above in Eq.~\ref{eq:previousterm}, we find
\begin{align}
A_{m',l'}^{(n+1)} \,T^{(n+1)}_{m',l'}  = &\, -(m'-(l'-1))\, A^{(n)}_{m',l'-1} \, \frac{\cos^{m'-(l'-1+1)}(\tau) \sin^{(l'-1+1)}(\tau)}{ (1- \cos(\tau))^{(Mm'+N)/M}} \nonumber \\
 + & \, (l'+1) \, A^{(n)}_{m',l'+1} \, \frac{\cos^{m'-(l'+1-1)}(\tau) \sin^{(l'+1-1)}(\tau)}{( 1 - \cos(\tau))^{(Mm'+N)/M}} \\
- & \, \frac{(M(m'-1)+N)}{M} \, A^{(n)}_{m'-1,l'-1} \frac{\cos^{(m'-1+1)-(l'-1+1)}(\tau) \sin^{(l'-1+1)}(\tau)}{(1-\cos(\tau))^{(M(m'-1)+N)/M}} ~~, \nonumber
\end{align}
or, rearranging the indices, 
\begin{equation}
\begin{aligned}\label{altra2}
A_{m',l'}^{(n+1)} \,T^{(n+1)}_{m',l'} & = \left[ -(m'-l'+1) \,
A^{(n)}_{m',l'-1} + \, (l'+1) \, A^{(n)}_{m',l'+1} \phantom{\Big(A \Big)}\right. \\
- & \, \left. \frac{M(m'-1)+N}{M}
\, A^{(n)}_{m'-1,l'-1} \right] 
\frac{\cos^{m'-l'}(\tau)
  \sin^{l'}(\tau)}{ (1- \cos(\tau))^{(2m'+1)/2}} ~.
\end{aligned}
\end{equation}
Then, the coefficients $A_{m',l'}^{(n+1)}$ are given by
\begin{equation}
A_{m',l'}^{(n+1)}  =  -(m'-l'+1) \,
A^{(n)}_{m',l'-1} + \, (l'+1) \, A^{(n)}_{m',l'+1} - \frac{M(m'-1)+N}{M}
\, A^{(n)}_{m'-1,l'-1}~~,
\end{equation}
which proves the preposition. \hfill $\blacksquare$

\subsection{Derivatives of $f$: evaluation}\label{deriv_f_eval}

The series of $f$ appearing in the expansion of the disturbing
function require derivatives for arbitrary $N$ but fixed $M =2$, and
they are evaluated at $\tau = \frac{\pi}{2}$.  Thus, the general
formula for $f^{(n)}$ is reduced to
\begin{equation}\label{formula_eval}
f^{(n)}\left( \pi/2 \right) = \sum_{m=1}^{n} A_{m,m}^{(n)} ~~,
\end{equation}
where the coefficients $A_{m,m}^{(n)}$ are given by
\begin{equation}\label{coefficient_eval}
A_{m,m}^{(n)} = -A_{m,m-1}^{(n-1)} - \left( \frac{2(m-1)+N}{2} \right) A_{m-1,m-1}^{(n-1)}~~,
\end{equation}
and
\begin{equation}\label{f1_eval}
f^{(1)}(\tau) = A_{1,1}^{(1)} \qquad \mathrm{with} \quad A_{1,1}^{(1)} = - \frac{N}{2}~~.
\end{equation}

\subsection{Derivatives of $F(\tau)$ and $G(\tau)$ at $\tau=\pi/2$}\label{FyGinpi}

Let $f(\tau)=\frac{1}{(1-\cos\tau)^{N/2}}$, and
\begin{equation}\label{laFgrande}
F(\tau)=\frac{\cos\tau}{(1-\cos\tau)^{N/2}} = \cos\tau \, f(\tau)~~,
\end{equation}
\begin{equation}\label{laGgrande}
G(\tau)=\frac{\sin\tau}{(1-\cos\tau)^{N/2}}= \sin\tau \, f(\tau)~~.
\end{equation}
Let $H(\tau)$ be a function whose expression is given by
$h_{1}(\tau)\, h_{2}(\tau)$, for two arbitrary functions $h_{1}$ and
$h_2$ $\in \mathbb{C}^{\infty}$.
Thus, according to Leibnitz's formula for derivatives, 
the $n$-th derivative of $H(\tau)$ with respect to $\tau$ is given
by
\begin{equation}\label{leibnitz} 
H^{(n)}(\tau) = \sum_{i=0}^{n} 
{i \choose k}
\, h_1^{(n-i)} \,h_2^{(i)}~~,
\end{equation}
where ${i \choose k}$ 
is the binomial coefficient. Replacing
$f(\tau)$ and $\cos \tau$ in ~\eqref{leibnitz},
we obtain the $n$-th derivative of $F$ in Eq.~\eqref{laFgrande}
\begin{displaymath}
F^{(n)}(\tau) = \sum_{i=0}^{n} 
{i \choose k}
\, \cos^{(i)}\tau \,f^{(n-i)}~~.
\end{displaymath}
Evaluating $\cos \tau$ at $\pi/2$, we find
\begin{equation}\label{laFgrande-deriv}
F^{(n)}(\pi/2) = \sum_{i=1}^{[\frac{n-1}{2}]} 
{i \choose k}
\, (-1)^{i} \,f^{(n-(2i-1))}(\pi/2)~~.
\end{equation}
Similarly,
\begin{equation}\label{laGgrande-deriv}
G^{(n)}(\pi/2) = \sum_{i=0}^{[\frac{n}{2}]}
{i \choose k}
\, (-1)^{i} \,f^{(n-2i)}(\pi/2)~~.
\end{equation}

\subsection{Newton's binomial formul\ae~for $(u -\pi/6)^i$}\label{newtform}

We decompose the term $(u -\pi/6)^i$ to construct a power series of $u$,
as follows:
\begin{displaymath}
\left( u - \frac{\pi}{6} \right)^i = 
\left( -\frac{\pi}{6} \right)^i \left( -\frac{ u}{\pi/6}  +1 \right)^i ~~.
\end{displaymath}
Let $X= \frac{-u}{\pi/6}$. Replacing in the expression
above, we obtain
\begin{displaymath}
\left( u - \frac{\pi}{6} \right)^i = \left(-\frac{\pi}{6}\right)^i (X+1)^i ~~.
\end{displaymath}
According to Newton's binomial formula, the previous expression 
reads
\begin{align}
\left(-\frac{\pi}{6}\right)^i (X+1)^i & = \left(-\frac{\pi}{6}\right)^i 
\sum_{k=0}^{i} 
{i \choose k}
\, X^k \nonumber \\
& = \left(-\frac{\pi}{6}\right)^i \sum_{k=0}^{i}
{i \choose k}
\, \left(\frac{-u}{\pi/6} \right)^k \nonumber \\
 & = \sum_{k=0}^{i}
{i \choose k}
\, \left(-\frac{\pi}{6}\right)^{i-k} 
\, u^k ~~, \label{eq:new_thing}
\end{align}
where ${i \choose k}$ 
is the binomial coefficient.
We include Eq.~\eqref{eq:new_thing} in the derivatives of
Eq.~\eqref{laFgrande-deriv} and Eq.~\eqref{laGgrande-deriv}

\subsection{Computation of $\frac{\cos \tau}{(2-2 \cos\tau)^{N/2}}$ and $\frac{\sin \tau}{(2-2 \cos\tau)^{N/2}}$}\label{complete}

The Taylor expansion of $\frac{\cos\tau}{(2-2 \cos\tau)^{N/2}}$ around $\pi/2$, is given by
\begin{align}
\frac{\cos\tau}{(2-2 \cos \tau)^{N/2}} & = \frac{\cos\tau}{\left[ 2 (1-\cos \tau)
 \right]^{N/2}} = \frac{1}{2^{N/2}} \, \frac{\cos\tau}{(1-\cos \tau)^{N/2}} \nonumber \\
& = \frac{1}{2^{N/2}} \, F(\tau ) = \frac{1}{2^{N/2}} \, \sum_{i=0}^{K}
\frac{F^{(i)}(\pi/2)}{i!}  \,\left(\tau-\frac{\pi}{2}\right)^i ~~. \nonumber
\end{align}
Considering $\tau = u + \pi/3$, we have
\begin{align*}
\frac{\cos\tau}{(2-2 \cos \tau)^{N/2}} & = \frac{1}{2^{N/2}} \,
\sum_{i=0}^{K} \frac{F^{(i)}(\pi/2)}{i!}
\,\left(u+\frac{\pi}{3}-\frac{\pi}{2}\right)^i = \frac{1}{2^{N/2}} \,
\sum_{i=0}^{K} \frac{F^{(i)}(\pi/2)}{i!}  \,\left(u-\frac{\pi}{6}\right)^i.
\end{align*}
Replacing Eq.~\eqref{eq:new_thing}
in the expression above, we get
\begin{displaymath}
\frac{\cos\tau}{(2-2 \cos \tau)^{N/2}} = \frac{1}{2^{N/2}} \,
\sum_{i=0}^{K} \frac{F^{(i)}(\pi/2)}{i!}  \, \sum_{k=0}^{i}
{i \choose k}
\, \left(- \frac{\pi}{6}\right)^{i-k} \, u^k ~~.
\end{displaymath}
Re-organizing the coefficients, we have
\begin{equation}\label{eq:cos_tau_betaN}
\frac{\cos\tau}{(2-2 \cos \tau)^{N/2}} = \frac{1}{2^{N/2}} \, \sum_{k=0}^{K} {\cal M}_{1}(k) \, u^k ~~,
\end{equation}
where
\begin{equation}\label{eq:M_cos_tau_betaN}
{\cal M}_{1}(k) =  \sum_{i=k}^{K} \frac{1}{i!} \,F^{(i)}(\pi/2)\,
{i \choose k}
\, \left( 
-\frac{\pi}{6} \right)^{i-k}~~,
\end{equation}
\vspace{0.5cm} which completes the generic expression of the expansion
for $\frac{\cos\tau}{(2-2 \cos \tau)^{N/2}}$. 

\noindent
The Taylor expansion of
$\frac{\sin\tau}{(2-2 \cos\tau)^{N/2}}$ around $\pi/2$, is given by
\begin{align*}
\frac{\sin\tau}{(2-2 \cos \tau)^{N/2}} & = \, \frac{\sin\tau}{\left[ 2 (1-\cos \tau) \right]^{N/2}}
= \frac{1}{2^{N/2}} \, \frac{\sin\tau}{(1-\cos \tau)^{N/2}} \\
& = \, \frac{1}{2^{N/2}} \, G(\tau ) = \frac{1}{2^{N/2}} \, \sum_{i=0}^{K}
\frac{G^{(i)}(\pi/2)}{i!}  \,(\tau-\frac{\pi}{2})^i ~~.
\end{align*}
Setting $\tau = u + \pi/3$, we have
\begin{displaymath}
\frac{\sin\tau}{(2-2 \cos \tau)^{N/2}} = \frac{1}{2^{N/2}} \,
\sum_{i=0}^{K} \frac{G^{(i)}(\pi/2)}{i!}
\,(u+\frac{\pi}{3}-\frac{\pi}{2})^i = \frac{1}{2^{N/2}} \,
\sum_{i=0}^{K} \frac{G^{(i)}(\pi/2)}{i!}  \,(u-\frac{\pi}{6})^i.
\end{displaymath}
Replacing~\eqref{eq:new_thing}
in the expression above, we get
\begin{displaymath}
\frac{\sin\tau}{(2-2 \cos \tau)^{N/2}} = \frac{1}{2^{N/2}} \,
\sum_{i=0}^{K} \frac{G^{(i)}(\pi/2)}{i!}  \, \sum_{k=0}^{i}
{i \choose k}
\left(- \frac{\pi}{6}\right)^{i-k} \, u^k ~~.
\end{displaymath}
Re-organizing the coefficients, we have
\begin{equation}\label{eq:sin_tau_betaN}
\frac{\sin\tau}{(2-2 \cos \tau)^{N/2}} = \frac{1}{2^{N/2}} \,
\sum_{k=0}^{K} {\cal M}_{2}(k) \, u^k ~~,
\end{equation}
where
\begin{displaymath}\label{eq:M_sin_tau_betaN}
{\cal M}_{2}(k) =  \sum_{i=k}^{K} \frac{1}{i!} \,G^{(i)}(\pi/2)\,
{i \choose k}
\, \left( 
-\frac{\pi}{6} \right)^{i-k}~~,
\end{displaymath}
which completes the generic expression of the expansion for
$\frac{\sin\tau}{(2-2 \cos \tau)^{N/2}}$.  Explicit formul\ae~for
$F^{(i)}(\pi/2)$ and $G^{(i)}(\pi/2)$ are given in
Sect.~\ref{FyGinpi}.

\section{Asymmetric expansion for $y(\tau)=\cos^M \tau$}\label{deriv_gen_cos}

\begin{lemma}\label{lemma2}
Let 
\begin{equation}\label{function_y}
y(\tau) = \cos^M \tau
\end{equation}
and $y^{(n)}$ be the $n$-th order derivative of $y$. Then, a general
formula for $y^{(n)}$ is given by
\begin{equation}\label{formula_yn}
y^{(n)}\left( \tau \right) = \sum_{i=0}^{\min(M,n)} B_{M,i}^{(n)} \, \cos^{M-i}\tau \, 
\sin^{i}\tau
\end{equation}
and the coefficients
$B_{M,i}^{(n)}$ are given by
\begin{equation}\label{coefficient_yn}
B_{M,i}^{(n)} = -\left( M - (i-1) \right) B_{M,i-1}^{(n-1)} + \left(i+1\right) B_{M,i+1}^{(n-1)}.
\end{equation}
The first derivative $y^{(1)}$ is given by
\begin{equation}\label{y1_fn}
y^{(1)}(\tau) =  B_{1,1}^{(1)}\, \cos^{m-1}\tau \,\sin \tau~~,  \: \mathrm{with} \: B_{1,1}^{(1)} = -M \quad \mathrm{and,\,by\,definition,\,} \, B_{1,0}^{(1)} = 0 ~~.
\end{equation}
\end{lemma}

Eqs.~\eqref{formula_yn}--\eqref{y1_fn} can be proven by induction as in 
Sect.~\ref{deriv_f_gen}. The series of $y$
appearing in the expansion of the disturbing function require
derivatives of arbitrary order $M$, but evaluated at $\tau = \frac{\pi}{2}$.
Thus, the formula for $y^{(n)}$ is reduced to
\begin{equation}\label{formula_eval_cos}
y^{(n)}\left( \pi/2 \right) = B_{M,M}^{(n)} ~~.
\end{equation}
Using the Newton binomial expansions of Sect.~\ref{newtform}, the
asymmetric expansion up to order $K$ for the function $\cos^M \tau$,
with $M\in\cal{N}$, is given by
\begin{equation}\label{formula_cosM}
\cos^M\tau= \sum_{k=0}^K {\cal M}_{3}(k)\, u^k~~,
\end{equation}
with
\begin{equation}\label{coeff_formula_cosM}
{\cal M}_{3}(k)= \sum_{i=k}^{K}\frac{1}{i!} B^{(i)}_{M,M}
{i \choose k}
\left(-\frac{\pi}{6} \right)^{i-k}~~.
\end{equation}

\section{Asymmetric expansion for $x(\tau)=\sin^M \tau$}\label{deriv_gen_sin}

\begin{lemma}\label{lemma3}
Let
\begin{equation}\label{function_x}
x(\tau) = \sin^M \tau
\end{equation}
and $x^{(n)}$ the $n$-th order derivative of $x$.  Then, a general formula
for $x^{(n)}$ is given by
\begin{equation}\label{formula_xn}
x^{(n)}\left( \tau \right) = \sum_{i=0}^{\min(M,n)} C_{M,i}^{(n)} \, \sin^{M-i}\tau \, 
\cos^{i}\tau
\end{equation}
and the coefficients
$C_{M,i}^{(n)}$ are given by
\begin{equation}\label{coefficient_xn}
C_{M,i}^{(n)} = \left( M - (i-1) \right) C_{M,i-1}^{(n-1)} - \left(i+1\right) C_{M,i+1}^{(n-1)}.
\end{equation}
The first derivative $x^{(1)}$ is given by
\begin{equation}\label{f1_xn}
x^{(1)}(\tau) =  C_{1,1}^{(1)}\, \sin^{m-1}\tau \,\cos \tau~~,  \qquad \mathrm{with} \quad C_{1,1}^{(1)} = M \quad \mathrm{and,\,by\,definition,\,} \, C_{1,0}^{(1)} = 0 ~~.
\end{equation}
\end{lemma}

Eqs.~\ref{formula_xn}--\ref{f1_xn} can be proven by induction, as 
in Sect.~\ref{deriv_f_gen}. The series of $x$
appearing in the expansion of the disturbing function require
derivatives of arbitrary order $M$ but evaluated at $\tau = \frac{\pi}{2}$.
Thus, the formula for $x^{(n)}$ is reduced to
\begin{equation}\label{formula_eval_sin}
x^{(n)}\left( \pi/2 \right) =  C_{M,0}^{(n)} ~~,
\end{equation}
Using the Newton binomial expansions of Sect.~\ref{newtform},
the asymmetric expansion up to order $K$ 
for the function $\sin^M \tau$, with $M\in\cal{N}$, is given by
\begin{equation}\label{formula_sinM}
\sin^M\tau= \sum_{k=0}^K {\cal M}_{4}(k)\, u^k~~,
\end{equation}
with
\begin{equation}\label{coeff_formula_sinM}
{\cal M}_{4}(k)= \sum_{i=k}^{K}\frac{1}{i!} C^{(i)}_{M,0}
{i \choose k}
\left(-\frac{\pi}{6} \right)^{i-k}~~.
\end{equation}


\chapter{Resonant normal form for the $1$:$5$ resonance}\label{Appe:res-norm-form}

We present a particular case of the resonant normal form construction
of Chapter 5 for $\mu= 0.0056$ and $e'=0.0$.

Starting from Eq.~\eqref{eq:hb_asym_expand_lambd_4}, we find
$\omega_f = 0.9811$, $\omega_s = -0.192232$. Since $Y_p$ represents an
integral of motion of the $H_b$, we do not carry $Y_p$
along with the normalization. Thus, the initial Hamiltonian reads
\begin{equation}
\begin{aligned}\label{eq:hb_asym_expand_lambd_5}
H_b^{(0)} (Y_s,{\cal Y},\phi_s,\phi_f,Y_p) = & \, {\cal Z}_0
+ \,\sum_{r=1}^{r_{max}} \lambda^r {\cal H}^{(0)}_r\\
= & \, -0.4972\, + Y_p -0.192232 \, Y_s \, +\, 
0.9811 \, {\cal Y} \,\\
& + \,\sum_{r=1}^{r_{max}} \lambda^r\,
\mathtt{c}_{(k_1,k_2,k_3,k_4)}\,  (\sqrt{Y_s})^{k_1} 
(\sqrt{{\cal Y}})^{k_2}\, 
{\textstyle {{\cos}\atop{\sin}}} (k_3 \phi_s + k_4\,\phi_f)~,
\end{aligned}
\end{equation}
where the coefficients of $\lambda^0$ are
\begin{equation*}
C = -0.4972 + Y_p~,\qquad \omega_s = -0.192232~,\qquad \omega_f = 0.9811~.
\end{equation*}
The coefficients $\mathtt{c}_{(k_1,k_2,k_3,k_3,c/s/-)}$ up to order
$\lambda^5$ are given in Tables~\ref{tab:lambda_1},
\ref{tab:lambda_2}, \ref{tab:lambda_3}, \ref{tab:lambda_4} and
\ref{tab:lambda_5} 
(the subscript $s$ or $c$ denotes a coefficient for sine or
cosine). These coefficients are found by using an asymmetric expansion
of $H_b^{(0)}$ up to order 20.

\begin{table}[h]
\begin{center}
\begin{tabular}{| l | l | l |}
\hline
{\color{white} --} & {\color{white} --} & {\color{white} --} \\
$\mathtt{c}_{(0,2,0,2,c)} = -3.15\snot[-3]$ & $\mathtt{c}_{(1,1,-1,1,c)} = 8.38243\snot[-2]$ &
$\mathtt{c}_{(1,2,1,0,c)} = -1.06144\snot[-2]$ \\ $\mathtt{c}_{(3,0,1,0,c)} = -7.62127\snot[-2]$ &
$\mathtt{c}_{(3,0,3,0,c)} = -1.606\snot[-2]$ & $\mathtt{c}_{(1,1,1,1,c)} = -6.45245\snot[-2]$ \\
$\mathtt{c}_{(0,2,0,2,s)} = 3.45544\snot[-2]$ & $\mathtt{c}_{(1,1,-1,1,s)} = -7.2366\snot[-2]$ &
$\mathtt{c}_{(1,2,1,0,s)} = -2.91364\snot[-1]$ \\ $\mathtt{c}_{(3,0,1,0,s)} = 6.83704\snot[-1]$ &
$\mathtt{c}_{(3,0,3,0,s)} = 2.41889\snot[-1]$ & $\mathtt{c}_{(1,1,1,1,2)} = 6.86516\snot[-2]$ \\
{\color{white} --} & {\color{white} --} & {\color{white} --} \\
\hline
\end{tabular}
\end{center}
\vspace{-0.5cm}
\caption{Coefficients for $\lambda^1$}\label{tab:lambda_1}
\vspace{0.5cm}
\begin{center}
\begin{tabular}{| l | l | l |}
\hline
{\color{white} --} & {\color{white} --} & {\color{white} --} \\
$\mathtt{c}_{(2,2,0,0,-)} = -1.38263\snot[0]$ & $\mathtt{c}_{(4,0,0,0,-)} = -1.51608\snot[0]$ & 
-------------------------------------- \\
$\mathtt{c}_{(2,1,0,1,c)} = 1.87829\snot[-1]$ & $\mathtt{c}_{(2,1,-2,1,c)} = -8.07313\snot[-1]$ 
& $\mathtt{c}_{(1,2,-1,2,c)} = 2.30056\snot[-1]$ \\  
$\mathtt{c}_{(2,2,2,0,c)} = 1.39878\snot[0]$ & $\mathtt{c}_{(4,0,2,0,c)} = 2.01083\snot[0]$ 
& $\mathtt{c}_{(0,4,0,4,c)} = -5.363794\snot[-1]$ \\  
$\mathtt{c}_{(1,2,1,2,c)} = 2.07311\snot[-1]$ & $\mathtt{c}_{(2,1,2,1,c)} = -1.09808\snot[-1]$ &
-------------------------------------- \\
$\mathtt{c}_{(2,1,0,1,s)} = 7.61681\snot[-1]$ & $\mathtt{c}_{(2,1,-2,1,s)} = -4.1622\snot[-1]$ 
& $\mathtt{c}_{(1,2,-1,2,s)} = -2.01581\snot[-1]$ \\  
$\mathtt{c}_{(2,2,2,0,s)} = -9.21549\snot[-2]$ & $\mathtt{c}_{(4,0,2,0,s)} = -4.70827\snot[-2]$ 
& $\mathtt{c}_{(0,4,0,4,s)} = 3.02004\snot[-2]$ \\  
$\mathtt{c}_{(1,2,1,2,s)} = 1.77943\snot[-1]$ & $\mathtt{c}_{(2,1,2,1,s)} = -3.5686\snot[-1]$ &
-------------------------------------- \\
{\color{white} --} & {\color{white} --} & {\color{white} --} \\
\hline
\end{tabular}
\end{center}
\vspace{-0.5cm}
\caption{Coefficients for $\lambda^2$}\label{tab:lambda_2}
\end{table}

\vspace{-0.1cm}
\begin{table}[t]
\begin{center}
\begin{tabular}{| l | l | l |}
\hline
{\color{white} --} & {\color{white} --} & {\color{white} --} \\
$\mathtt{c}_{(2,2,0,2,c)} = 2.11849\snot[0]$ & $\mathtt{c}_{(3,1,-3,1,c)} = -1.2371\snot[0]$ &
$\mathtt{c}_{(2,2,-2,2,c)} = -1.1784\snot[0]$ \\  
$\mathtt{c}_{(3,1,-1,1,c)} = 3.4132\snot[0]$ & $\mathtt{c}_{(3,2,1,0,c)} = -5.18809\snot[-1]$ &
$\mathtt{c}_{(5,0,1,0,c)} = -2.25109\snot[-1]$ \\  
$\mathtt{c}_{(3,2,3,0,c)} = 5.40089\snot[-1]$ & $\mathtt{c}_{(5,0,3,0,c)} = 3.33405\snot[-1]$ &
$\mathtt{c}_{(5,0,5,0,c)} = -1.26818\snot[-1]$ \\  
$\mathtt{c}_{(3,1,1,1,c)} = -3.2061\snot[0]$ & $\mathtt{c}_{(2,2,2,2,c)} = -9.67646\snot[-1]$ &
$\mathtt{c}_{(3,1,3,1,c)} = 1.02408\snot[0]$ \\  
$\mathtt{c}_{(2,2,0,2,s)} = -8.49661\snot[-1]$ & $\mathtt{c}_{(3,1,-3,1,s)} = 5.65338\snot[-2]$ &
$\mathtt{c}_{(2,2,-2,2,s)} = 4.77707\snot[-1]$ \\  
$\mathtt{c}_{(3,1,-1,2,s)} = -2.01249\snot[-1]$ & $\mathtt{c}_{(3,2,1,0,s)} = -1.84924\snot[1]$ &
$\mathtt{c}_{(5,0,1,0,s)} = -1.3411\snot[1]$ \\  
$\mathtt{c}_{(3,2,3,0,s)} = 6.32667\snot[0]$ & $\mathtt{c}_{(5,0,3,0,s)} = 6.892495\snot[0]$ &
$\mathtt{c}_{(5,0,5,0,s)} = -1.47242\snot[0]$ \\  
$\mathtt{c}_{(3,1,1,1,s)} = 2.5525\snot[-1]$ & $\mathtt{c}_{(2,2,2,2,s)} = 3.93525\snot[-1]$ &
$\mathtt{c}_{(3,1,3,1,s)} = -1.15715\snot[-1]$ \\  
{\color{white} --} & {\color{white} --} & {\color{white} --} \\
\hline
\end{tabular}
\end{center}
\vspace{-0.5cm}
\caption{Coefficients for $\lambda^3$}\label{tab:lambda_3}
\vspace{0.5cm}
\begin{center}
\begin{tabular}{| l | l | l |}
\hline
{\color{white} --} & {\color{white} --} & {\color{white} --} \\
$\mathtt{c}_{(4,2,0,0,-)} = -7.336\snot[1]$ & $\mathtt{c}_{(6,0,0,0,-)} = -3.5544\snot[1]$ &
-------------------------------------- \\
$\mathtt{c}_{(4,1,0,1,c)} = 2.56466\snot[0]$ & $\mathtt{c}_{(4,1,-4,1,c)} = 4.05496\snot[-1]$ &
$\mathtt{c}_{(3,2,-3,2,c)} = 1.89832\snot[0]$ \\  
$\mathtt{c}_{(4,1,-2,1,c)} = -1.6044\snot[0]$ & $\mathtt{c}_{(3,2,-1,2,c)} = -4.63742\snot[0]$ &
$\mathtt{c}_{(4,2,2,0,c)} = 9.88183\snot[1]$ \\  
$\mathtt{c}_{(6,0,2,0,c)} = 5.4184\snot[1]$ & $\mathtt{c}_{(4,2,4,0,c)} = -2.58252\snot[1]$ &
$\mathtt{c}_{(6,0,4,0,c)} = -2.27375\snot[1]$ \\  
$\mathtt{c}_{(6,0,6,0,c)} = 4.0902\snot[0]$ & $\mathtt{c}_{(3,2,1,2,c)} = 3.93398\snot[0]$ &
$\mathtt{c}_{(4,1,2,1,c)} = -1.97847\snot[0]$ \\  
$\mathtt{c}_{(3,2,3,2,c)} = 1.17857\snot[0]$ & $\mathtt{c}_{(4,1,4,1,c)} = 6.12665\snot[-1]$ &
-------------------------------------- \\
$\mathtt{c}_{(4,1,0,1,s)} = 1.99487\snot[1]$ & $\mathtt{c}_{(4,1,-4,1,s)} = 4.06227\snot[0]$ &
$\mathtt{c}_{(3,2,-3,2,s)} = 4.14059\snot[0]$ \\  
$\mathtt{c}_{(4,1,-2,1,s)} = -1.45283\snot[1]$ & $\mathtt{c}_{(3,2,-1,2,s)} = -1.10837\snot[1]$ &
$\mathtt{c}_{(4,2,2,0,s)} = -4.71074\snot[0]$ \\  
$\mathtt{c}_{(6,0,2,0,s)} = -1.48538\snot[0]$ & $\mathtt{c}_{(4,2,4,0,s)} = 2.50468\snot[0]$ &
$\mathtt{c}_{(6,0,4,0,s)} = 1.2897\snot[0]$ \\  
$\mathtt{c}_{(6,0,6,0,s)} = -3.65832\snot[-1]$ & $\mathtt{c}_{(3,2,1,2,s)} = 9.9994\snot[0]$ &
$\mathtt{c}_{(4,1,2,1,s)} = -1.24769\snot[1]$ \\  
$\mathtt{c}_{(3,2,3,2,s)} = -3.03302\snot[0]$ & $\mathtt{c}_{(4,1,4,1,s)} = 2.99649\snot[0]$ &
-------------------------------------- \\
{\color{white} --} & {\color{white} --} & {\color{white} --} \\
\hline
\end{tabular}
\end{center}
\vspace{-0.5cm}
\caption{Coefficients for $\lambda^4$}\label{tab:lambda_4}
\vspace{0.5cm}
\begin{center}
\begin{tabular}{| l | l | l |}
\hline
{\color{white} --} & {\color{white} --} & {\color{white} --} \\
$\mathtt{c}_{(4,2,0,2,c)} = 8.3806\snot[1]$ & $\mathtt{c}_{(5,1,-5,1,c)} = 1.40493\snot[1]$ &
$\mathtt{c}_{(4,2,-4,2,c)} = 1.86122\snot[1]$\\
$\mathtt{c}_{(5,1,-3,1,c)} = -6.22934\snot[1]$ & $\mathtt{c}_{(4,2,-2,2,c)} = -6.46295\snot[1]$ &
$\mathtt{c}_{(5,1,-1,1,c)} = 1.13157\snot[2]$\\
$\mathtt{c}_{(5,2,1,0,c)} = -1.88076\snot[1]$ & $\mathtt{c}_{(7,0,2,0,c)} = -4.05322\snot[0]$ &
$\mathtt{c}_{(5,2,3,0,c)} = 2.94141\snot[1]$\\
$\mathtt{c}_{(7,0,3,0,c)} = 7.68892\snot[0]$ & $\mathtt{c}_{(5,2,5,0,c)} = -1.06258\snot[0]$ &
$\mathtt{c}_{(7,0,5,0,c)} = -4.72853\snot[0]$\\
$\mathtt{c}_{(7,0,7,0,c)} = 1.08978\snot[0]$ & $\mathtt{c}_{(5,1,1,1,c)} = 1.06258\snot[2]$ &
$\mathtt{c}_{(4,2,2,2,c)} = -5.08451\snot[1]$\\
$\mathtt{c}_{(5,1,3,1,c)} = 5.0474\snot[1]$ & $\mathtt{c}_{(4,2,4,2,c)} = 1.15009\snot[1]$ &
$\mathtt{c}_{(5,1,5,1,c)} = -9.90992\snot[0]$\\
$\mathtt{c}_{(4,2,0,2,s)} = 4.06877\snot[1]$ & $\mathtt{c}_{(5,1,-5,1,s)} = -1.22433\snot[0]$ &
$\mathtt{c}_{(4,2,-4,2,s)} = -9.48024\snot[0]$\\
$\mathtt{c}_{(5,1,-3,1,s)} = 6.2547\snot[0]$ & $\mathtt{c}_{(4,2,-2,2,s)} = 3.28216\snot[1]$ &
$\mathtt{c}_{(5,1,-1,1,s)} = -1.33605\snot[1]$\\
$\mathtt{c}_{(5,2,1,0,s)} = -9.04863\snot[2]$ & $\mathtt{c}_{(7,0,2,0,s)} = -3.28218\snot[2]$ &
$\mathtt{c}_{(5,2,3,0,s)} = 4.65471\snot[2]$\\
$\mathtt{c}_{(7,0,3,0,s)} = 1.93595\snot[2]$ & $\mathtt{c}_{(5,2,5,0,s)} = -9.89688\snot[1]$ &
$\mathtt{c}_{(7,0,5,0,s)} = -7.22133\snot[1]$\\
$\mathtt{c}_{(7,0,7,0,s)} = 1.07427\snot[1]$ & $\mathtt{c}_{(5,1,1,1,s)} = 1.51191\snot[1]$ &
$\mathtt{c}_{(4,2,2,2,s)} = 2.44049\snot[1]$\\
$\mathtt{c}_{(5,1,3,1,s)} = -9.05332\snot[0]$ & $\mathtt{c}_{(4,2,4,2,s)} = -5.79586\snot[0]$ &
$\mathtt{c}_{(5,1,5,1,s)} = 2.26686\snot[0]$\\
{\color{white} --} & {\color{white} --} & {\color{white} --} \\
\hline
\end{tabular}
\end{center}
\vspace{-0.5cm}
\caption{Coefficients for $\lambda^5$}\label{tab:lambda_5}
\end{table}

We identify the kernel of our homological equation as ${\cal Z}_0 =
-0.192231\,Y_s\,+\,0.9811\,{\cal Y}$. 
The two frequencies are close to the commensurability
\begin{equation}\label{eq:res_condit_1to5}
1\,\times\,(0.9811) \,+\, 5\,\times\,(-0.192232) \sim 0~,
\end{equation}
i.e. a resonant condition $m_f\,\omega_f\,+\,m_s\,\omega_s$ with
$m_f=1$ and $m_s=5$. This suggests to consider the integers
$\ell_f=5$ and $\ell_s=-1$, such that
the resonant module is defined by
\begin{equation}\label{eq:res_modul_1to5}
{\cal M} = \{ \mathbf{k} = (k_f,k_s): k_f\, \ell_f + k_s\, \ell_s = 0 \:\land\:
\ell_f = 5,\,\ell_s = -1\}~.
\end{equation}

At the first normalization step, we isolate the components to normalize from
the term of order $\lambda^1$, ${\cal H}^{(0)}_{1}$, given by
{\small
\begin{equation}
\begin{aligned}\label{eq:h1-for1to5}
{\cal H}^{(0)}_{1} = & -0.00315\,{\cal Y}\, \cos(2 \phi_f) 
+ 0.0838243\, \sqrt{{\cal Y}}\, \sqrt{Y_s}\, \cos(\phi_f - \phi_s) 
-  0.0106144\, {\cal Y}\, \sqrt{Y_s}\, \cos(\phi_s) \\
&-  0.0762127\, Y_s^{3/2}\, \cos(\phi_s)\, 
- 0.01606\, Y_s^{3/2}\, \cos(3 \phi_s) 
- 0.0645245\, \sqrt{{\cal Y}}\, \sqrt{Y_s}\, \cos(\phi_f + \phi_s) \\
&- 0.0345544\, {\cal Y}\, \sin(2 \phi_f)
- 0.072366\, \sqrt{{\cal Y}}\, \sqrt{Y_s}\, \sin(\phi_f - \phi_s)  
- 0.291364\, {\cal Y}\, \sqrt{Y_s}\, \sin(\phi_s) \\
&- 0.683704\, Y_s^{3/2}\, \sin(\phi_s) 
+ 0.241889\, Y_s^{3/2}\, \sin(3 \phi_s) 
+  0.0686516\, \sqrt{{\cal Y}}\, \sqrt{Y_s}\, \sin(\phi_f + \phi_s)~.
\end{aligned}
\end{equation}}
In this case, none of the components is in normal form, so we must
normalize them all. In complex Fourier terms:
\begin{align}
{\cal H}^{(0)}_{1} = & (-0.001575\,-\,\mathrm{i}\, 0.0172772) {\cal Y} \mathrm{e}^{-\mathrm{i}2\phi_f}
- (0.001575 -\,\mathrm{i}\, 0.0172772) {\cal Y} \mathrm{e}^{\mathrm{i}2\phi_f} \nonumber\\
& - (0.0322622 -\,\mathrm{i}\, 0.0343258)  \sqrt{{\cal Y}} \sqrt{Y_s} \mathrm{e}^{\mathrm{i}(-\phi_f -\phi_s)}  
+ (0.041912 +\,\mathrm{i}\, 0.036183) \sqrt{{\cal Y}} \sqrt{Y_s} \mathrm{e}^{\mathrm{i}(\phi_f-\phi_s)} \nonumber\\
& + (0.041912 -\,\mathrm{i}\, 0.036183) \sqrt{{\cal Y}} \sqrt{Y_s} \mathrm{e}^{\mathrm{i}(-\phi_f+\phi_s)}  
- (0.0322622 +\,\mathrm{i}\, 0.0343259) \sqrt{{\cal Y}} \sqrt{Y_s} \mathrm{e}^{\mathrm{i}(\phi_f+\phi_s)} \nonumber\\
& - (0.00530718 + \,\mathrm{i}\, 0.14582) {\cal Y} \sqrt{Y_s}  \mathrm{e}^{-\mathrm{i}\phi_s} 
- (0.00530718 -\,\mathrm{i}\, 0.14582) {\cal Y} \sqrt{Y_s} \mathrm{e}^{\mathrm{i}\phi_s} \nonumber\\
& - (0.0381064 +\,\mathrm{i}\, 0.341852) Y_s^{3/2} \mathrm{e}^{-\mathrm{i}\phi_s}  
- (0.0381064 -\,\mathrm{i}\, 0.341852) Y_s^{3/2} \mathrm{e}^{\mathrm{i}\phi_s} \nonumber\\
& - (0.00803 -\,\mathrm{i}\, 0.120944) Y_s^{3/2} \mathrm{e}^{-\mathrm{i}3\phi_s}  
- (0.00803 +\,\mathrm{i}\, 0.120944) Y_s^{3/2} \mathrm{e}^{\mathrm{i}3\phi_s} ~.\nonumber
\end{align}
Using Eq.~\eqref{eq:gene_func_res_nf}, we obtain the form of the first generating function,
{\footnotesize
\begin{equation}
\begin{aligned}
\chi_{1} = &\, \lambda \left( (0.00880502 -\mathrm{i}\, 0.00080271) {\cal Y} \mathrm{e}^{-\mathrm{i}\,2 \phi_f}    
+ (0.00880502 +\mathrm{i}\, 0.00080271) {\cal Y} \mathrm{e}^{\mathrm{i}\,2 \phi_f} \right. \\    
& - (0.0435127 +\mathrm{i}\, 0.0408969) \sqrt{{\cal Y}} \sqrt{Y_s} \mathrm{e}^{\mathrm{i}\,(-\phi_f -\phi_s)}
+ (0.0308377 -\mathrm{i}\, 0.0357206) \sqrt{{\cal Y}} \sqrt{Y_s} \mathrm{e}^{\mathrm{i}\,(\phi_f -\phi_s)} 
 \\ 
& + (0.0308377 +\mathrm{i}\, 0.0357206) \sqrt{{\cal Y}} \sqrt{Y_s} \mathrm{e}^{\mathrm{i}\,(-\phi_f + \phi_s)}
- (0.0435127 -\mathrm{i}\, 0.0408969) \sqrt{{\cal Y}} \sqrt{Y_s} \mathrm{e}^{\mathrm{i}\,(\phi_f +\phi_s)}  
 \\
& - (0.757841 - \mathrm{i}\, 0.0276082) {\cal Y} \sqrt{Y_s} \mathrm{e}^{-\mathrm{i}\,\phi_s}  
- (0.757841 + \mathrm{i}\, 0.0276082) {\cal Y} \sqrt{Y_s} \mathrm{e}^{\mathrm{i}\, \phi_s}  \\
& - (1.77833 - \mathrm{i}\, 0.198231) Y_s^{3/2} \mathrm{e}^{-\mathrm{i}\, \phi_s}  
- (1.77833 + \mathrm{i}\, 0.198231) Y_s^{3/2} \mathrm{e}^{\mathrm{i}\, \phi_s} \\
&\left. + (0.209719 + \mathrm{i}\, 0.0139241) Y_s^{3/2} \mathrm{e}^{-\mathrm{i}\,3  \phi_s}  
+ (0.209719 - \mathrm{i}\, 0.0139241) Y_s^{3/2} \mathrm{e}^{\mathrm{i}\,3\phi_s} \right)~.
\end{aligned}
\end{equation}}
or in trigonometric form,
\begin{align}\label{eq:chi_1}
\chi_{1} = &\, \lambda \left(0.01761 {\cal Y} \cos(2 \phi_f) 
+ 0.0616755 \sqrt{{\cal Y}} \sqrt{Y_s} \cos(\phi_f - \phi_s) 
- 1.51568 {\cal Y} \sqrt{Y_2} \cos(\phi_s) \right. \nonumber \\
& - 3.55666 Y_s^{3/2} \cos(\phi_s) 
+ 0.419438 Y_s^{3/2} \cos(3 \phi_s) 
- 0.0870255 \sqrt{{\cal Y}} \sqrt{Y_s} \cos(\phi_f + \phi_s) \nonumber \\
& - 0.00160534 {\cal Y} \sin(2 \phi_f) 
+ 0.0714412 \sqrt{{\cal Y}} \sqrt{Y_s} \sin(\phi_f - \phi_s) 
+ 0.0552163 {\cal Y} \sqrt{Ys} \sin(\phi_s) \nonumber \\
& \left. + 0.396461 Y_s^{3/2} \sin(\phi_s) 
+ 0.027848 Y_s^{3/2} \sin(3 \phi_s) 
- 0.0817938 \sqrt{{\cal Y}} \sqrt{Y_s} \sin(\phi_f + \phi_s)\right)~.\nonumber
\end{align}
Finally, we apply the Lie operator
\begin{equation}\label{eq:Hb1-after-nf}
H_b^{(1)} = \mathrm{exp}({\cal L}_{\chi_{1}}) H_b^{(0)}~,
\end{equation}
where we notice that the term of order $\lambda^1$ in the transformed
Hamiltonian vanishes, since there were no terms in normal form to keep
in the original ${\cal H}^{(0)}_{1}$ in Eq.~\eqref{eq:h1-for1to5}.

We repeat this procedure at consecutive orders in $\lambda$. As an
example, after 5 steps, the Hamiltonian is given by
\begin{equation}
H_b^{(5)} = {\cal Z}_0 + \lambda {\cal Z}_1 + \ldots + \lambda^{5} {\cal Z}_{5} 
+ \lambda^{6}{\cal H}^{(0)}_6 + \lambda^{7}{\cal H}^{(0)}_7 + \ldots
\end{equation}
where the normal form terms up to $\lambda^5$ are
\begin{align}
{\cal Z}_0 &=  -0.4972 + 0.9811 {\cal Y} - 0.192232 Y_s \\
{\cal Z}_1 &=  0 \\
\begin{split}
{\cal Z}_2 &=  -8.13668\snot[-4]\, {\cal Y} 
+ 1.10548\snot[-1] \,{\cal Y}^2 
- 5.42596\snot[-3]\, Y_s \\  
&\quad - 3.42146\snot[-1] \,{\cal Y}\, Y_s 
+ 5.59639\snot[-1]\, Y_s^2 
\end{split}\\
{\cal Z}_3 & =  -3.7977\snot[-5]\,{\cal Y} -1.93824\snot[-4]\, Y_s\\
\begin{split}
{\cal Z}_4 & =  -2.56596\snot[-6]\,{\cal Y} 
+ 1.344\snot[-2]\, {\cal Y}^2 
- 1.198\snot[0] \,{\cal Y}^3\\
&\quad + 6.52028\snot[-5]\, Y_s 
- 4.91754\snot[-2]\, Y2\, Y_s 
+ 1.20226\snot[1]\, {\cal Y}^2 Y_s\\ 
&\quad - 3.30968\snot[-2]\, Y_s^2 
+ 3.7839\snot[0]\, {\cal Y}\, Y_s^2  
+ 6.71463\snot[0]\, Y_s^3 
\end{split}\\
\begin{split}
{\cal Z}_5 &  =  -1.82594\snot[-7]\, {\cal Y}
+ 1.04172\snot[-3]\, {\cal Y}^2 
+ 4.69972\snot[-6]\, Y_s \\
& - 1.0448\snot[-2]\,{\cal Y}\, Y_s 
- 2.15687\snot[-2]\, Y_s^2 
- 5.15683\snot[1]\, \sqrt{{\cal Y}}\, Y_s^{5/2} \,\cos(\phi_f + 5 \phi_s) \\
& + 3.23298\snot[1] \, \sqrt{{\cal Y}}\, Y_s^{5/2}\, \sin(\phi_f + 5 \phi_s) ~.
\label{hola-A}
\end{split}
\end{align}
Eq.~\eqref{hola-A} contains the first resonant terms $\cos(\phi_f + 5 \phi_s)$,
$\sin(\phi_f + 5 \phi_s)$. In general for the resonance $1$:$n$, the
first resonant terms appear in the normal form at the book-keeping
order $n$, and at all subsequent orders.

\backmatter

\chapter{Acknowledgements}\label{cha:thanks}

\vspace{1.5cm}

I have been thinking about how to write this
part of the thesis since quite long. But even today,
I am not sure whether I have chosen the right words. 
I hope that the following transmits my gratitude towards
people that certainly deserve it.

\vspace{0.5cm} In the very first place, I would like to thank my
supervisor Dr. Ugo Locatelli, not only for his guidance during my
Ph.D., but also for his company and friendship during three years.
For the former I hope to be able to reward him with many
collaborations, future papers and joint works, but the latter is no
doubt priceless.

\vspace{0.5cm} I would also like to thank the referees of my
thesis, Prof. Angel Jorba and Prof. Antonio Giorgilli, for their
careful reading of the manuscript with constructive comments.

\vspace{0.5cm} I had the opportunity to develop my Ph.D. research in
the framework of a valuable project, the Astronet-II network, that
allowed me to combine new knowledge and work with something I love:
meeting people and traveling. I want to thank Prof. Alessandra
Celletti for giving me such an oportunity, and I truly hope to have
fulfilled hers as well as Ugo's expectations when they selected me as
Early Stage Researcher. Additionally, I want to thank all the people
involved in the management of this big network, which was a tedious
and hard work. And, of course, all the nice people I had the chance to
meet during these three years, in particular the Astronet-II fellows
(Zubin, Fabrizio, Mattia, Albert, Elisabetta, Andrea, Claudiu,
Mohammad, Marta, Willem, Leon, Lily, Luca, Pedro, Pawel and Alex) and
the Astronet-II Add-Ons (Dani, Ariadna, Marco, Christoph, Chris,
several Stardusters and many many others).

\vspace{0.5cm} An important part of the experience of living so far
from home for so long is the ``\emph{d\'epaysement}''. It was truly
difficult in some moments. Fortunately, I was not alone. I would like
to thank Yannis, Fabien, Nicole, Maryam and all the people that for
one reason or another were found at the University of ``Tor Vergata''
and with whom I shared nice moments, lunches in the office, dinners
out, complains, laughs and working hours. A particular line here for
Carolina, who made me a lot of company even being far, with the excuse
of 'Tenemos que terminar el poster!', so many times along these years.

\vspace{0.5cm}
Thanks to my Buenos Aires friends: Sofia, Nadia, Ioni, Santiago and
Fran. With them I learned that the concept of distance is quite
relative, and some friendships are made to last, independently of 
the circumstances. 

\vspace{0.5cm} I would like to thank my wonderfully big family, full
of uncles, aunts and cousins, for all the unconditional love and
support they have provided me with ever since I exist. They have been
motivating me to follow the paths I chose, and to remain persistent
when the things get darker. My brother Bruno, my sister Flavia and my
nieces Isabella and Olivia make the world a brighter place to me.
Above all, I thank my parents Maria and Sergio, for what should be
properly called 'everything'.

\vspace{0.5cm}
And Christos. I tried to put it in words and I could not.
His presence in this thesis, in my past and in my future,
is so important that I cannot find the way to express it. 
I hope he realizes that.

\vspace{1cm}
\flushright
Roc\'io Isabel P\'aez\\
Roma, February 19th 2016


\end{document}